\documentclass[onecolumn]{aastex631}

\newcommand\codename{\texttt{GWFAST}}
\newcommand\wfname{\texttt{WF4Py}}
\newcommand\trickfootnote[1]{}

\usepackage[english]{babel}
\usepackage[utf8]{inputenc}
\usepackage{savesym}
\usepackage{comment}
\usepackage{nccmath}
\usepackage{bbold, amsmath}
\usepackage[caption=false]{subfig}
\usepackage{enumitem}
\usepackage{xcolor}
\definecolor{neonpurple}{RGB}{176, 38, 255}
\definecolor{lightblue}{RGB}{114, 189, 212}
\definecolor{brightgreen}{RGB}{72, 179, 0}
\usepackage{nicefrac}
\usepackage{physics}
\usepackage{mathtools}
\usepackage{booktabs}
\usepackage{multirow}
\usepackage{letltxmacro}
\savesymbol{tablenum}
\usepackage{siunitx}
\restoresymbol{SIX}{tablenum}
\DeclareSIUnit \parsec {pc}
\DeclareSIUnit \arcsecondfull {arcsec}
\DeclareSIUnit \year{yr}
\DeclareSIUnit \day{day}
\DeclareSIUnit \hour{hr}
\DeclareSIUnit \radiant{rad}
\DeclareSIUnit \degfull{deg}
\DeclareSIUnit \erg {erg}
\DeclareSIUnit \Lsun {L_\odot}
\DeclareSIUnit \Msun {M_\odot}
\DeclareSIUnit \AstroUnit {au}
\DeclareSIUnit \steradian {sr}

\LetLtxMacro{\originaleqref}{\eqref}
\renewcommand{\eqref}{Eq.~\originaleqref}
\renewcommand{\eq}{\eqref}

\addto\extrasenglish{%
}

\def\stopsqrt{\mathpalette\DHLhksqrt}
\def\DHLhksqrt#1#2{\setbox0=\hbox{$#1\sqrt{#2\,}$}\dimen0=\ht0
	\advance\dimen0-0.2\ht0
	\setbox2=\hbox{\vrule height\ht0 depth -\dimen0}%
	{\box0\lower0.4pt\box2}}

\shorttitle{Forecasting the detection capabilities of third--generation GW detectors using \codename{}}
\shortauthors{Iacovelli, Mancarella, Foffa, Maggiore}

\begin{document}

\reportnum{ET-0141A-22}

\title{Forecasting the detection capabilities of third--generation  gravitational--wave detectors using \codename{}}

\correspondingauthor{Francesco Iacovelli}
\email{Francesco.Iacovelli@unige.ch}

\author[0000-0002-4875-5862]{Francesco Iacovelli}
\affiliation{D\'epartement de Physique Th\'eorique and Gravitational Wave Science Center,\\ Universit\'e de Gen\`eve, 24~quai Ernest Ansermet, 1211~Gen\`eve~4, Switzerland}

\author[0000-0002-0675-508X]{Michele Mancarella}
\affiliation{D\'epartement de Physique Th\'eorique and Gravitational Wave Science Center,\\ Universit\'e de Gen\`eve, 24~quai Ernest Ansermet, 1211~Gen\`eve~4, Switzerland}

\author[0000-0002-4530-3051]{Stefano Foffa}
\affiliation{D\'epartement de Physique Th\'eorique and Gravitational Wave Science Center,\\ Universit\'e de Gen\`eve, 24~quai Ernest Ansermet, 1211~Gen\`eve~4, Switzerland}

\author[0000-0001-7348-047X]{Michele Maggiore}
\affiliation{D\'epartement de Physique Th\'eorique and Gravitational Wave Science Center,\\ Universit\'e de Gen\`eve, 24~quai Ernest Ansermet, 1211~Gen\`eve~4, Switzerland}

\begin{abstract}
We introduce  \codename{}, a novel Fisher--matrix code for gravitational--wave studies, tuned toward third--generation gravitational--wave detectors such as Einstein Telescope (ET) and Cosmic Explorer (CE). We use it to perform a comprehensive study of the capabilities of ET alone, and of a network made by  ET and two CE detectors, as well as to  provide forecasts for the forthcoming O4 run of the LVK collaboration. We consider  binary neutron stars, binary black holes and neutron star--black hole binaries, and compute basic metrics such as the distribution of signal--to--noise ratio (SNR), the accuracy in the reconstruction of various parameters (including distance, sky localization, masses, spins and, for neutron stars, tidal deformabilities), and  the redshift distribution of the detections for different thresholds in SNR and different levels of accuracy in localization and distance measurement. We examine the expected distribution and properties of `golden events', with especially large values of the SNR. We also pay special attention to the dependence of the results on astrophysical uncertainties and on various technical details (such as choice of waveforms, or the threshold in SNR), and we compare with other Fisher codes in the literature. 
In the companion paper \cite{Iacovelli:2022mbg} we discuss the technical aspects of the code.
Together with this paper,  we publicly release the code \codename{} \raisebox{-1pt}{\href{https://github.com/CosmoStatGW/gwfast}{\includegraphics[width=10pt]{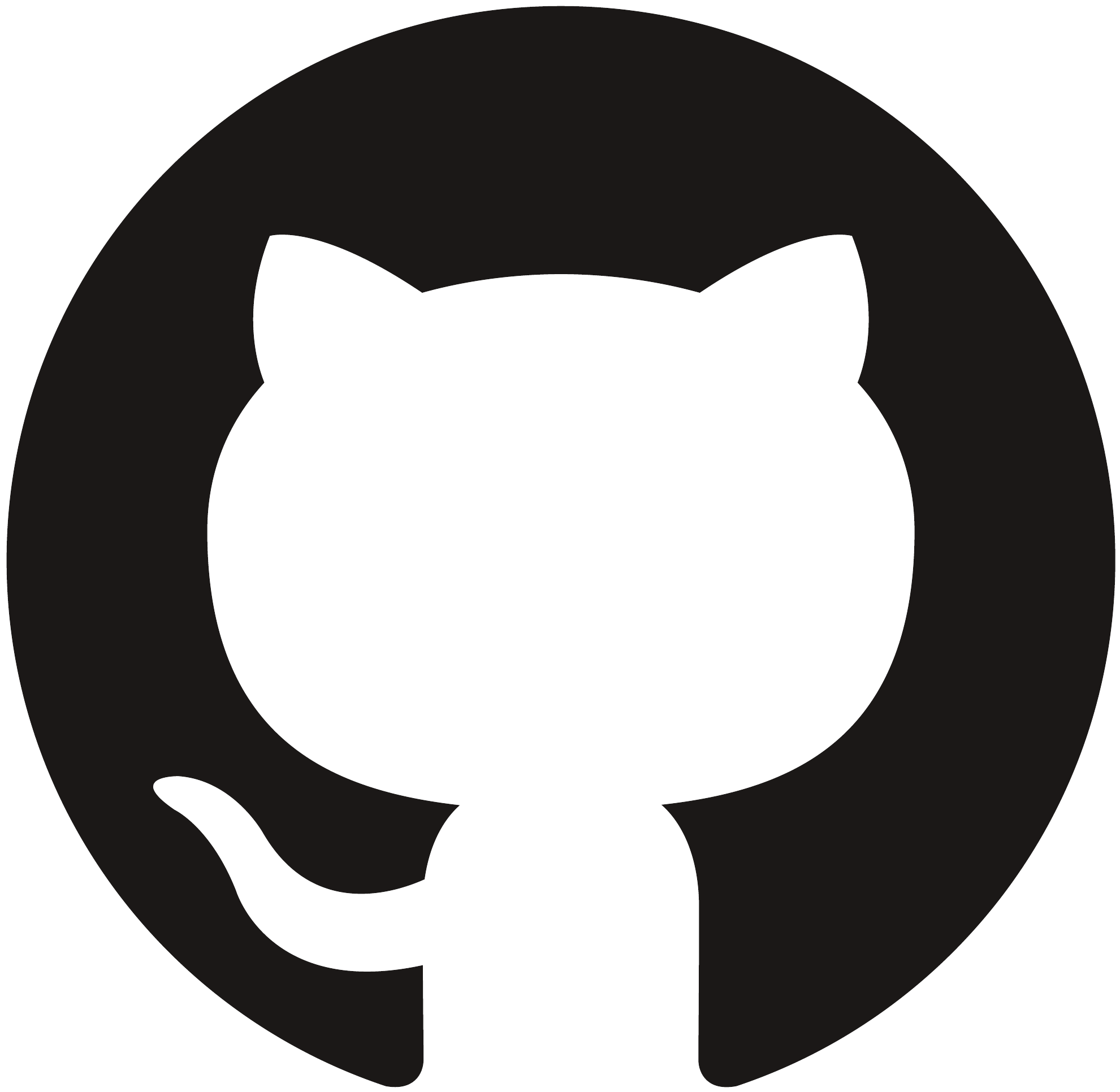}}}~\footnote{\url{https://github.com/CosmoStatGW/gwfast}}, and the library \wfname{} \raisebox{-1pt}{\href{https://github.com/CosmoStatGW/WF4Py}{\includegraphics[width=10pt]{GitHub-Mark.pdf}}}~\footnote{\url{https://github.com/CosmoStatGW/WF4Py}} implementing state--of--the--art gravitational--wave waveforms in pure \texttt{Python}.
\\

\end{abstract}

\section{Introduction} \label{sec:intro}

The discoveries made in the last few years by  the LIGO and Virgo gravitational wave (GW) detectors have   opened a new window on the Universe. After the first historic detection of the GWs from a binary black hole (BBH) coalescence in September~2015 \citep{Abbott:2016blz},  
and the first detection in 2017 of a binary neutron star (BNS)  coalescence and the identification and follow--up of its electromagnetic counterpart \citep{TheLIGOScientific:2017qsa,Monitor:2017mdv}, which opened the field of multi--messenger astronomy, GW observations have become a routine, with detections made on a weekly basis. After three observing runs (O1--O3), the current catalog of GW detections contains about 90 events, mostly BBHs, but include also two BNSs  and  two neutron star--black hole (NSBH) binaries \citep{LIGOScientific:2020ibl,LIGOScientific:2021djp}. These discoveries are already starting to have a remarkable impact on astrophysics, cosmology and fundamental physics including, e.g., studies of population properties of compact 
binary systems \citep{LIGOScientific:2021psn}, tests of General Relativity \citep{LIGOScientific:2021sio}, the determination of the speed of GWs to one part in $10^{15}$
\citep{Monitor:2017mdv}, first constraints on the expansion history of the Universe using GWs
\citep{LIGOScientific:2021aug}, or 
constraints on the neutron star equation of state \citep{HernandezVivanco:2019vvk}.

Currently LIGO and Virgo, now joined by KAGRA and forming  the LVK collaboration, are further improving their sensitivity and preparing for their fourth observational run, O4, scheduled for 2023, and eventually for the O5 run, currently scheduled for 2026--2028. However,  the community has long been preparing the jump toward `third--generation' (3G) detectors, Einstein Telescope (ET) in Europe~\citep{Punturo:2010zz,Hild:2010id}  and Cosmic Explorer (CE) in the US~\citep{Reitze:2019iox,Evans:2021gyd}.
Thanks to an increase by one order of magnitude in sensitivity and a significant enlargement of the  bandwidth toward both  low and high frequencies, these detectors will have an extraordinary potential for discoveries. In particular, for coalescing binaries, they will
allow us to make a huge jump in the distances reached, in the number of detections,  in the accuracy of signal reconstructions,  and in the range of masses  that can be explored,  providing data that
have the potential of triggering revolutions in fundamental physics, in cosmology and in astrophysics. 
Building on many previous works, the science case for 3G detectors is being systematically investigated (see \cite{Maggiore:2019uih,Kalogera:2021bya} and references therein). The activity around 3G detectors recently received  a significant boost from the inclusion of  ET in the ESFRI Roadmap, the roadmap of large European scientific  infrastructures,\footnote{See \url{https://www.esfri.eu/latest-esfri-news/new-ris-roadmap-2021}.} resulting in an acceleration of activities around ET and, very recently, the ET Collaboration has  formally been created at the XII ET Symposium.\footnote{See \url{https://indico.ego-gw.it/event/411/}.} On the US side, activity around Cosmic Explorer has started more recently, and has led to the Cosmic Explorer Horizon Study document~\citep{Evans:2021gyd}.

A crucial aspect, in order to assess the scientific potential of 3G detectors, is the development of codes that allow us to forecast the performance of these detectors, in terms of some rather general ``metrics'', such as 
the distribution of signal--to--noise ratio (SNR) for coalescing binaries, and the accuracy in the reconstruction of  their parameters;  in particular,  distance, sky localization, masses, spins and, for neutron stars, quantities (such as the tidal deformability) that are sensitive to their inner structure. It is also very important to understand the reach in redshift of these detectors and  how the distribution in SNR and parameter reconstruction depends on redshift. The reference tool for this kind of investigations is an approach based on the Fisher matrix. Despite some of its well--known limitations, that we will  review in \autoref{sec:formalism}, this is at the moment the only computationally feasible way of performing parameter estimation  on large populations, such as the $10^4-10^5$ BBHs and BNSs  per year that, as we will discuss below,  3G detectors are expected to detect.
This tool  also allows us to compare  the performance of different  configurations of a given detector, and of different detector networks,  
which is necessary in order  to  take informed decisions on  individual detector designs and on optimal network configurations.
 
For this reason, a number of parameter estimation codes tuned to 3G detectors have been developed  recently,  in particular \texttt{GWBENCH}~\citep{Borhanian:2020ypi} and   \texttt{GWFISH} \citep{Harms:2022ymm} [see also  \cite{Chan:2018csa,Grimm:2020ivq,Nitz:2021pbr,Li:2021mbo,Pieroni:2022bbh}]. In this paper we present a novel parameter estimation code,  \codename{}, also meant  mainly for application to 3G detectors. 
The existence of several different parameter estimation codes is a welcome, and in fact necessary, feature;   these codes can contribute to taking decisions on detectors which are meant to dominate the scientific landscape for decades (and require the huge financial and human resources typical of Big Science), so they must be extremely reliable and have undergone cross--checks between different groups, that developed different codes independently. In this spirit,  we have undergone a  process of cross--checking between 
 \texttt{GWBENCH},  \texttt{GWFISH} and \codename{}, which are, arguably, the most complete and advanced codes currently available, finding broad consistency.\footnote{These checks are being performed in the context of the activities of the Observational Science Board (OSB) of ET, which is in charge of developing the Science Cases and the technical tools relevant for ET. See \url{https://www.et-gw.eu/index.php/observational-science-board} for a repository of papers relevant for ET, produced in the context of the OSB activities.} 
Each of these codes has different technical implementations. We discuss the technical aspects of our code,  together with reliability and performance tests, in the companion paper 
\cite{Iacovelli:2022mbg}, while here we briefly summarize them in \autoref{sec:technicalities}. One relevant aspect is that the problem of computing Fisher matrices for a large catalog of independent events is clearly parallel. Parallelization techniques can be used, but one limitation of current \texttt{Python}--based implementations is that they still usually compute the SNRs and Fisher matrices serially on each node/CPU.  
In our implementation, we are able to  ‘vectorize' the evaluation 
of the  Fisher matrices even on a single CPU (\emph{on top of} the use of parallel computing), resulting in a gain in computational speed, which is at the origin of the name \codename{}.
This is also due to the implementation of the waveforms in \texttt{Python}, which motivated the development and release of the library \wfname{}.  Furthermore, this allows us to make use of automatic differentiation to compute derivatives, which is a technique alternative to finite differences leading to faster and more robust evaluations. Note that \codename{} also supports an interface with the LIGO Algorithm Library, \texttt{LAL} \citep{lalsuite}, which allows the use of all waveforms available in this library.

In this paper we discuss the results obtained for  BBHs, BNSs and NSBHs with our code, and we compare them, in particular, with those reported for BBHs and BNSs in \cite{Borhanian:2022czq} (using  \texttt{GWBENCH}), and for BNSs in
\cite{Ronchini:2022gwk} (which uses \texttt{GWFISH}). The comparison also needs to take care
of different assumptions on astrophysical populations, that reflect our current uncertainties, as well as of different technical choices (waveforms, thresholds on SNR, detector network configurations, etc.). Our results will then contribute to giving an overall picture of the capabilities of 3G detectors.
Compared in particular to \cite{Borhanian:2022czq}, 
our work is more focused on ET. To avoid a proliferation of plots, and of lines in each plot, in this work we will only report the results for two 3G configurations: ET alone, and a network of ET together with two CE detectors (ET+2CE), and we will compare them with the expectations for the forthcoming LVK--O4 run. Studying ET alone allows us to assess the strength of the ET Science Case, independently of decisions that will be taken by different funding agencies on CE. On the other hand, combining ET with two CE detectors allows us to examine the full strength of a 3G detector network.  Our code, however, can be used  to study a large variety of networks.

The paper is organized as follows. In \autoref{sec:formalism} we recall the Fisher matrix formalism, as well as its limitations. In \autoref{sect:modeling} we discuss the modelization of the GW signal. In particular, since BNSs can stay in the bandwidth of 3G detectors for  as long as hours or a day, it is important to include the effect of the Earth's rotation on the signal, which can be exploited to improve the sky localization of the source. In \autoref{sec:detectors} we discuss the assumptions that we make on the astrophysical populations of BBHs, BNSs and NSBHs, and we present the detectors that we will study, computing their horizon (and the distance at which $50\%$ of the events are detected) for different type of sources. \autoref{sect:Results} contains the bulk of our results for parameter estimation of BBHs, BNSs and NSBHs. Further results are included in the Appendices, including, in \autoref{app:comparison}, a comparison with the results presented in \cite{Borhanian:2022czq} and in \cite{Ronchini:2022gwk}.

\section{Formalism for parameter estimation} \label{sec:formalism}

\subsection{Fisher Information Matrix}
In this section we recall the basic formalism and interpretation of the use of the Fisher matrix formalism in GW parameter estimation. We refer to \cite{Cutler:1994ys,Vallisneri:2007ev,Rodriguez:2013mla} for comprehensive treatments.
We assume that the time--domain signal in a GW detector can be written as the superposition of an expected signal $h_0$ and stationary, Gaussian noise $n$ with zero mean:
\begin{equation}\label{eq:signalNoise}
s(t)=h_0(t)+n(t)\; .
\end{equation}
The statistical properties of the noise are encoded in the one--sided power spectral density, defined by
\begin{equation}\label{eq:specDens}
\langle \tilde{n}^*(f)\, \tilde{n}(f^{\prime})\rangle =\frac{1}{2} \delta(f-f^{\prime})  S_{n}(f)\; ,
\end{equation}
where the tilde denotes a temporal Fourier transform.
This determines an inner product for any two time--domain signals $a(t)$, $b(t)$:
\begin{equation}\label{eq:scProd}
\left( a \, | \, b \right) =  4 \Re{ \int_0^{\infty} \dd{f} \, \frac{\tilde a^*(f) \, \tilde b(f) }{S_{n}(f)}} \; .
\end{equation}
Under the assumption of stationarity and Gaussianity, from \eqref{eq:specDens} we have that the variance of a Fourier mode with frequency $f$ is $S_{n}(f)/2$, so the probability that the noise has a given distribution $n_0$ can be written in terms of \eqref{eq:scProd} as 
$p(n_0) \propto \exp{-\left( n_0 \, | \, n_0 \right) /2}.$ 
Together with \eqref{eq:signalNoise}, this results in the following likelihood for a data realisation $s$ conditioned on the waveform parameters $\vb*{\theta}$:
\begin{equation}\label{eq:likelihood}
\mathcal{L}(s \;|\; \vb*{\theta}) \propto \exp{-\left( s -h(\vb*{\theta})  \, | \, s -h(\vb*{\theta}) \right) /2} \; .
\end{equation}
Note that $\vb*{\theta}$ refers to the parameters of the template waveform model, which may be different from the ones of the actual signal $h_0$, that we denote by $\vb*{\theta}_0$.
Using \eqref{eq:scProd} we can also express the signal--to--noise ratio (SNR) of the true signal as 
\begin{equation}\label{eq:SNR}
\text{SNR} =  \left(h_0 \, | \, h_0\right)^{\nicefrac{1}{2}}\; .
\end{equation}
The Fisher Information Matrix (FIM) for the likelihood in \eqref{eq:likelihood} is defined as:
\begin{equation}\label{eq:Fisher_def}
\Gamma_{ij} \equiv -\eval*{\langle \partial_{i} \partial_{j} \log \mathcal{L}(s \,|\, \vb*{\theta})  \rangle_n}_{\vb*{\theta}=\vb*{\theta}_0} =\left(h_i \, | \, h_j\right)   \; ,
\end{equation}
where $h_i \equiv \partial_{i}h$, and the notation $\langle\,\dots\rangle_n$ denotes an average over noise realizations with fixed parameters. The last equality is a consequence of the property $\langle \left(a \, | \, n\right) \left(n \, | \, b\right)  \rangle_n = \left(a \, | \, b\right)$ that follows from \eqref{eq:specDens}.
Intuitively, near a maximum of the likelihood, the latter is approximated by a multivariate Gaussian with covariance $\Gamma_{ij}^{-1}$. To understand the correct interpretation of this statement it is useful to consider an expansion of the template signal around the waveform with true parameters' values $\vb*{\theta}_0$ as
\begin{equation}\label{eq:LSA}
h(\vb*{\theta}) = h_0 +  h_i \, \delta \theta^{i}
+\dots  \qquad\; ,
\end{equation}
where $ \delta \theta^{i} \equiv \theta^i - \theta_{0}^{i} $. 
This approximation, where only first derivatives of the signal are included, is known as the \emph{linearized signal approximation} (LSA). The LSA likelihood is obtained by inserting \eqref{eq:LSA} in \eqref{eq:likelihood}:
\begin{equation}\label{eq:LSAlikelihood}
\mathcal{L}(s \,|\, \vb*{\theta}) \propto \exp \Big[ -\frac{1}{2} \left(n \, | \, n\right)
+ \delta\theta^{i} \left(n \, | \, h_i\right)
-\frac{1}{2} \delta\theta^{i}\delta\theta^{j}  \left(h_i \, | \, h_j\right) \Big] \; .
\end{equation}
It can be shown that the LSA is equivalent to the limit of large SNR~\citep{Vallisneri:2007ev}.\footnote{Formally, the LSA likelihood is equivalent to the leading order term in a series expansion in $1/\text{SNR}$.}
Two interpretations of the meaning of the FIM appearing in \eqref{eq:LSAlikelihood} are possible:

\begin{itemize}
    \item From a \emph{frequentist} point of view, one can compute the maximum likelihood (ML) estimator $\hat{\vb*{\delta \theta}}$ and its covariance over noise realizations at fixed parameters, which yields
    \begin{equation}
        \begin{aligned}
            \hat{\delta \theta}^i &= {\left(h_i \, | \, h_j\right)}^{-1} \left(n \, | \, h_j\right)\,, \\
       \langle  \hat{\delta \theta}^i \hat{\delta \theta}^j \rangle_n  &= \left(h_i \, | \, h_j\right)^{-1}\,.
        \end{aligned}
    \end{equation}
    This shows that \emph{the inverse of the FIM is equal to the covariance of the frequentist maximum--likelihood estimator in the LSA/large SNR limit and for Gaussian noise}.\footnote{It can also be shown that the ML estimator is unbiased, i.e. the expectation value of $\vb*{\theta}$ is $\vb*{\theta}_0$.}
    \item 
    From a \emph{Bayesian} perspective one can compute the mean and variance of the posterior probability $p(\vb*{\theta} \,|\, s )\propto p(\vb*{\theta}) \mathcal{L}(\vb*{\theta} \,|\, s)$. Using a flat prior $p(\vb*{\theta}) \propto \text{const.}$, we obtain
    \begin{equation}\label{eq:BayesFisher}
        \begin{aligned}
       \langle {\delta \theta}^i \rangle &= \frac{\int \dd\vb*{\theta}\,{\delta \theta}^i \,p(\vb*{\theta} \,|\, s )}{\int \dd\vb*{\theta} \, p(\vb*{\theta} \,|\, s )} = {\left(h_i \, | \, h_j\right)}^{-1} \left(n \, | \, h_j\right)\,, \\
       \langle  {\delta \theta}^i {\delta \theta}^j \rangle  &= \langle \big({\delta \theta}^i-\langle {\delta \theta}^i \rangle \big)\big({\delta \theta}^j-\langle {\delta \theta}^j \rangle \big) \rangle = \left(h_i \, | \, h_j\right)^{-1}\,.
    \end{aligned}
    \end{equation}
    In the above equations, $\langle\,\dots\rangle$ denotes an average over parameter realizations at fixed data.
    Hence, \emph{the inverse of the FIM is also equal to the covariance of the Bayesian posterior probability distribution of the true parameters $\vb*{\theta}_0$ for a given experiment with data $s$, assuming: a flat prior, the LSA/large SNR limit and Gaussian noise.}\footnote{In this case, note that an additional uncertainty might be present, i.e. the fact that the posterior might not peak at the true parameters $\vb*{\theta}_0$ due to contribution of the noise. In this sense, one should talk about uncertainty rather than error~\citep{Vallisneri:2007ev,Rodriguez:2013mla}. However, under the LSA, this bias has expectation value zero under repeated realizations of the noise with covariance $\Gamma_{ij}^{\mkern-4mu -1}$. Note also that this coincidence between frequentist ML estimator and posterior mean is in general not true beyond the LSA~\citep{Vallisneri:2007ev}.}
\end{itemize}
The Bayesian point of view is the most useful to compare to results of an actual parameter estimation as well as to incorporate the effects of priors on the source parameters. In this case, one can explicitly re--write the LSA likelihood (omitting factors that do not depend on $\vb*{\delta \theta}$) as
\begin{equation}\label{eq:LSAlikelihoodBayes}
\mathcal{L}(s \,|\, \vb*{\theta}) \propto \exp \left[ -\frac{1}{2} \left(h_i \, | \, h_j\right)  \big( \delta\theta^{i}- \langle {\delta \theta}^i \rangle\big) \big( \delta\theta^{j}- \langle {\delta \theta}^j \rangle\big) \right] \, .
\end{equation}
On the other hand, in absence of real data, as is the case when forecasting parameter estimation capabilities of future experiments, it makes sense also to consider the frequentist approach which makes statements about ensembles of possible data realizations~\citep{Cutler:1994ys}.

Finally, another important interpretation of the FIM concerns the Cramer--Rao bound: the inverse FIM gives the lower bound for the covariance of any unbiased estimator of the true parameters $\vb*{\theta}_0$ under different noise realization (hence, it is a frequentist error estimate). However, its interpretation is subtle \citep{Vallisneri:2007ev} and this bound does not translate in a bound on the variance of the Bayesian posterior~\citep{Rodriguez:2013mla}.

\subsection{Singularities}\label{sec:singularities}
The most relevant issues when computing the FIM are the accurate computation of the derivatives and the presence of singularities, which can result in instability of the numerical inversion. The computational aspects as implemented in \codename{} are described in \autoref{sec:technicalities}, while here we discuss the interpretation of singular or ill--conditioned matrices. 

If the FIM is exactly singular, the likelihood has one or more directions in parameter space along which it is exactly constant, corresponding to null eigenvalues. If the LSA approximation was exact, one could conclude that one or more combination of parameters are impossible to constrain and discard them. In particular, one can discard the combination of parameters corresponding to the null singular values of a singular--value decomposition. However, this is true only in the LSA, as higher--order contributions may cure the singularity, and using this approach in the FIM does not necessarily lead to realistic forecasts of the parameter estimation performance, since only truly degenerate combinations should be discarded.

A more common situation to face in practice is the case where the matrix is not exactly singular but \emph{ill conditioned}, i.e. the inverse of the condition number (defined as the ratio of the largest to the smallest eigenvalues) is comparable or smaller than machine precision. This can lead to large amplifications of inversion errors, resulting in a covariance that may be inaccurate even at the 100\% level. Condition numbers might be improved by numerical techniques (see \autoref{sec:technicalities} for the implementation in \codename{}), but the physical interpretation is that the directions in parameter space corresponding to small eigenvalues need large variations of the parameters to produce relevant changes in the waveform (at least comparable to the noise level). Hence, the LSA might no longer be sufficient to accurately describe the likelihood over all the range of interest~\citep{Vallisneri:2007ev}, even if the SNR at the true parameters' values is high enough. We will discuss this point in \autoref{sec:technicalities}. One can resort to some regularization of the singularities, for example truncating the singular values of the singular--value decomposition to the minimum allowed numerical precision; however this has not a direct link to actual generalizations of the FIM that might cure the singularity, such as the inclusion of higher order terms (or of priors, see \autoref{sec:priors}).
In general, when using the FIM to forecast parameter estimation, one must be aware that the bad conditioning of the matrix is pointing to some possible breakdown of the LSA on the likelihood surface irrespective of the specific regularization technique used.

In this work we prefer not to resort to regularization of the singular values nor to discard combinations of parameters corresponding to small eigenvalues. 
The reason is that this might lead to underestimating the errors on the waveform parameters that are most affected by degeneracies, unless careful checks and comparisons with other methods are performed.

One possible alternative approach corresponds to discarding all the matrices with condition number larger than the inverse machine precision.
This is the most conservative option, which might on the other hand lead to discarding many events for which a sensible inversion of the Fisher matrix could still be obtained.
A somewhat intermediate possibility is to discard all the events for which the inversion error of the FIM is larger than a given threshold. In this work [as, e.g., in   \cite{Berti:2004bd}, see their App.~B], we adopt the latter strategy and quantify the inversion error as 
\begin{equation}\label{eq:invError}
    \epsilon = {|| \Gamma \cdot \Gamma^{-1} - \mathbb{1} ||}_{\rm max} = \max_{ij} |(\Gamma \cdot \Gamma^{-1}- \mathbb{1} )_{ij} | \; . 
\end{equation}
The chosen threshold for this work is $\epsilon_{\rm max} = \num{5e-2}$.\footnote{Note that the v1 arXiv version of this paper we used $\epsilon_{\rm max} = \num{1e-3}$. This is at the origin of some minor differences in the results.} In \autoref{sec:appendix_comparisonInv} we detail the choice of this value and compare the results with other methods to obtain forecasts on the uncertainty, including a direct sampling from the likelihood in \eqref{eq:LSAlikelihoodBayes} which does not rely on the inversion of the Fisher matrix. In \autoref{app:comparison} we compare to other choices in the literature and discuss the impact of different approaches on the population of sources studied in this paper.

In a Bayesian context, the use of priors can cure singularities and is the most realistic way to proceed, as priors are actually used in parameter estimation.
We discuss the role of priors in the next section, while their effect is also studied in \autoref{sec:appendix_comparisonInv}fv.

\subsection{Role of priors}\label{sec:priors}
In a Bayesian parameter estimation problem, prior distributions are expected to have a role whenever their information content is more restrictive than the one in the likelihood, typically either because the prior changes significantly in the region of non--vanishing likelihood or because hard boundaries are present. 
If the LSA is valid, the presence of priors can play an important role in curing possible singularities of the FIM.
In practice, the FIM formalism allows to treat in a simple way only the case of a Gaussian prior $p(\mathbf{\theta}) \propto \exp{-P_{ij} \delta \theta^i \delta \theta^j}$, in which case adding the prior amounts to substitute $\left(h_i \, | \, h_j\right) \mapsto \left(h_i \, | \, h_j\right) + P_{ij}$ in \eqref{eq:BayesFisher}--\originaleqref{eq:LSAlikelihoodBayes}, i.e. to add the prior matrix to the FIM. 

In a GW parameter estimation problem, we have to deal with the fact that some parameters might require priors that are far from Gaussian, in particular to incorporate information on the physical range of some of them, such as the symmetric mass ratio $\eta$ (which is constrained to be in the range $[0;\, 0.25]$), the luminosity distance $d_L$ (constrained to be positive), and the angles.
In \textit{Markov chain Monte Carlo} (MCMC) analyses, priors on the angles might not be used if not for increasing speed, since the full likelihood carries information on the periodicity of these variables. In contrast, the Fisher formalism does not have any information about the periodicity of the likelihood with respect to some parameters, hence the use of a prior can be a more realistic choice. For angular variables, a crude but simple approximation can be the use of a Gaussian prior of width $2\pi$.
The situation is more complicated for other parameters, in particular $\eta$ and $d_L$. When using a FIM analysis to forecast overall trends in future experiments rather than concentrating on single events, one can check that the majority of the events have a predicted $1\sigma$ contour that does not exceed the physical boundary. For more realistic estimates, an exact prior can be included by explicitly drawing samples from the likelihood in \eqref{eq:LSAlikelihoodBayes}, using rejection sampling to account for the prior, and estimating the posterior covariance from the remaining samples. We show the effect of adopting this procedure, and compare it to different inversion methods, in  \autoref{sec:appendix_comparisonInv}.

\subsection{Limits of applicability}\label{sec:limits}

\begin{figure}[t]
\hspace{-1.95cm}
    \begin{tabular}{c@{\hskip -4mm}c@{\hskip -5mm}c}
  \includegraphics[width=70mm]{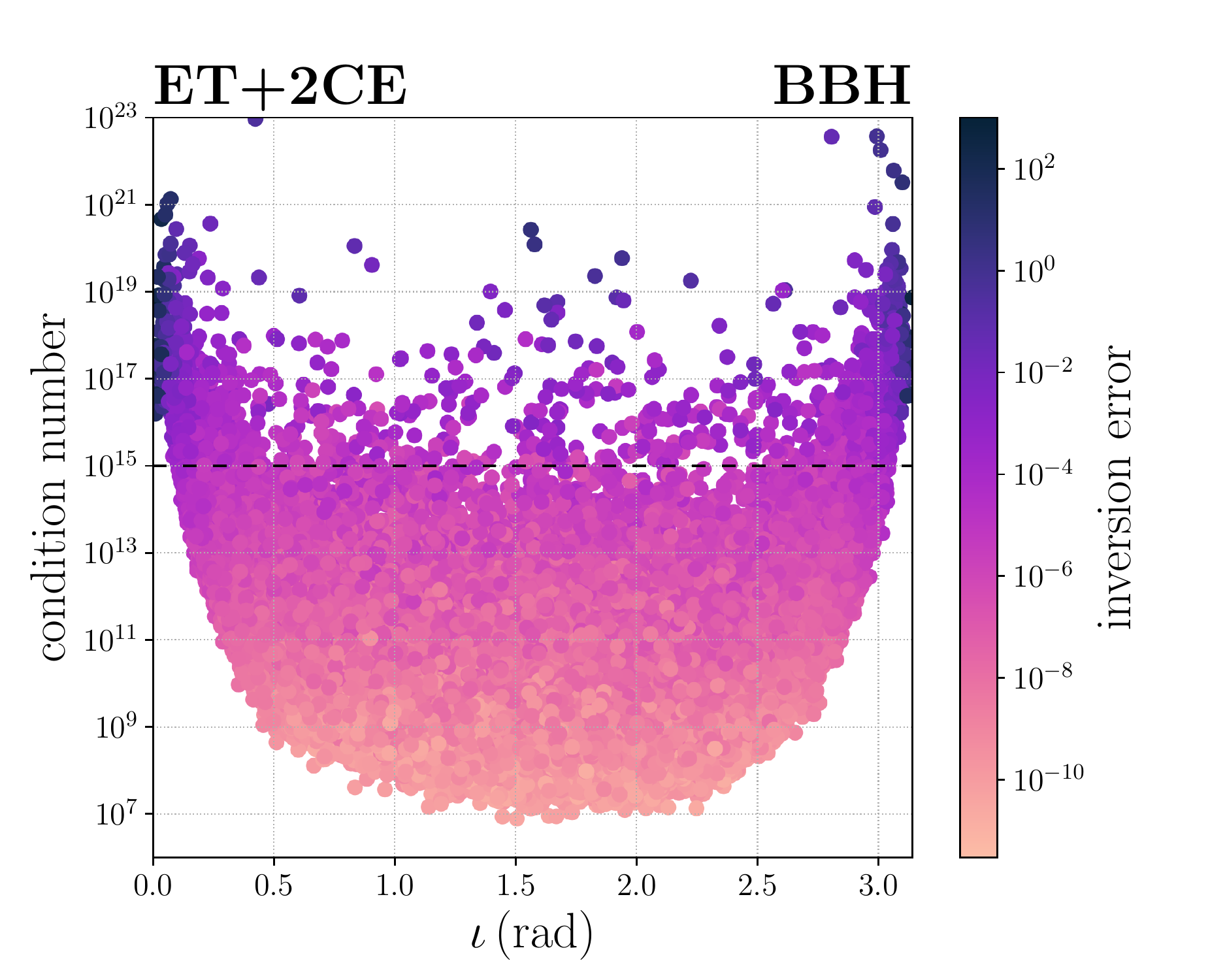} &   \includegraphics[width=70mm]{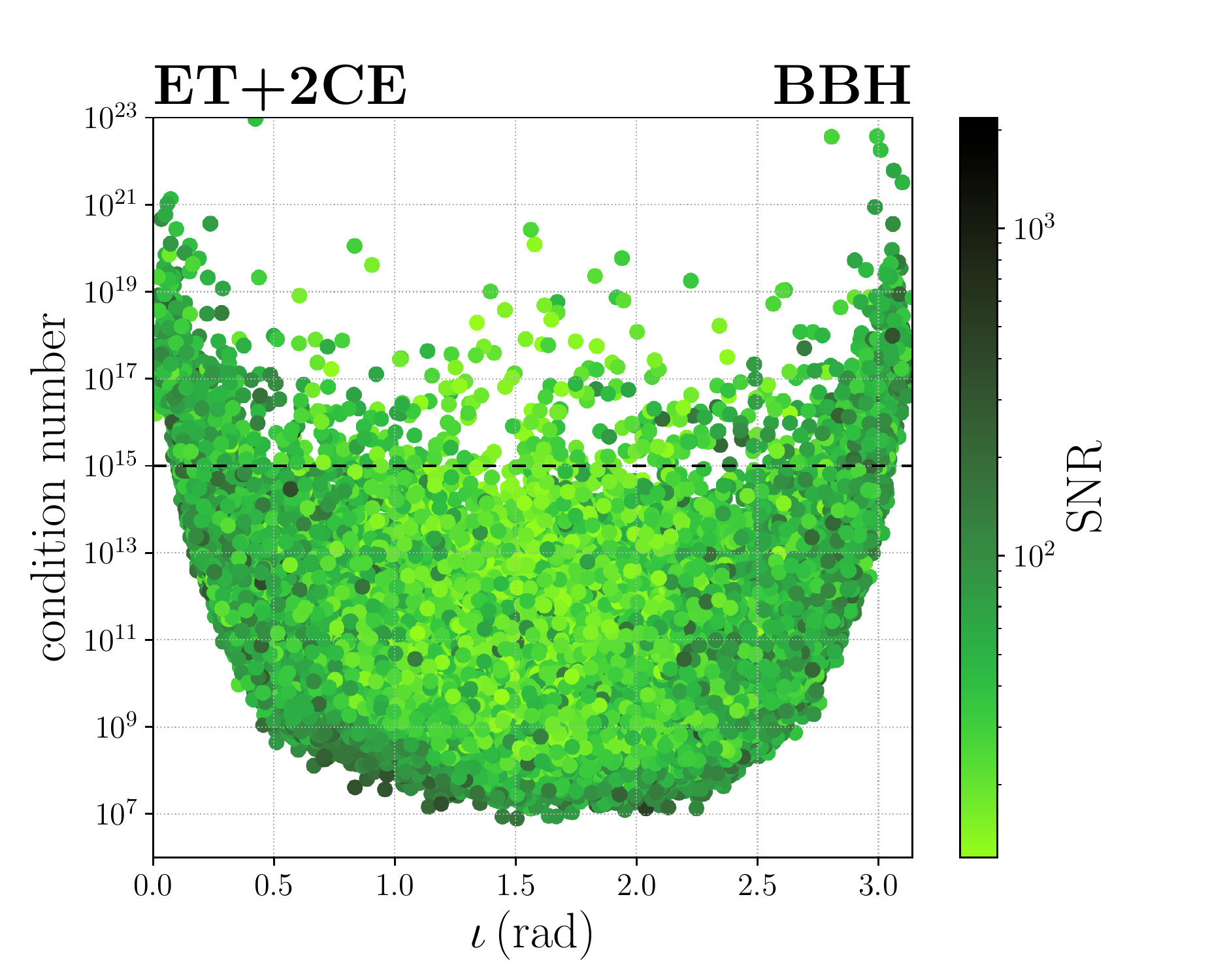} & \includegraphics[width=70mm]{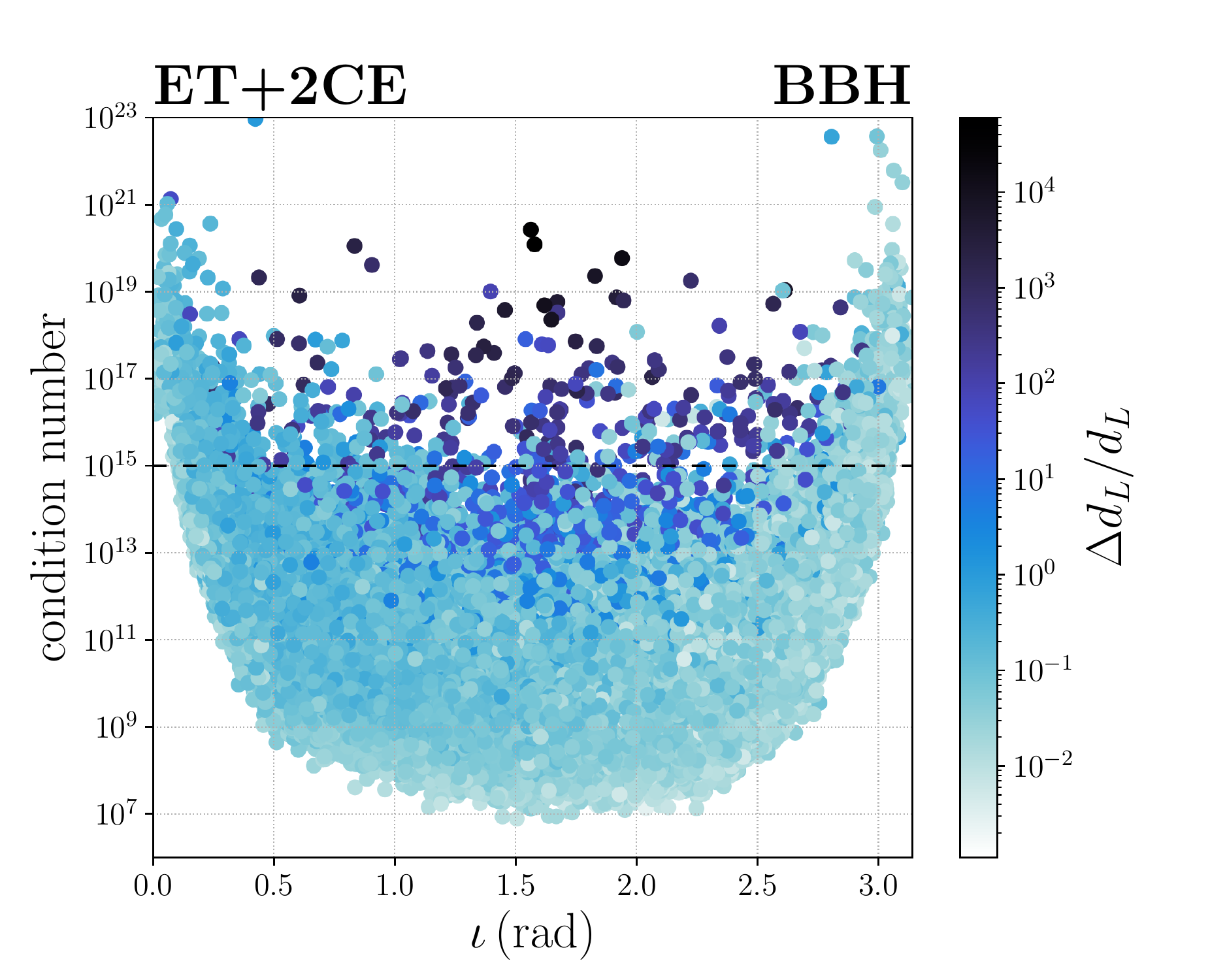}\\[-.5cm]
\end{tabular}
    \caption{Scatter plot of the condition numbers of the Fisher matrices as a function of the system inclination angle, computed for BBH events detected at a network consisting of ET+2CE, using the \texttt{IMRPhenomHM} waveform model. In the left panel we report on the color bar the inversion errors associated with the events and in the central panel the SNRs and in the right panel the relative errors on the luminosity distance. The horizontal dashed line shows the inverse machine precision limit, of \num{e15}.}
    \label{fig:CondNumbsVSiota}
\end{figure}

We have seen that the applicability of the FIM relies on the LSA/high SNR limit. From the discussion in \autoref{sec:singularities}, however, it is clear that such an approximation should hold within all the likelihood region of interest -- for example, the 1$\sigma$ or 2$\sigma$ contours. For this to be the case, not only the SNR at the true values should be high, but of particular relevance is the problem of conditioning: a high condition number might signal a breakdown of the LSA in the region of interest~\citep{Finn:1992wt, Vallisneri:2007ev}. 

One instructive example concerns the inclination angle of the source with respect to the observer, $\iota$. The amplitudes of the two GW polarisations depend differently on this parameter and their joint measurement would allow disentangling $\iota$ from the distance to the source. However, it is for nearly face--on binaries ($\iota \sim 0$) that the signal is louder, but in this case the two polarisations are nearly equal~\citep{Nissanke:2009kt, Schutz:2011tw, Usman:2018imj}, which leads to a strong degeneracy of the inclination angle with the distance to the source. Thus, an event observed face--on will be louder than the same system observed in an inclined configuration, but its FIM will be ill--conditioned. In the limit of exactly face--on systems, $\iota = 0$, and ignoring the presence of higher modes in the GW signal, the FIM will even be exactly degenerate, since the first derivative of the signal with respect to $\iota$ vanishes exactly in this limit. We refer to \autoref{fig:CondNumbsVSiota} for an illustration based on the population studied in this work. This shows that, for events with nearly face--on configuration, we can be in the situation where the SNR is large but the FIM is ill--conditioned \citep{Cutler:1994ys}.\footnote{In these cases, distance estimates will be particularly affected, since the luminosity distance has a strong correlation with the inclination, and the FIM can yield inaccurate predictions \citep{Cutler:1994ys}. It is however possible to extend the approximation of the full posterior beyond linear order for the marginal posterior in the subspace $(\iota, d_L)$ \citep{Cutler:1994ys,Chassande-Mottin:2019nnz}.}

In practise, assessing the validity of the LSA on all the likelihood surface without explicitly calculating higher--order corrections or resorting to explicit Monte Carlo analyses is a subtle problem. \cite{Vallisneri:2007ev} proposed a ‘‘maximum mismatch criterion'' to determine the validity of the LSA, which consists in sampling the $1\sigma$ likelihood surface, compute the difference between the waveforms on this surface and at the true value, $\Delta h = h(\vb*{\theta}_{1\sigma})-h_0$ and computing the ratio $r$ of the LSA likelihood to the full likelihood, which can be shown to be
\begin{equation}
    |\log(r)| = \frac{1}{2}\,\left(\theta^i h_i -\Delta h\, | \theta^j h_j -\Delta h \, h_j\right) \; .
\end{equation}
Then, a signal can be considered linear if some threshold is satisfied by this mismatch, for example by requiring that $|\log(r)|< 0.1$ for $90\%$ of the points on the $1\sigma$ likelihood surface~\citep{Vallisneri:2007ev, Rodriguez:2013mla}. This is an internal consistency criterion rather than a proof of validity. However, explicit comparison with MCMC showed that, while this is indeed  a sufficient condition for the LSA, many systems showing good agreement between the FIM and a full MCMC analysis failed this test~\citep{Rodriguez:2013mla}. 
In summary, there seems not to be a conclusive, flexible and computationally cheap test to assess the validity of the FIM over all the surface of interest. In this work we apply the threshold on the inversion error of the FIM described in \autoref{sec:singularities} and \autoref{sec:appendix_comparisonInv}.
However, one must be aware of such limitations of the FIM approach. We believe these issues to be a further motivation for the development and comparison of different implementations of the FIM technique, in order to assess the robustness of the predictions to different choices.

\section{Modeling the gravitational--wave signal}\label{sect:modeling}

The response of a detector to a GW signal emitted by a coalescing binary system is given by a linear combination of the two polarisations $(h_+, h_{\times})$, obtained by their projection on the detector arms by suitable ‘‘antenna pattern functions'' $(F_+, F_{\times})$ that depend on the source position and polarisation angle, as well as the location, orientation and shape of the detector, which we denote collectively by $\vb*{\lambda}$.
In full generality, we denote the parameters of the waveform by $\vb*{\theta} = \{{\cal M}_c, \eta, d_L, \theta, \phi, \iota, \psi, t_c, \Phi_c, \chi_{1,x}, \chi_{2,x}, \chi_{1,y}, \chi_{2,y}, \chi_{1,z}, \chi_{2,z}, \Lambda_1, \Lambda_2\}$ [see e.g. \cite{Maggiore:2007ulw}], where ${\cal M}_c$ denotes the detector--frame chirp mass, $\eta$ the symmetric mass ratio, $d_L$ the luminosity distance to the source, $\theta$ and $\phi$ are the sky position coordinates, defined as $\theta=\pi/2-\delta$ and $\phi=\alpha$ (with $\alpha$ and $\delta$ right ascension and declination, respectively), $\iota$ the inclination angle of the binary with respect to the line of sight, $\psi$ the polarisation angle, $t_c$ the time of coalescence, $\Phi_c$ the phase at coalescence, $\chi_{i,c}$ the dimensionless spin of the object $i=\{1,2\}$ along the axis $c = \{x,y,z\}$ and $\Lambda_i$ the dimensionless tidal deformability of the object $i$ (which is present only for systems containing a NS). 
Instead of $\Lambda_1, \Lambda_2$, we will actually use~\citep{PhysRevD.89.103012}
\begin{subequations}\label{eq:LamTdelLam_def}
\begin{align}
    \tilde{\Lambda} &= \dfrac{8}{13} \left[(1+7\eta-31\eta^2)(\Lambda_1 + \Lambda_2) + \sqrt{1-4\eta}(1+9\eta-11\eta^2)(\Lambda_1 - \Lambda_2)\right]\, ,\\
    \delta\tilde{\Lambda} &= \dfrac{1}{2} \left[\sqrt{1-4\eta} \left(1-\dfrac{13272}{1319}\eta + \dfrac{8944}{1319}\eta^2\right)(\Lambda_1 + \Lambda_2) + \left(1 - \dfrac{15910}{1319}\eta + \dfrac{32850}{1319}\eta^2 + \dfrac{3380}{1319}\eta^3\right)(\Lambda_1 - \Lambda_2)\right]\, ,
\end{align}
\end{subequations}
which have the advantage that $\tilde{\Lambda}$ is the combination that enters the inspiral waveform at 5\,PN, while $\delta\tilde{\Lambda}$ first enters at 6\,PN.

In the time domain, the signal of the quadrupole mode is given by
\begin{equation}\label{hTime}
h(t,\vb*{\theta},\vb*{\lambda}) = A_{+}(t,\vb*{\theta})F_{+}(t, \vb*{\theta}, \vb*{\lambda}) \cos{\Phi(t, \vb*{\theta})} + A_{\times}(t,\vb*{\theta}) F_{\times}(t, \vb*{\theta}, \vb*{\lambda}) \sin{\Phi(t, \vb*{\theta})} \, .
\end{equation}
The amplitudes $A_{+,\times}$ and the phase $\Phi$ are obtained from a waveform model. We describe the models used in \autoref{sect:waveforms}.
In particular, in this work and in \codename{} we work in the frequency domain. 
In this case, when computing the full signal in 
\eq{hTime}, an important point is that the low--frequency sensitivity of 3G detectors, and in particular of ET, makes it possible to observe the inspiral phase of low--mass events, such as BNSs, for several hours to possibly one day. In those cases the  pattern functions evolve in time during the detection due to the change in the relative position of the source and the detector. 
Since we work in frequency rather than time domain, it is important to correctly account for this time evolution when Fourier transforming the signal. 
We describe this in detail in \autoref{sec:detResp}.
\subsection{Waveform models}\label{sect:waveforms}

We adopt Fourier domain full inspiral--merger--ringdown models, tuned on Numerical Relativity (NR) simulations. Our code can be adapted to a large variety of waveforms. As reference waveforms, for BBH, BNS and NSBH systems, we will use, respectively:
\begin{description}[align=left]
    \item[\texttt{IMRPhenomHM}] \citep{PhysRevLett.120.161102, PhysRevD.101.103004} this is a recent model, tuned for BBH systems (in quasi--circular orbits) with non--precessing spins, which takes into account the quadrupole of the signal and the sub--dominant modes $(l,m) = (2,1),\ (3,2),\ (3,3),\ (4,3),\ {\rm and}\ (4,4)$. The  contribution of these higher modes is of fundamental importance for parameter estimation, since they can break the degeneracy between the luminosity distance and inclination angle. In fact, each mode depends on a different combination of sines and cosines of the inclination angle $\iota$, through the spin--weighted spherical harmonics $ _{-2}Y^{lm}$ \citep{1967JMP.....8.2155G}, while the fundamental mode (2,2), only depends on the cosine, leading to a degeneracy with $d_L$;
    \item[\texttt{IMRPhenomD\_NRTidalv2}] \citep{PhysRevD.100.044003} this model is an extension of \texttt{IMRPhenomD} \citep{PhysRevD.93.044006, PhysRevD.93.044007}, which also  accounts for tidal effects in BNS systems. In fact, the two neutron stars in a coalescing binary, differently from black holes, will deform when getting closer to each other, and this leaves clear signatures on the waveform, which have to be accurately modelled and tuned to specific NR simulations;
    \item[\texttt{IMRPhenomNSBH}] \citep{PhysRevD.92.084050, PhysRevD.100.044003}, this model can describe the signals coming from the merger of a NS and a BH, which have very distinctive features. In particular, the mass ratios of these systems can be much higher than that of BNSs and BBHs, so this model is tuned up to $q\sim 100$, and it also accounts for tidal effects, since the NS will deform getting closer to the BH. Moreover, this model is built and tuned to account for the fact that, when it is close enough to the BH, the NS can either plunge into it or get disrupted forming an accretion torus\footnote{A torus can be formed also if the NS plunges into the BH, resulting in a mildly disruptive merger, which is also a case taken into account in the tuning of \texttt{IMRPhenomNSBH}.} \citep{1976ApJ...210..549L}, two scenarios that result in very different features on the waveforms.
\end{description}
We will then perform a comparison between the results obtained using different waveform models. Note that, in addition to the above waveform models, \codename{} includes \texttt{Python} implementations of the models \texttt{TaylorF2\_RestrictedPN} and \texttt{IMRPhenomD}, as well as a wrapper to all waveforms contained in the LIGO Algorithm Library, \texttt{LAL} (note, however, that when using the latter, which are implemented in \texttt{C}, it is no longer possible  to exploit \texttt{Python} vectorization).
\subsection{Detector response and effect of Earth's rotation}\label{sec:detResp}

In this work and in \codename{} we use the expression of the ‘pattern functions' which takes into account the rotation of the Earth given in \cite{PhysRevD.58.063001}. We collectively denote the parameters characterizing the detector by $\vb*{\lambda} = \{\lambda,\ \varphi,\ \gamma,\ \zeta\}$, where $\zeta$ is the angle between the detector's arms (e.g. \SI{90}{\degree} for an L--shaped detector), $\lambda$ and $\varphi$ are the detector's latitude and longitude, respectively, and $\gamma$ is the angle formed by the arms bisector and East. We then have for the pattern functions:
\begin{equation}\label{eq:patt_func_expr}
    \begin{aligned}
        F_{+}(t; \theta, \phi, \psi, \vb*{\lambda}) &= \sin{\zeta}\, \big[a(t; \theta, \phi, \vb*{\lambda})\cos{2\psi} + b(t;  \theta, \phi, \vb*{\lambda})\sin{2\psi}\big] \, ,\\
        F_{\times}(t; \theta, \phi, \psi, \vb*{\lambda}) &= \sin{\zeta}\, \big[b(t;  \theta, \phi, \vb*{\lambda})\cos{2\psi} - a(t;  \theta, \phi, \vb*{\lambda})\sin{2\psi}\big]\, ,
    \end{aligned}
\end{equation}
with (omitting from now on the explicit dependence on the parameters)
\begin{equation}
    \begin{aligned}
        a(t) &= \dfrac{1}{16}\sin{2\gamma}(3-\cos{2\lambda})(3+\cos{2\theta})\cos{[2(\phi-\varphi-2\pi f_{\oplus}t)]} - \dfrac{1}{4} \cos{2\gamma}\sin{\lambda}(3+\cos{2\theta})\sin{[2(\phi-\varphi-2\pi f_{\oplus}t)]} \\
        & \quad\, +\dfrac{1}{4}\sin{2\gamma}\sin{2\lambda}\sin{2\theta}\cos{(\phi-\varphi-2\pi f_{\oplus}t)} -\dfrac{1}{2}\cos{2\gamma}\cos{\lambda}\sin{2\theta}\sin{(\phi-\varphi-2\pi f_{\oplus}t)} \\ & \quad\, + \dfrac{3}{4}\sin{2\gamma}\cos^2{\lambda}\sin^2{\theta} \, ,\\ \\
        b(t) &= \cos{2\gamma}\sin{\lambda}\cos{\theta}\cos{[2(\phi-\varphi-2\pi f_{\oplus}t)]} + \dfrac{1}{4}\sin{2\gamma}(3-\cos{2\lambda})\cos{\theta}\sin{[2(\phi-\varphi-2\pi f_{\oplus}t)]} \\
        & \quad\, +\cos{2\gamma}\cos{\lambda}\sin{\theta}\cos{(\phi-\varphi-2\pi f_{\oplus}t)} + \dfrac{1}{2} \sin{2\gamma}\sin{2\lambda}\sin{\theta}\sin{(\phi-\varphi-2\pi f_{\oplus}t)}\, ,
    \end{aligned}
\end{equation}
where $f_{\oplus} \simeq\SI{1}{\per\day}$ is the Earth's rotational frequency.

\begin{figure}[t]
    \centering
    \includegraphics[width=.8\textwidth]{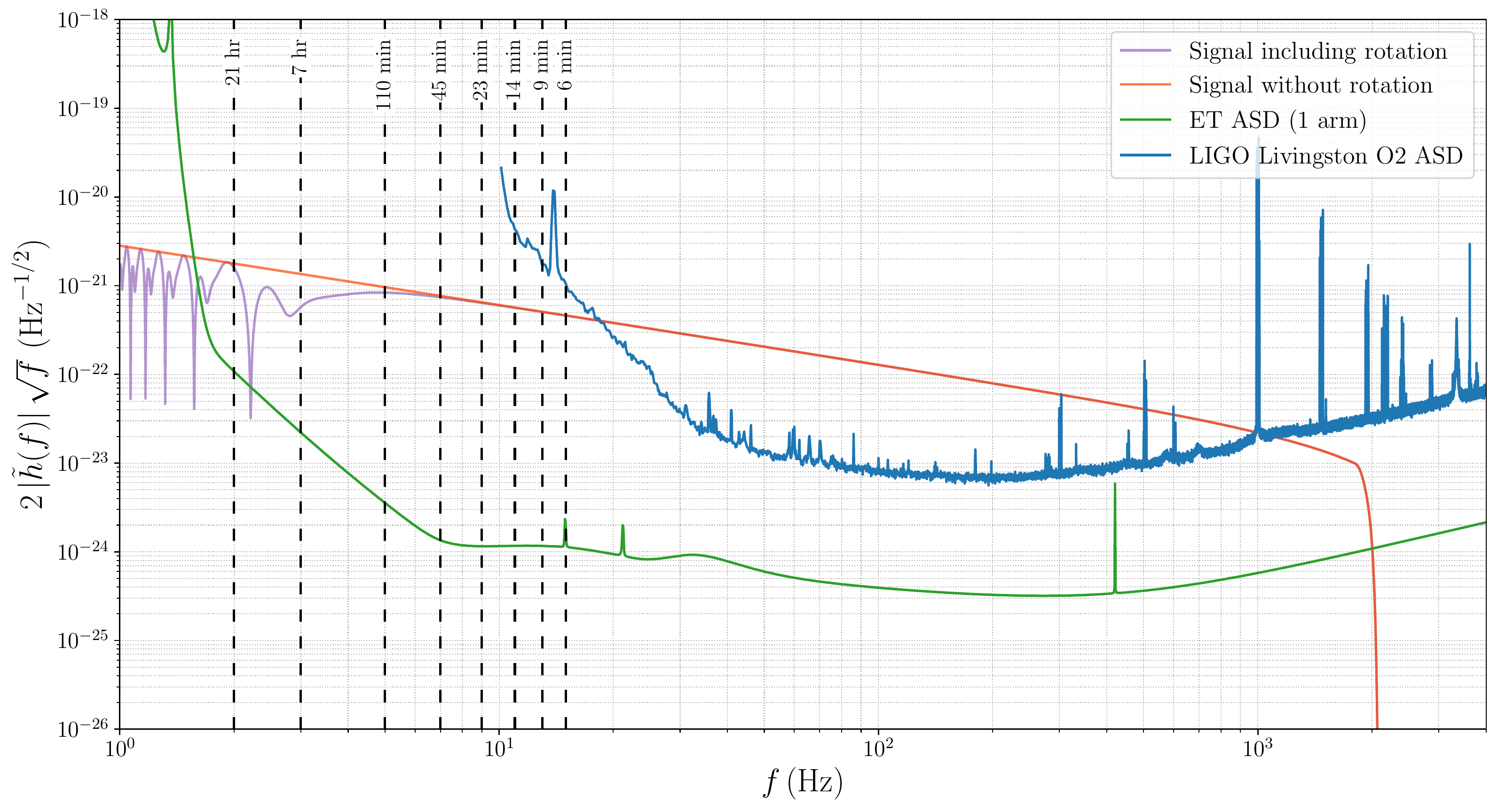}
    \caption{Comparison between the amplitude of  a BNS signal analogous to GW170817, as seen by a single instrument of the ET triangular detector (whose sensitivity curve is shown in green)    accounting for the effect of Earth's rotation (violet curve) and without  accounting for it (orange curve). The dashed vertical lines indicate the amount of time left before coalescence and the amplitude is computed using the waveform model \texttt{IMRPhenomD\_NRTidalv2}. For comparison, we also report in blue a representative LIGO Livingston sensitivity curve during the second observing run, available at \url{https://dcc.ligo.org/LIGO-G1801952/public}.}
    \label{fig:singalinband_change_rot}
\end{figure}

The effect of the Earth's rotation on the signal consists of an amplitude modulation, due to the variation of the pattern functions with time as in \eqref{eq:patt_func_expr}, a phase modulation arising for the same reason, due to the fact that the pattern functions relative to the two polarisations evolve differently in time,\footnote{The phase modulation is implicit when writing the two components of the signal separately, as in \eqref{hTime}, but becomes apparent by re--writing it as $h(t) = A(t) \cos\{\Phi(t) - \arctan{[{A_{\times}F_{\times}(t)}/{(A_{+}F_{+}(t))}}]\}$ with $A(t) = [(A_{\times}F_{\times}(t))^2+(A_{+}F_{+}(t))^2]^{\nicefrac{1}{2}}$.} 
and a Doppler contribution to the phase due to the relative motion between the source and the detector \citep{PhysRevD.57.7089, PhysRevD.67.103001}. In time domain, the Doppler contribution can be conveniently expressed as a time--dependent shift of the time variable [see, e.g. Sect.~7.6.2 of \cite{Maggiore:2007ulw} and \cite{Wen:2010cr}].
We work with  signals in  frequency domain.
To compute the Fourier transform one can adopt the Stationary Phase Approximation,  that applies if the change in the amplitude during a cycle is much slower than the corresponding change in the phase. This is the case for
each of the two terms in the sum in \eqref{hTime},\footnote{The Stationary Phase Approximation is usually adopted when the pattern functions do not depend on time, in which case the condition is satisfied, as shown e.g. in \cite{Maggiore:2007ulw}. This remains true even including the time dependence in \eqref{eq:patt_func_expr}, given that the frequency of Earth's rotation is much smaller than the frequency of the gravitational wave when it enters the detectors band.}  so, neglecting for the moment the time shift due to the Doppler effect, we get
\begin{equation}\label{statPhase}
\begin{aligned}
    \tilde{h}_{+,\times}(f) &= \int A_{+,\times}(t)F_{+,\times}(t)\cos{\Phi(t)} \, e^{2\pi i f t} \dd{t} \\
    &\simeq \dfrac{1}{2} A_{+,\times}(t^\ast) F_{+,\times}(t^\ast) \left(\dfrac{2\pi}{\ddot{\Phi}(t^\ast)}\right)^{\nicefrac{1}{2}}\exp{i\left[2 \pi f t^\ast \, -\, \Phi (t^\ast) - \dfrac{\pi}{4}\right]} \, ,
\end{aligned}
\end{equation}
where the stationary point $t^\ast (f)$ is determined by the condition $2\pi f = \dot{\Phi}(t^\ast)$.
To lowest order in the post--Newtonian (PN) expansion,
this gives
\begin{equation}
    t^\ast (f) = t_c - \dfrac{5}{256} \left(\dfrac{G{\cal M}_c}{c^3}\right)^{-\nicefrac{5}{3}} (\pi f)^{-\nicefrac{8}{3}} \times \left[1 + \order{(\pi f G M_{\rm tot}/c^3)^{\nicefrac{2}{3}}}\right]\,,
\end{equation}
where $t_c$ is the time of coalescence and $M_{\rm tot}$ the total mass of the system. Note, however, that in our analysis we compute the time to coalescence at 3.5\,PN order, as in \cite{PhysRevD.80.084043}, Eq. (3.8b).
The amplitude modulation is reflected in the fact that the Fourier transformed signal contains the expressions of $F_+$ and $F_{\times}$ evaluated at the stationary point $t^\ast (f)$, according to \eqref{statPhase} (see also \cite{Zhao:2017cbb}). An illustration of this effect is given in \autoref{fig:singalinband_change_rot}, for a BNS signal analogous to GW170817. 

We next consider the Doppler effect due to the rotation of the Earth around its axis. Its contribution to the calculation of $t^\ast(f)$ and to the amplitude of the Fourier transform on the right--hand side of \eqref{statPhase} is totally negligible, since the frequency scale $1\,  {\rm day}^{-1}$ is very small compared to the frequencies relevant for ground--based detectors, so we only need to consider its effect in the phase. This is encoded in the time dependence of the time delay corresponding to the travel time of the signal from the origin of the reference frame, i.e. the center of the Earth, to the detector. This time delay in time domain results in a phase shift in Fourier domain, known as location phase factor, given by

\begin{equation}\label{locPhase}
    \phi_{L} = 2\pi f \Delta t_{L} = -2\pi f\dfrac{R_{\oplus}}{c} \vu*{s}(\theta,\phi) \dotproduct \vu*{d}(\lambda,\varphi, t^\ast (f))  \, ,
\end{equation}
where $R_{\oplus} \simeq \SI{6371}{\kilo\meter}$ is the Earth's radius, $\vu*{s}(\theta,\phi)$ and $\vu*{d}(\lambda,\varphi)$ are the unit vectors pointing to the source and the detector, respectively, with the second one being evaluated at the stationary point $t^\ast (f)$ when including Earth's rotation, i.e.
\begin{equation}
\begin{aligned}
   \vu*{s}(\theta,\phi) &= \Big( \sin{\theta}\cos{\phi},\ \sin{\theta}\sin{\phi},\ \cos{\theta} \Big) \, , \\
    \vu*{d}(\lambda,\varphi, t^\ast (f))  &=  \Big( \cos{\lambda} \cos{(\varphi+2\pi f_\oplus t^\ast)},\ \cos{\lambda} \sin{(\varphi+2\pi f_\oplus t^\ast)},\ \sin{\lambda} \Big)\, .
   \end{aligned} 
\end{equation}

\noindent
The effect of this location phase is of particular relevance in the case of a network of detectors, since the difference in the arrival time of the signal between different detectors gives a fundamental information to localize the source. The Doppler effect due to the rotation of the Earth around the Sun, which is relevant for instance for LISA, is negligible for the signals expected at ground--based detectors.

In summary, the signal in the frequency domain takes the form

\begin{equation}
\begin{split}
\tilde{h} (f) = \Big[ F_{+}\big(t^\ast (f); \vb*{\theta}, \vb*{\lambda}\big)\, A^{(F)}_{+}(f, \vb*{\theta}) & + F_{\times}\big(t^\ast (f); \vb*{\theta}, \vb*{\lambda}\big)\,A^{(F)}_{\times}(f,\vb*{\theta})\Big] \times \\ & \exp{i \left[2\pi f t_c  + \phi_{L}\left(t^\ast (f)\right) - \Phi_c + \Phi^{(F)}(f; \vb*{\theta}) -\frac{\pi}{4}\right]} \, ,
\end{split}
\end{equation}
where the quantities
$ A^{(F)}_{+, \times}(f, \vb*{\theta})$ and $\Phi^{(F)}(f; \vb*{\theta})$ are the output of the waveform model in frequency domain [which already includes the factor $(2\pi/\ddot{\Phi}(t^\ast))^{\nicefrac{1}{2}}$ in \eqref{statPhase}], and $\Phi_c$ the value of the phase at coalescence, which we explicitly separated from $\Phi^{(F)}(f; \vb*{\theta})$ for clarity.

When including the contribution of higher modes, the signal can be expressed as a superposition of multipoles of spin--weighted  spherical harmonics \citep{1967JMP.....8.2155G} of weight $-2$, whose Fourier transform can be computed independently. To compute the frequency--domain signal one can thus proceed as before, obtaining
\begin{equation}
\begin{split}
    \tilde{h}(f, \vb*{\theta},\vb*{\lambda}) = \Big[F_+(t^\ast (f), \vb*{\theta},\vb*{\lambda})\, \tilde{h}_+ (f, \vb*{\theta}) & + F_\times(t^\ast (f), \vb*{\theta},\vb*{\lambda})\, \tilde{h}_\times (f, \vb*{\theta})\Big] \times \\ & \exp{i\left[ 2\pi f t_c  + \phi_{L}\left(t^\ast (f)\right) - \Phi_c -\frac{\pi}{4}\right]}\,,
\end{split}
\end{equation}
where the quantities $\tilde{h}_{+,\times}$ are now the output of the waveform model, computed in frequency domain [see e.g. \cite{Garcia-Quiros:2020qpx}] as
\begin{equation}
\begin{aligned}
    \tilde{h}_+(f, \vb*{\theta}) & = \sum_{\ell\geq 2}\sum_{m=0}^\ell \dfrac{1}{2} \left[_{-2}Y_{\ell m} (\iota) + (-1)^\ell _{-2}Y_{\ell -m}^\ast(\iota)\right] A^{(F)}_{\ell m}(f, \vb*{\theta}) \exp{i\Phi^{(F)}_{\ell m} (f, \vb*{\theta})}, \\ 
    \tilde{h}_\times(f, \vb*{\theta}) & = \sum_{\ell\geq 2}\sum_{m=0}^\ell \dfrac{i}{2} \left[_{-2}Y_{\ell m}(\iota) - (-1)^\ell _{-2}Y_{\ell -m}^\ast(\iota)\right] A^{(F)}_{\ell m}(f, \vb*{\theta}) \exp{i\Phi^{(F)}_{\ell m} (f, \vb*{\theta})}\,,
\end{aligned}
\end{equation}
with $A^{(F)}_{\ell m}$ and $\Phi^{(F)}_{\ell m}$ being the amplitude and phase of the $(\ell, m)$ mode in frequency domain.

Another noteworthy aspect is that, when using a triangular--shaped detector, like ET (see \autoref{sect:detnetworks} for a brief description of its design), it is not needed to compute the signal in each interferometer, since the sum of all the signals will vanish by geometrical reasons \citep{Freise_2009}. In fact, the difference among the signals observed in each instrument will arise only from the pattern functions, that depend on the orientation of the detector as
\begin{equation}\label{eq:dep_pattfun_orient_triang}
    F_n \propto C_1 \sin\left(2\gamma + n\dfrac{2\pi}{3}\right) + C_2 \cos\left(2\gamma + n\dfrac{2\pi}{3}\right); \qquad n\in \{1,2,3\},
\end{equation}
 where $\gamma$ denotes the orientation of the first interferometer (the angle between east and the bisector of the first arm in the expressions we use) and the terms $C_1$ and $C_2$ collect all the other terms appearing in the pattern functions. We here assumed the three interferometers to be equal and co--located, which is a sensible approximation for ground--based detectors. It can then be trivially shown that $F_1 + F_2 + F_3 = 0 $, thus the signal in one of the instruments can simply be obtained from the signals in the other two, reducing the amount of needed calculations by one third.\footnote{
 This is true in general for a detector forming a closed loop of generic shape, again assuming the various interferometers to be equal and co--located: the generalisation of \eqref{eq:dep_pattfun_orient_triang} to a detector with $N_{\rm ifo}$ interferometers is given by
\(
    F_n \propto C_1 \sin\left(2\gamma + {2\pi n}/{N_{\rm ifo}}\right) + C_2 \cos\left(2\gamma + {2\pi n}/{N_{\rm ifo}}\right)\) 
with 
\( n\in \{1,2,\dots, N_{\rm ifo}\},\)
and computing the total signal one obtains
\[
        \sum_{n=1}^{N_{\rm ifo}} F_n = (C_1 \sin2\gamma + C_2 \cos2\gamma) \sum_{n=1}^{N_{\rm ifo}} \cos\left(n\dfrac{2\pi}{N_{\rm ifo}}\right) + (C_1 \cos2\gamma - C_2 \sin2\gamma) \sum_{n=1}^{N_{\rm ifo}} \sin\left(n\dfrac{2\pi}{N_{\rm ifo}}\right) = 0\,,
\]
 since the two sums trivially vanish.} This is the basis of the so--called null--stream, that is the data stream obtained by summing the signals in the instruments of a closed--loop detector, and offers outstanding capabilities for analysing data \citep{PhysRevD.40.3884, Wen:2005ui, Ajith:2006qk, Chatterji:2006nh, Rakhmanov:2006qm, Harry:2010fr, 2012PhRvD..86l2001R, Schutz:2020hyz, Wong:2021cmp, Wong:2021eun}. In particular, as it has recently be shown in \cite{Goncharov:2022_NullSt} , it can be used to eliminate glitches contaminating the detector data stream, and to get unbiased estimates of the power spectral density (PSD). Notice also that the null--stream in ET corresponds to the so--called T channel of the LISA space interferometer \citep{Prince:2002hp}.
\subsection{Technical aspects}\label{sec:technicalities}

We here briefly outline some technical choices made to implement the above features in \codename{}, which are  described in more detail, and tested, in the companion paper \cite{Iacovelli:2022mbg}. \codename{} is a pure \texttt{Python} code that fully exploits the vectorization capabilities of this language, and is able to rapidly get signal--to--noise ratios and Fisher matrices for large catalogs of events. In particular, we entirely re--wrote in \texttt{Python}, in fully vectorized form,  the waveform models \texttt{IMRPhenomD}, \texttt{IMRPhenomD\_NRTidalv2}, \texttt{IMRPhenomHM} and \texttt{IMRPhenomNSBH}, available in \texttt{C} in the LIGO Algorithm Library, \texttt{LAL} \citep{lalsuite}. Together with this paper and \codename{}, we also release the open--source library \wfname{} containing the \texttt{Python} implementation of the waveforms. The agreement with \texttt{LAL} is excellent, at the level of $\sim 10^{-14}$ in the inspiral part and $\sim 10^{-5}$ in the worst case in the merger--ringdown. The difference in the latter case is entirely due to interpolation routines needed to compute the waveform in the ringdown phase. \codename{} anyway also implements an interface with \texttt{LAL}, which makes possible to easily use all the waveforms available in that library.

When using pure \texttt{Python} waveforms, derivatives in \codename{} are computed using a mixture of analytical differentiation and automatic differentiation. The latter is a technique alternative to finite--difference, that allows us to compute derivatives up to machine precision without issues of convergence due to the step--size choice in finite difference, based on a decomposition of the function in elementary functions, see \cite{DBLP:journals/corr/abs-1811-05031} for a review. In particular, \codename{} makes use of the implementation in the \texttt{JAX} package \citep{jax2018github}.
This further has the advantage of allowing vectorization of the computation of derivatives, thus fully exploiting the capabilities of \texttt{Python}. For this to be efficient, a vectorized implementation of the waveforms is needed, which motivated the development of \wfname{}.
Finally, differentiation with respect to the parameters $d_L, \ \theta,\ \phi, \ \iota, \ \psi, \ t_c $ and $\Phi_c$, which do not depend on the waveform model, is implemented analytically, to further gain in speed and accuracy.
As additional checks of reliability, we verified that, for these parameters, the analytical results and the result obtained by \texttt{JAX} agree at machine precision ($10^{-15}$ in our case). We also implemented the FIM formalism in an independent code in \texttt{Wolfram Mathematica} in the case of the \texttt{TaylorF2\_RestrictedPN} waveform model, for which the calculation of all derivatives can be more easily performed analytically. The agreement on the diagonal elements is never worse than $10^{-4}$, including differences in the integration routines in the two languages. We refer to~\cite{Iacovelli:2022mbg} for more details of these tests. 

When resorting instead to the \texttt{LAL} waveforms, the computation of the derivatives is performed using finite difference techniques, as implemented in the \texttt{numdifftools} library,\footnote{\url{https://pypi.org/project/numdifftools/}.} with an adaptive step--size, while  the  differentiation with respect to the parameters $d_L, \ \theta,\ \phi, \ \iota, \ \psi, \ t_c $ and $\Phi_c$,  is still performed  analytically.

Coming to the inversion of the FIM, each row and column is normalized to the square root of the diagonal entries, so that the resulting matrix has ones on the diagonal and the remaining elements in the interval $[-1,\, 1]$~\citep{Harms:2022ymm}.
This transformation is applied again after the inversion of the resulting matrix to obtain the inverse of the original FIM. The inversion itself is done by means of the Cholesky decomposition, which amounts to express a (hermitian, positive--definite) matrix as a product of a lower triangular matrix and its conjugate transpose. The inversion of a triangular matrix is an easier task than that of a full matrix, which improves the inversion.\footnote{There is a small sub--sample of matrices which may be not positive--definite due to the presence of very small eigenvalues that can assume small negative values due to numerical fluctuations, in which case the Cholesky decomposition cannot be found. For those matrices, \codename{} resorts by default to a singular--value decomposition for the inversion. In any case, the inversion error for these events is always larger than the threshold adopted, so they are discarded. } Other methods supported by \codename{} are discussed in \cite{Iacovelli:2022mbg}. The inversion makes use of the \texttt{Python} library \texttt{mpmath} for precision arithmetic. 

\section{Applications to current and future ground--based detectors} \label{sec:detectors}

In this section, we use \codename{} to study the detection and parameter estimation capabilities of current and future observatories for the three kind of GW sources that have been detected so far, namely, BBHs, BNSs, and NSBHs. The goal is to give realistic forecasts based on updated population models. For the short term, we focus on the  forthcoming O4 run of the LIGO--Virgo--KAGRA (LVK) collaboration. For the long term, we consider a single Einstein Telescope (ET) observatory, and a network ``ET+2CE'', made by ET, located in Europe, and two Cosmic Explorers (CE), located in the US.

\subsection{Populations}\label{sect:populations}
We here describe our baseline assumptions for the populations of the three kinds of compact binary systems considered in the analysis, (astrophysical) BBHs, BNSs and NSBHs, also summarised in \autoref{tab:popParamsTable}. The functional form and numerical values of the parameters for all distributions are reported in \autoref{sec:appendix_adoptedDistr}.

\begin{description}[align=left]
    \item[BBH]    We adopt a source--frame mass and spin distribution calibrated on the latest LVK results \citep{LIGOScientific:2021psn}, using the \textsc{Power Law + Peak} profile for the former,  and the \textsc{Default} model for the latter, and assume that the mass and spin distributions do not evolve with redshift. For the local rate, the value inferred  from the GWTC--3 catalog is ${\cal R}_{0, {\rm BBH}}=17^{+10}_{-6.7}~\si{\per\cubic\giga\parsec\per\year}$.\footnote{This is the median value with $90\%$ c.l. error, inferred using the \textsc{Power Law + Peak} distribution. It is not explicitly given in \cite{LIGOScientific:2021psn} (a plot of  ${\cal R}_{{\rm BBH}}(z)$ is anyhow shown in their Fig.~13), but is available in the associated data release at \url{https://zenodo.org/record/5655785\#.YnUnPS8QN70}, inside the file \texttt{PowerLawPeakObsOneTwoThree.json}.} We will then adopt ${\cal R}_{0, {\rm BBH}}=\SI{17}{\per\cubic\giga\parsec\per\year}$ as our reference value. Given the still limited redshift range of the detected events, the rate distribution  in redshift has large  uncertainties, and only its \textsc{power-law} behavior at low redshift has been constrained. 
    Since 3G detectors will cover a much broader redshift range, for which a power--law behavior is not realistic, we choose to adopt a Madau--Dickinson profile \citep{Madau:2014bja} in which the low--end slope is fixed to the LVK value and the other parameters assume typical values used in the literature \citep{Madau:2016jbv}, see \autoref{sec:appendix_adoptedDistr} for details. 
    With these choices, we find that the number of BBHs coalescing in  one year, out to $z=20$, is
    $N_{\rm BBH}\simeq\num{7.5e4}$.
    
    Several caveats are here in order. Beside the uncertainty in the local rate (see \cite{Mandel:2021smh} for a comprehensive review of  the uncertainties from both observation and theory),  there is an uncertainty on the BBH mass function, for which the LVK data already provide some information, but which can still vary significantly; even more important are the uncertainties on the redshift evolution of the merger rate 
    \citep{Dominik:2013tma,Santoliquido:2020axb,Rozner:2022ydm,Chruslinska:2022ovf}
    and of the mass and spin distributions. For the redshifts of interest at 3G detectors, these distributions cannot be significantly constrained by current  data, while   theoretical modelizations  still have large uncertainties. Our choice of neglecting any redshift dependence in the mass and spin distributions is the simplest one, and is consistent with  \cite{LIGOScientific:2021psn}, that find no evidence for redshift dependence in the range of redshifts currently explored by 2G detectors. For the broader range of redshifts that will be accessible to 3G detectors, this is not expected to continue to hold, see e.g.
    \cite{Fishbach:2021yvy,vanSon:2021zpk,Belczynski:2022wky} for the redshift dependence of the mass distribution, 
    and \cite{Qin:2018vaa, Biscoveanu:2022qac,Bavera:2022mef}  for the redshift dependence of the spin distribution.  Our choice of  redshift--independent mass and spin distributions should therefore be considered only as  dictated by simplicity, and will likely have to be modified as the observational and theoretical understanding improve. It should also be stressed that,  already for BBHs of astrophysical origin, different formation channels have merger rates with different redshift dependence; in particular, BBHs whose progenitors were population~III stars have a  redshift dependence of the merger rate which is sensibly different from the one that we have assumed, and can extend to redshift $z\sim 20$ and beyond, see  \cite{Kinugawa:2014zha,Ng:2020qpk} and references therein. Furthermore, BBHs with a primordial origin would have a completely different merger rate,  that increases monotonically with redshift as $R_{\rm PBH}\propto [t(z)]^{-\nicefrac{34}{37}}$ up to $z=\order{\num{e3}}$ \citep{Raidal:2018bbj,DeLuca:2020qqa,DeLuca:2021wjr}, 
    see also \cite{Franciolini:2021nvv} for recent review.
    
    \item[BNS] The knowledge of the  population of merging BNS binaries is still limited compared to BBH systems, given the very small number of  GW detections of this class of sources. Following \cite{LIGOScientific:2020kqk,LIGOScientific:2021psn}, we assume that the (source--frame)  masses of neutron stars in merging binaries have a flat
    distribution  in the interval $[1,2.5]\, \si{\Msun}$.  
    The other distribution commonly used in the literature is a Gaussian for each of the two NSs, such as  $\mathcal{N}(\mu=1.33, \sigma=0.09)$ (with masses in units of \si{\Msun}), which  comes from galactic electromagnetic (EM) observations \citep{Farrow:2019xnc}. The local rate for BNS mergers inferred from GW observations is quite sensitive to the assumptions made for the mass function distribution, and it is therefore important to choose a rate consistent with the assumed mass distribution. The value inferred from the GWTC--3 catalog, assuming a flat mass distribution, is 
    ${\cal R}_{0, \rm BNS} = 105.5^{+190.2}_{-83.0}~\si{\per\cubic\giga\parsec\per\year}$~\citep{LIGOScientific:2021psn}.
    In the following, we will then use  as reference value ${\cal R}_{0, \rm BNS} = \SI{105.5}{\per\cubic\giga\parsec\per\year}$. However,  the uncertainty on this number is  still quite large and, depending on the astrophysical modelization used, values of  ${\cal R}_{0, \rm BNS}$ in the range $(10-1700)~\si{\per\cubic\giga\parsec\per\year}$ are consistent with current observations~\citep{LIGOScientific:2021psn}. We will  emphasize this uncertainty whenever we quote numbers that depend on the local BNS rate. With our choices,  the number of BNSs coalescing in  one year, out to $z=20$, is
    $N_{\rm BNS}\simeq\num{1e5}$.
    
    Given the expected small values for the spins of these objects, we sample their components aligned with the orbital angular momentum independently and uniformly in the range $[-0.05, \, 0.05]$. Another parameter characterising NSs is the adimensional tidal deformability, $\Lambda$, which strongly depends on the equation of state (EoS) of dense matter above the nuclear density, and is still largely unknown \citep{2010PhRvD..81l3016H}. We thus make the agnostic choice of sampling $\Lambda$ for each component uniformly in the range $[0,\,2000]$, which is compatible with current observations \citep{Abbott:2018wiz, Abbott_2020GW190425}. 
    In \autoref{sec:appendix_comparisonNSEoS} we compare with the results obtained by using some specific NS equations of state.
    For the redshift distribution of these systems we assume a Madau--Dickinson profile convolved with a time delay distribution $P(t_d)\propto 1/t_d$ with a minimum time delay of $\SI{20}{\mega\year}$, which is again a standard choice in the literature [see e.g. \cite{2012PhRvD..86l2001R,Belgacem:2019tbw}] and, again, no redshift evolution for the mass distribution.
    
    \item[NSBH] The population of this class of sources is the most uncertain, with only two GW observations by the time of writing \citep{Abbott_2021:NSBHdetection}, and no electromagnetic observation. For the NS we adopt  a Gaussian distribution for the masses, $\mathcal{N}(1.33, 0.09)$ (with masses in units of \si{\Msun}), which seems a good approximation to realistic astrophysical scenarios  [see e.g. Fig. 14 of \citep{10.1093/mnras/stab2716}],  an aligned spin uniformly distributed in the interval $[-0.05, \, 0.05]$, and an adimesional tidal deformability parameter uniformly sampled in $[0,\,2000]$.  Both from observations and astrophysical simulations, there seems to be a preference for low--mass BHs appearing in this systems, also with small spins. We thus adopt for the BH mass distribution the fitting function provided in Eq. (6) of \cite{Zhu_2021_NSBH}, which is tuned on the astrophysical simulations of \cite{10.1093/mnras/sty1999}, and shows a peak around $m_{\rm BH}\sim \SI{8}{\si{\Msun}}$, while for the BH aligned spin component we assume a Gaussian distribution $\mathcal{N}(0, 0.15)$ as suggested in \cite{Zhu_2021_NSBH}, which is consistent with the values found in the analysis of GW200105 and GW200115. The rate distribution is the most uncertain, as can be seen for example from the various plots and discussion in \cite{10.1093/mnras/stab280, 10.1093/mnras/stab2716}. We thus adopt the same distribution used in the BNS case, i.e. a Madau--Dickinson profile convolved with a time delay distribution $P(t_d)\propto 1/t_d$ with a minimum time delay of $\SI{20}{\mega\year}$.
    With these choices, the number of merger per year out to $z=20$ is $N_{\rm NSBH}\simeq\num{4.5e4}$. We will then study the dependence of our results on these choices for the BH mass distribution and the event rate, comparing with some of the models presented in \cite{10.1093/mnras/stab2716}.
\end{description}
The assumptions for the remaining parameters are common to all the three kinds of sources: the non--aligned spin components are set to 0, the sky position parameters $\theta$ and $\phi$ are sampled uniformly over the whole sphere, the cosine of the inclination angle is sampled uniformly in the interval $[-1,\, 1]$, while the polarisation angle in the interval $[0,\, \pi]$, and coalescence phase in $[0,\, 2\pi]$, and the time of coalescence is sampled uniformly in time over a 10 year period.\footnote{What really matters in the analysis is anyway the Greenwich Mean Sidereal Time, GMST, associated to the GPS time, which, expressed in days, can only range between 0 and 1.} The luminosity distances are computed from the redshifts assuming a flat $\Lambda$CDM cosmology with \textsc{Planck18} parameters \citep{Aghanim_Planck18}, using the \texttt{astropy.cosmology} package \citep{price2018astropy}.
\begin{table}[t]
    \centering
    \begin{tabular}{!{\vrule width .09em}c|c|c|c!{\vrule width .09em}}
        \toprule
        \midrule
        Parameter & BBH & BNS & NSBH \\
        \midrule
        \midrule
        $m_1$ & \multirow{2}{*}{\textsc{Power Law + Peak} \citep{LIGOScientific:2021psn}} & \multirow{2}{*}{uniform in $[1,\,2.5]$  \si{\Msun}} & \citep{Zhu_2021_NSBH} Eq. (6)\\
        \cmidrule{0-0}\cmidrule{4-4}
        $m_2$ & & & $\mathcal{N}(1.33, 0.09)$ in \si{\Msun}\\
        \midrule
        $z$ & Madau--Dickinson \citep{Madau:2014bja} & \multicolumn{2}{c!{\vrule width .09em}}{Madau--Dickinson + $P(t_d) \propto1/t_d$, $t_{d, {\rm min}} = \SI{20}{\mega\year}$} \\
        \midrule
        $d_L$ & \multicolumn{3}{c!{\vrule width .09em}}{computed from $z$ assuming \textsc{Planck18} flat $\Lambda$CDM \citep{Aghanim_Planck18}} \\
        \midrule
        $\chi_{1,z}$ & \multirow{2}{*}{\textsc{Default} \citep{LIGOScientific:2021psn}} & \multirow{2}{*}{uniform in $[-0.05,\,0.05]$} & $\mathcal{N}(0, 0.15)$ \\
        \cmidrule{0-0}\cmidrule{4-4}
        $\chi_{2,z}$ & & & uniform in $[-0.05,\,0.05]$\\
        \midrule
        $\chi_{x}, \chi_{y}$ & \multicolumn{3}{c!{\vrule width .09em}}{0} \\
        \midrule
        $\Lambda_1$ & \multirow{2}{*}{0} & \multirow{2}{*}{uniform in $[0,\,2000]$} & 0 \\
        \cmidrule{0-0}\cmidrule{4-4}
        $\Lambda_2$ & & & uniform in $[0,\,2000]$\\
        \midrule
        $\theta$ & \multicolumn{3}{c!{\vrule width .09em}}{$\cos(\theta)$ uniform in $[-1,\, 1]$} \\
        \midrule
        $\phi$ & \multicolumn{3}{c!{\vrule width .09em}}{uniform in $[0,\,2\pi]$} \\
        \midrule
        $\iota$ & \multicolumn{3}{c!{\vrule width .09em}}{$\cos(\iota)$ uniform in $[-1,\, 1]$} \\
        \midrule
        $\psi$ & \multicolumn{3}{c!{\vrule width .09em}}{uniform in $[0,\,\pi]$} \\
        \midrule
        $t_c$ & \multicolumn{3}{c!{\vrule width .09em}}{uniform in 10 yr} \\
        \midrule
        $\Phi_{c}$ & \multicolumn{3}{c!{\vrule width .09em}}{uniform in $[0,\,2\pi]$} \\
        \midrule
        \bottomrule
    \end{tabular}
    \caption{Summary of the distributions assumed for all the parameters of the three classes of compact binary systems considered, refer to the the text for more details. For BBHs and BNSs the index 1 refer to the heaviest component and the index 2 to the lightest. For NSBH systems the index 1 refers to the BH and the index 2 to the NS.}
    \label{tab:popParamsTable}
\end{table}

\subsection{Detector networks}\label{sect:detnetworks}

\begin{table}[t]
    \centering\hspace{-2.5cm}
    \begin{tabular}{!{\vrule width .09em}c|c|c|c|c|c|c|c!{\vrule width .09em}}
    \toprule\midrule
    Detector & arms length & latitude $\lambda$ & longitude $\varphi$ & orientation $\gamma$ & arms aperture $\zeta$ & shape & duty cycle \\
    \midrule\midrule
    CE 1 & \SI{40}{\kilo\meter} & \SI{46.5}{\degree} & \SI{-119.4}{\degree} & \SI{171}{\degree} & \SI{90}{\degree} & L & 85\%\\
    \midrule
    CE 2 & \SI{20}{\kilo\meter} & \SI{30.6}{\degree} & \SI{-90.8}{\degree} & \SI{242.7}{\degree} & \SI{90}{\degree} & L & 85\%\\
    \midrule
    ET & \SI{10}{\kilo\meter} & \SI{40.5}{\degree} & \SI{9.4}{\degree} & \SI{0}{\degree} & \SI{60}{\degree} & Triangle & 85\%\\
    \midrule
    LIGO H1 & \SI{4}{\kilo\meter} & \SI{46.5}{\degree} & \SI{-119.4}{\degree} & \SI{171}{\degree} & \SI{90}{\degree} & L & 70\%\\
    \midrule
    LIGO L1 & \SI{4}{\kilo\meter} & \SI{30.6}{\degree} & \SI{-90.8}{\degree} & \SI{242.7}{\degree} & \SI{90}{\degree} & L & 70\%\\
    \midrule
    Virgo & \SI{3}{\kilo\meter} & \SI{43.6}{\degree} & \SI{10.5}{\degree} & \SI{115.6}{\degree} & \SI{90}{\degree} & L & 70\%\\
    \midrule
    KAGRA & \SI{3}{\kilo\meter} & \SI{36.4}{\degree} & \SI{137.3}{\degree} & \SI{15.4}{\degree} & \SI{90}{\degree} & L & 70\%\\
    \midrule\bottomrule
    \end{tabular}
    \caption{Summary of the positions, orientations, angle between arms, shapes and duty cycles of the detectors used in our analysis. The orientation denotes the angle between the bisector of the arms (the first arm in the case of a triangle) and East.}
    \label{tab:detectorsData}
\end{table}

Our analysis is carried out for three different networks of detectors, which are also summarised in \autoref{tab:detectorsData}:
\begin{description}[align=left]
    \item[LVK O4] This network consists of the four  L--shaped ground--based GW detectors which are operative at the time of writing, namely the two LIGO detectors, in the U.S., in the sites of Hanford and Livingston, Virgo, in Italy, near Cascina, and KAGRA, in Japan, near Hida. The expected PSDs for all the detectors can be downloaded from \url{https://dcc.ligo.org/LIGO-T2000012/public}.  To forecast the capabilities of the O4 observational run we  use the Advanced LIGO sensitivity with a BNS range of \SI{190}{\mega\parsec}, the Advanced Virgo sensitivity with a BNS range of \SI{120}{\mega\parsec}, and the KAGRA sensitivity with a BNS range of \SI{10}{\mega\parsec}, which corresponds to current  expectations for the best sensitivities that could be reached during O4.\footnote{See \url{https://observing.docs.ligo.org/plan/}. For Virgo, current expectations are rather of a maximum range of \SI{115}{\mega\parsec}, but we use the publicly available PSD, that still corresponds to \SI{120}{\mega\parsec}. It should also  be stressed that initial O4 sensitivities will be much lower, with expected ranges of \SI{160}{\mega\parsec} for LIGO, \SI{80}{\mega\parsec} for Virgo and 1--\SI{3}{\mega\parsec} for KAGRA.} To be more realistic, we produce our results assuming an uncorrelated 70\% duty cycle for each detector, as suggested in \cite{AbbottLivingRevGWobs}.
    
    \item[ET] Einstein Telescope (ET) is a proposed third generation GW detector, to be built in Europe. A candidate site  is in the municipality of Lula, in Sardinia,  Italy,  and we use this location for definiteness. Very similar results would be obtained choosing the candidate site in the Meuse--Rhine Euroregion, across the borders of the Netherlands, Belgium and Germany (basically, the only difference is in the effect of the Earth's rotation on the localization capability of BNSs, that slightly improves increasing the latitude).
    Differently from current detectors, ET will be an underground detector and has a triangular design, consisting of three nested interferometer  with \SI{10}{\kilo\meter} long arms forming angles of \SI{60}{\degree}. Each interferometer actually has a xylophone design, meaning that each arm will consist of two separate instruments, one optimised for high frequencies and one for low frequencies. For our purposes they can be treated effectively as three equal co--located detectors rotated by \SI{120}{\degree} with respect to each other, neglecting the distance between each arm.\footnote{See \url{http://www.et-gw.eu/index.php/relevant-et-documents} for an collection of documents on the ET design study and  Science Case.} The official ET PSD (corresponding to what was formerly called the \textsc{ET--D} design) can be downloaded 
    at \url{https://apps.et-gw.eu/tds/?content=3&r=14065}. 
    We  also assume an uncorrelated duty cycle of 85\% for each arm of the detector, as in \cite{Ronchini:2022gwk}.
    
    \item[ET + 2CE] Beside ET alone,  we will study  a network of three  3G  ground--based detectors, made by  ET in Europe, and  by two Cosmic Explorer detectors in the U.S. The CE detectors are planned to be  L--shaped interferometers, on the surface (rather than underground, as ET).   
    Two main  configurations  are being investigated, one in which both  detectors have  \SI{40}{\kilo\meter}  arms, and one that consists of a detector with \SI{40}{\kilo\meter}  arms and  another with  \SI{20}{\kilo\meter}  arms
    \citep{Evans:2021gyd}. The latter, beside its baseline design (that, for the given length, maximizes the range to compact binaries), can also be occasionally tuned to have a better sensitivity to the post--merger phase of BNSs. The network with a \SI{40}{\kilo\meter} detector and a tunable \SI{20}{\kilo\meter} detector appears to maximize the science output and is currently the reference CE configuration 
    \citep{Evans:2021gyd}. We will then 
    study the case in which one CE detector has \SI{40}{\kilo\meter} arms and the other has \SI{20}{\kilo\meter} arms (which we will set in its baseline configuration, that maximizes the range to compact binaries), and we  will refer to the network made by ET and these two CE detectors as ``ET+2CE".\footnote{The most recent PSDs of CE
can be found at \url{https://dcc.cosmicexplorer.org/CE-T2000017/public}.} Given the current uncertainty about their construction sites, for definiteness  we assume the \SI{40}{\kilo\meter} interferometer to be located and oriented as the LIGO Hanford detector while the \SI{20}{\kilo\meter} instrument as LIGO Livingston; these are not expected to be the actual locations but, at the level of the present analysis, this will not be very important, as long as their relative distance is comparable to  the one assumed here. Another option under consideration, in which one of the two CE detectors could rather be placed in Australia, would of course lead to better angular localization. In our analysis we will not consider the two CE detectors alone (except in the plots showing the range, in \autoref{fig:Detector_Horizons_BBH} and \ref{fig:Detector_Horizons_NSBH} below), but always a network consisting of them and ET. Again, we assume an uncorrelated 85\% duty cycle for the two CE detectors and each arm of ET.
\end{description}

\begin{figure}[t]
    \centering
    \includegraphics[width=.8\textwidth]{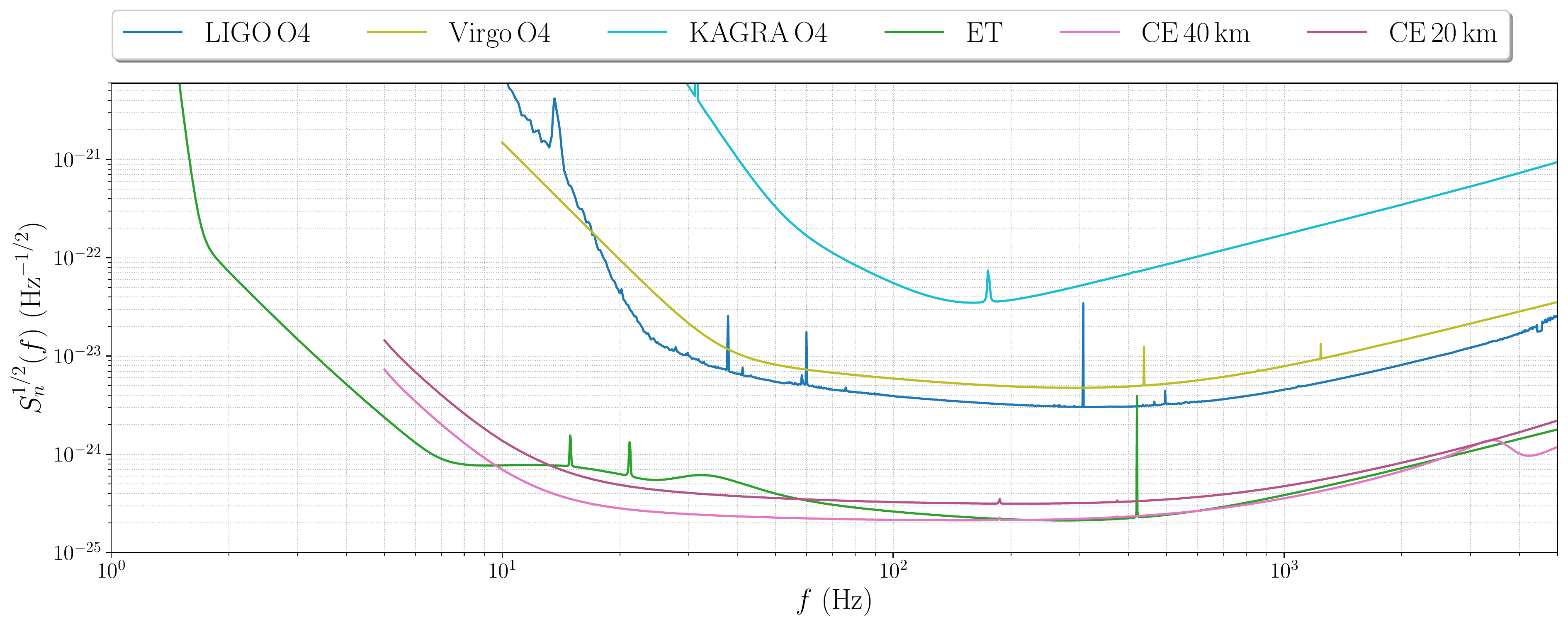}
    \caption{Noise Amplitude Spectral Densities, ASDs, of the various detectors considered. To have a uniform comparison between different geometries, the  single--detector ET sensitivity curve  has been rescaled by a factor 2/3 to take into account that, having a triangular design, ET actually consists of 3 nested instruments with an opening angle of $\SI{60}{\degree}$.}
    \label{fig:Detector_ASDs}
\end{figure}

The noise spectral densities of the various detectors considered are shown in \autoref{fig:Detector_ASDs}. We take into account that ET is made of three nested interferometers, with an opening angle of $\SI{60}{\degree}$. In the ${\rm SNR}^2$, this gives a factor $3\times (\sqrt{3}/2)^2=9/4$, i.e. a factor $3/2$ in the  SNR. Therefore, to compare  with the sensitivity of a single L--shaped interferometer, we multiply the ET sensitivity by a factor $2/3$.
In \autoref{fig:Detector_Horizons_BBH} we show, for  LVK during the O4 run, for ET, and for a network of two CE (without ET),  the corresponding horizon distance to BBHs  (i.e., the maximum distance to which a BBH optimally oriented and with optimal sky location can be detected),  requiring  ${\rm SNR}\geq 8$ for detection, where SNR is the signal--to--noise ratio of the detector network. 
The lower edge of the shaded bands gives the distance to which about $50\%$ of the BBHs can be detected. This has been obtained requiring that the SNR, averaged over sky position and inclination, is above our threshold value, which is a proxy for a more accurate computation obtained performing a sampling of a population of events for each mass bin. The left panel shows the case of equal mass and non--spinning binaries, and is analogous to a well--known plot presented in \cite{Hall:2019xmm}.\footnote{Actually, the plot in \cite{Hall:2019xmm}  refers to two CEs both of \SI{40}{\kilo\meter}, while we consider the \SI{40}{\kilo\meter}+\SI{20}{\kilo\meter} configuration. Note also that we  use the  \texttt{IMRPhenomHM} waveform, while the plot in \cite{Hall:2019xmm} was obtained using \texttt{IMRPhenomD} (we thank Evan Hall for providing this information). However, in the equal--mass case, higher modes give a negligible contribution.} In the central panel we show the result for non--spinning binaries with a mass ratio $q=m_1/m_2=5$, so in this case the fact that we include higher modes matters for the $50\%$ detection range (but not for the horizon, which is obtained for $\iota=0$, in which case the higher modes vanish). Note that, for a given total mass, the horizon decreases by increasing the mass ratio, simply because, for fixed  total mass, the chirp mass ${\cal M}_c$ decreases as the mass ratio moves away from $q=1$, and the amplitude in the inspiral phase is proportional to ${\cal M}_c^{\nicefrac{5}{3}}$. In the right panel we also turn on the spins, taking aligned spins for the two BHs, choosing for definiteness $\chi_{1,z}= \chi_{2,z}=0.8$, while still keeping $q=5$. We see that the parallel spins have the effect of raising the horizon distance again, because of their repulsive effect in the inspiral phase, which delays the merger (conversely, anti-parallel spins accelerate the merger and lower the horizon distance).
\begin{figure}[t]
    \centering
    \includegraphics[width=1.\textwidth]{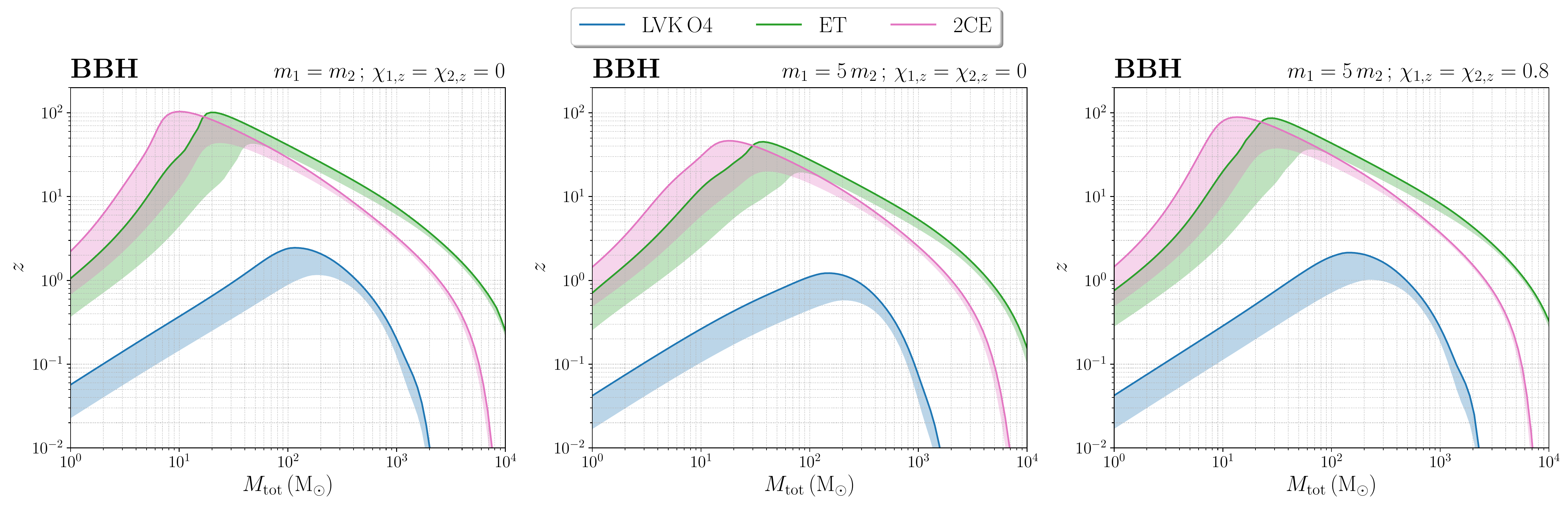}
    \caption{Detection horizons for different monochromatic populations of BBH sources for the LVK detector network during O4, for ET alone, and for a network of two CE detectors (without ET), taken one with 40~km arms and one with 20~km arms. The solid lines show the maximum redshift out to which a binary with optimal sky location and inclination can be detected, while the shaded bands represent the maximum $z$ out to which a binary  could be observed after averaging the SNR over the sky,
    which is a proxy for the redshift out to which $50\%$ of the  population with given masses and spins could be detected. 
    We set a detection threshold ${\rm SNR}=8$.
    In the left panel we show the result for equal mass non--spinning binaries, in the central panel for non--spinning binaries with a mass ratio $q=5$, and in the right panel for binaries with a mass ratio $q=5$ and aligned spin components $\chi_{1,z} = \chi_{2,z} = 0.8$. We use the \texttt{IMRPhenomHM} waveform.}
    \label{fig:Detector_Horizons_BBH}
\end{figure}
\begin{figure}[t]
    \centering
    \includegraphics[width=1.\textwidth]{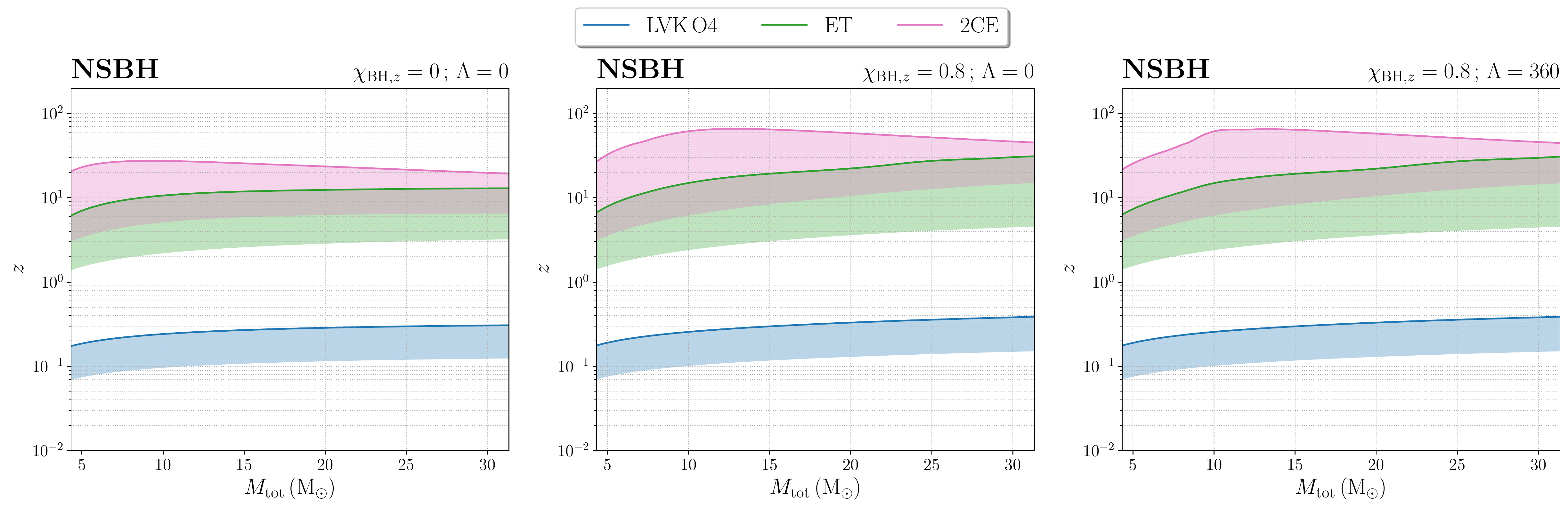}
    \caption{As in \autoref{fig:Detector_Horizons_BBH} for NSBH sources. The NS mass is fixed to $m_{\rm NS} = \SI{1.33}{\Msun}$ while we vary the BH mass in the astrophysically motivated range $[3,\, 30]\,\si{\Msun}$ \citep{10.1093/mnras/stab2716}. In the left panel we show the results for non--spinning binaries neglecting tidal effects; in the central panel for binaries having a BH with aligned spin component $\chi_{{\rm BH},z}=0.8$, and in the right panel we consider both the effect of a spinning BH with $\chi_{{\rm BH},z}=0.8$ and a tidal deformability parameter for the NS $\Lambda=360$, which is consistent with the APR4 \citep{PhysRevC.58.1804} soft equation of state. We use the waveform model \texttt{IMRPhenomNSBH}.}
    \label{fig:Detector_Horizons_NSBH}
\end{figure}

For BNS,  taking them of  equal mass and spinless, which is a very good approximation for actual systems, and neglecting the small effect of tidal deformability, the horizon and $50\%$ detection range  can also be read from the left panel of \autoref{fig:Detector_Horizons_BBH}, using a value of $M_{\rm tot}$, such as $M_{\rm tot}\simeq 2.7$, appropriate to typical BNSs. 

For NSBHs, instead, the mass ratio is very different from one, and the BH spin is not necessarily small.
In \autoref{fig:Detector_Horizons_NSBH} we show the results for NSBH 
using a horizontal scale for the total mass appropriate to current expectations for these systems,
obtained fixing the NS mass  to $m_{\rm NS} = \SI{1.33}{\Msun}$ and varying the BH mass in the astrophysically motivated range $[3,\, 30]\,\si{\Msun}$ \citep{10.1093/mnras/stab2716},
and using the
\texttt{IMRPhenomNSBH} waveform. In the left panel we show the case of a spinless BH, and we set  to zero the parameter $\Lambda$ that describes the tidal deformability of the NS. In the central panel we turn on the BH spin (setting, for definiteness, $\chi_{{\rm BH},z}=0.8$) and, in the right panel, we also turn on the NS tidal deformability. We see that, for NSBHs, over the relevant range of masses, the horizon curves are much more flat than for BBHs. This, first of all, is simply  due to the much smaller range that we have taken for $M_{\rm tot}$, based on current expectations for NSBH systems.
Furthermore, the chirp mass is related to the total mass $M_{\rm tot}$ and to the mass ratio $q=m_1/m_2$  by ${\cal M}_c=[q/(1+q)^2]^{\nicefrac{3}{5}} M_{\rm tot}$. Therefore, for $q\gg 1$, as is the case for NSBHs (recall that, with our conventions, $m_1\geq m_2$, so $q\geq 1$), we have ${\cal M}_c\simeq q^{-\nicefrac{3}{5}}M_{\rm tot}$ and
a given change in $M_{\rm tot}$ results in a  smaller change in ${\cal M}_c$, which is the mass scale that characterizes the amplitude. We also see that, for CE, increasing $M_{\rm tot}$, beyond some value in the considered range the horizon starts to decrease; this is due to the fact that, increasing  $M_{\rm tot}$, the  merger takes place at lower frequencies, so the signal is moved toward the region where  the sensitivity of CE degrades faster, compared to ET.

\section{Results}\label{sect:Results}

We now show the results of our analysis for the three populations of sources, as seen by the different detector networks and with the assumptions discussed above. We first compute, for each event, the  network SNR, defined by
\begin{equation}
{\rm SNR}^2=\sum\nolimits_{i} {\rm SNR}^2_i\, ,
\end{equation}
where the sum runs over the detectors, and the matched filter signal--to--noise ratios ${\rm SNR}_i$ of the individual detectors are given in \eqref{eq:SNR} or, more explicitly, 
\begin{equation}
    {\rm SNR}^2_i = 4 \int_{f_{\rm min}}^{f_{\rm cut}} \dfrac{|\tilde{h}_{(i)} (f)|^2}{S_{n,i}(f)} \dd{f} \, . 
\end{equation}
Here $\tilde{h}_{(i)}(f)$ denotes the GW signal as detected by the $i^{\rm th}$ detector in the network (thus including the contribution of the pattern function of the detector and the location phase factor), $S_{n,i}(f)$ is the one--sided noise PSD of the $i^{\rm th}$ detector, $f_{\rm min}$ denotes the minimum frequency of the adopted frequency grid,  which we set at \SI{2}{\hertz} for ET, \SI{5}{\hertz} for CE and \SI{10}{\hertz} for 2G detectors, and $f_{\rm cut}$ is the maximum frequency of the grid, which depends on the source characteristics and the adopted waveform model.\footnote{For \texttt{TaylorF2} this is set to twice the the binary Innermost Stable Circular Orbit frequency, $f_{\rm ISCO} = 1/(2 \pi \, 6\sqrt{6} \, G M_{\rm tot}/c^3)$, while for the chosen full inspiral--merger--ringdown waveforms, as in \texttt{LALSimulation}, we set the cut frequency at $(G M_{\rm tot}/c^3)\,f_{\rm cut} = 0.2$.} After the SNRs have been determined, we perform a Fisher analysis restricting to the events having a network ${\rm SNR} \geq 12$ (in the case of ET, the network SNR is obtained combining the contributions of the three arms).  We will also compare the results for the number of detections and horizon distances with the results obtained with a network ${\rm SNR} \geq 8$, while not performing a  Fisher matrix analysis in this case, since it becomes less reliable for such low values of the SNR.

We further discard the signals with an inversion error of the Fisher matrix bigger than $ 5 \times 10^{-2}$ (see \autoref{sec:singularities}). The Fisher matrices are computed according to \eqref{eq:Fisher_def}. For a network of detectors, the total Fisher matrix is just the sum of the Fisher matrices computed for each detector. For BBHs  the set of parameters is $\vb*{\theta} = \{{\cal M}_c,\ \eta,\ d_L,\ \theta,\ \phi,\ \iota,\ \psi,\ t_c,\ \Phi_c,\ \chi_{1},\ \chi_{2}\}$, where we denoted the aligned spin components, $\chi_{1,z}$ and $\chi_{2,z}$, simply  as $\chi_{1}$ and $\chi_{2}$, to simplify the notation. For BNS and NSBH systems, $\vb*{\theta}$ also includes the tidal deformability parameters $\tilde{\Lambda}$ and $\delta\tilde{\Lambda}$, defined in \eqref{eq:LamTdelLam_def},
where, in the case of NSBH, the parameter $\Lambda_1$ corresponding to the BH must be set to zero (recall, from \autoref{tab:popParamsTable}, that for NSBH we use the convention that the index $i=1$ always refers to the BH).
After the inversion of the Fisher matrix, we compute the sky localization area for the events according to the definition \citep{Barack:2003fp, Wen:2010cr}
\begin{equation}\label{eq:skyLoc}
    \Delta\Omega_{{\rm X}\%} = -2\pi |\sin\theta|\stopsqrt{\left(\Gamma^{-1}\right)_{\theta\theta}\, \left(\Gamma^{-1}\right)_{\phi\phi} - \left(\Gamma^{-1}\right)_{\theta\phi}^2}\ \ln{\left(1 - \dfrac{{\rm X}}{100}\right)}\,,
\end{equation}
where X denotes the confidence level.  
The result of this expression is in units of steradian, so a further multiplication by $(\SI{180}{\degree}/\pi)^2$ is needed to get the estimation in the usual \si{\square\degfull} units. We will give our results in terms of $\Delta\Omega_{90\%}$.

\subsection{Binary black holes}

We first focus on BBH systems, which are on average the loudest signals that can be observed in the frequency band of current and future ground--based detectors. As recalled in \autoref{sect:populations}, the most recent estimate for the local rate from the GWTC--3 catalog  is ${\cal R}_{0, {\rm BBH}}=17^{+10}_{-6.7}~\si{\per\cubic\giga\parsec\per\year}$ \citep{LIGOScientific:2021psn}, so we use the  value 
${\cal R}_{0, {\rm BBH}}=17~\si{\per\cubic\giga\parsec\per\year}$ as our reference value for normalizing the redshift distribution of the merger rate
discussed in \autoref{sec:appendix_adoptedDistr}. In particular, this means that  our merger rate as a function of redshift, ${\cal R}_{{\rm BBH}}(z)$, will be consistent with that shown in Fig.~13 of 
\cite{LIGOScientific:2021psn}.
For these systems we simulate a population of $N_{\rm BBH}=\num{7.5e4}$ sources out to $z=20$ that, as mentioned in \autoref{sect:populations}, using this value for ${\cal R}_{0, {\rm BBH}}$ and our choice for the redshift distribution of the merger rate, corresponds to the number of systems coalescing in  one year. 

\begin{figure}
    \centering
    \includegraphics[width=.9\textwidth]{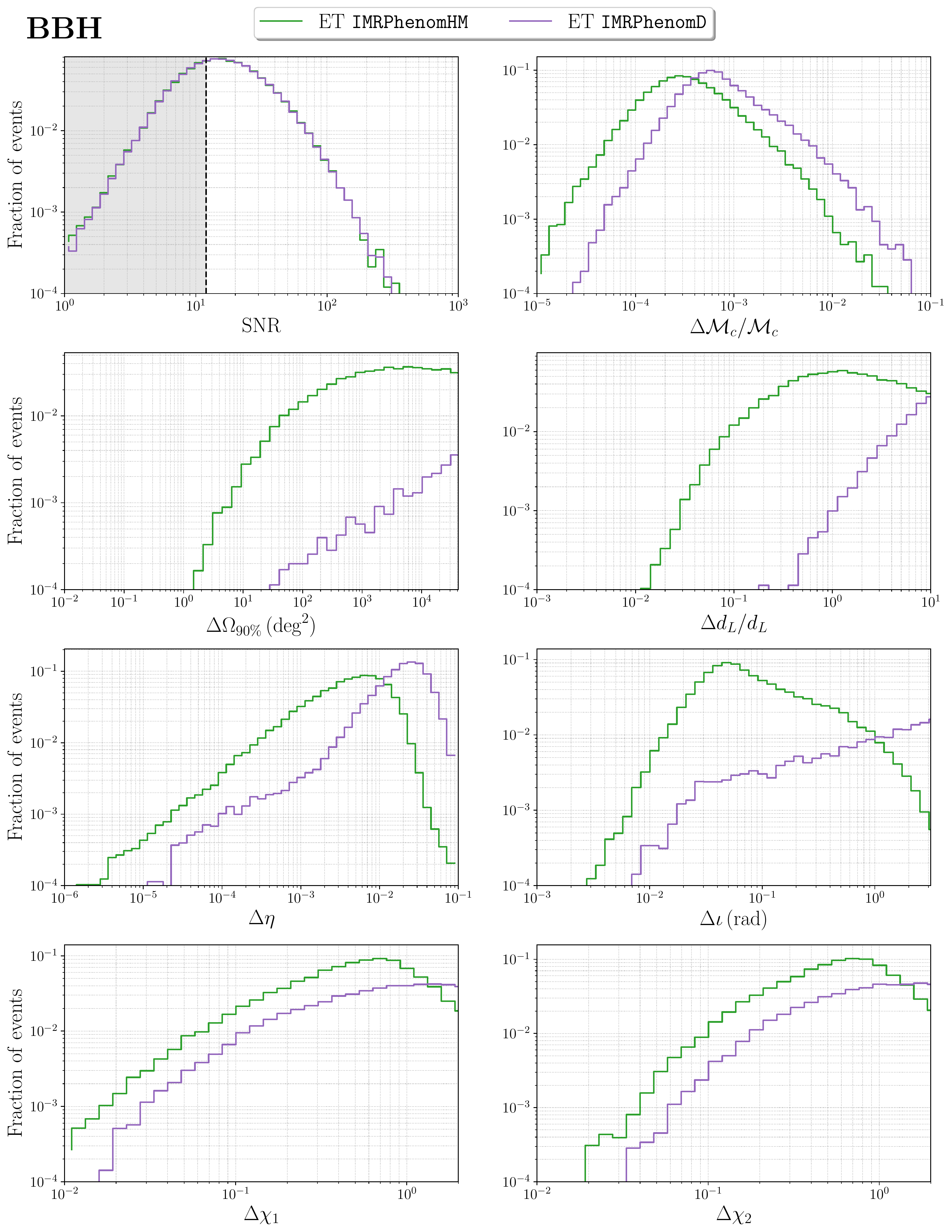}
    \caption{Histograms of the SNR and parameter errors for BBHs for   ET alone, using the waveform model \texttt{IMRPhenomHM}, which includes sub--dominant modes of the signal, and \texttt{IMRPhenomD}, which only contains the quadrupole. For the SNR, in each bin we show the fraction of events, normalized  to the total number of sources in our sample. For the parameters of the waveform, we limit to the $\tilde{N}_{\rm det}$ events that pass the cut ${\rm SNR}\geq 12$, and that furthermore admit a reliable inversion of the Fisher matrix, and we
    show the fraction of events, normalized to $\tilde{N}_{\rm det}$.}
    \label{fig:HistBBHcomp_PhenDvsHM_ET}
\end{figure}

\begin{figure}
    \centering
    \includegraphics[width=.9\textwidth]{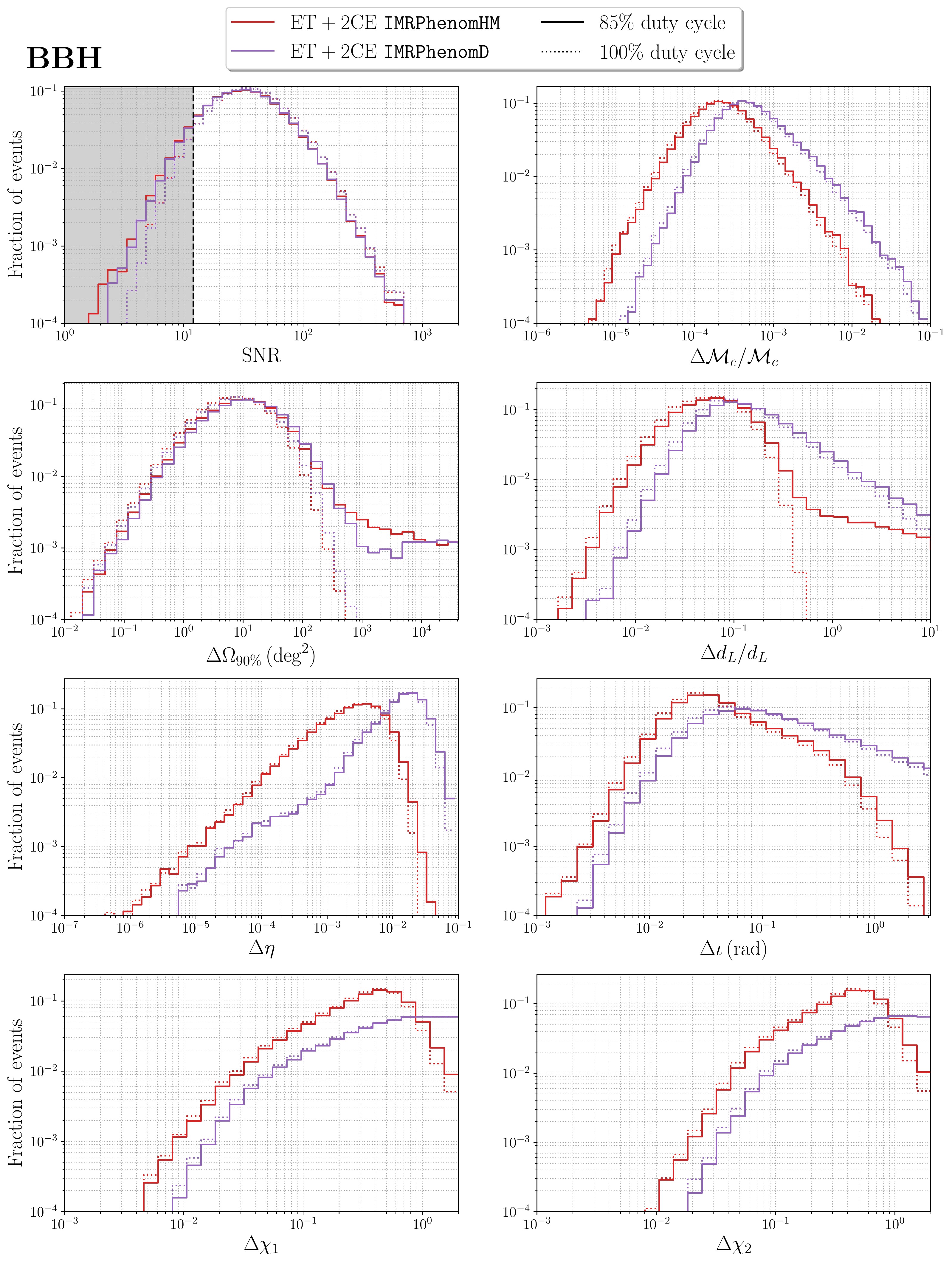}
    \caption{As in \autoref{fig:HistBBHcomp_PhenDvsHM_ET}, for the ET+2CE network. We here further show the results obtained without imposing the 85\% uncorrelated duty cycle for the detectors.}
    \label{fig:HistBBHcomp_PhenDvsHM}
\end{figure}

At the methodological level, it is important first of all to understand the effect of the waveform model used and, in particular, for heavy systems such as BBHs, the role of higher modes.
We then begin by comparing  the results 
obtained using \texttt{IMRPhenomHM}, which is our baseline waveform for BBHs   and, as discussed in \autoref{sect:waveforms}, includes several higher modes,
with the results that we get analysing the same catalog of sources with the waveform model \texttt{IMRPhenomD} \citep{PhysRevD.93.044006, PhysRevD.93.044007}. This is a full inspiral--merger--ringdown model, tuned to binaries with non--precessing spins but, differently from \texttt{IMRPhenomHM}, it only contains the dominant quadrupole mode of the signal. The results are shown in \autoref{fig:HistBBHcomp_PhenDvsHM_ET} for ET alone and in
\autoref{fig:HistBBHcomp_PhenDvsHM} for ET+2CE. In the upper left panels of these figures
we show, for the two waveforms, the distribution in SNR of the whole sample of events, displayed as a fraction of events with respect to the total number of events in the sample, i.e. the \num{7.5e4} sources that we have simulated. We then restrict parameter estimation to the events that pass the cut ${\rm SNR}\geq 12$, which is our criterion for detection and, for these, we perform the Fisher matrix analysis. As discussed in \autoref{sec:singularities}, for some of these events the Fisher matrix is ill--conditioned and its inversion can lead to large amplification of numerical errors, resulting in a covariance matrix that can be inaccurate even at the $100\%$ level, and our strategy is to simply discard these events, accepting this as a limitation of the Fisher  matrix formalism.  We denote by $N_{\rm det}$ the number of events detected in our sample, i.e. those that pass the cut ${\rm SNR}\geq 12$, and by $\tilde{N}_{\rm det}$ the number of detected event that, furthermore, have a Fisher matrix that can be inverted reliably, according to our criterion discussed in \autoref{sec:singularities}. In the plots showing the distribution of the errors on the parameters  we only  include these $\tilde{N}_{\rm det}$ events. The corresponding panels in \autoref{fig:HistBBHcomp_PhenDvsHM_ET} and \ref{fig:HistBBHcomp_PhenDvsHM}, as well as all similar plots in the following, show
the distribution  of these events, as a fraction normalized to   $\tilde{N}_{\rm det}$.

From \autoref{fig:HistBBHcomp_PhenDvsHM_ET} we see that, for ET alone, while the distribution of SNR is unaffected by the inclusion of higher modes, parameter estimation is significantly improved. This is especially remarkable for the angular resolution and for the error on the luminosity distance, where the  inclusion of higher--order modes allows us to break the distance--inclination degeneracy, but the estimation of all parameters is significantly improved,  thanks to the better description offered by the waveform model. For all parameters shown, the inclusion of higher modes has the effect of increasing the fraction of events for which accurate parameter reconstruction is possible, and also of cutting the  long tails corresponding to events with large errors, as especially evident in $\Delta\iota$ and $\Delta\eta$.  
For ET+2CE we see from \autoref{fig:HistBBHcomp_PhenDvsHM} that a rather similar pattern emerges, except for the angular localization which, in this case, is completely dominated by the triangulation, rather than by the accuracy of the waveform.
Another interesting information that emerges from \autoref{fig:HistBBHcomp_PhenDvsHM}  is the role of the duty cycle. The dashed lines show the results obtained assuming a $100\%$ duty cycle, while the solid line use a more realistic independent duty cycle of $85\%$ for each detector. We see that the duty cycle has a large effect on the tails of the distribution of $\Delta d_L/d_L$, and of
$\Delta\Omega_{90\%}$. This is natural, since these are the quantities that are more directly sensitive to the network SNR, which decrease when one or more detectors in the network is down. This highlights the importance of having detectors with a high duty cycle, or of having a network of more detectors.

\begin{figure}
    \centering
    \includegraphics[width=.9\textwidth]{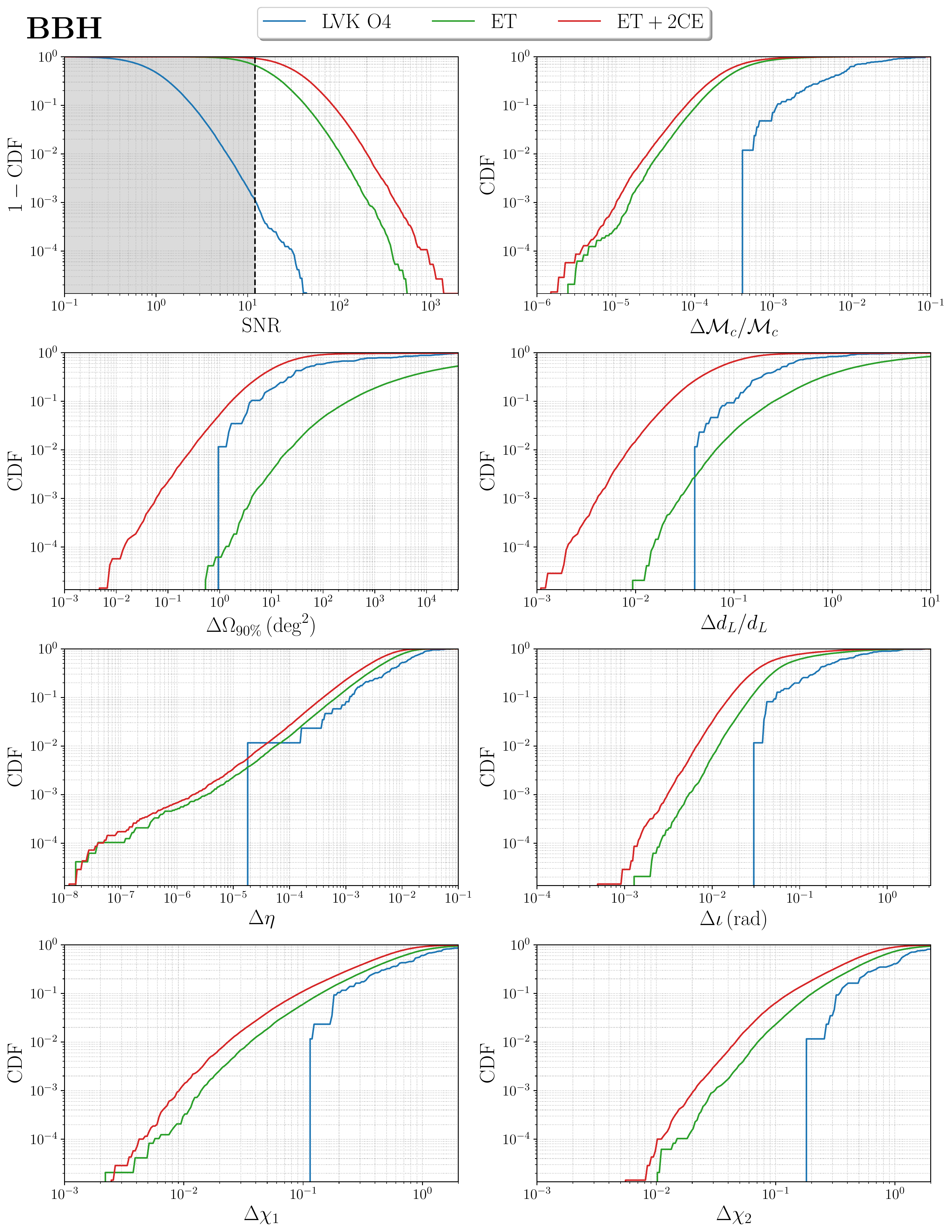}
    \caption{Cumulative distributions of the detection efficiency, for the SNRs and for the error on the parameters, for BBH signals with the three considered network configuration, using the waveform model \texttt{IMRPhenomHM}.}
    \label{fig:BBH_cumuldist}
\end{figure}

In the following, we therefore focus only on the results 
obtained using \texttt{IMRPhenomHM}. While a histogram of detection fractions contains the information in its most raw form, the corresponding cumulative distributions give information that is more condensed and often easier to interpret (and less sensitive to the specific random realization used, when small numbers are involved). 
In \autoref{fig:BBH_cumuldist} we then show, for LVK--O4, ET alone, and ET+2CE, 
the corresponding cumulative distributions function (CDF). For the SNR, we show the cumulative distribution of the fraction of events, where the fraction is obtained normalizing the events in a given bin  to the total number of events that we have generated;  for the other panels, involving parameter estimation, we show the cumulative distribution of
the fraction of events obtained normalizing the events in a given bin to $\tilde{N}_{\rm det}$, i.e. to the total number of {\em detected} events which, furthermore, admit a reliable inversion of the Fisher matrix. 
For the SNR we actually plot $1-{\rm CDF}$, so the vertical axis  gives the  fraction of events with signal--to--noise ratio {\em larger} than a given value, while for the parameters we show the CDF, so the  fraction of events with error {\em smaller} than a given value. 

We find that ET alone can detect $67\%$ of the BBH population, while a network of ET and 2CE could detect $93\%$   of the full population of sources. With our reference value 
for  the local rate, ${\cal R}_{0, {\rm BBH}}=\SI{17}{\per\cubic\giga\parsec\per\year}$, a  threshold in signal--to--noise ratio ${\rm SNR} \geq 12$, and
our choices for the BBH mass function and merger rate distribution in redshift, discussed in \autoref{sect:populations},
this means, 
for  ET alone,  that the number of events detected in one year is $N_{
\rm det}\simeq\num{5.1e4}$ while,  
for ET+2CE, the number of BBH events detected per year is $N_{\rm det}\simeq\num{7.0e4}$.  Using, more generally, the currently $90\%$ c.l. allowed range ${\cal R}_{0, {\rm BBH}}=[10.3,27]~\si{\per\cubic\giga\parsec\per\year}$ for the local rate,  the number of detections per year for ET is in the range
$[\num{3.1e4}, \num{8.0e4}]$ and, for ET+2CE, is in the range $[\num{4.2e4},\num{1.1e5}]$.
For LVK in the O4 observational run, we find instead  86 detections per year (which raise to 141 assuming a duty cycle of $100\%$). Our result for the O4 run is perfectly consistent with the forecast in \cite{AbbottLivingRevGWobs}. 
Given the logarithmic scale, the lines
in  \autoref{fig:BBH_cumuldist}
(in particular, on this scale, the blue lines that refers to LVK--O4) drop vertically to zero when the accuracy required on a parameter becomes so small
that no more system, in our sample of detections, satisfies it. The precise value of the parameters when this happens, as well as the discrete steps apparent in these curves, depend of course on the specific random realization of our sample. 

When performing parameter estimation, we restrict as usual to the $\tilde{N}_{\rm det}$ detected events that pass our criterion on the inversion of the Fisher matrix; for ET we find that, 
with $N_{\rm det}\simeq \num{5.1e4}$ detection, we have  $\tilde{N}_{\rm det}\simeq\num{4.8e4}$, corresponding to $96\%$ of the detected events; for ET+2CE,  with $N_{\rm det}\simeq\num{7.0e4}$, $99.2\%$ of the detected events pass the criterion on the inversion, so we have
$\tilde{N}_{\rm det}\simeq\num{6.9e4}$; for LVK--O4, $N_{\rm det}=\tilde{N}_{\rm det}=86$. 
We expect that most of the events that we have discarded because their Fisher matrix is ill--conditioned correspond to cases in which an analysis based on the full likelihood, rather than on the Fisher matrix approximation, would anyhow return large errors on the parameters, so most of the detected events that we have discarded should only contribute to the tails of the distributions, corresponding to large parameter errors.

\begin{figure}
    \centering
    \includegraphics[width=.9\textwidth]{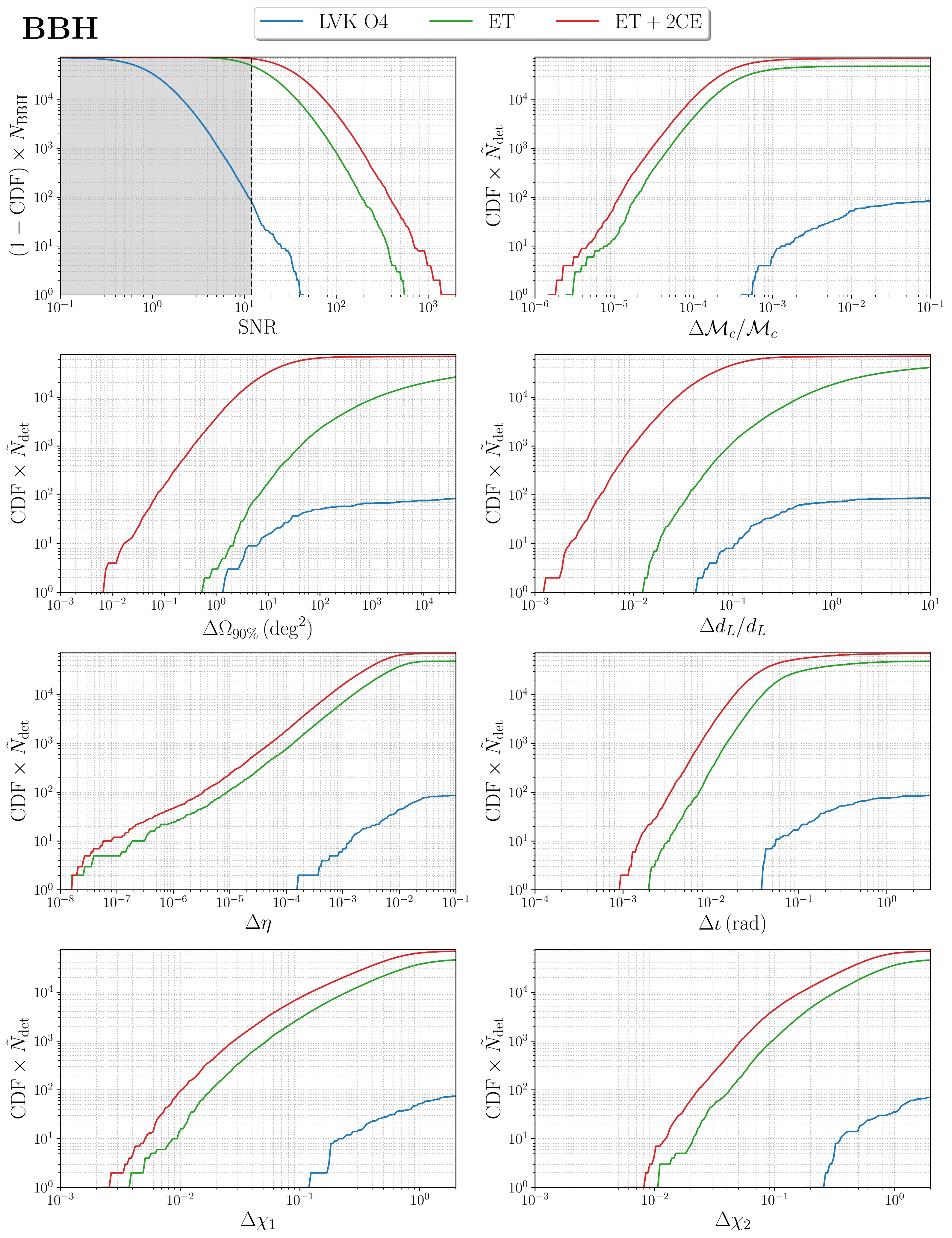}
    \caption{Cumulative distributions of the number of detections per year, for the SNRs and for the error on the parameters, for BBH signals with the three considered network configuration, using the waveform model \texttt{IMRPhenomHM}, and  the value ${\cal R}_{0, {\rm BBH}}=\SI{17}{\per\cubic\giga\parsec\per\year}$ for the local rate. Except for the panel showing  the distribution of the SNR, only the events whose Fisher matrix could be reliably inverted are included.}
    \label{fig:BBH_cumuldistNdet}
\end{figure}

The cumulative detection fraction is a useful metric to evaluate the potential of GW detectors,  in particular to appreciate how a sample of detections is representative of the whole population, and also has the advantage of being independent of the value chosen for the local rate, which  makes easier the comparison between different papers in the literature. However,
the full  potential of 3G detectors is better  appreciated by showing, instead, the cumulative distribution of the absolute number of detected events, since the overall normalizations, in particular between 2G and 3G detectors, are very different.
Indeed, in some of the panels of \autoref{fig:BBH_cumuldist}, such as that for $\Delta d_L/d_L$, the blue curve for LVK--O4 partly stays above the green curve for ET alone. This, of course, does not mean that LVK--O4 has a better sensitivity to $d_L$ than ET; rather, it is simply due to  the fact that ET sees many more events, much further away, and some of these events have worse resolution, so have the effect of decreasing the fraction of detected events which are accurately measured.
In \autoref{fig:BBH_cumuldistNdet} we show the same plots as in \autoref{fig:BBH_cumuldist}, but now in terms of  the cumulative number of events, rather than the cumulative fraction of events; for the panel on parameter estimation this is  obtained multiplying  the cumulative detection fraction curves for  LVK--O4, ET and
ET+2CE, by the respective values of  $\tilde{N}_{\rm det}$, while, for the SNR, we multiply by number of simulated events, $N_{\rm BBH}$.
We can then better appreciate  how significantly even ET alone improves on LVK--O4. In particular, now, for $\Delta d_L/d_L$, the green curve for ET alone is well above the blue curve for LVK--O4: for instance, while LVK--O4 is expected to detect only $\order{10}$ events per year with an accuracy on $d_L$ better than $10\%$, ET alone will detect  $\order{10^3}$ events per year with $\Delta d_L/d_L<10\%$, of which  $\order{10^2}$ per year will have  $\Delta d_L/d_L<4\%$.
It is also important, in particular for applications to multi--messenger astronomy and to cosmology,  that a single ET improves  over LVK--O4 even on sky localization accuracy. This was not obvious {\em a priori} since a single detector, compared to  a network of four widely separated detectors,  cannot exploit triangulation. Nevertheless, the increase in sensitivity of ET, and therefore its capability to reconstruct all parameters of the signal, provides a significant improvement, compared to LVK--O4, even  in the number of detected events with a given angular resolution, as we see from the panel for $\Delta\Omega_{90\%}$ in \autoref{fig:BBH_cumuldistNdet}: for instance, we see that ET  alone can reach an accuracy on $\Delta\Omega_{90\%}$ better than \SI{100}{\square\degfull} on about 2000 events per year (compared to about 45 events/yr for LVK--O4); 
better than
\SI{10}{\square\degfull} on about 160 events per year (compared to about 15 events/yr for LVK--O4); and can even reach  
$\Delta\Omega_{90\%}\leq \SI{1}{\square\degfull}$ on a few events per year. An ET+2CE network, combining the sensitivity of 3G detectors with the long baselines for triangulation, provides  further remarkable improvement on source localization, with about 3400~BBH/yr localized to better than \SI{1}{\square\degfull}, and the very best events localized to less than   \SI{e-2}{\square\degfull}.

\begin{figure}[t]
    \hspace{-2cm}
    \begin{tabular}{c@{\hskip 3mm}c@{\hskip 3mm}c}
  \includegraphics[width=64mm]{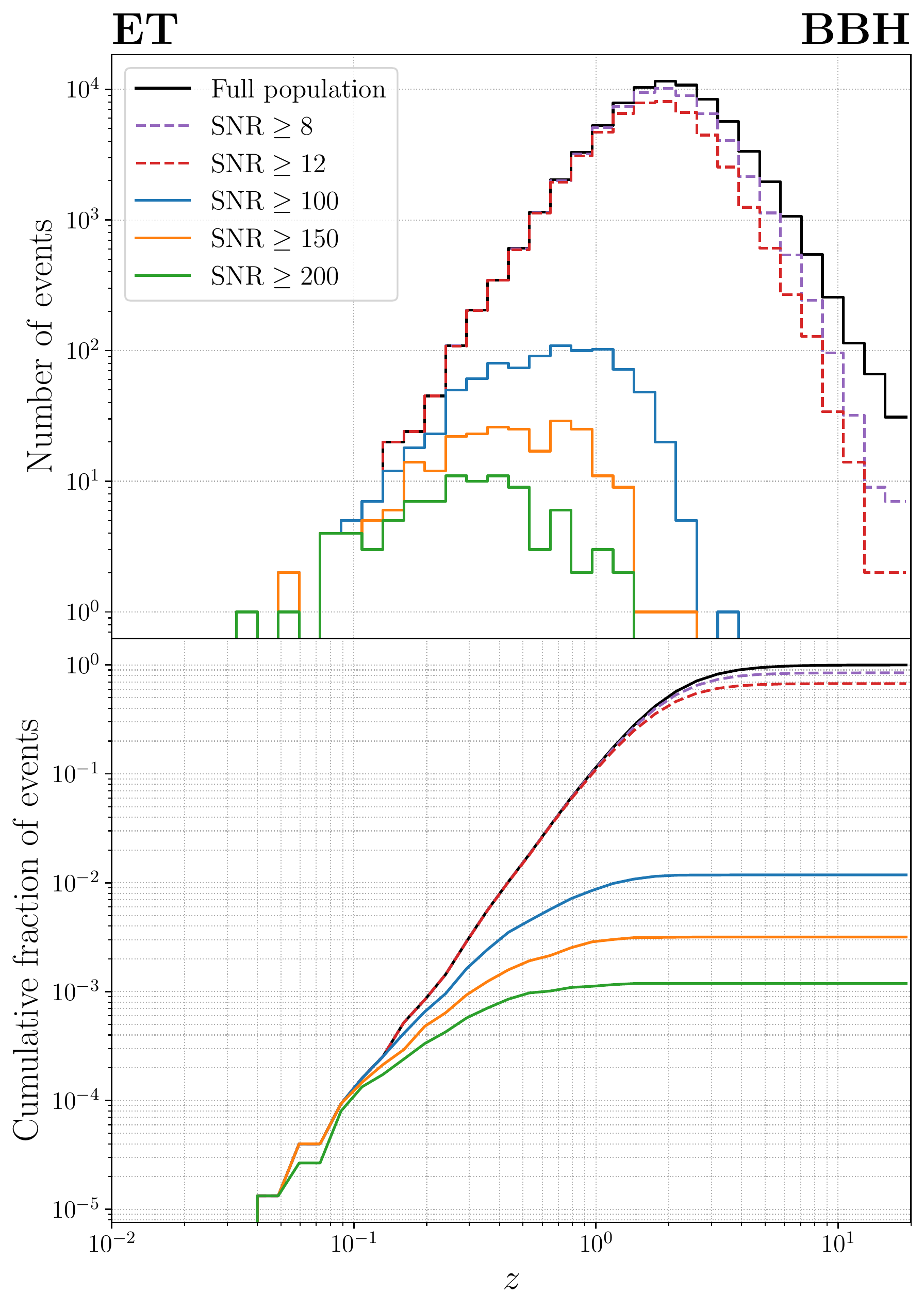} &   \includegraphics[width=64mm]{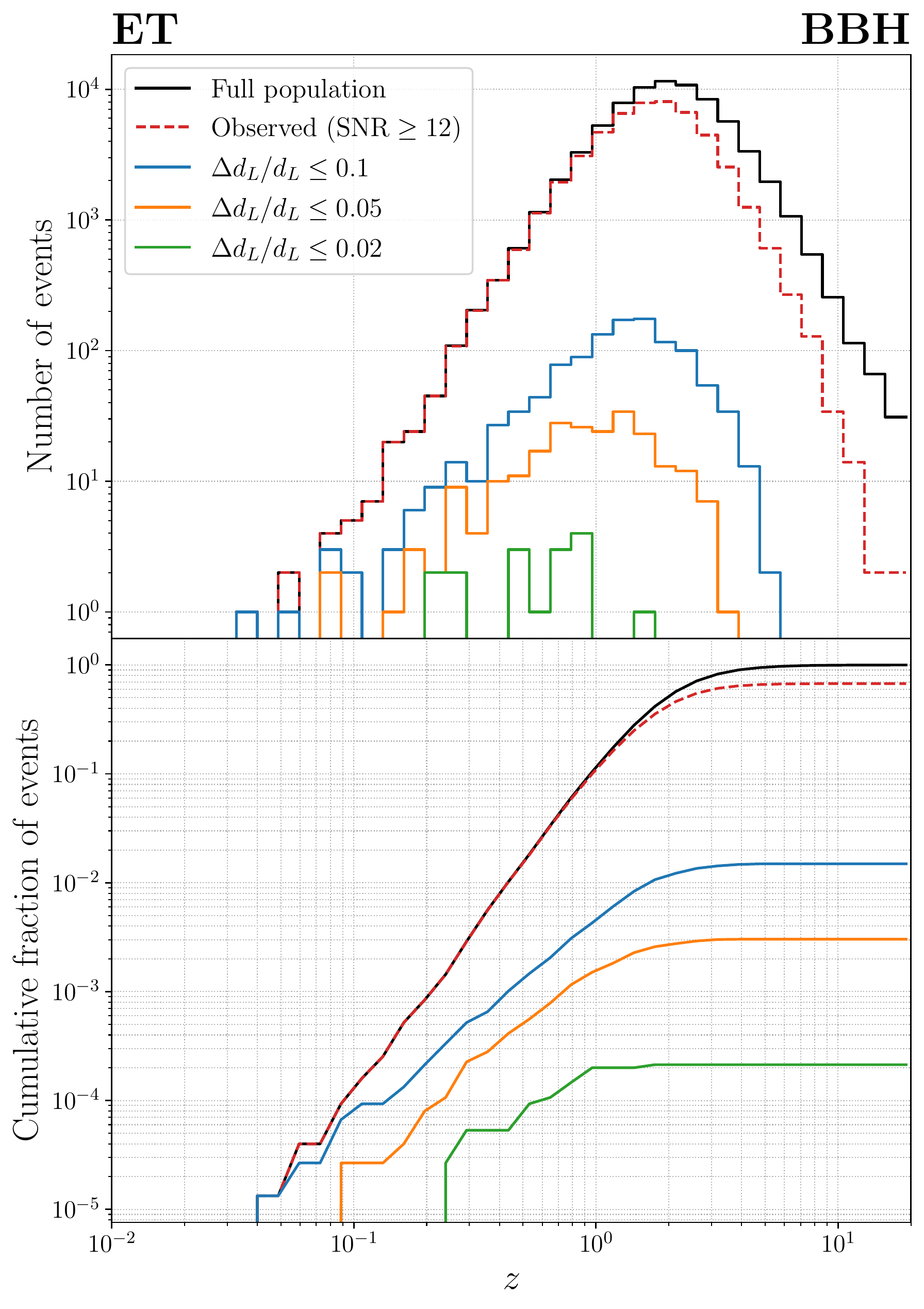} & \includegraphics[width=64mm]{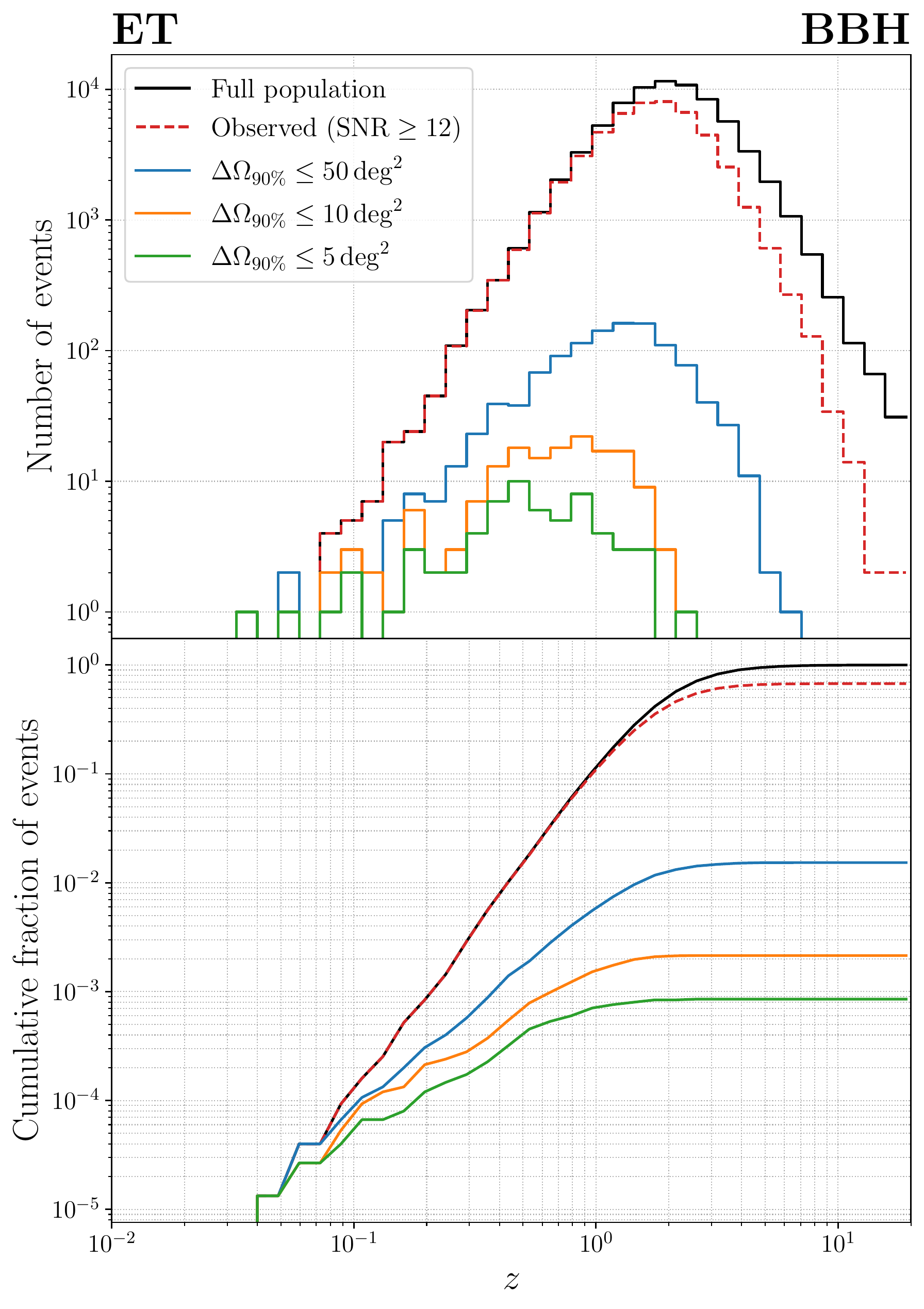}
\end{tabular}
    \caption{Redshift distributions of the BBHs observed  at ET alone in 1~yr,
    selected on the basis of different thresholds for the SNR (left panel), or setting ${\rm SNR}\geq 12$ and applying further cuts on $\Delta d_L/d_L$ (central panel), or  on  $\Delta\Omega_{90\%}$ (right panel).  The black solid line corresponds to the total BBH population in the astrophysical model that we have assumed.
    We set  ${\cal R}_{0, {\rm BBH}}=\SI{17}{\per\cubic\giga\parsec\per\year}$, and show the results for the detections in one year  (taking into account our assumptions of the duty cycle). In each column, the upper panel shows the number of events per redshift bin, while the lower panel shows the corresponding cumulative distributions, normalized to the number of BBH events in our sample, $N_{\rm BBH}=7.5\times 10^4$. }
    \label{fig:ET_BBH_zhists}
\end{figure}

\begin{figure}[t]
    \hspace{-2cm}
    \begin{tabular}{c@{\hskip 3mm}c@{\hskip 3mm}c}
  \includegraphics[width=64mm]{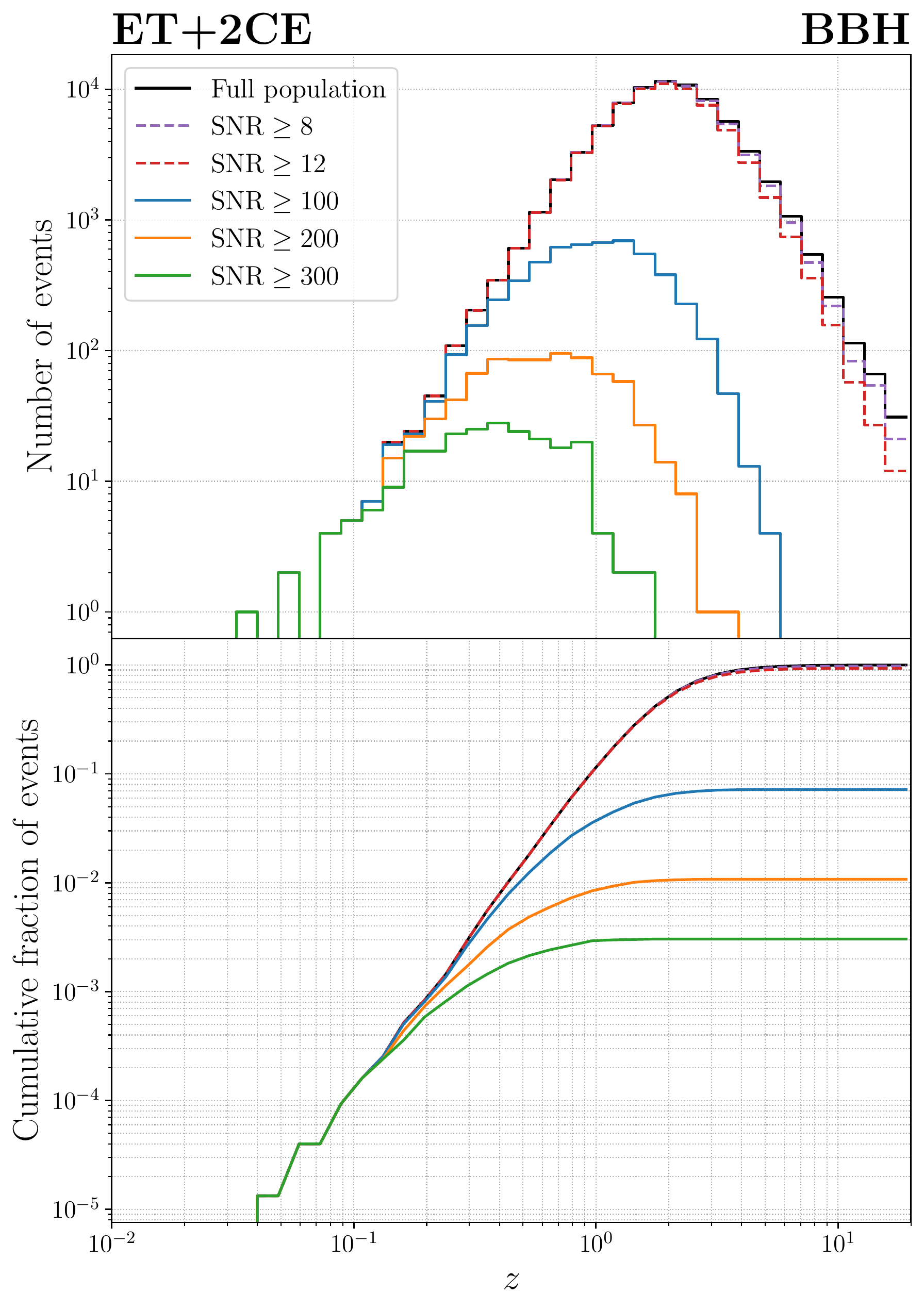} &   \includegraphics[width=64mm]{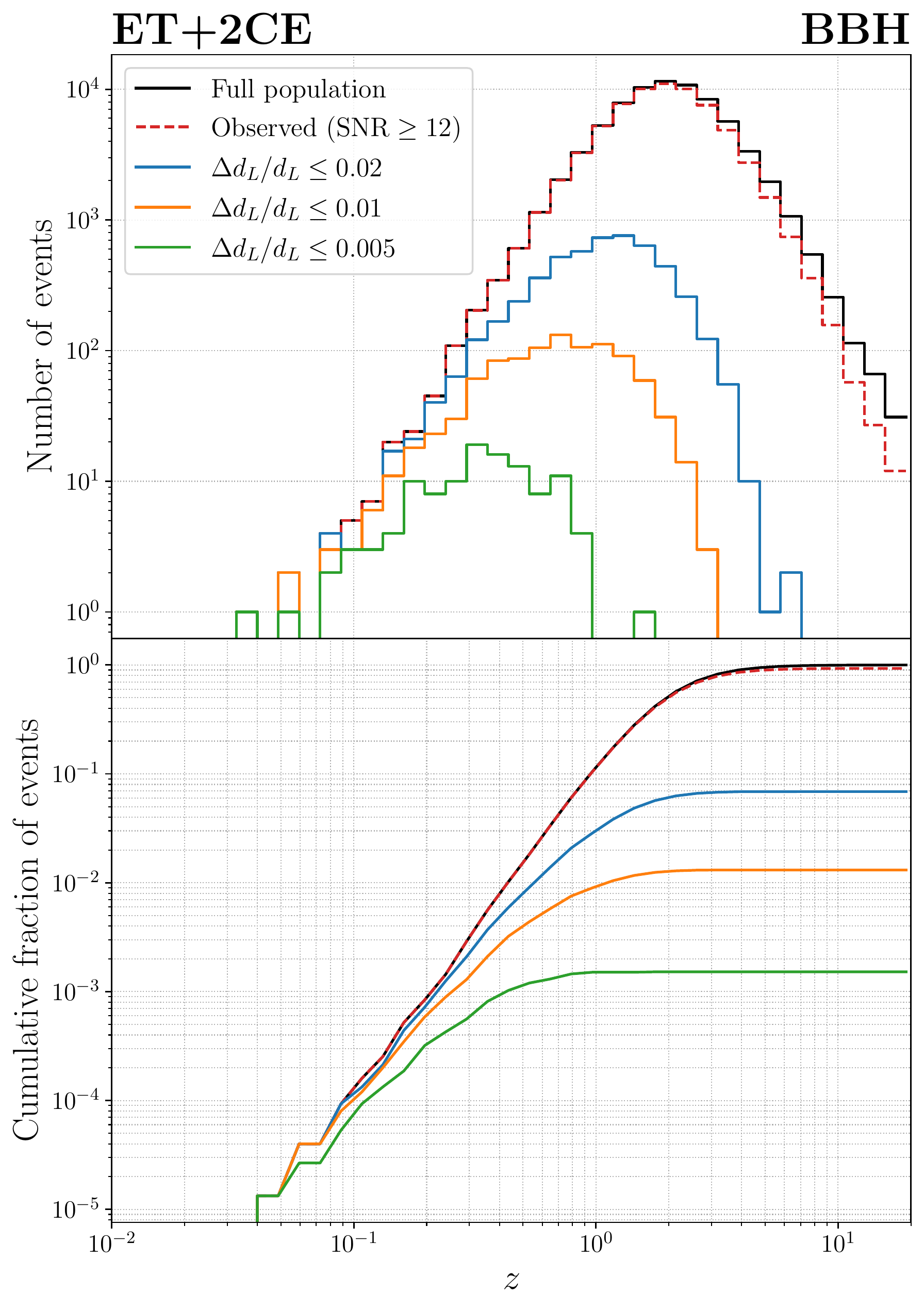} & \includegraphics[width=64mm]{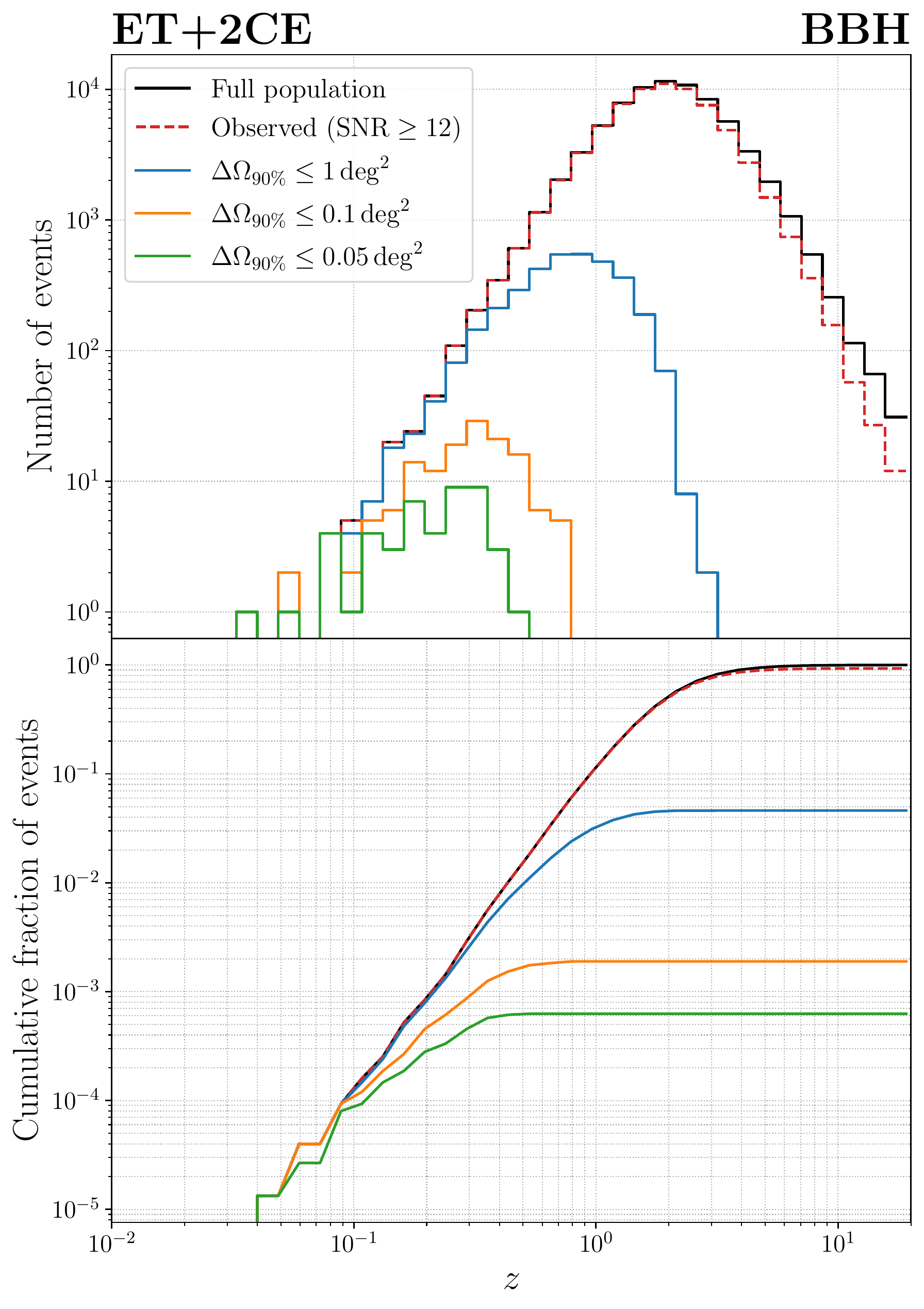}
\end{tabular}
    \caption{As in \autoref{fig:ET_BBH_zhists}, for ET+2CE.}
    \label{fig:ET2CE_BBH_zhists}
\end{figure}

Another crucial information, for understanding the scientific potential of 3G detectors, is provided by the redshift distribution of the detected events, for different values of the threshold in SNR, or for different selection cuts on the accuracy that can be obtained on some especially important parameters. In the left panel of \autoref{fig:ET_BBH_zhists} we show the redshift distributions of the detected BBHs  at ET alone, selected on the basis of the SNR, and we compare it to the redshift distribution of the total population. Beside showing the distribution of detected events, with detection defined from the criterion ${\rm SNR}\geq12$, we also show the distribution in redshift of the events that satisfy a less stringent cut on the SNR, that we take for definiteness ${\rm SNR}\geq8$, as well as the events that pass some very high thresholds on the SNR, while in the central and right panels we show the redshift distribution of the observed events (defined from the condition ${\rm SNR}\geq12$), which furthermore pass stringent cuts  on the error on the luminosity distance (central panel) or sky localization (right panel). In this figure, the upper panels give the fraction of events per redshift bin, and the lower panels give the corresponding cumulative distributions.
A lower threshold on the SNR can be useful in multi--messenger observations. In general, BBHs are not expected to have an electromagnetic counterpart; however, should a counterpart be present for any of those events, the temporal association between the gravitational and electromagnetic signals could be used to lower the SNR, while maintaining  a high statistically significance.\footnote{BBH systems could trigger detectable EM emissions if their merger happen in dense environments \citep{Perna:2016jqh,Mink:2017npg}, but these are expected to be more difficult to detect, see \cite{Palmese:2021wcv} and references therein for a recent discussion. For instance,
a mechanism  that has been investigated, for producing an electromagnetic counterpart to a BBH,  is that two BHs could merge in the accretion disk of an AGN. The final BH then receives a kick  and moves at high speed through the accretion disk, emitting a flare~\citep{McKernan:2019hqs}.
In \cite{Graham:2020gwr} has been proposed that the 
flare ZTF19abanrhr, detected by the Zwicky Transient Facility, could indeed be associated to the event GW190521, described in ~\cite{Abbott:2020tfl,Abbott:2020mjq}. See also 
\cite{Mastrogiovanni:2020mvm,Finke:2021aom} for the possibility of testing modified GW propagation with the association of GW190521 with ZTF19abanrhr, which could also remove possible objections to the association, related to the difference in their luminosity distance.}  Another reason for using a lower SNR, specific to ET, is that, as we have discussed in \autoref{sect:waveforms}, the three--arms configuration of ET allows us to construct a null stream, where the GW signal cancels. This can provide a powerful veto against glitches and other non--Gaussian noise  \citep{Goncharov:2022_NullSt}, that allows us to dig more confidently into events with a lower  SNR. However,
one must keep in mind that the Fisher matrix analysis becomes less reliable for such low values of the SNR.
On the opposite side, events with very large SNR, such as ${\rm SNR}\geq100$, or events with especially good accuracy on parameter reconstruction, are obviously interesting for precision physics, such as tests of GR or  measurements of $H_0$. In 
\autoref{sect:golden}
we will discuss   these `golden binaries' further.   For the moment, we just observe from \autoref{fig:ET_BBH_zhists} that, with the detection criterion ${\rm SNR}\geq12$, ET alone can detect basically $100\%$ of the population up to $z\sim 1$, and about $67\%$ of the BBH population out to $z\sim 20$; furthermore, about $1\%$ of the population up to $z\sim 10$ is detected with ${\rm SNR}\geq100$.
\autoref{fig:ET2CE_BBH_zhists} shows the same results for the ET+2CE network. In this case, basically $100\%$ of the BBH population out to $z\simeq2$ is detected, and $93\%$  up to $z\sim 20$.\footnote{Individual BBHs can be detected to much higher redshifts. In particular, observing a BBH at  $z\gtrsim 30$ , with sufficient accuracy on the redshift reconstruction, would allow us to establish with good confidence that the event is due to a BBH with a primordial, rather than  astrophysical, origin
\citep{DeLuca:2021hde,DeLuca:2021wjr,Ng:2021sqn,Ng:2022agi}.}
The central and right panels show the redshift distribution of events that pass the detection cut ${\rm SNR}\geq12$, and furthermore satisfy more stringent requirement on the accuracy that can be obtained on $d_L$ or on sky localization. We will further comment on them in \autoref{sect:golden}.

\begin{figure}
    \hspace{-2cm}
    \begin{tabular}{c@{\hskip -3mm}c@{\hskip -4mm}c}
  \includegraphics[width=70mm]{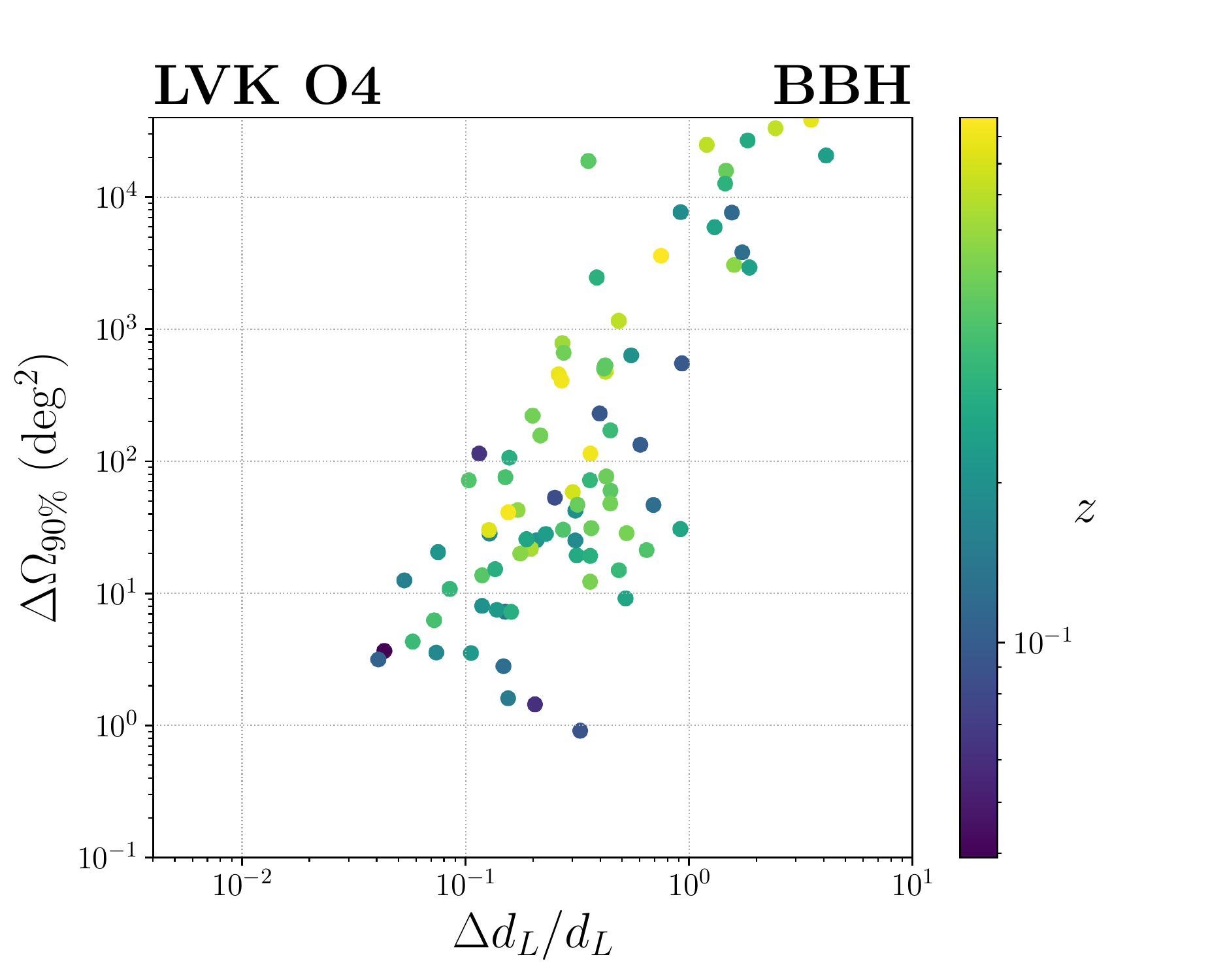} &   \includegraphics[width=70mm]{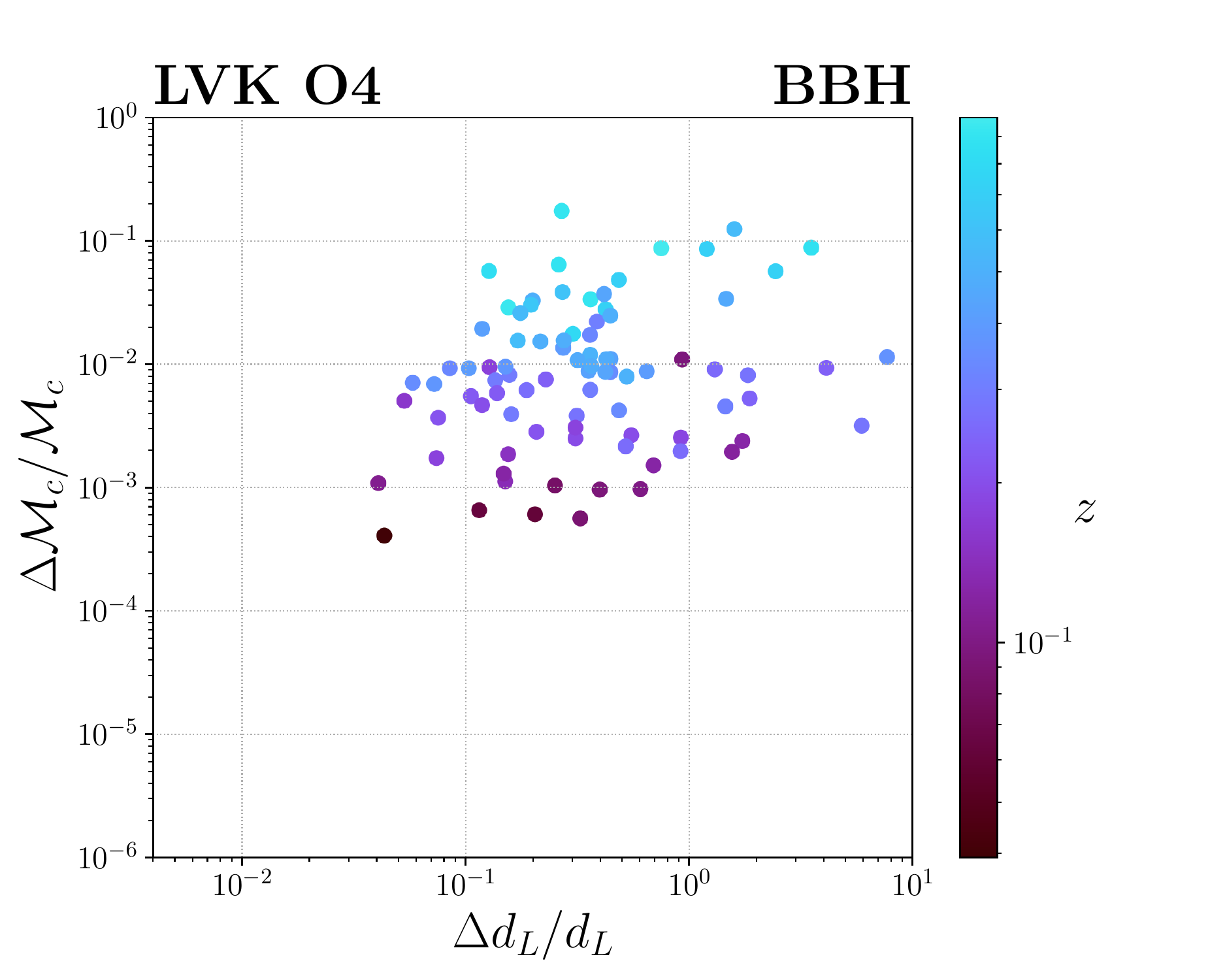} & \includegraphics[width=70mm]{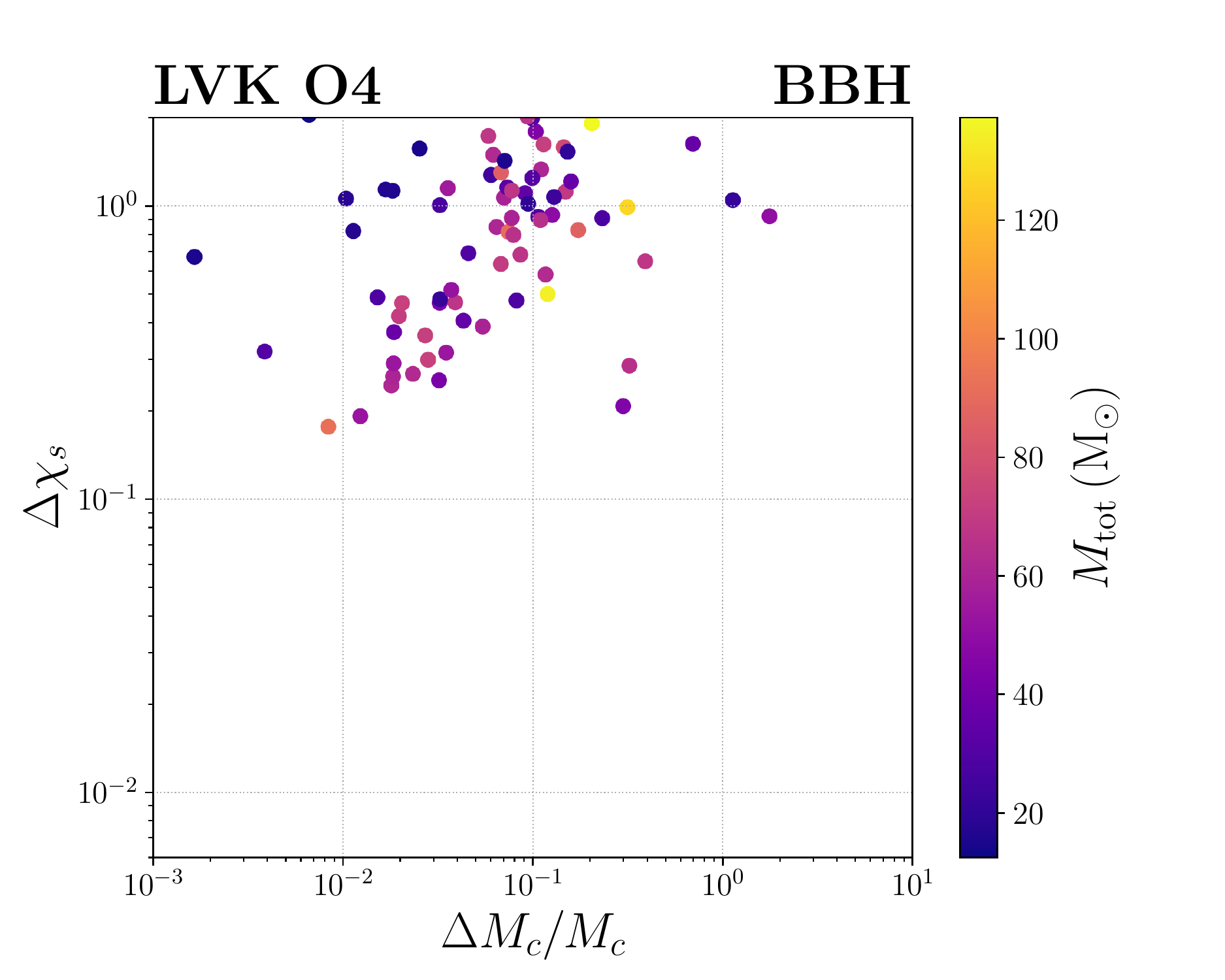}\\[-.5cm]
  \includegraphics[width=70mm]{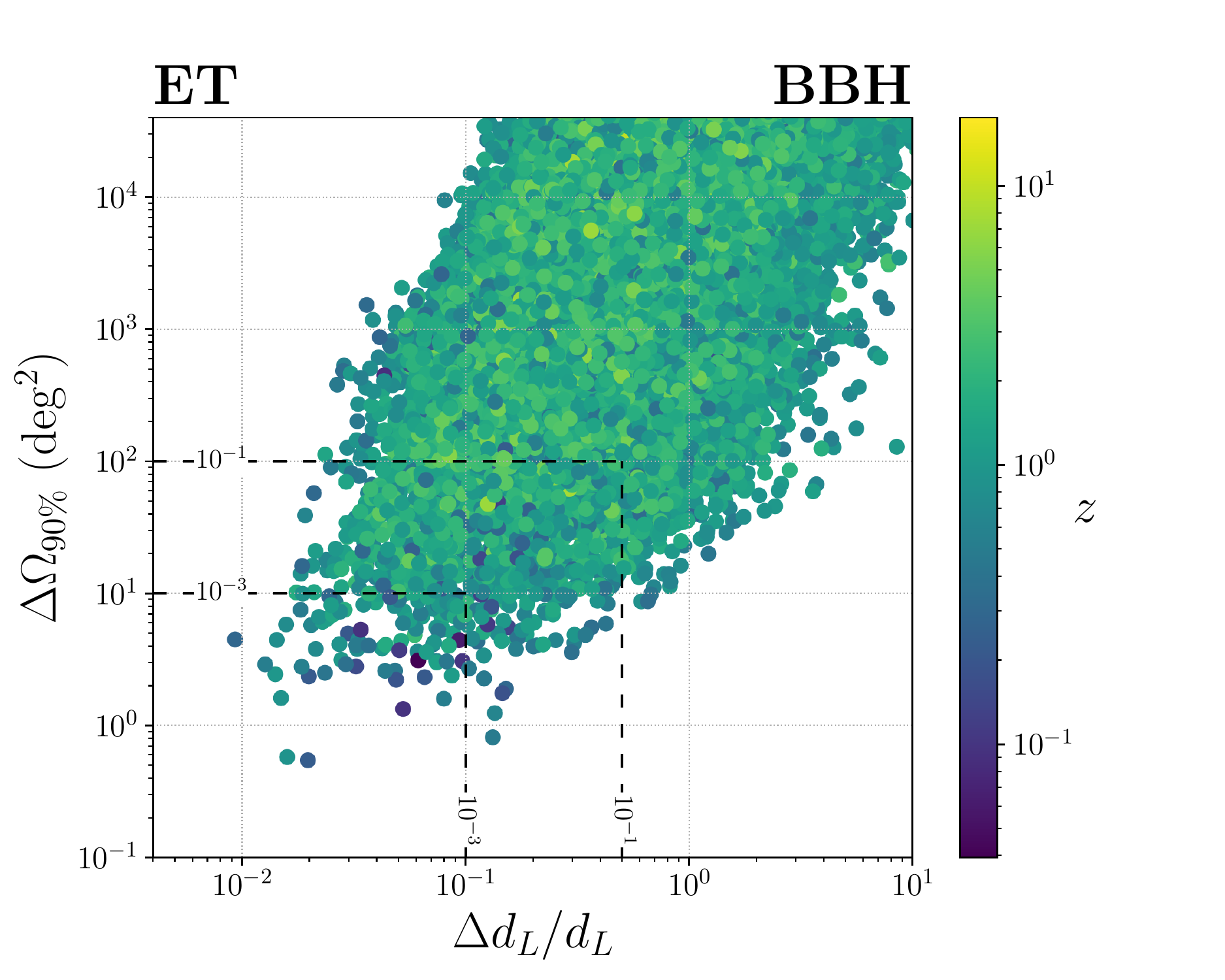} &   \includegraphics[width=70mm]{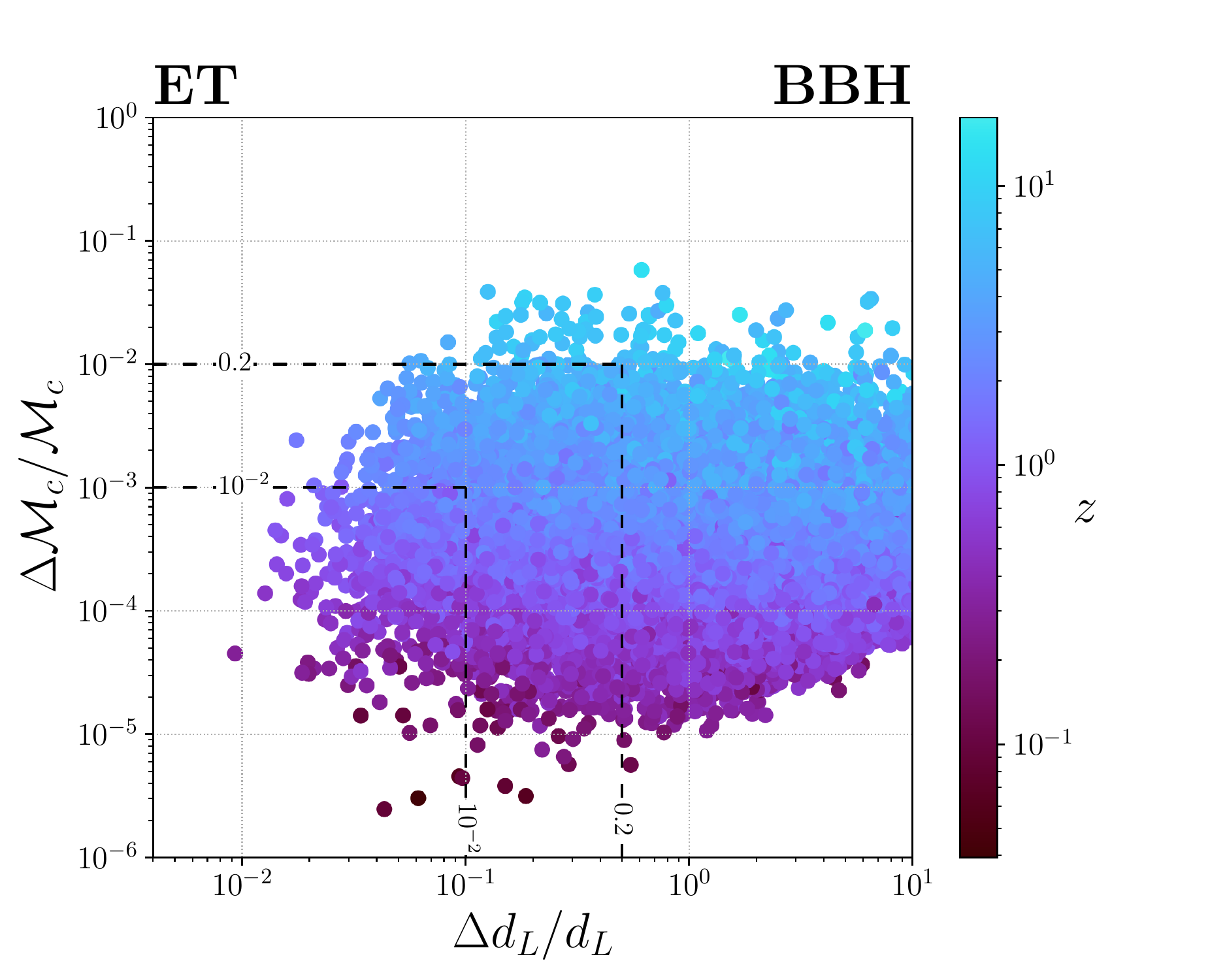} & \includegraphics[width=70mm]{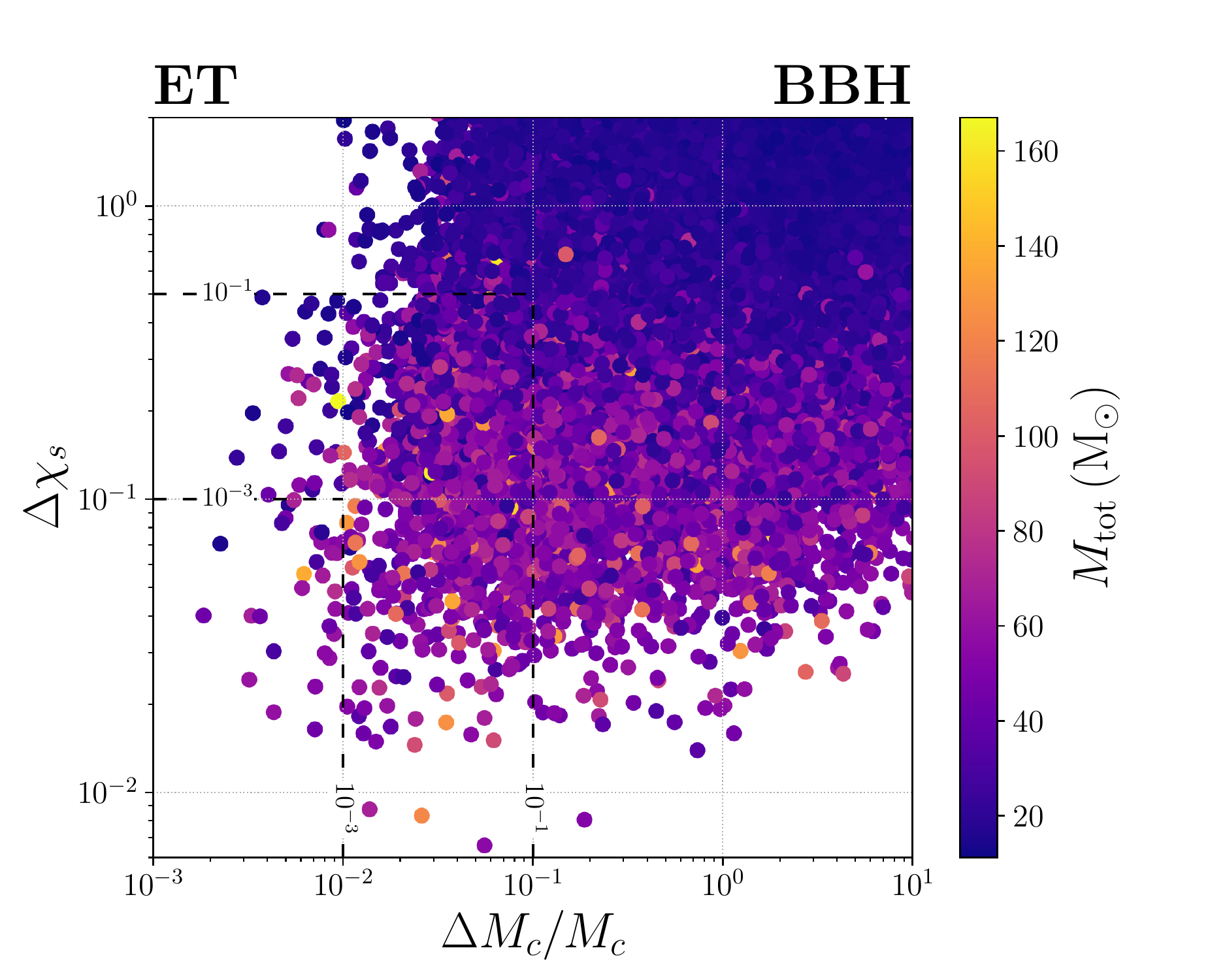}\\[-.5cm]
  \includegraphics[width=70mm]{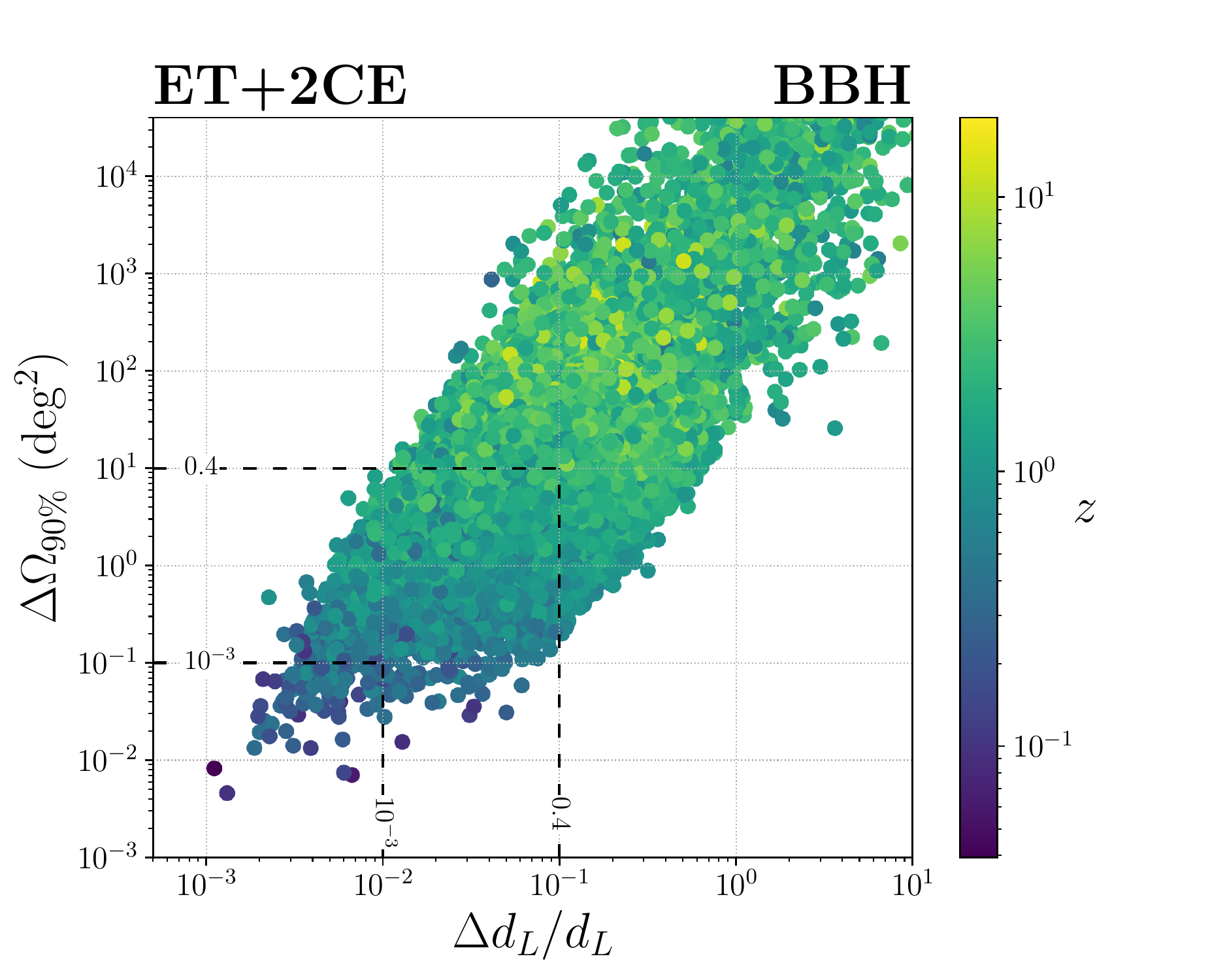} & \includegraphics[width=70mm]{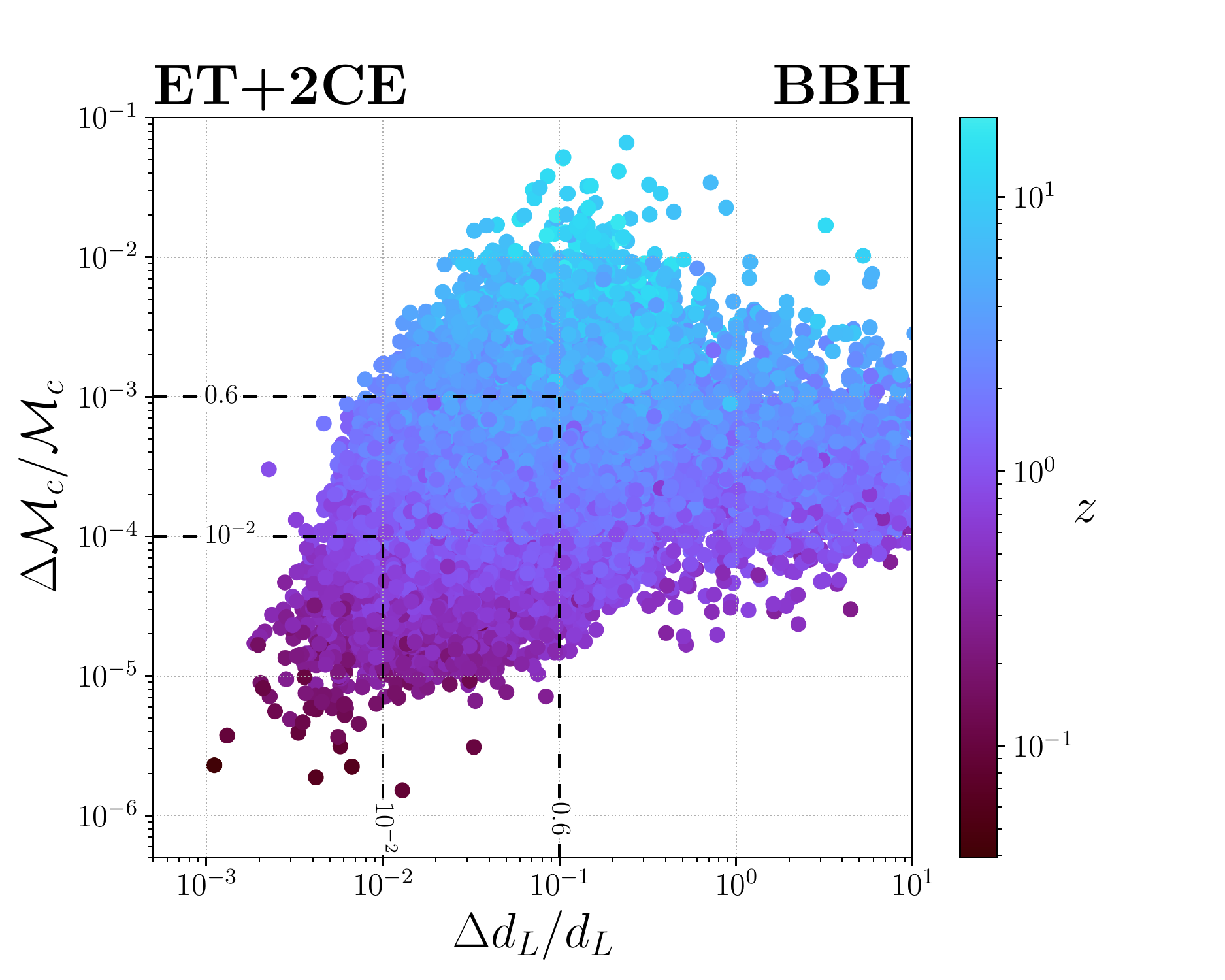} & \includegraphics[width=70mm]{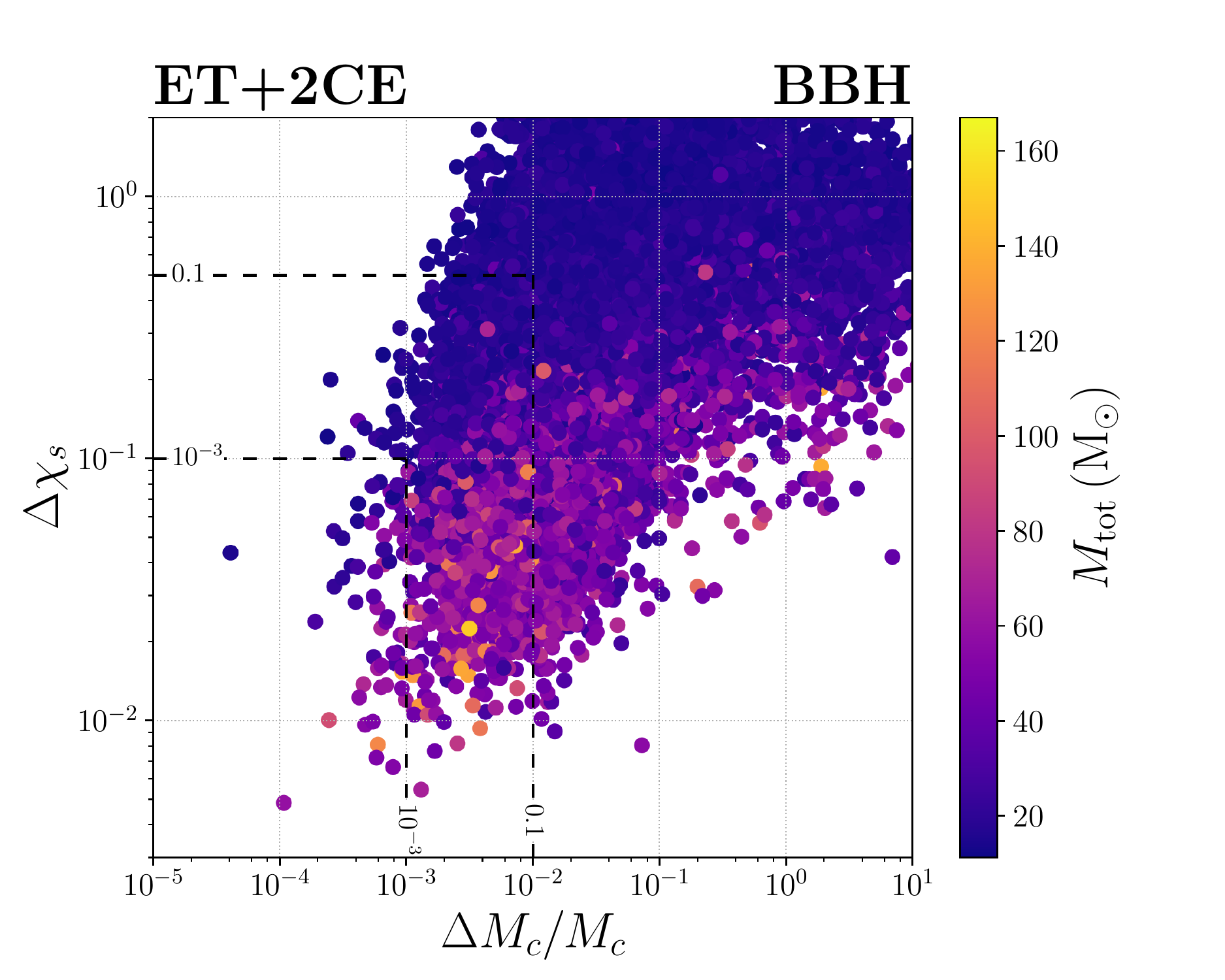}
\end{tabular}
    \caption{Scatter plots of the observed BBH population at the various networks. In the first row we show the results for the LVK network in O4, in the middle for ET alone and in the bottom for the ET + 2CE network. In each row, the left panel shows the distribution of errors on the luminosity distance and sky location, with the color code representing the  redshift, the central panel the distribution of errors on the luminosity distance and detector--frame chirp mass, again  with the color code representing the  redshift, and the right panel the distribution of errors on the source--frame chirp mass and symmetric spin $\chi_s = (\chi_1 + \chi_2)/2$, with the color code  representing the total source--frame mass. The numbers reported on the dashed lines refer to the fraction of observed events lying inside the corresponding region.}
    \label{fig:BBH_scatter}
\end{figure}

In \autoref{fig:BBH_scatter} we present scatter plots  showing the correlations of the errors on quantities that are relevant in cosmological and astrophysical applications of GWs. In particular, in the left panels we show the scatter of detections in the plane $(\Delta\Omega_{90\%},\,\Delta d_L/d_L)$
(with a color scale giving information on the redshift). We see that, for an ET + 2CE network, about 36\%  of the events will have a localization region smaller than \SI{10}{\square\degfull}, while also having an accuracy on the luminosity distance $\Delta d_L/d_L\lesssim 10^{-1}$, even at $z\gtrsim1$.
This would have an important impact on late--time cosmology, in particular for the possibility of determining the Hubble constant and constraining the phenomenon of ``modified GW propagation'' \citep{Belgacem:2017ihm, Belgacem:2018lbp, Belgacem:2019pkk} by correlating GWs and galaxy catalogs [see \cite{Finke:2021aom, LIGOScientific:2021aug, Palmese:2021mjm} for the most recent applications].
A small fraction of the events, of order \num{e-3}  (which still, for ET+2CE, and with our assumptions on  the population and the local rate, corresponds to about 90  events per year), could even have a localization region $\Delta\Omega_{90\%}\lesssim\SI{e-1}{\square\degfull}$ and $\Delta d_L/d_L\lesssim\SI{e-2}{}$, which could allow a unique host galaxy identification even in absence of a direct EM counterpart \citep{Borhanian:2020vyr}, either from a galaxy catalog or from a dedicated follow--up.

Very significant results  are obtained also with ET alone, despite the limitations in the localization with a single detector, with a fraction of events of order $ 10^{-3}$  (which, in the case of ET alone, and with our reference value for the local rate, corresponds to about 90 events per year) having at the same time a sky localization  better than $\SI{10}{\square\degfull}$ and an error on the luminosity distance $\Delta d_L/d_L\lesssim\num{e-1}$. As we mentioned above, ET alone will see  about $10^3$ BBH events with $\Delta d_L/d_L\lesssim\num{e-1}$ and about $160$ BBH events with  $\Delta\Omega_{90\%}\leq \SI{10}{\square\degfull}$. We see that, among these, about $90$ satisfy both criteria, $\Delta d_L/d_L\lesssim\num{e-1}$ and  $\Delta\Omega_{90\%}\leq \SI{10}{\square\degfull}$, and will therefore be particularly useful for cosmological studies.

For the current network of detectors in the O4 run we find that it should observe  $\sim 5$ BBHs per year with localization $\Delta\Omega_{90\%}\lesssim\SI{10}{\square\degfull}$ and $\Delta d_L/d_L\lesssim\num{e-1}$, including some sources at $z\gtrsim 0.3-0.4$. This would allow substantial improvements in the constraints on the Hubble constant within $\sim 2$ years of data taking, using the correlation  with galaxy catalogs.
By comparison, in the GWTC--2 and  GWTC--3 catalogs, which include the detections up to the O3 run, the best localized BBH event, GW190814, has a sky localization of $\SI{19}{\square\degfull}$, and only two more BBH events, GW190412 ($\SI{21}{\square\degfull}$) and GW200208$\_$130117 ($\SI{30}{\square\degfull}$) have  $\Delta\Omega_{90\%}\leq \SI{30}{\square\degfull}$~\citep{LIGOScientific:2020ibl,LIGOScientific:2021djp}.

In the central panels of \autoref{fig:BBH_scatter} we show our forecasts for the joint errors on the detector--frame chirp mass and  the luminosity distance, again with a color scale giving information on the redshift. An accurate measurement of detector--frame masses, together with information on the distribution of source--frame masses, can be used to break the mass--redshift degeneracy in the GW waveform and constrain 
the underlying cosmological model together with the compact binary population \citep{Farr:2019twy, Ezquiaga:2021ayr, Mastrogiovanni:2021wsd, LIGOScientific:2021aug, Mancarella:2021ecn, Leyde:2022orh, Karathanasis:2022rtr}. 

Finally, in the right column of \autoref{fig:BBH_scatter}, we show the expected errors attainable on the source--frame chirp mass and symmetric spin parameter, 
$\chi_s = (\chi_1 + \chi_2)/2$,
with a color code that provides information on the total source--frame mass. These three  parameters are of fundamental importance to characterise the astrophysical BBH population \citep{LIGOScientific:2021psn} and also to disentangle it from a possible primordial component \citep{DeLuca:2020qqa}. Again, for both combinations of parameters, these results are very promising, showing the potential of the next generation of GW detectors.

For ease of readability, some  selected results presented in this section are summarised in \autoref{tab:BBH_Summary}.

\begin{table}[htp!]
    \centering\hspace{-2.5cm}
    \begin{tabular}{!{\vrule width .09em}c|c|c||c|c|c!{\vrule width .09em}}
    \toprule\midrule
    \multicolumn{6}{!{\vrule width .09em}c!{\vrule width .09em}}{\bf BBH}\\
    \midrule\midrule
    Network & Detected & Analysed & $\rm SNR\geq 100$ & $\Delta d_L/d_L \leq 10\%$ & $\Delta\Omega_{90\%}\leq\SI{10}{\square\degfull}$\\
    \midrule\midrule
    \textbf{LVK--O4} & 86 & 86 & 0 & 8 & 15\\
    \midrule
    \textbf{ET} & 50607 & 48456 & 885 & 1120 & 161\\
    \midrule
    \textbf{ET+2CE} & 69799 & 69610 & 5384 & 45331 & 30889 \\
    \midrule\bottomrule
    \end{tabular}
    \caption{A selection of  results from the analysis of the \num{7.5e4} BBHs (corresponding to the full population in about \SI{1}{\year} with our choices for the parameters) at the considered networks. Here, and in similar tables below, the column labeled ``Analysed'' reports the number of events that passed our detection threshold (${\rm SNR}\geq 12$) and for which, furthermore, we could get a reliable inversion of the Fisher matrix, according to our criterion explained in \autoref{sec:singularities}.}
    \label{tab:BBH_Summary}
\end{table}
\newpage
\subsection{Binary neutron stars}\label{sect:resBNS}

We next present our results for the population of BNS systems. The  current best estimates for the local merger rate, ${\cal R}_{0, \rm BNS}$, is obtained  from the GWTC--3 catalog \citep{LIGOScientific:2021psn}, but still suffers from a large uncertainty.  Using a rather specific set of assumptions (in particular, a mass distribution for NS in binaries flat between $1\si{\Msun}$ and $2.5\si{\Msun}$), \cite{LIGOScientific:2021psn} finds
${\cal R}_{0, \rm BNS} = 105.5^{+190.2}_{-83.0}~\si{\per\cubic\giga\parsec\per\year}$, consistent with the value ${\cal R}_{0, \rm BNS} = 320^{+490}_{-240}~\si{\per\cubic\giga\parsec\per\year}$ inferred from the GWTC--2 catalog using the same mass distribution~\citep{LIGOScientific:2020kqk}, but still with a median value lower by a factor $\sim 3$. However, these numbers are quite model dependent. Three different models have been investigated in \citep{LIGOScientific:2021psn} and, taking the lowest $5\%$ and highest $95\%$ credible interval out of all three models, ${\cal R}_{0, \rm BNS}$ is inferred to be in the range $(10-1700)~\si{\per\cubic\giga\parsec\per\year}$.
In the following, we will use as reference value ${\cal R}_{0, \rm BNS} = \SI{105.5}{\per\cubic\giga\parsec\per\year}$, which is the median value obtained for the flat mass distribution that we are using, but it is important to keep in mind  the uncertainty on this number, which could be up to a factor $\sim 10$ lower, or up to a factor $\sim 16$  higher.
We then simulate a population of $N_{\rm BNS}=\num{e5}$ BNS out to $z=20$, which,  for ${\cal R}_{0, \rm BNS} = \SI{105.5}{\per\cubic\giga\parsec\per\year}$, and our assumed redshift dependence for the merger rate, corresponds to the number of coalescences in about  one year. 

\begin{figure}
    \centering
    \includegraphics[width=.9\textwidth]{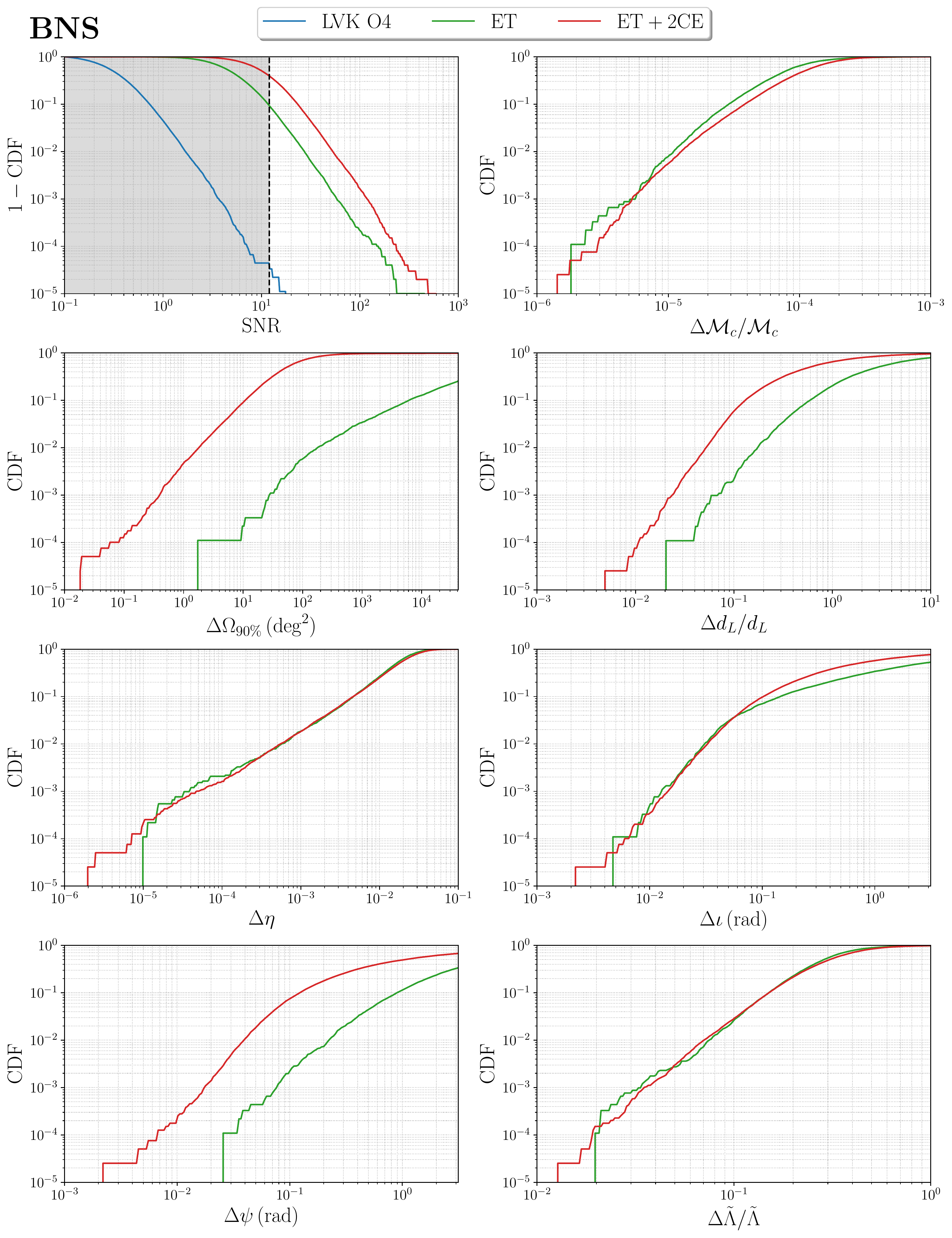}
    \caption{Cumulative distributions of the event fraction for BNS signals as function of SNR (for LVK--O4,  ET and ET+2CE) and as a function of parameter errors (for ET and ET+2CE), using the waveform model \texttt{IMRPhenomD\_NRTidalv2}.}
    \label{fig:BNS_cumuldist}
\end{figure}

\begin{figure}
    \centering
    \includegraphics[width=.9\textwidth]{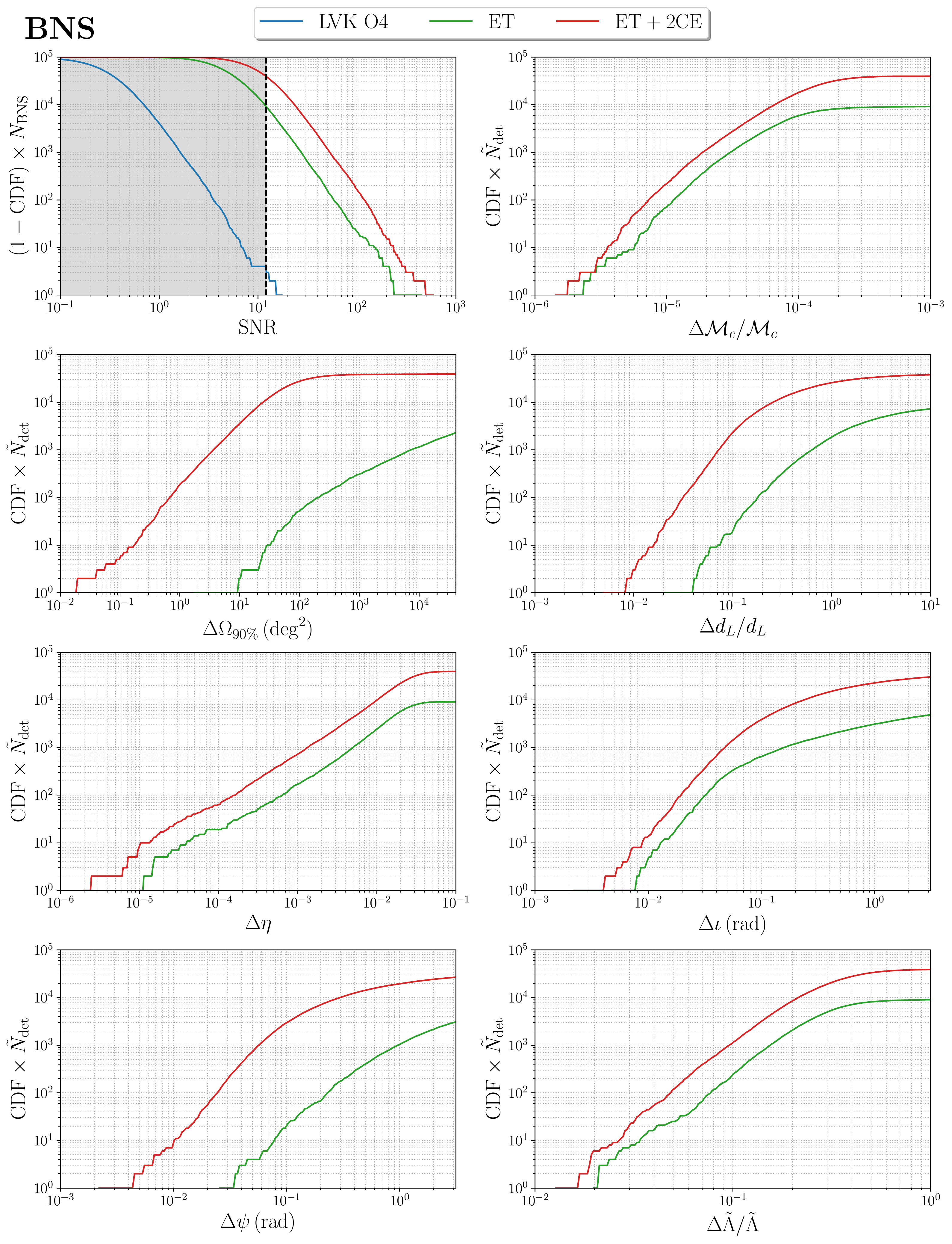}
    \caption{As in \autoref{fig:BNS_cumuldist}, in terms of the total number of events rather than detection fraction, using ${\cal R}_{0, \rm BNS} = 105.5\,  {\rm Gpc}^{-3}{\rm yr}^{-1}$. Except for the panel showing  the distribution of the SNR, only the events whose Fisher matrix could be reliably inverted are included.}
    \label{fig:BNS_cumuldistNdet}
\end{figure}

For this value of the local rate, for the current LVK network of detectors during the O4 run and accounting for the duty cycle, we find that only about 4 events per year should detected,  consistently with the forecast in \cite{AbbottLivingRevGWobs}. Even assuming the current detectors to be operational 100\% of the time during O4, the number of detections just raises to 5. Given the smallness of this sample, for LVK--O4 we do not plot the results for the parameter estimation of the corresponding events, since they would strongly depend  on the particular random realization considered. Therefore, except for the cumulative distribution of the SNR, we only show the results for ET and for ET+2CE.

\begin{figure}[t]
    \hspace{-2cm}
    \begin{tabular}{c@{\hskip 3mm}c@{\hskip 3mm}c}
  \includegraphics[width=64mm]{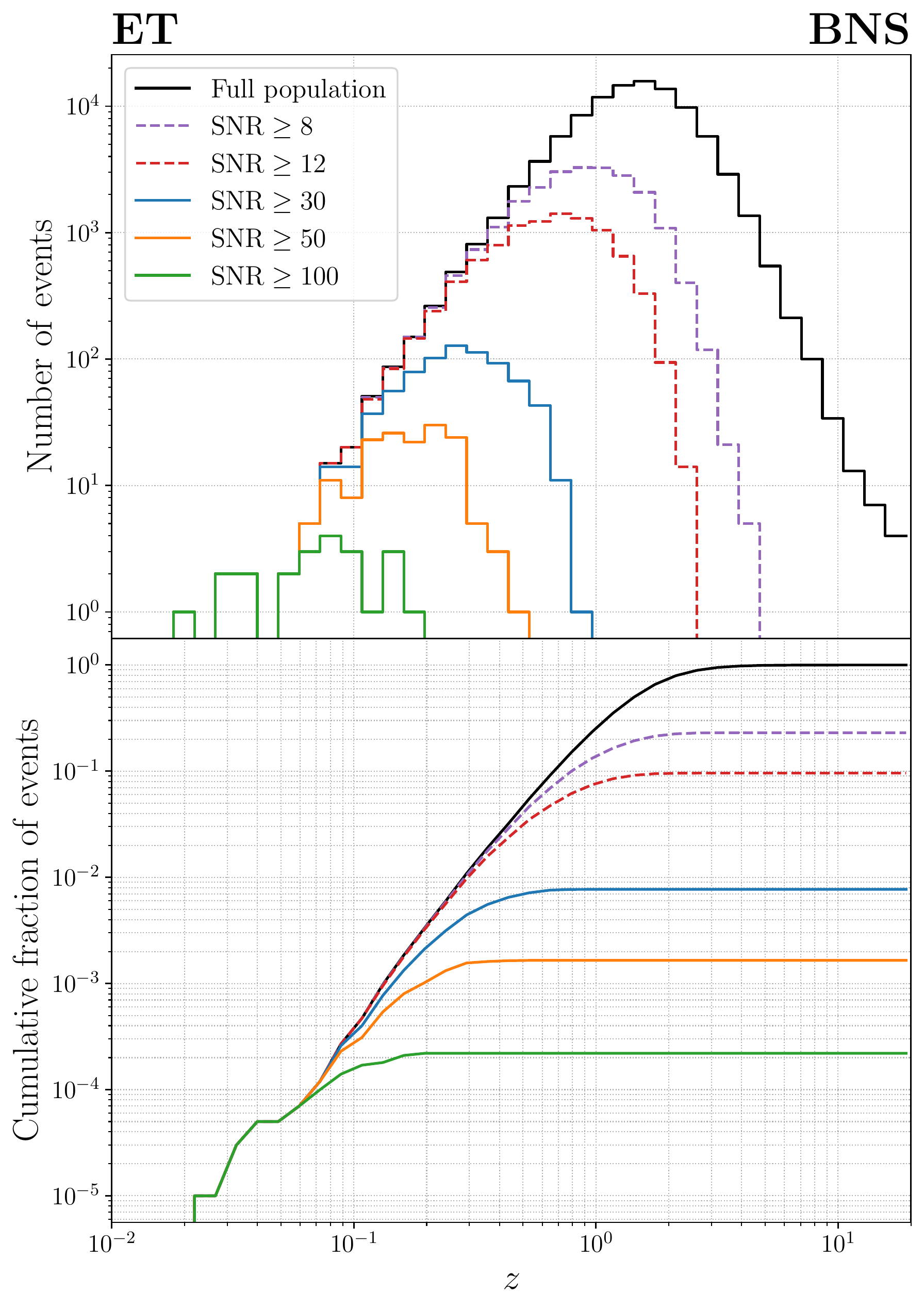} &   \includegraphics[width=64mm]{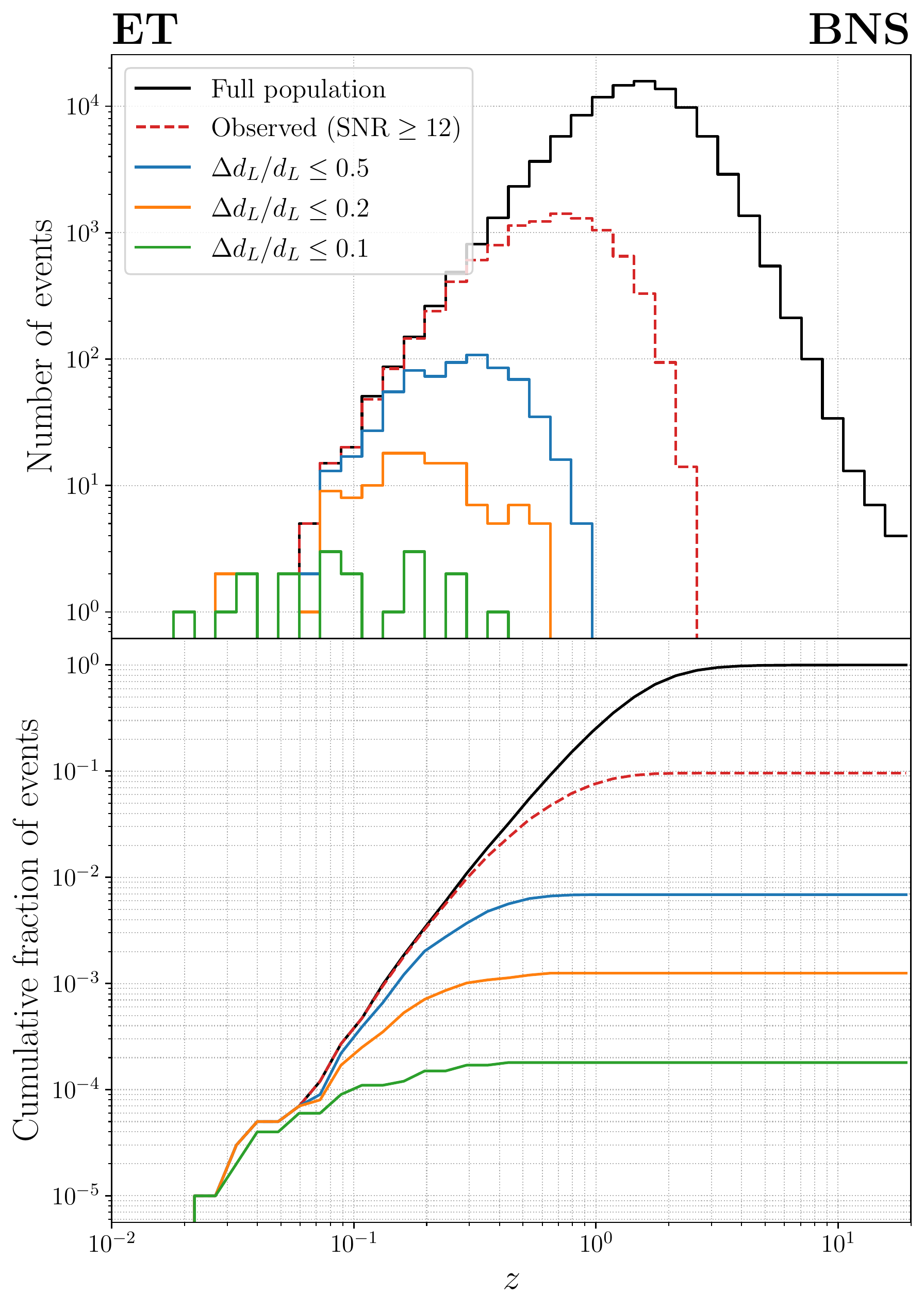} & \includegraphics[width=64mm]{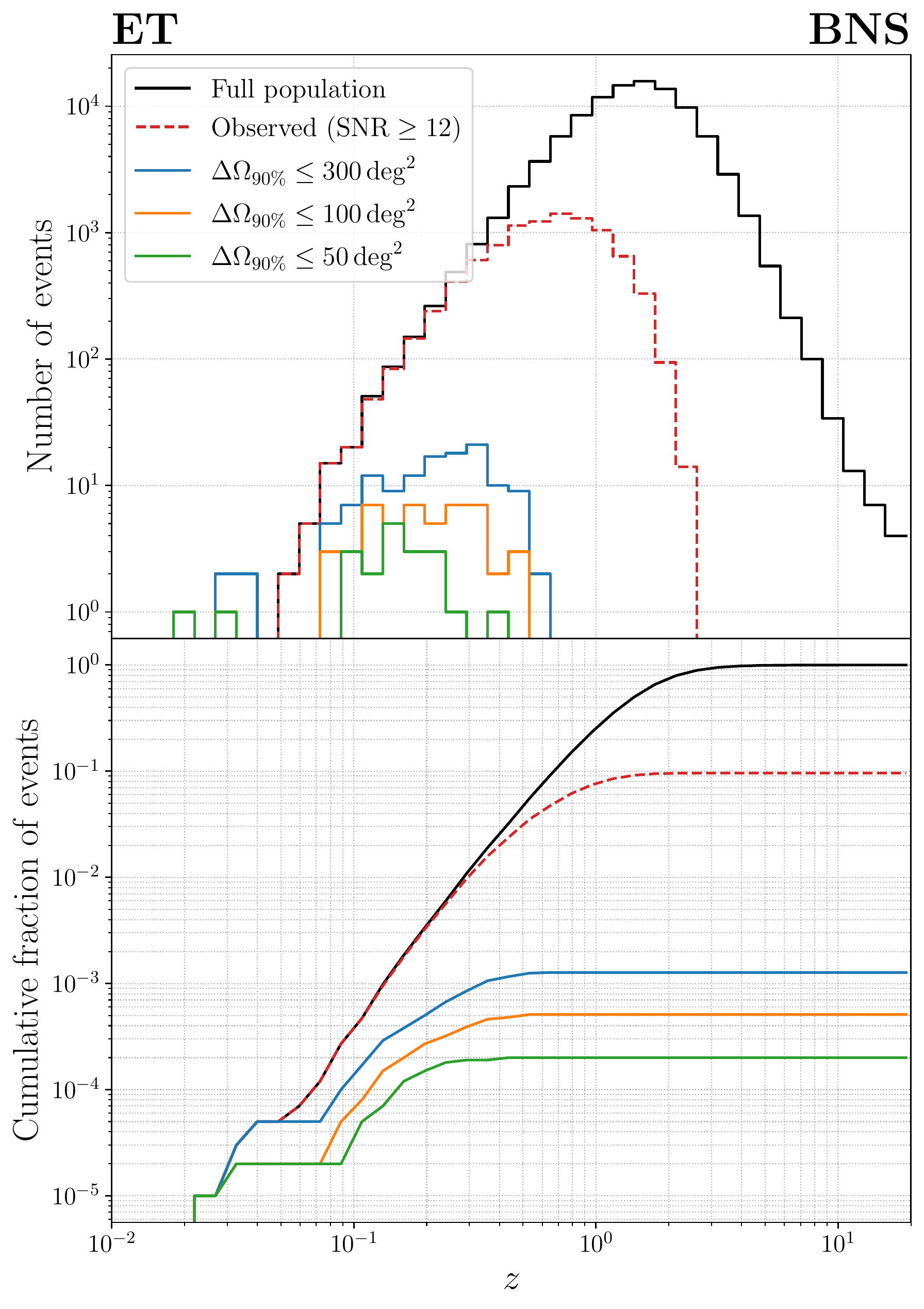}
\end{tabular}
    \caption{Redshift distributions of the BNSs observed  at ET alone, selected on the basis of different thresholds for the SNR (left panel), or setting ${\rm SNR}\geq 12$ and applying further cuts on $\Delta d_L/d_L$ (central panel), or  on  $\Delta\Omega_{90\%}$ (right panel). The black solid line corresponds to the total BNS population in the astrophysical model that we have assumed.
    We set  ${\cal R}_{0, {\rm BNS}}=\SI{105.5}{\per\cubic\giga\parsec\per\year}$, and show the results for the detections in one year  (taking into account our assumptions of the duty cycle).  Given the uncertainly on ${\cal R}_{0, {\rm BNS}}$, which currently can be in the range $(10-1700)~\si{\per\cubic\giga\parsec\per\year}$, one should keep in mind that the absolute number can still change by a factor $\order{10}$ or more. In each column, the upper panel shows the number of events per redshift bin, while the lower panel shows the corresponding cumulative distributions, normalized to the number of BNS events in our sample, $N_{\rm BNS}=\num{e5}$.}
    \label{fig:ET_BNS_zhists}
\end{figure}

\begin{figure}[t]
    \hspace{-2cm}
    \begin{tabular}{c@{\hskip 3mm}c@{\hskip 3mm}c}
  \includegraphics[width=64mm]{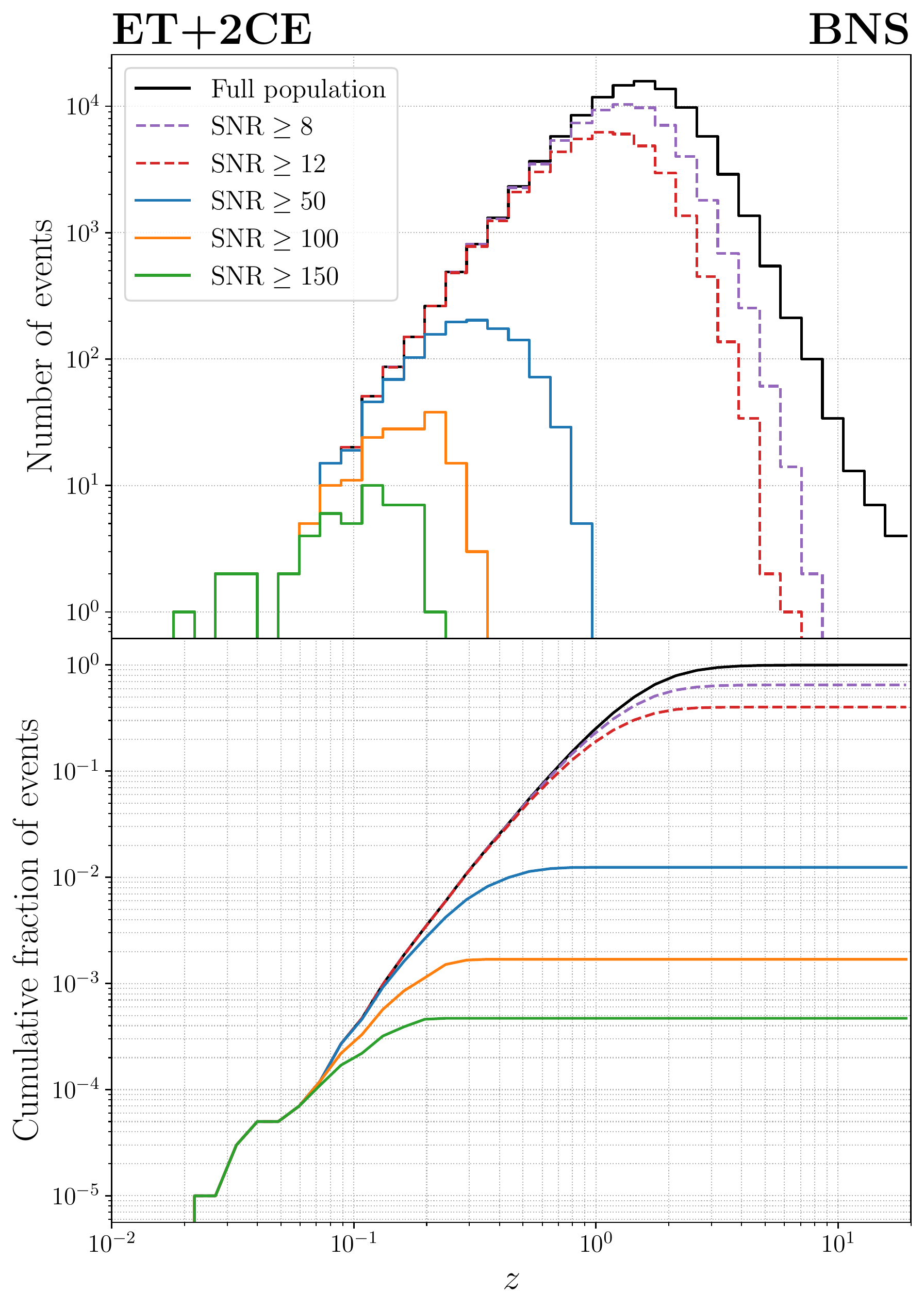} &   \includegraphics[width=64mm]{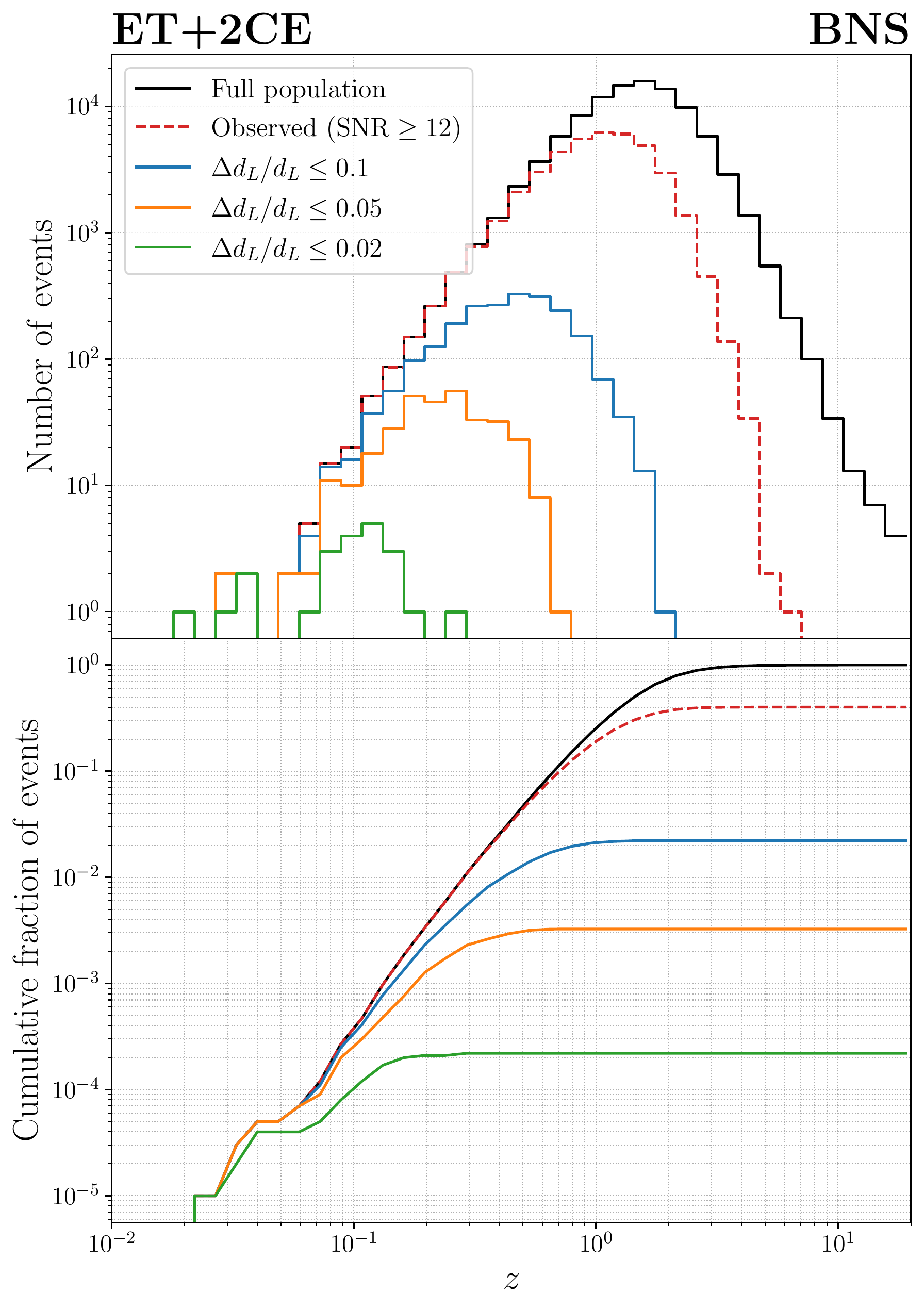} & \includegraphics[width=64mm]{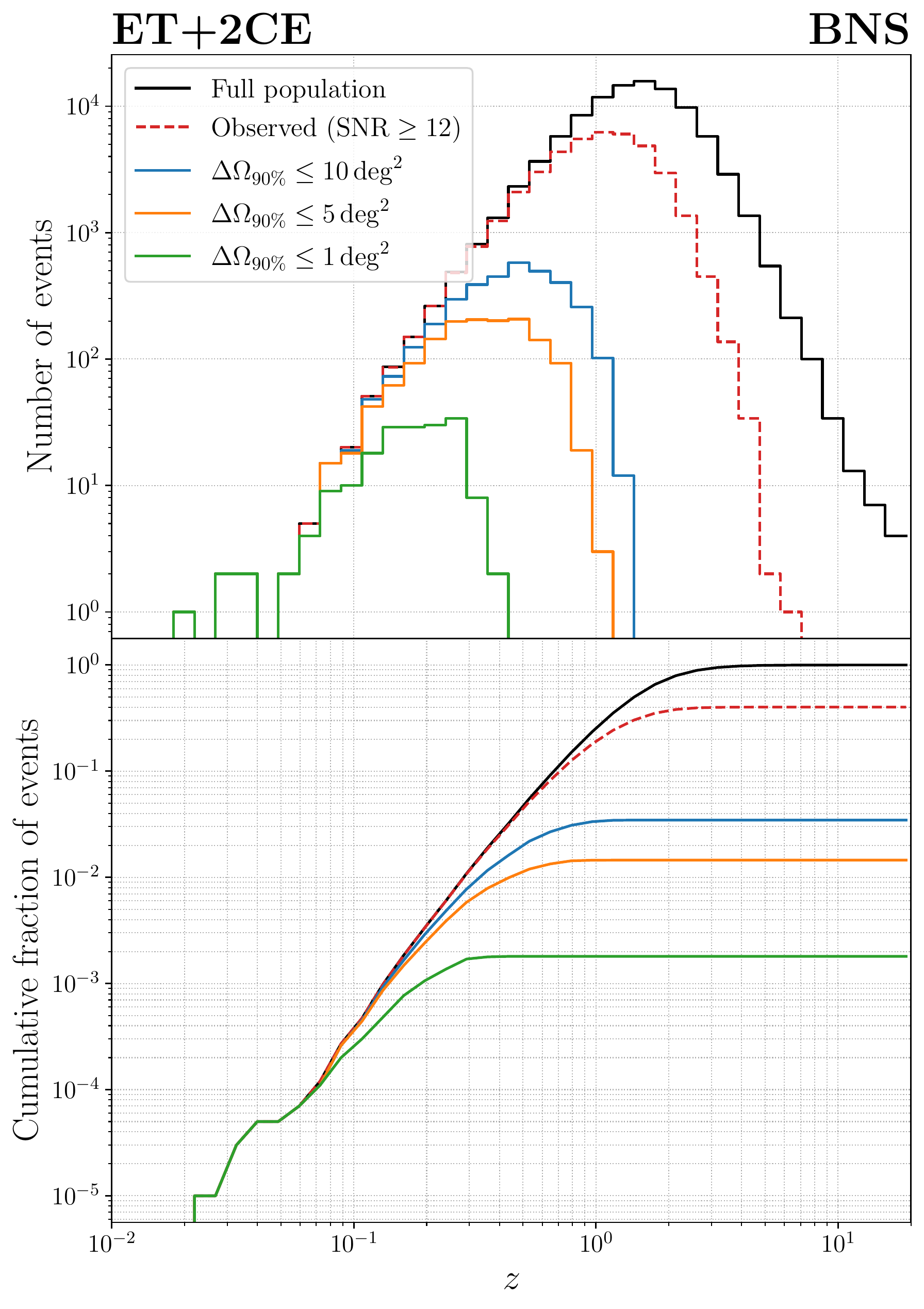}
\end{tabular}
    \caption{As in \autoref{fig:ET_BNS_zhists}, for ET+2CE.}
    \label{fig:ET2CE_BNS_zhists}
\end{figure}

For 3G detectors,  with our threshold ${\rm SNR}\geq 12$ we find   $N_{\rm det}\simeq 
\num{9.6e3}$ detections per year for ET alone (corresponding to $9.6\%$ of the population) and 
$N_{\rm det}\simeq \num{4.0e4}$ detections per year for ET + 2CE (corresponding to $40.1\%$ of the population).
Varying ${\cal R}_{0, \rm BNS}$ in the currently allowed range $(10-1700)\,  {\rm Gpc}^{-3}{\rm yr}^{-1}$, while keeping fixed the redshift distribution of our population model, the number of detections per year at LVK--O4 ranges between zero and 65, for
ET varies in the range $[\num{9.1e2}, \num{1.6e5}]$, and for
ET+2CE varies in the range $[\num{3.8e3}, \num{6.5e5}]$. When performing parameter estimation, we will restrict again to the $\tilde{N}_{\rm det}$ detected events that also pass our criterion for having a reliable inversion of the Fisher matrix. We find $\tilde{N}_{\rm det}\simeq\num{9.2e3}$ for ET (corresponding to about $96\%$ of the detected events), and
$\tilde{N}_{\rm det}\simeq \num{4.0e4}$ (corresponding to about $99\%$  of the detected events) for ET+2CE.

Our results for the SNR and for parameter estimation are shown in
\autoref{fig:BNS_cumuldist} (where  we show the detection fractions)  and in \autoref{fig:BNS_cumuldistNdet} (where, for parameter estimation, we multiply the detection fractions at the different detector networks  by the respective values of $\tilde{N}_{\rm det}$, and for the SNR we multiply by $N_{\rm BNS}$), while in \autoref{fig:ET_BNS_zhists} we show the redshift dependence of various distributions at ET, similarly to \autoref{fig:ET_BBH_zhists} for the BBH case; in 
\autoref{fig:ET2CE_BNS_zhists} we show the corresponding results for ET+2CE.

Compared to BBHs, the smaller mass of BNSs allows them to stay longer in the detector band, but makes them less loud, thus more difficult to see, especially at high redshifts.  
As we discussed in \autoref{sec:detResp}, thanks to the long time spent in the detector bandwidth at ET, the BNS  localization can be improved by exploiting  the rotation of the Earth. From  the panel for  $\Delta\Omega_{90\%}$ in \autoref{fig:BNS_cumuldist} and \ref{fig:BNS_cumuldistNdet} we see that, for BNS, with our fiducial values of the rate and our assumptions on the population, ET as a single detector can reach an error of order $10\, \si{\square\degfull}$ for the very best localized sources, and about $0.6\%$ of the detected BNS will have 
$\Delta\Omega_{90\%} <\SI{e2}{\square\degfull}$.\footnote{Observe that the CDF for  $\Delta\Omega_{90\%}$ in \autoref{fig:BNS_cumuldist}, in the case of ET, does not saturate to one as $\Delta\Omega_{90\%}$ approaches $4\pi\, {\rm sr}\simeq \SI{4.1e4}{\square\degfull}$. This is due to the fact that the Fisher matrix does not know about the physical constraint that $\Delta\Omega$ must be smaller or equal than $4\pi$ and, for poorly constrained events, can return a $90\%$ c.l. value larger than that. Indeed, we see from the plot that, for BNSs at ET alone, only about $25\%$ of the detected events (and that, furthermore, have  passed our criterion on the inversion of the Fisher matrix) satisfy $\Delta\Omega_{90\%}\leq4\pi$. The effect was already present, although less pronounced, for BBHs at ET alone, as we can see from the panel for $\Delta\Omega_{90\%}$ in \autoref{fig:BBH_cumuldist}. Similar considerations hold for $\iota$ and $\psi$, for which the Fisher matrix does not know  the constraints $\Delta\iota\leq \pi $ and $\Delta\psi\leq \pi$, and for the dimensionless spin variables $\chi_i$, that range between $-1$ and $1$, and so should satisfy $\Delta\chi_i\leq 2$. As a consequence, the corresponding CDF, computed with the Fisher matrix, do not  necessarily saturate to one when these boundaries are reached.}
When comparing with other results in the literature, it is important to observe that, given the current large uncertainty on the BNS local rate and population properties, different papers can  use very different assumptions, which can easily lead to results  differing by about one order of magnitude. In particular, in \autoref{app:comparison} we compare with the results found in
\cite{Borhanian:2022czq} and in 
\cite{Ronchini:2022gwk}, where we find that our results are consistent, once the differences in the assumptions  made (all legitimate, within the current uncertainties), are taken into account.

It is interesting to observe  that, despite the improvement brought by the inclusion of the Earth's rotation (which is only relevant for BNSs, given the long duration of their signal)  still, for   ET as a single detector, the  localization accuracy of BNSs is not comparable to that of BBHs,  that was shown in the corresponding panel of \autoref{fig:BBH_cumuldistNdet}. For instance, with  a single ET  detector  a fraction of about \num{3e-3} of the detections (corresponding, for our fiducial value of the local rate and our population model, to $\order{150}$ BBH/yr)  can be localized to better than \SI{10}{\square\degfull}, while, with our assumptions on the BNS local rate and BNS population model, we only find 2 BNS that can be localized to such an accuracy.
This shows that the loudness of typical BBH signals is more important, for sky localization, than the long time spent in the bandwidth by BNS signals.

For a  network ET+2CE, thanks to triangulation, we see again from  \autoref{fig:BNS_cumuldistNdet} that we can reach an accuracy below \SI{e-1}{\square\degfull} for the best localized BNS systems, to be compared with the localization below \SI{e-2}{\square\degfull} that ET+2CE can reach for the best localized BBHs, as we saw in  \autoref{fig:BBH_cumuldistNdet}. 
However, it is important to notice that, for multi--messenger observations of BNSs and for the identification and follow--up of the associated kilonova, already the angular resolution of a single ET detector  can be adequate. For instance, to understand the nucleosynthesis spectra of the kilonova, for sources up to $z\simeq 0.3-0.4$  the best instrument is the Extremely Large Telescope (ELT),\footnote{\url{https://elt.eso.org}. planned to start observations in 2027, and that could still be operational by the time of 3G GW detectors \citep{2008Msngr.133....2S, Rossi:2019fnm}}. However, a direct pointing with  ELT would require arcsec localization, which is anyhow out of question for GW detectors. The actual, strategy for sources at these moderate redshifts, is rather to use telescopes with large field--of--view (FOV): these instruments, with a localization of the GW events of the order of tens to hundreds of square degrees, can indeed localize the source. Notice that such a GW resolution can be given already by ET alone.
In particular, the Vera Rubin Observatory's LSST \citep{LSST:2008ijt} has a FOV of $\SI{9.6}{\square\degfull}$ and can observe kilonovae up to $z\sim 0.1$ (with $5\%$ of the kilonovae observable up to $z\simeq 0.4$) and 
several other instruments, such as  ULTRASAT \citep{Sagiv:2013rma},  can reach $z\simeq 0.1$; the Nancy Roman Space Telescope (formerly WFIRST) \citep{2015arXiv150303757S} has the highest reach, being able to observe $50\%$ of the kilonovae up to $z\sim 0.2-0.3$, with $5\%$ of the kilonovae observable up to $z\simeq 1$, although its FOV, \SI{0.28}{\square\degfull}, is not as large, so it requires sub--degree localization, or an earlier localization by instruments with a larger FOV
(see \cite{Cowperthwaite:2018gmx,Chase:2021ood,Ronchini:2022gwk} for recent discussions). Once localized the kilonova with these  large FOV instruments, telescopes such as the ELT can perform a more detailed follow--up of the source. For BNSs at larger redshifts, the identification of the electromagnetic counterpart can only be made by ${\rm X}/\gamma$--ray satellites with 
large FOV;  a sky localization of order \SI{100}{\square\degfull}, as can be provided by ET alone, is about $1/10$ of
the typical FOV of wide--field X--ray telescope, which in general  is larger than \SI{1}{\steradian},
so such a localization can already provide sufficient information for the search and localization of the $\gamma$-- or X--ray counterpart; then, the ${\rm X}/\gamma$--ray satellites
can  provide the arcmin  localization needed to drive the ground--based follow--up~\citep{Ronchini:2022gwk}.
For these multi--messenger studies, a lower threshold on the SNR, such as ${\rm SNR}=8$, can be appropriate, since the statistical significance will be enhanced by the temporal coincidence of the gravitational and electromagnetic signals. From the upper--left panel of \autoref{fig:ET_BNS_zhists} we see that, for ET alone and a threshold ${\rm SNR}=8$, the farthest detected  BNSs in our sample are at $z\sim 4.5$ (the farthest one happens to be at $z\simeq 4.77$ and has $M_{\rm tot}=\SI{4.15}{\Msun}$ and nearly equal masses). Higher values of the rate, or longer observation times, would lead to the gradual appearance of more and more rare events. For ${\rm SNR}=8$, at ET alone, an equal mass non--spinning BNS, with $m_1=m_2=\SI{1.35}{\si{\Msun}}$, optimal inclination and sky location could be detected up to $z_{\rm max}\simeq 3.5$, while 
a heavier equal mass non--spinning BNS, say with $m_1=m_2=\SI{2.1}{\si{\Msun}}$, optimal inclination and sky location could be detected up to $z_{\rm max}\simeq 6.9$.\footnote{Because of the long duration of the BNS signal and the Earth movement, the sky location changes in time. In practice, we find that the optimal sky location is such that the BNS is near a maximum of the pattern function at the time of merger.}
These estimates are in agreement with those in \cite{Ronchini:2022gwk}. For ET+2CE, the farthest detection in our sample, with ${\rm SNR}\geq8$, is at $z\simeq8.8$ and has $M_{\rm tot}=\SI{3.73}{\Msun}$ and a mass ratio of $q\simeq 1.33$.
An ideally oriented BNS with the same (rather extreme) characteristics could in principle be detected up to $z\simeq 20$  (where, of course, NSs are not even expected to exist), while a more realistic 
equal--mass non--spinning BNS, with $m_1=m_2=\SI{1.35}{\si{\Msun}}$, could be detected up to  $z\simeq11$.

\begin{figure}
    \hspace{-2cm}
    \begin{tabular}{c@{\hskip -3mm}c@{\hskip -4mm}c}
  \includegraphics[width=70mm]{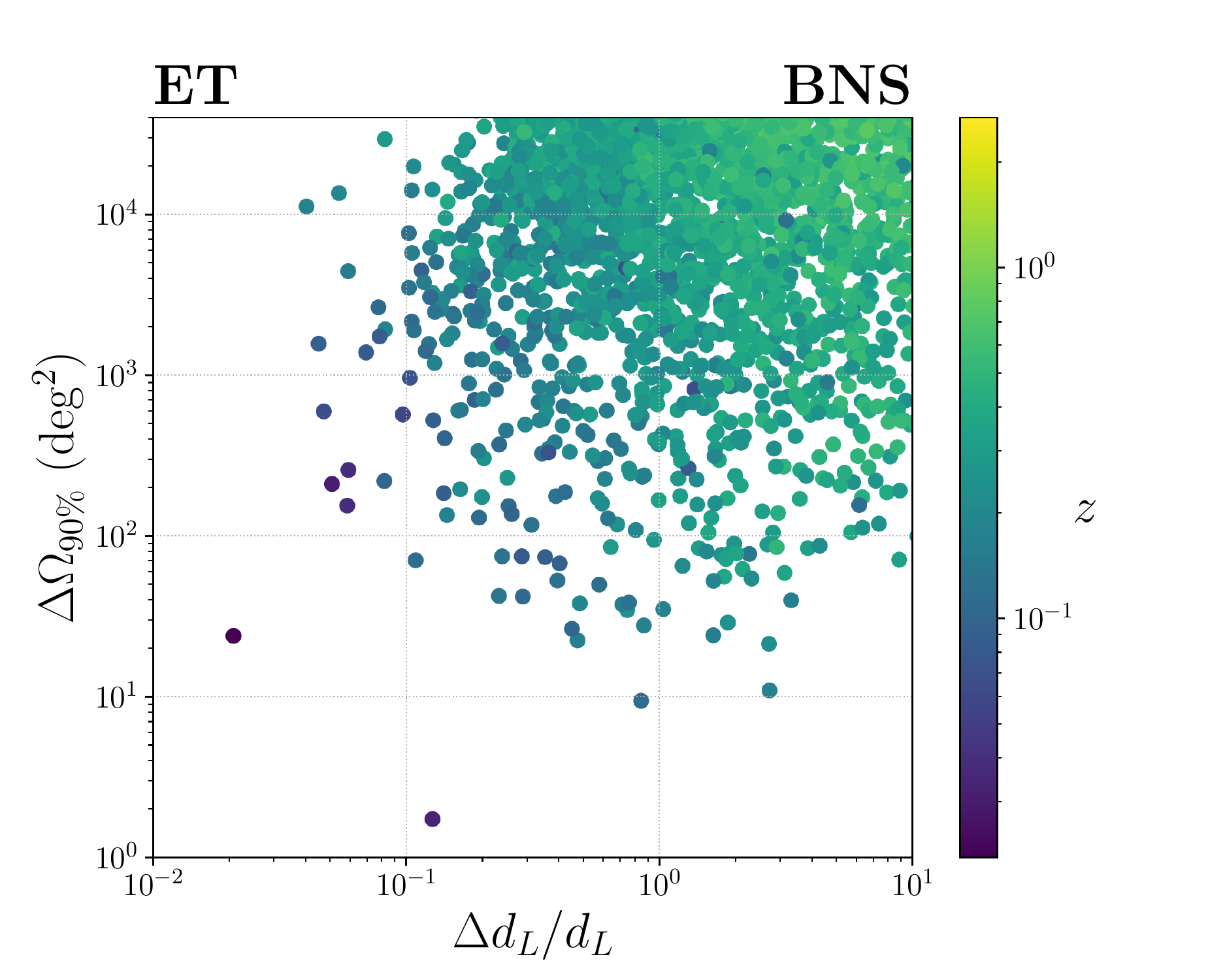} &   \includegraphics[width=70mm]{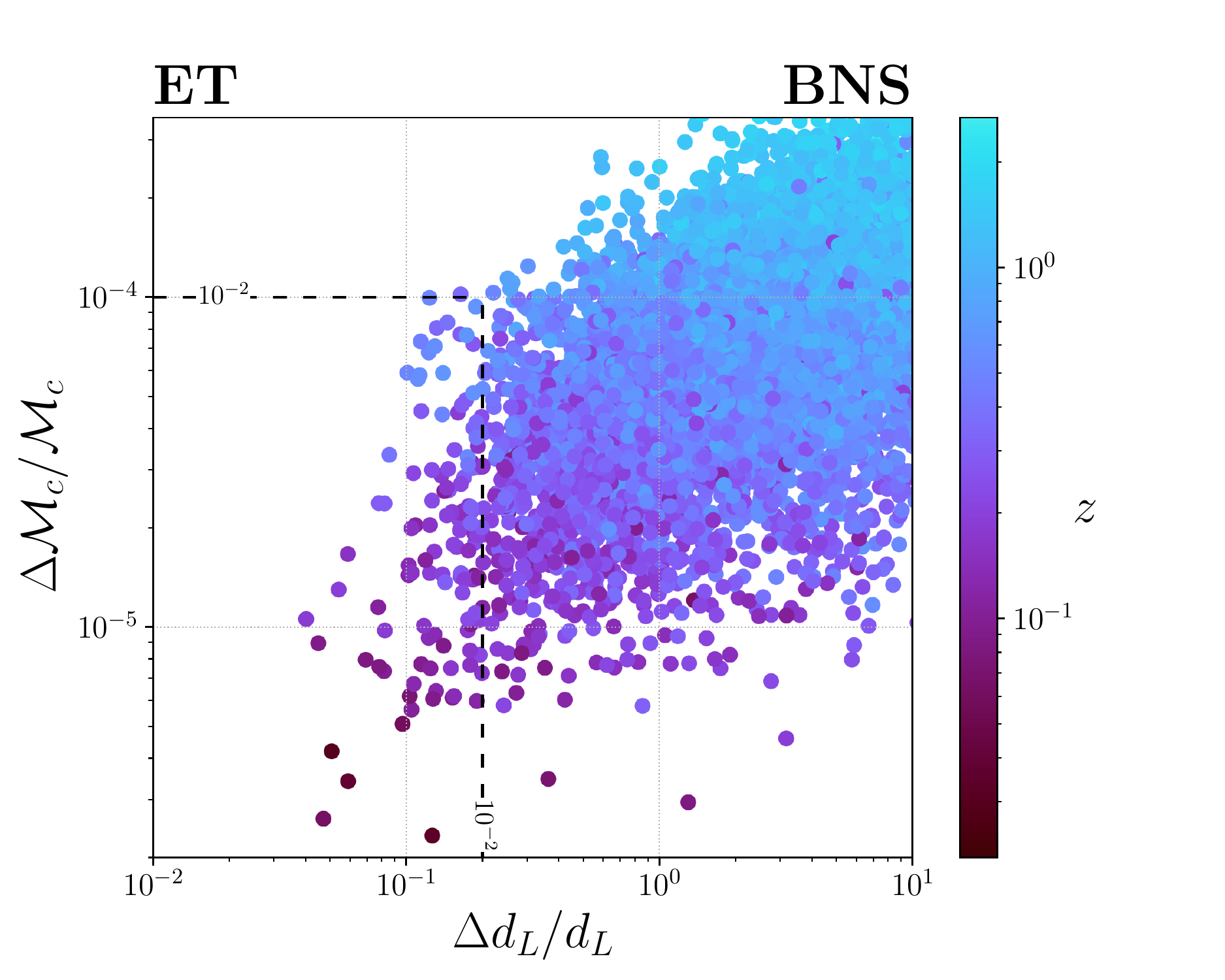} & \includegraphics[width=70mm]{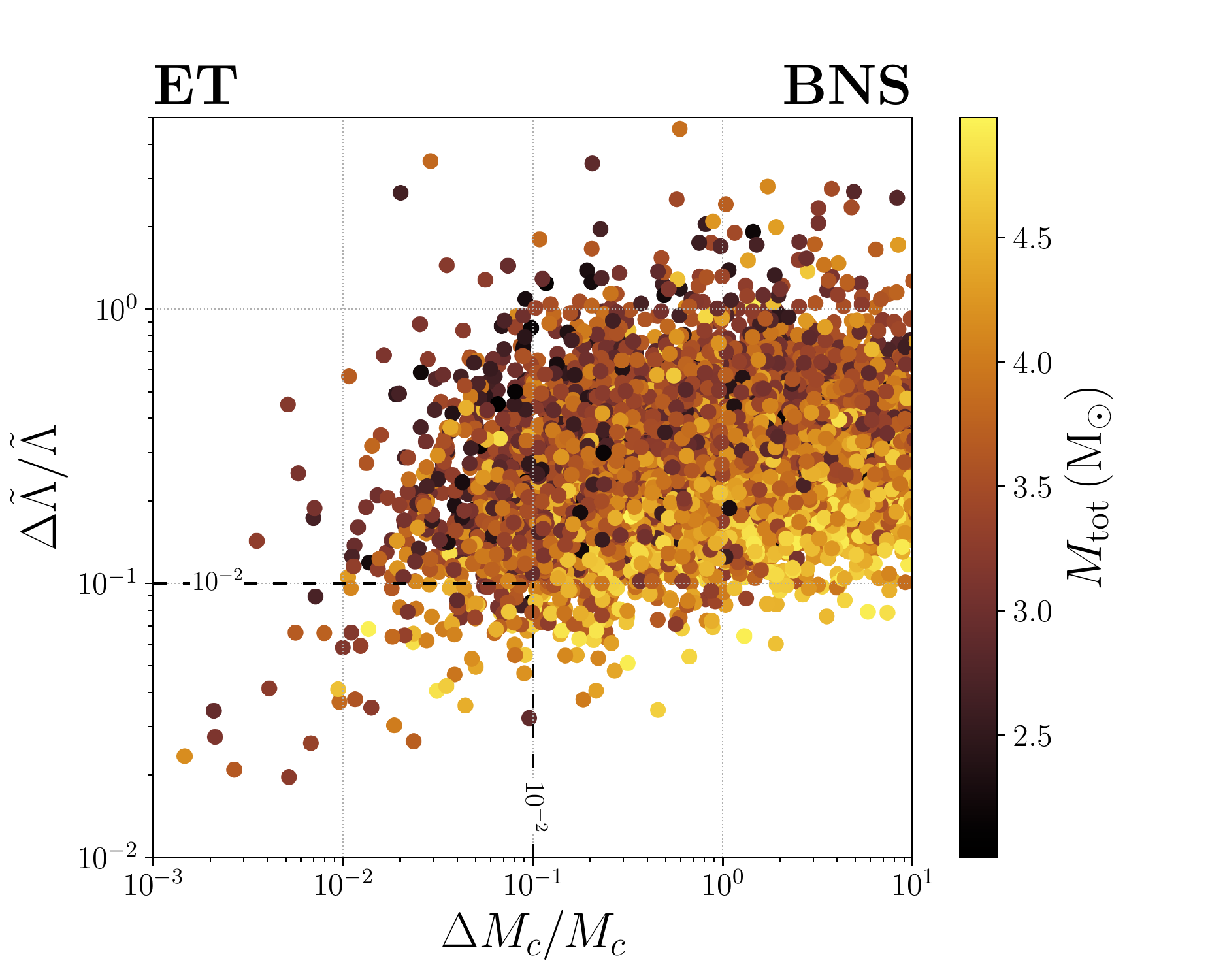}\\[-.5cm]
  \includegraphics[width=70mm]{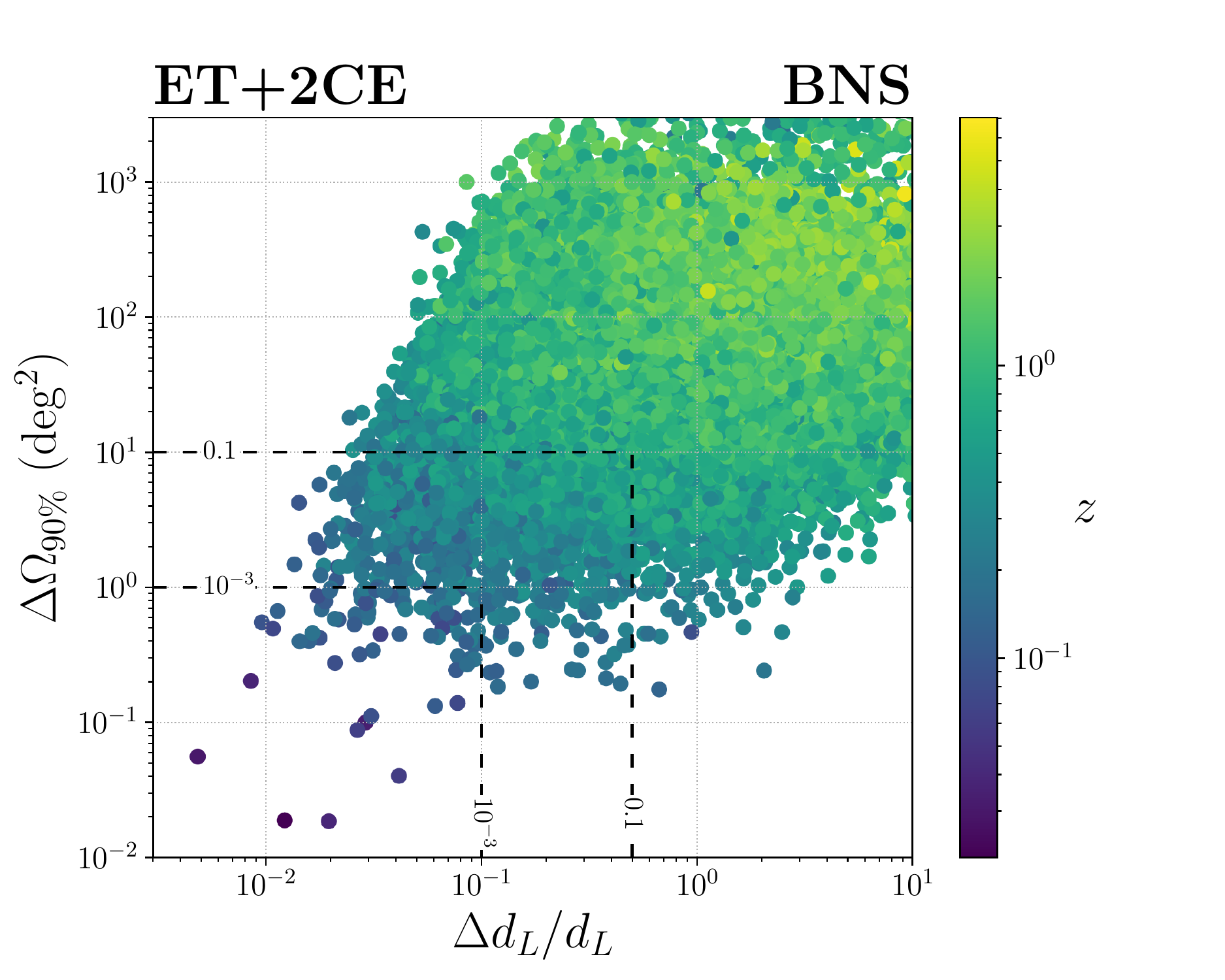} & \includegraphics[width=70mm]{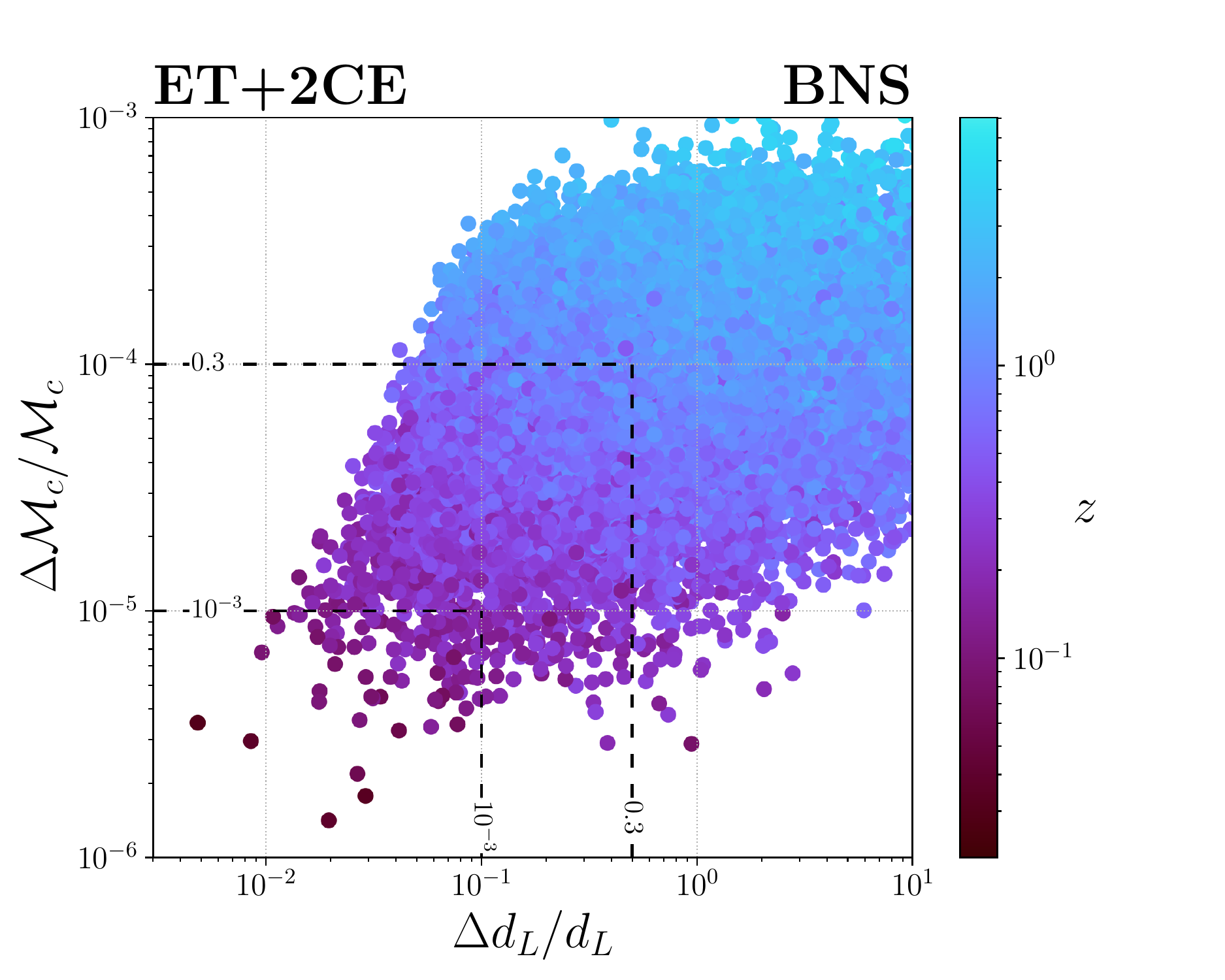} & \includegraphics[width=70mm]{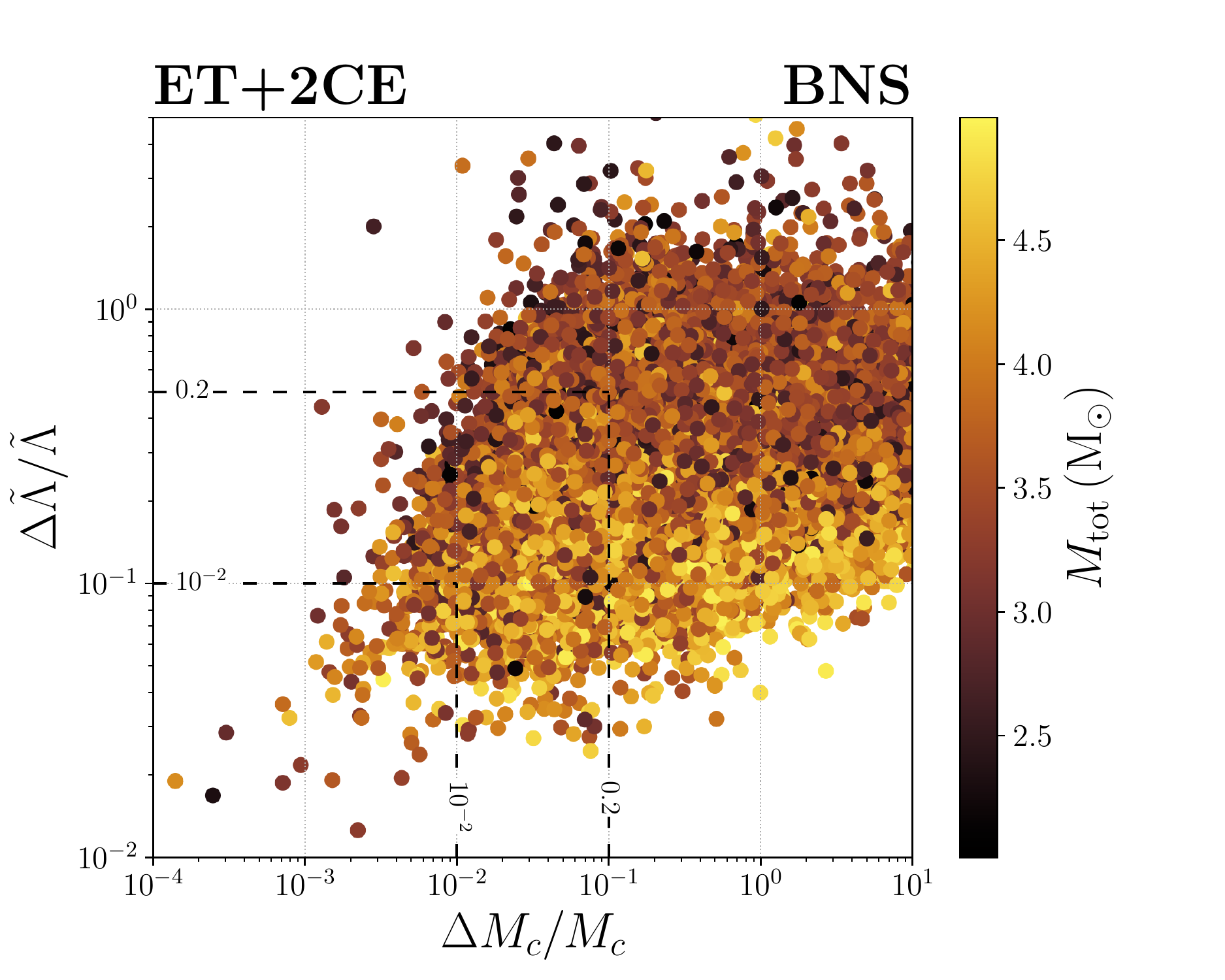}
\end{tabular}
    \caption{Scatter plots of the observed BNS population at 3G detectors. On the top row we report the results for ET alone and in the bottom for the ET + 2CE network. In each row, the left panel shows the distribution of errors on the luminosity distance and sky location as a function of redhsift, the central panel the distribution of errors on the luminosity distance and detector--frame chirp mass as a function of redshift, and the right panel the distribution of errors on the source--frame chirp mass and adimensional tidal deformability as a function of the total source--frame mass. The numbers reported on the dashed lines refer to the fraction of observed events lying inside the corresponding region.}
    \label{fig:BNS_scatter}
\end{figure}

When drawing conclusions from  plots featuring the absolute number of BNS detections,  it is important to keep in mind  that the estimate of 
${\cal R}_{0, {\rm BNS}}$ from current observations is still very uncertain, given the limited sample of detected BNS coalescences, much more than for BBHs, for which we already have a much larger sample of detections.
As we have mentioned above, current data, depending on the model used (e.g. on the assumed mass, spin and redshift distributions) are consistent with values of 
${\cal R}_{0, {\rm BNS}}$ in the range $(10-1700)~\si{\per\cubic\giga\parsec\per\year}$.  Therefore, as more and more BNS detections will accumulate with future runs, and the BNS local rate will become better constrained, the absolute scale on the vertical axis  in \autoref{fig:BNS_cumuldistNdet} could still change, possibly by as much as a factor $\order{10}$. Furthermore, a change in the rate is not just reabsorbed in a rescaling of the absolute vertical scale of these plots. For a fixed observation time, such as the standard reference choice of one year, the tails of the distributions, i.e. rare and particularly interesting  events, change. As an obvious example, if, with a given choice of rate, in one year there is no event with, say, $z>4$, this does not mean that, with a rate three times larger, there will be three times zero events, which is still zero. Rarer and rarer events will appear as we increase the rate or the observation time. Therefore, given the current large  uncertainty, when comparing different results for BNSs in the literature, one must first of all  check the value of ${\cal R}_{0, {\rm BNS}}$ used.

Finally, similarly to the plots shown  for BBH signals, in the left and central panels of \autoref{fig:BNS_scatter} we  show scatter plots of the joint estimation errors for the sky location and luminosity distance,  and for detector--frame chirp mass and luminosity distance, with a color code showing the redshift dependence. These are again the basis for the statistical methods based on correlation with galaxy catalogs and the mass function, the latter being particularly promising given the expected narrowness of the BNS mass distribution \citep{Taylor:2011fs, Finke:2021eio}. Exploiting statistical techniques is of fundamental importance also for BNS systems even though they can have an associated electromagnetic counterpart, since the detection of such EM emissions is expected only for a  fraction of the sources \citep{Belgacem:2019tbw, Ronchini:2022gwk}. In the right panels of \autoref{fig:BNS_scatter} we instead report the forecast of the  relative errors attainable on the source--frame chirp mass and the combination of the adimensional tidal deformability parameters $\tilde{\Lambda}$ defined in \eqref{eq:LamTdelLam_def} (with a color code giving information  on the total source--frame mass). These are the basic quantities needed to study the equation of state of dense matter above the nuclear density. We find a large amount of detections with percent--level accuracy attainable both at ET alone and at ET+ 2CE detectors. This would lead to dramatic improvements in our understanding of the equation of state of NSs [see e.g. \cite{Gupta:2022qgg} for a recent discussion]. We again show a selection of some of the main results presented in this section in \autoref{tab:BNS_Summary}.
\begin{table}[tp!]
    \vspace{-.2cm}
    \centering\hspace{-2.5cm}
    \begin{tabular}{!{\vrule width .09em}c|c|c||c|c|c!{\vrule width .09em}}
    \toprule\midrule
    \multicolumn{6}{!{\vrule width .09em}c!{\vrule width .09em}}{\bf BNS}\\
    \midrule\midrule
    Network & Detected & Analysed & $\rm SNR\geq 30$ & $\Delta d_L/d_L \leq 10\%$ & $\Delta\Omega_{90\%}\leq\SI{100}{\square\degfull}$\\
    \midrule\midrule
    \textbf{LVK--O4} & 4 & 4 & 0 & 1 & 2\\
    \midrule
    \textbf{ET} & 9577 & 9153 & 770 & 18 & 51\\
    \midrule
    \textbf{ET+2CE} & 40107 & 39584 & 5289 & 2225 & 27331 \\
    \midrule\bottomrule
    \end{tabular}
    \caption{A selection of  results from the analysis of the \num{1e5} BNSs (corresponding to the full population in about \SI{1}{\year} with our choices for the parameters) at the considered networks.}
    \label{tab:BNS_Summary}
    \vspace{-.2cm}
\end{table}

\subsection{Neutron star--black hole binaries}

Finally, we present our results for the population of NSBH systems, for which, currently, there are  only two detections, which  took place in the second part of the O3 run. 
In \cite{Abbott_2021:NSBHdetection}, the local rate has been estimated to be
${\cal R}_{0, {\rm NSBH}} = 45^{+75}_{-33}~\si{\per\cubic\giga\parsec\per\year}$ (assuming that the observed NSBH are representative of the underlying population), or
${\cal R}_{0, {\rm NSBH}} = 130^{+112}_{-69}~\si{\per\cubic\giga\parsec\per\year}$ (assuming a broad NSBH population),
while the re--analysis in  \cite{LIGOScientific:2021psn}, using  different modelizations, produce broadly consistent rates, between $7.8$ and $\SI{140}{\per\cubic\giga\parsec\per\year}$. In the following, as a reference value, we will assume for definiteness the value ${\cal R}_{0, {\rm NSBH}} = \SI{45}{\per\cubic\giga\parsec\per\year}$, keeping in mind that the current  uncertainty on the local rate can result in an increase or a decrease in the number of detections per year, by a factor $\sim 3$ in either direction. 

\begin{figure}[t]
    \centering
    \includegraphics[width=.9\textwidth]{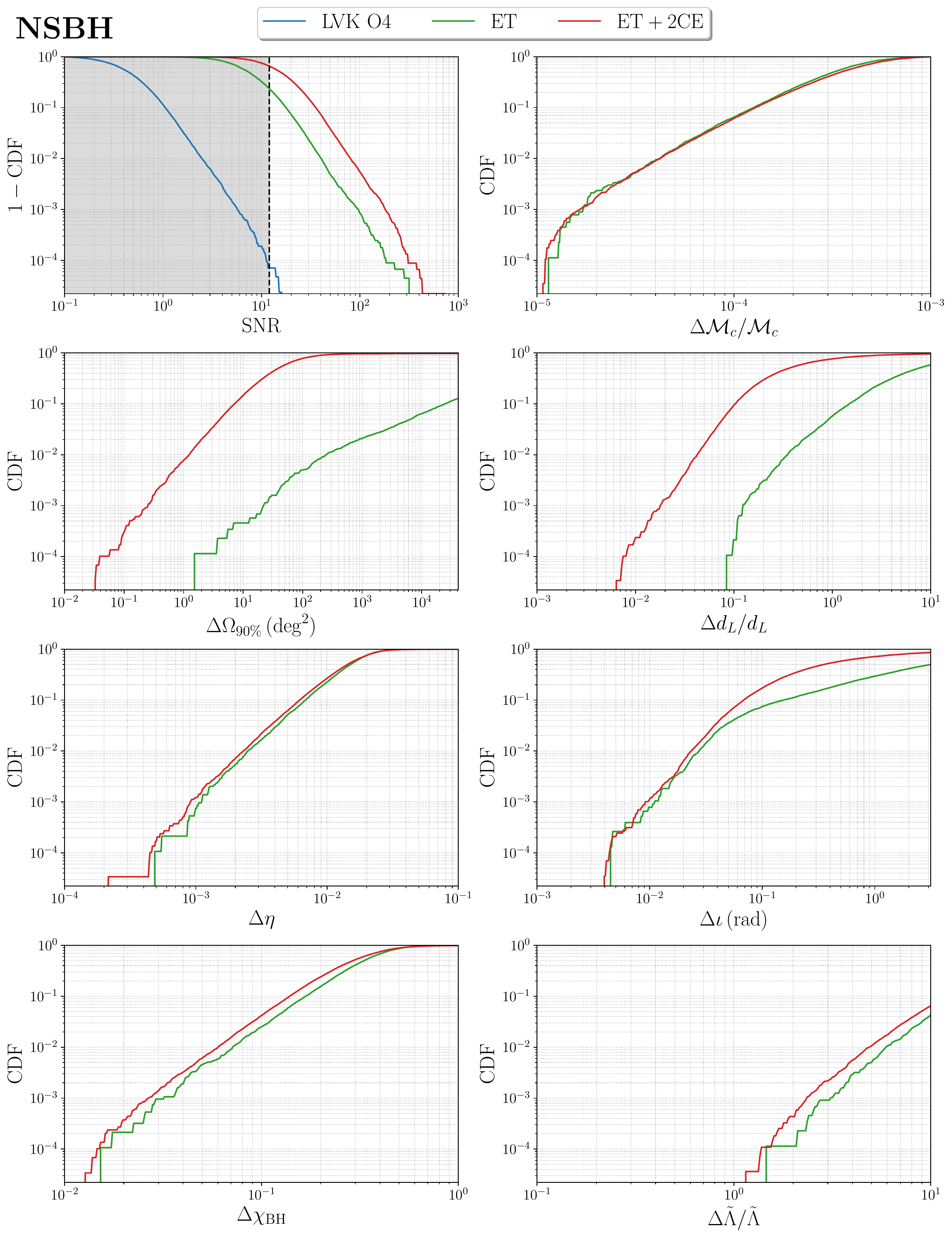}
    \caption{Cumulative distributions of the SNRs (for LVK--O4,  ET and ET+2CE) and parameter errors (for ET and ET+2CE) for NSBH signals, using the waveform model \texttt{IMRPhenomNSBH}.}
    \label{fig:NSBH_cumuldist}
    \vspace{1.cm}
\end{figure}

\begin{figure}[t]
    \centering
    \includegraphics[width=.9\textwidth]{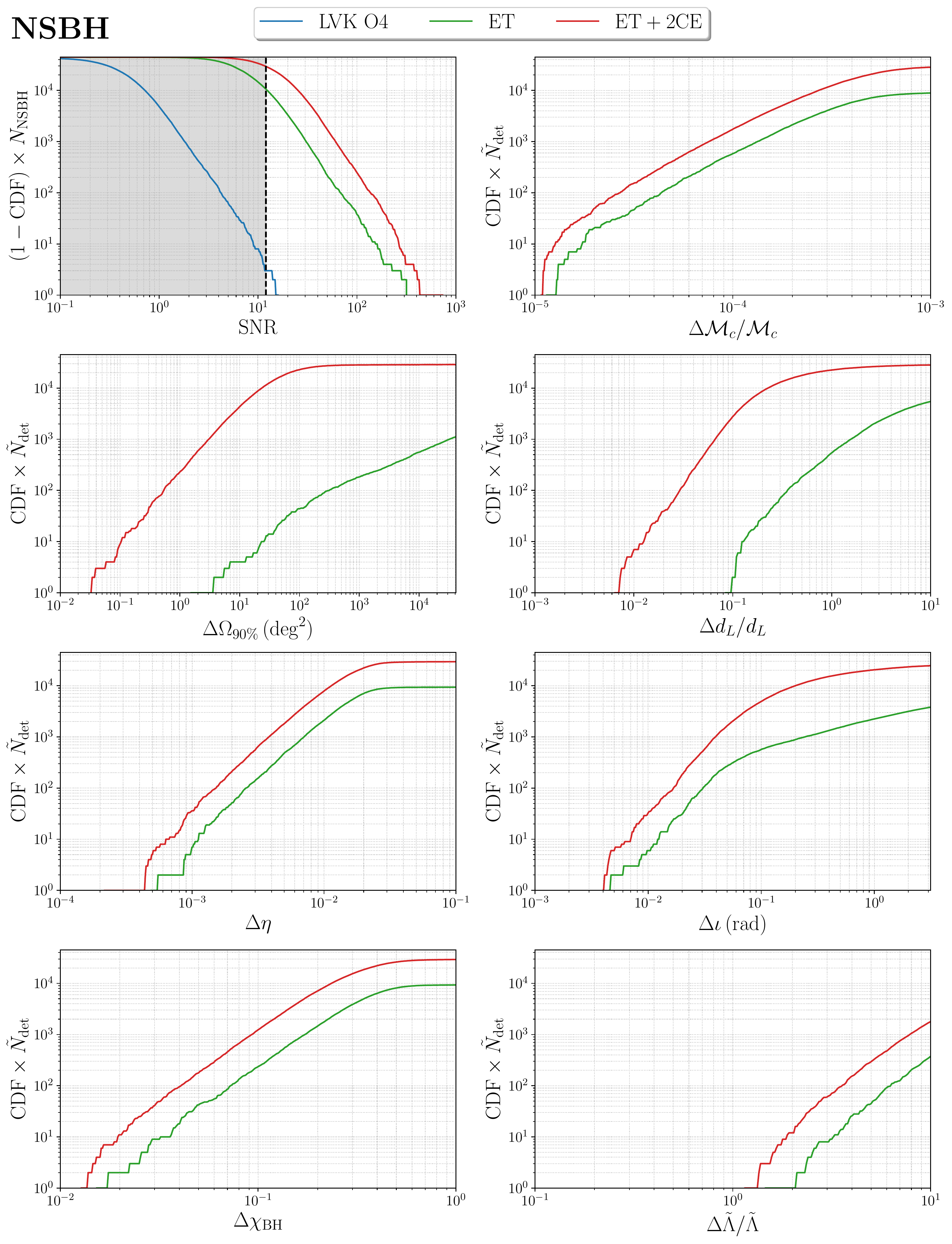}
    \caption{As in \autoref{fig:NSBH_cumuldist}, in terms of the total number of events rather than detection fraction.}
    \label{fig:NSBH_cumuldistNdet}
    \vspace{1.cm}
\end{figure}

\begin{figure}[t]
    \hspace{-2cm}
    \begin{tabular}{c@{\hskip 3mm}c@{\hskip 3mm}c}
  \includegraphics[width=64mm]{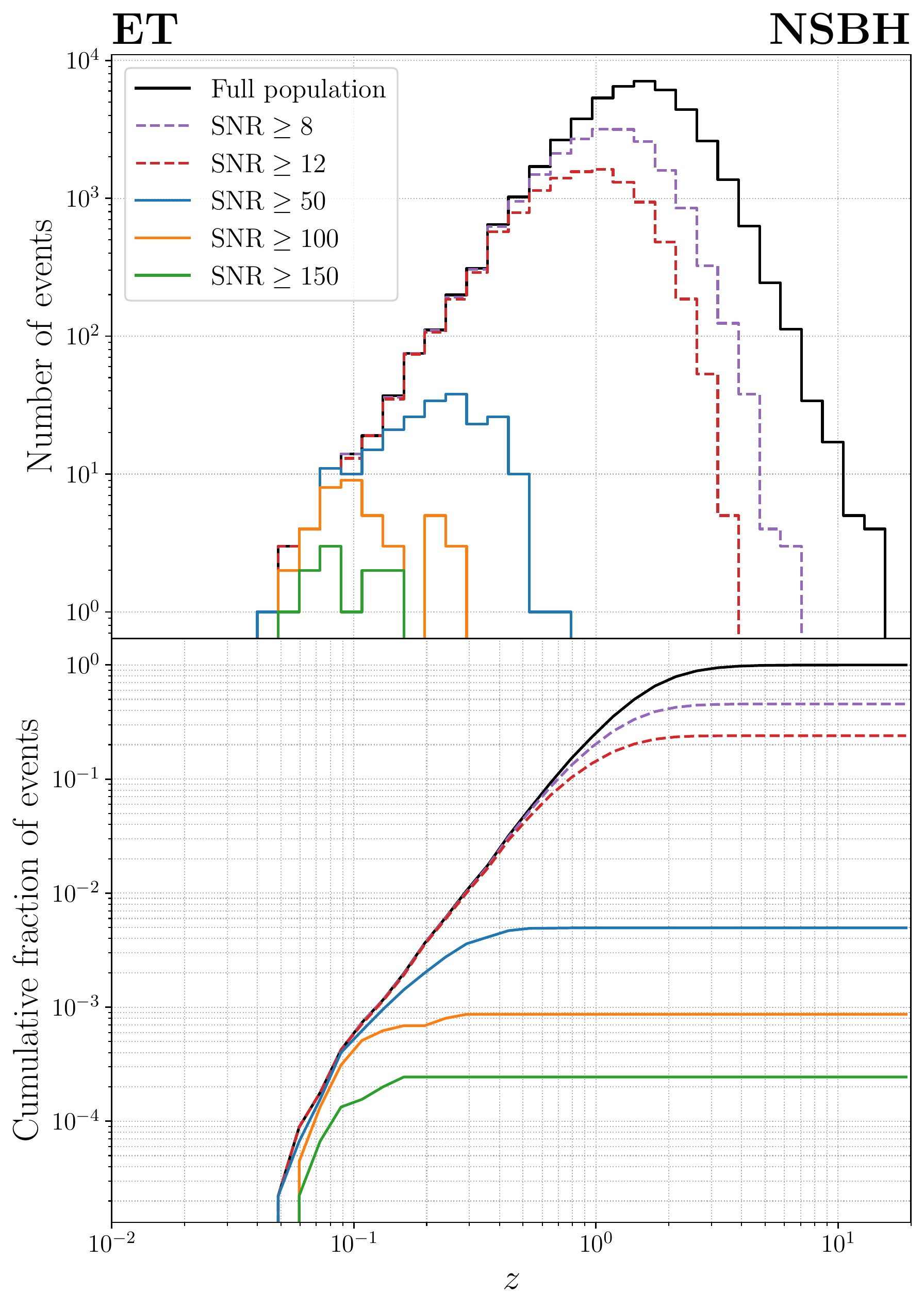} &   \includegraphics[width=64mm]{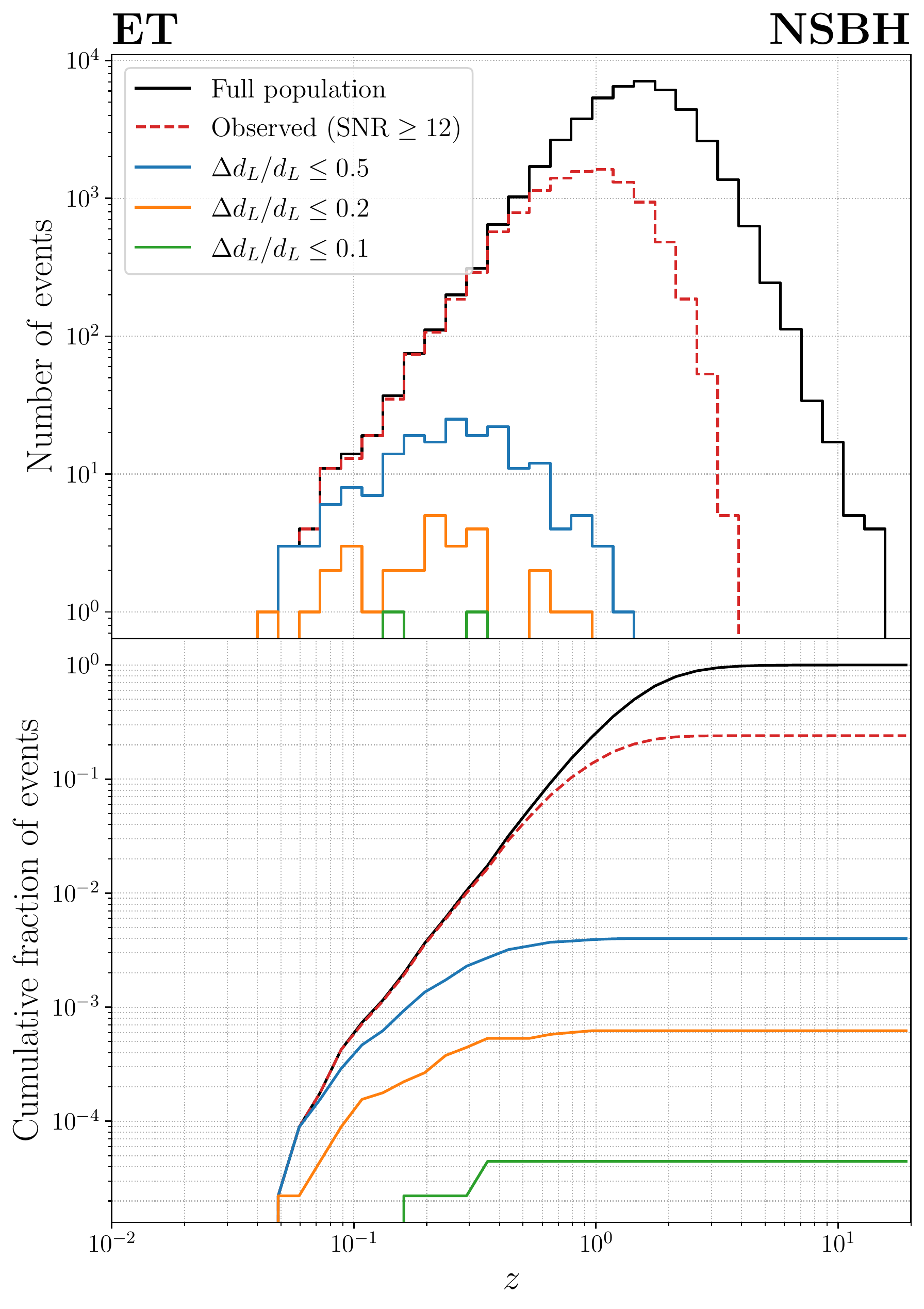} & \includegraphics[width=64mm]{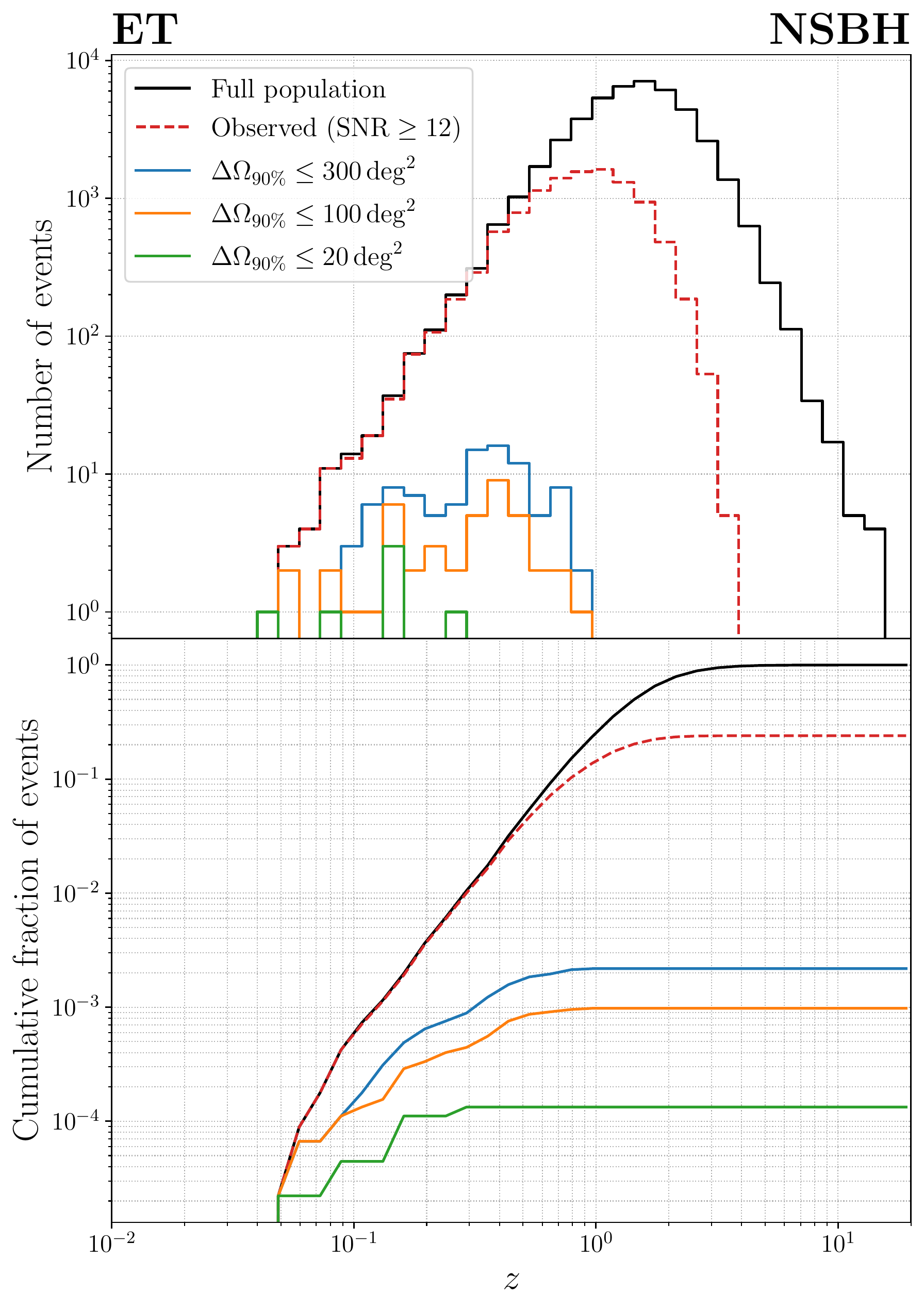}
\end{tabular}
    \caption{Redshift distributions of the NSBHs observed  at ET alone, selected on the basis of different thresholds for the SNR (left panel), or setting ${\rm SNR}\geq 12$ and applying further cuts on $\Delta d_L/d_L$ (central panel), or  on  $\Delta\Omega_{90\%}$ (right panel). The black solid line corresponds to the total NSBH population in the astrophysical model that we have assumed.
    We set  ${\cal R}_{0, {\rm NSBH}} = \SI{45}{\per\cubic\giga\parsec\per\year}$, and show the results for the detections in one year  (taking into account our assumptions of the duty cycle).  Given the current uncertainly on ${\cal R}_{0, {\rm NSBH}}$, one should keep in mind that the absolute number can still change by a factor $\order{3}$. In each column, the upper panel shows the number of events per redshift bin, while the lower panel shows the corresponding cumulative distributions, normalized to the number of BNS events in our sample, $N_{\rm NSBH}=\num{4.5e4}$.}
    \label{fig:ET_NSBH_zhists}
\end{figure}

\begin{figure}[t]
    \hspace{-2cm}
    \begin{tabular}{c@{\hskip 3mm}c@{\hskip 3mm}c}
  \includegraphics[width=64mm]{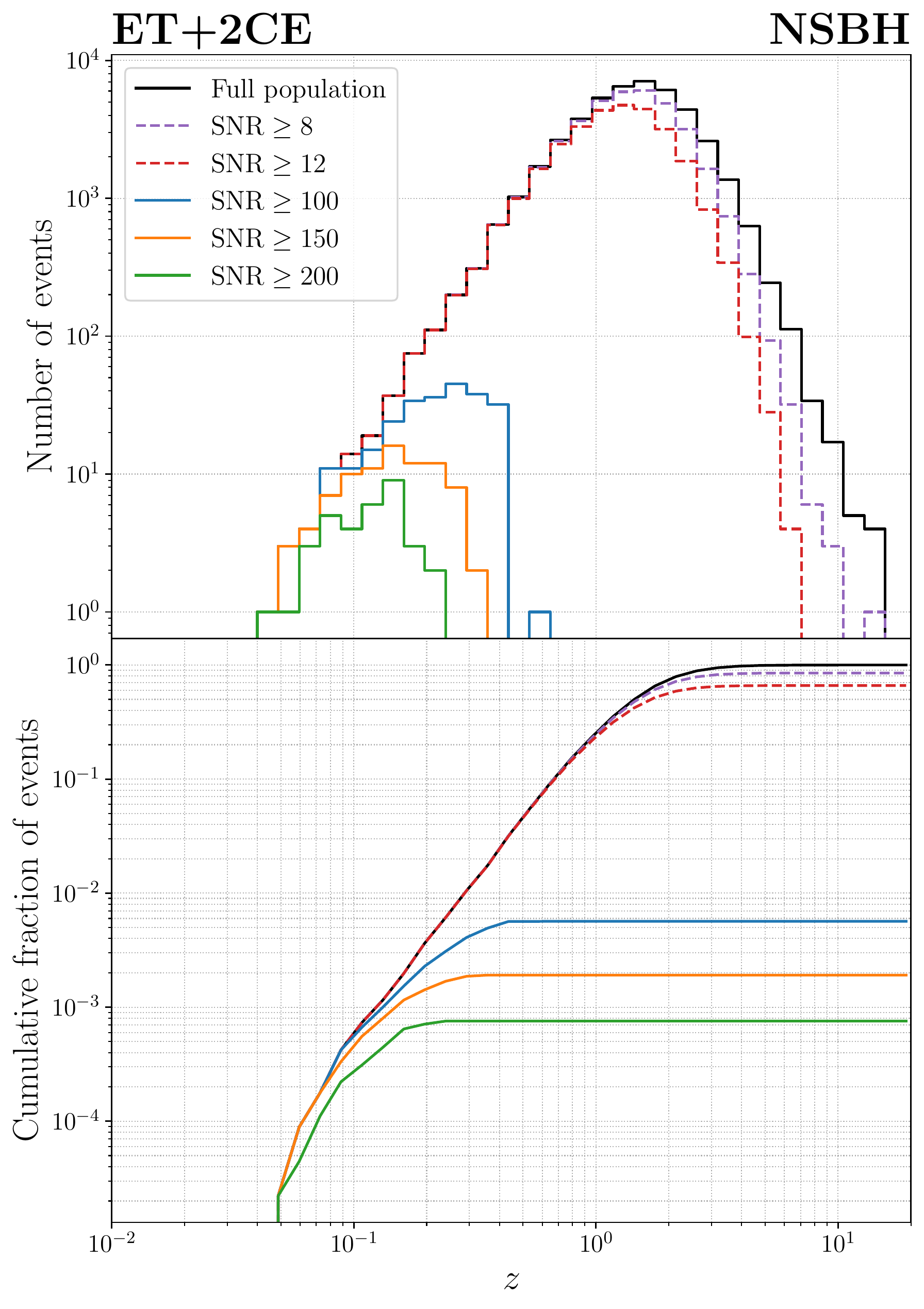} &   \includegraphics[width=64mm]{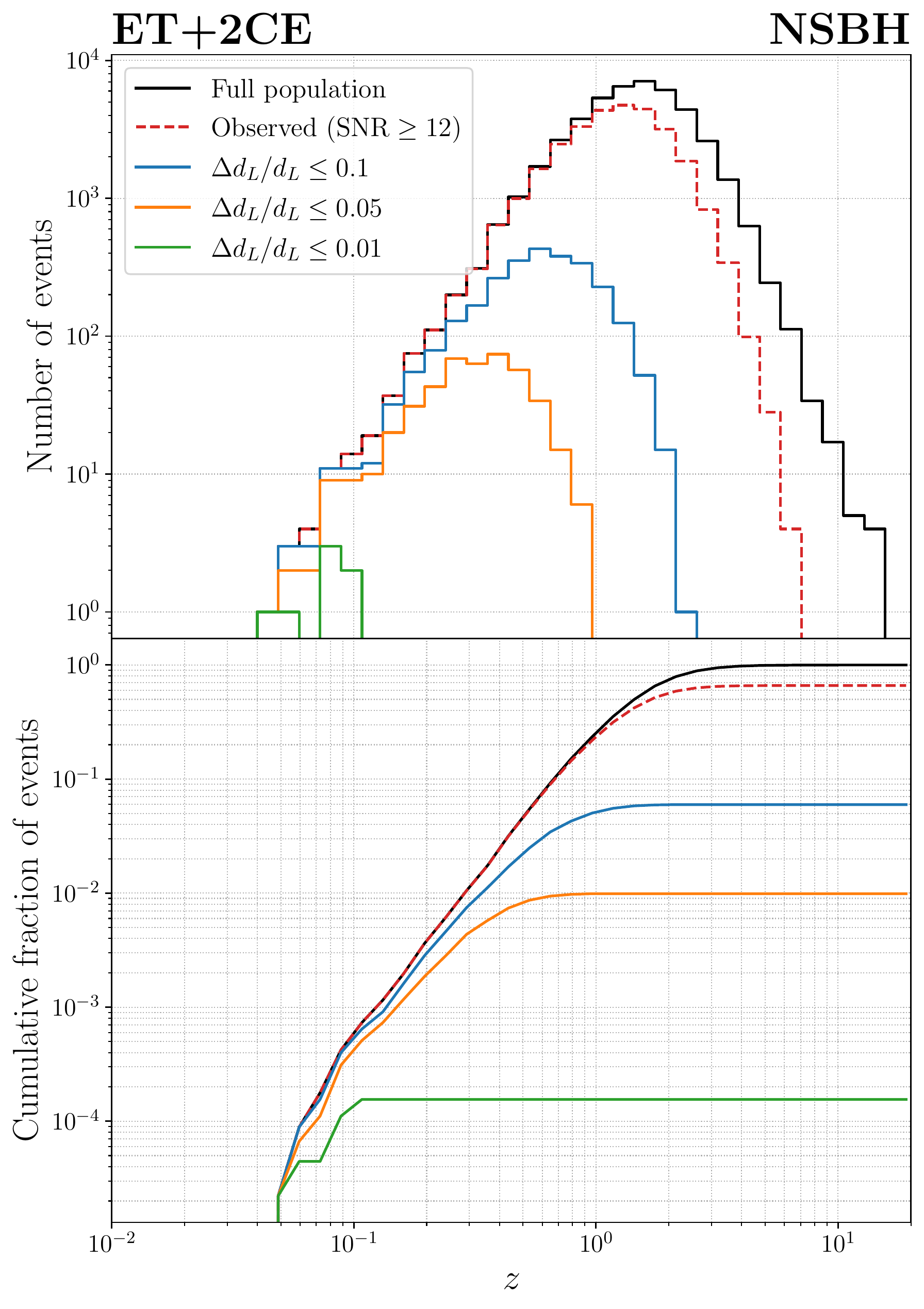} & \includegraphics[width=64mm]{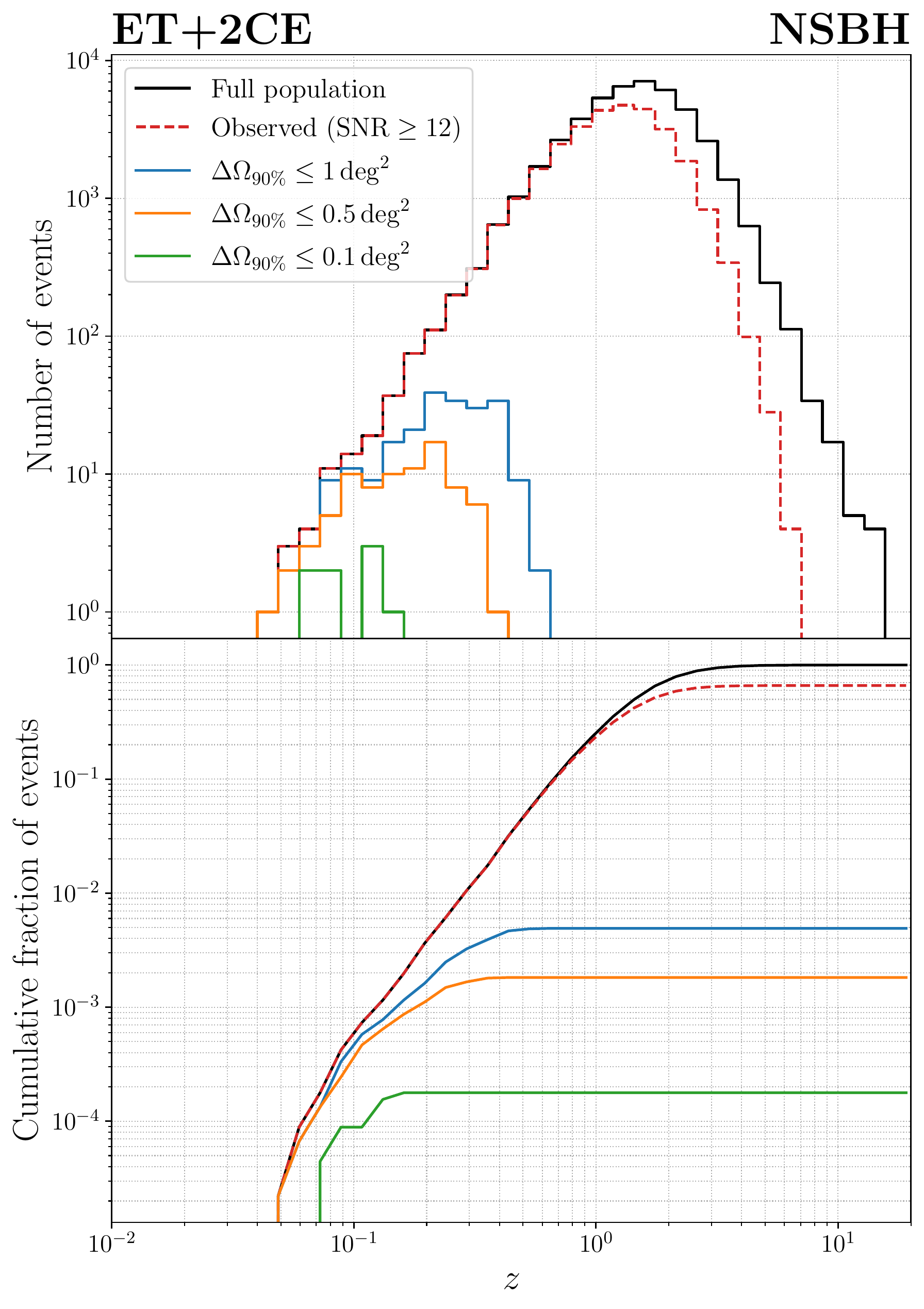}
\end{tabular}
    \caption{As in \autoref{fig:ET_NSBH_zhists}, for ET+2CE.}
    \label{fig:ET2CE_NSBH_zhists}
\end{figure}

For these systems, we then simulate a population of $N_{\rm NSBH}=\num{4.5e4}$ sources out to $z=20$ that, with our reference value of  ${\cal R}_{0, {\rm NSBH}}$ and the redshift dependence of the merger rate that we have assumed, described in \autoref{sect:populations}, would correspond to the number of coalescences happening in about one year.  
The cumulative distributions for the SNRs and for the parameter errors are shown in \autoref{fig:NSBH_cumuldist} and \ref{fig:NSBH_cumuldistNdet}, for the relative fractions and the absolute number of detection, respectively, while the redshift distributions are shown in \autoref{fig:ET_NSBH_zhists} for ET and in \autoref{fig:ET2CE_NSBH_zhists} for ET+2CE.  We find that, at ${\rm SNR}\geq12$, ET alone can detect up to $24\%$ of the sources (corresponding, with our assumptions on the population, to about \num{1.1e4} detections per year), while a network consisting of ET + 2CE could be able to detect 66\%  of the full population of sources (corresponding to about \num{3.0e4}  detections per year).
In contrast, for the current network of detectors  during O4, accounting for the duty cycle, we get 3 events, again in agreement with the broad expectations of \cite{AbbottLivingRevGWobs}. Assuming the current detectors to be operational 100\% of the time during O4 the number of detections only raises to 4, thus again we do not plot the corresponding results for the parameter estimation. Varying ${\cal R}_{0, \rm NSBH}$ in the currently allowed range $(7.8-140)~\si{\per\cubic\giga\parsec\per\year}$ (while keeping fixed the redshift dependence of the merger rate and the other features of our population model), the number of detection per year at LVK--O4 ranges between one and 10, for 
ET varies in the range $[\num{1.9e3}, \num{3.4e4}]$ and, for
ET+2CE, varies in the range $[\num{5.2e3}, \num{9.2e4}]$. For parameter estimation, using ${\cal R}_{0, {\rm NSBH}} = \SI{45}{\per\cubic\giga\parsec\per\year}$, the corresponding values of $\tilde{N}_{\rm det}$ are
$\num{9.4e3}$ for ET ($87\%$ of the detected events) and $\num{2.9e4}$  for ET+2CE ($99\%$).

\begin{figure}
    \hspace{-2cm}
    \begin{tabular}{c@{\hskip -3mm}c@{\hskip -4mm}c}
  \includegraphics[width=70mm]{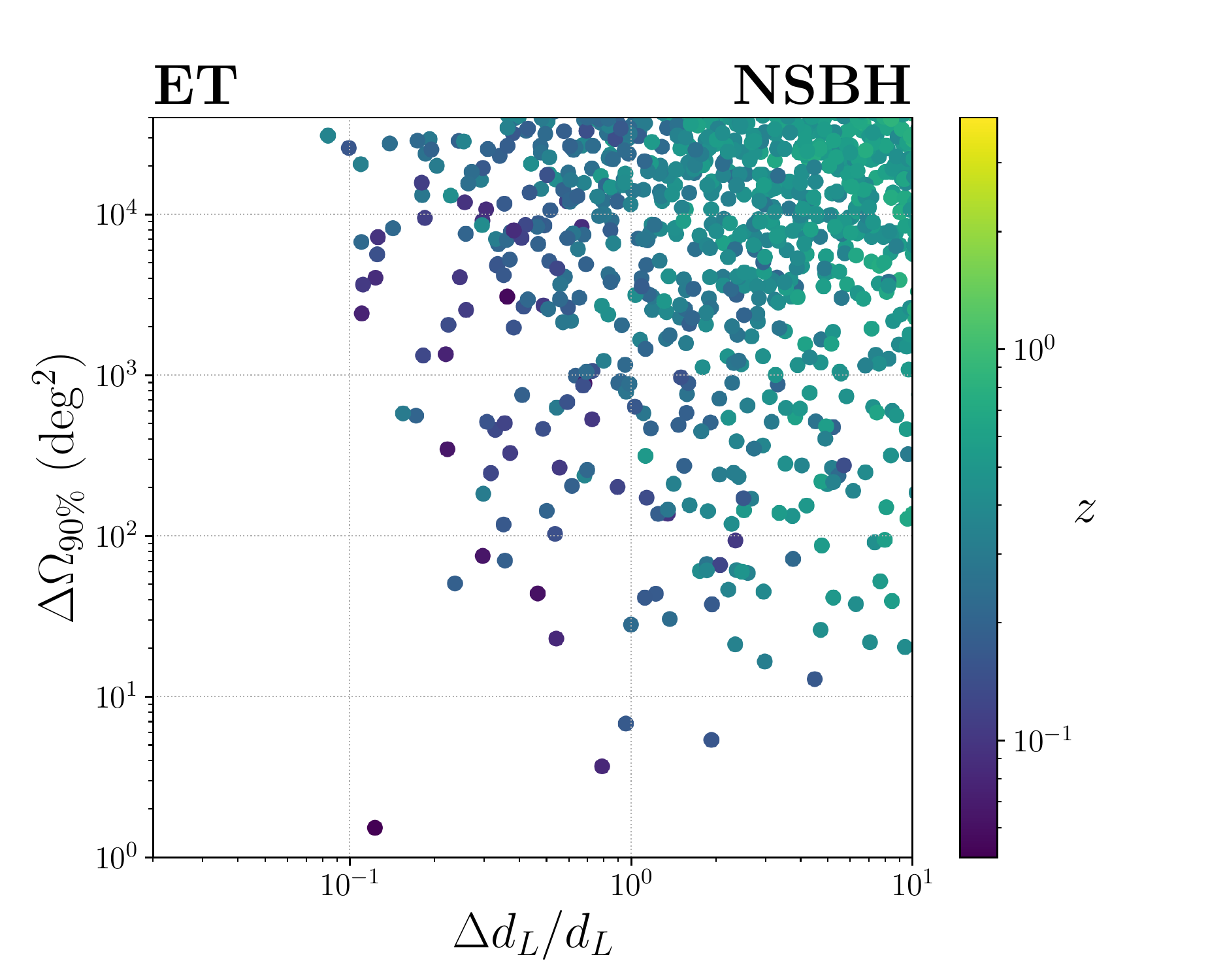} &   \includegraphics[width=70mm]{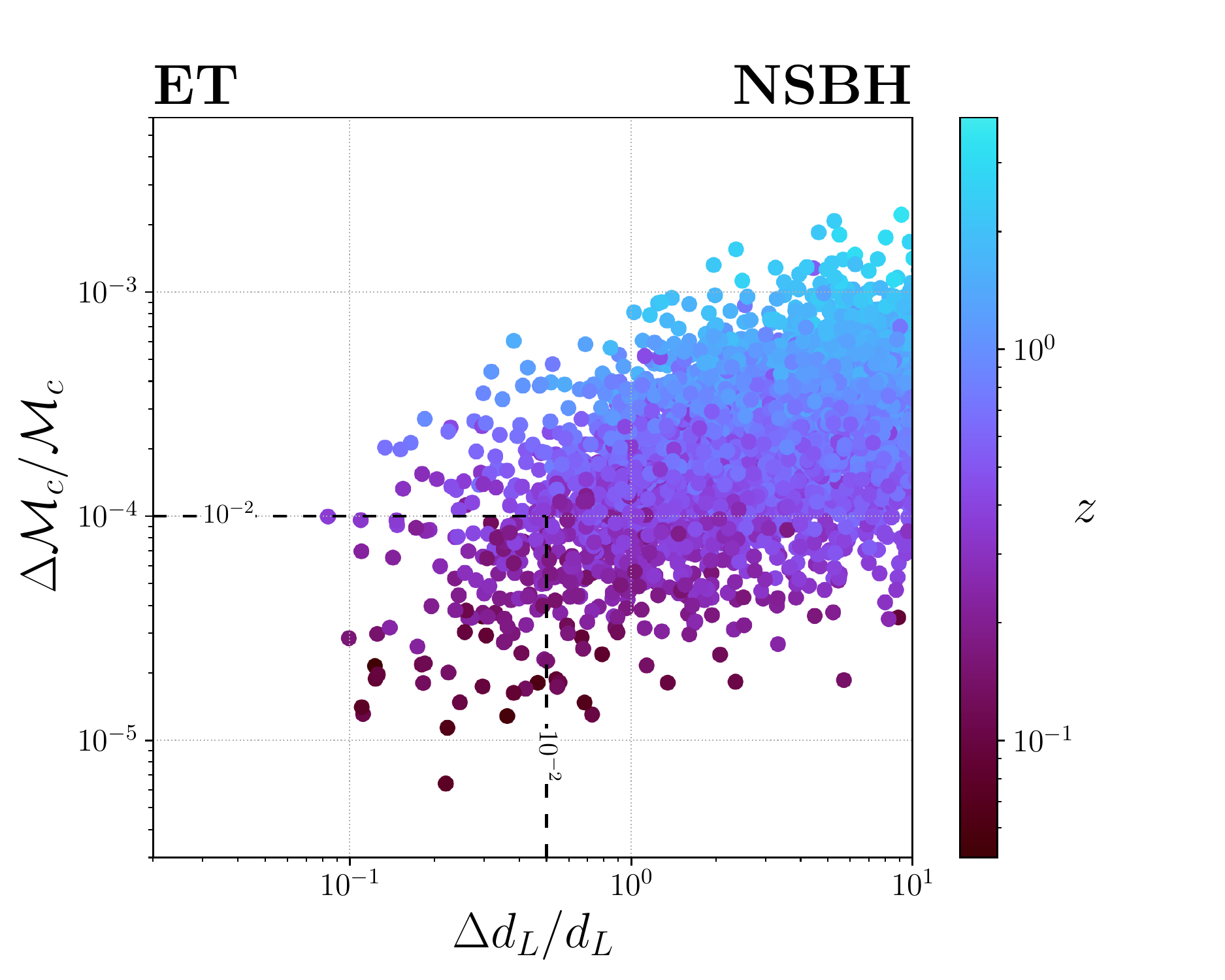} & \includegraphics[width=70mm]{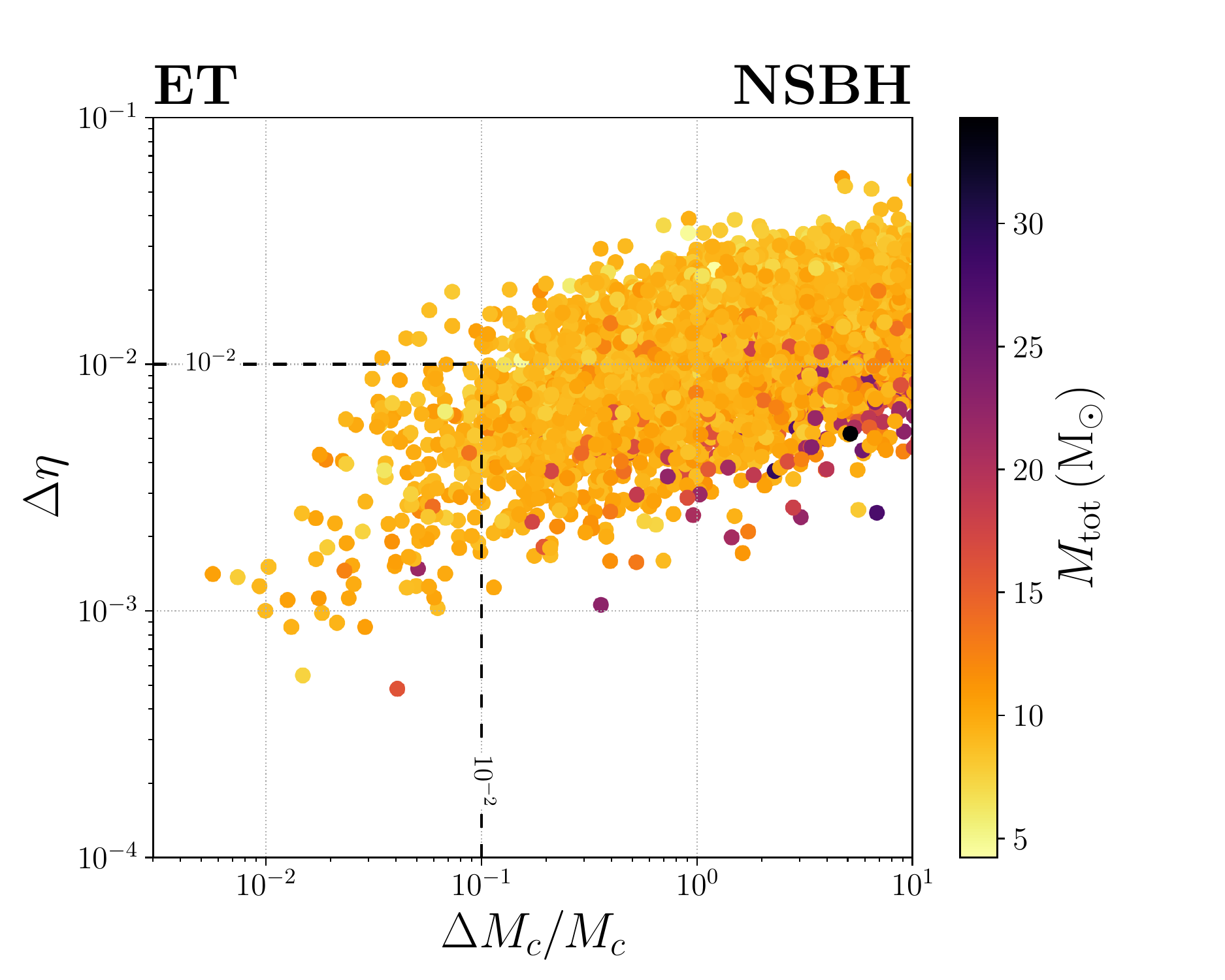}\\[-.5cm]
  \includegraphics[width=70mm]{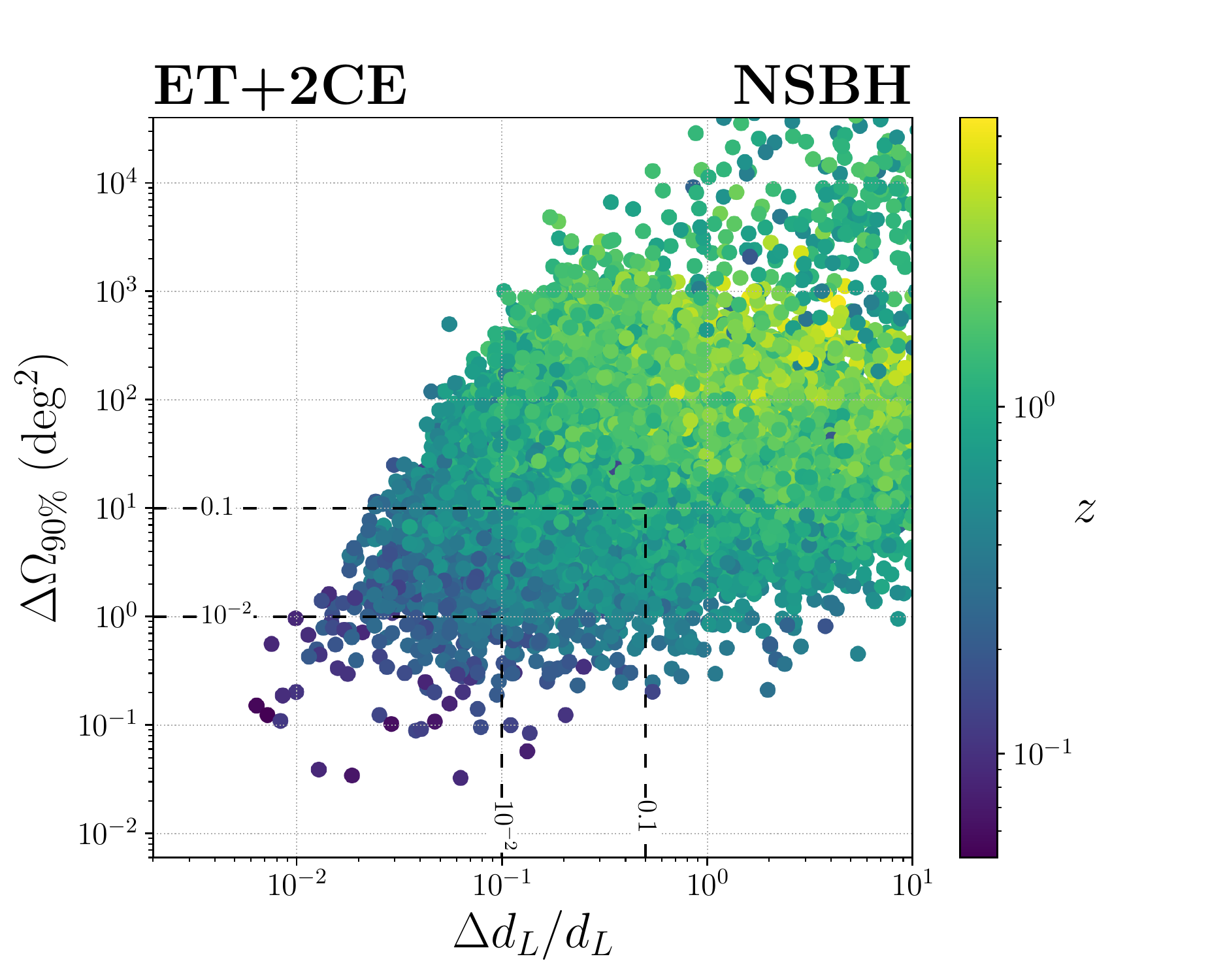} & \includegraphics[width=70mm]{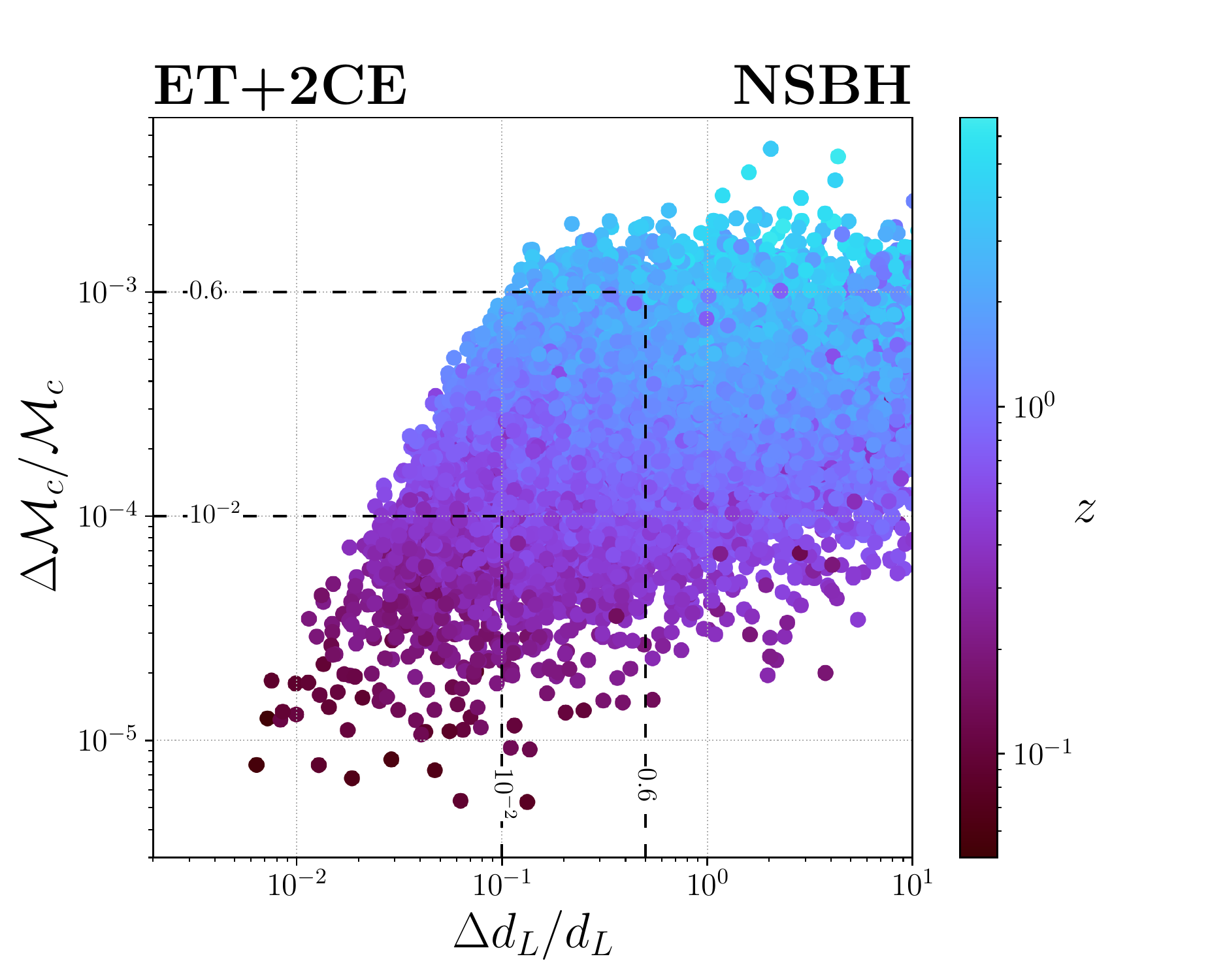} & \includegraphics[width=70mm]{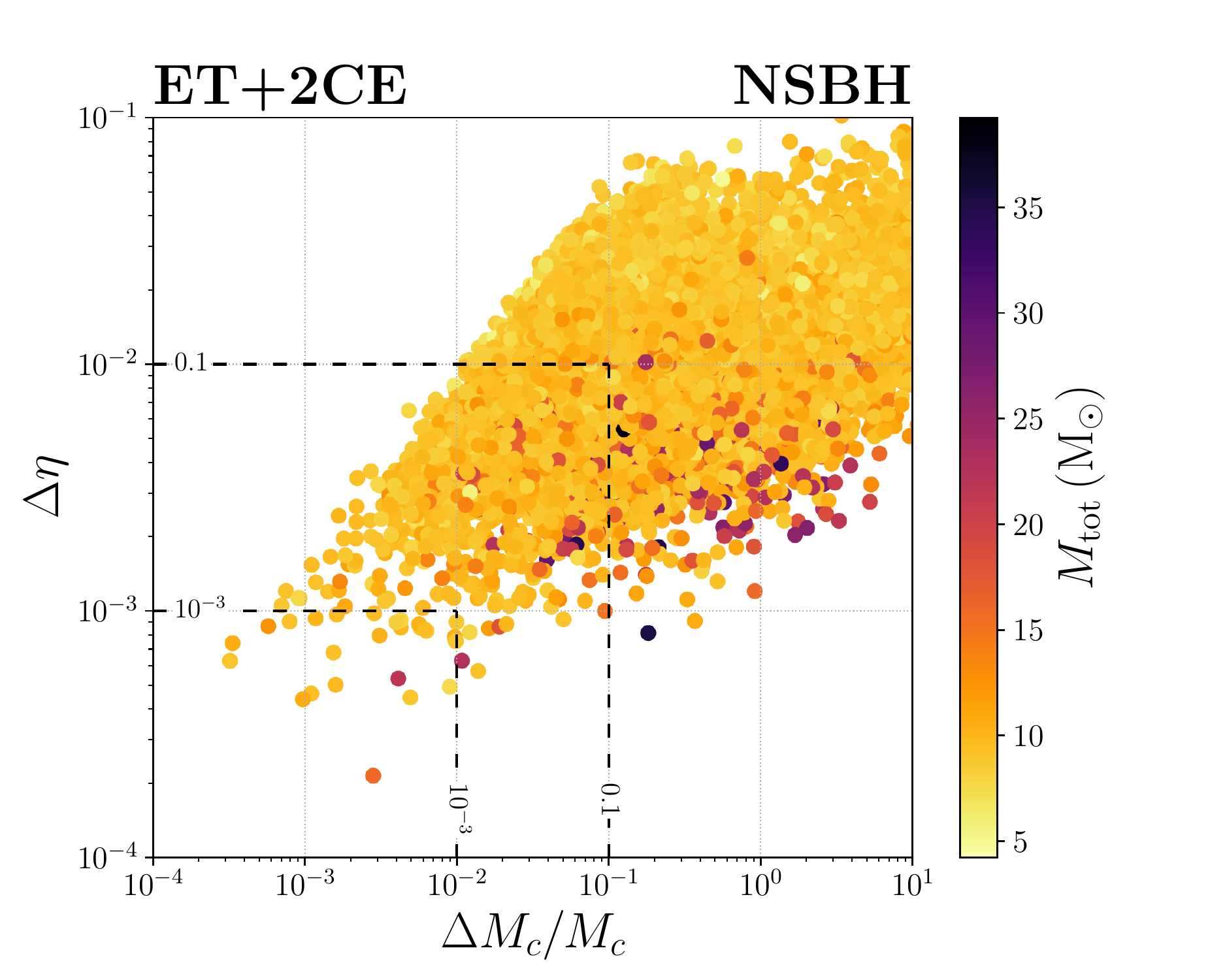}
\end{tabular}
    \caption{Scatter plots of the observed NSBH population at 3G detectors. On the top row we report the results for ET alone and in the bottom for the ET + 2CE network. In each row, the left panel shows the distribution of errors on the luminosity distance and sky location as a function of redhsift, the central panel the distribution of errors on the luminosity distance and detector--frame chirp mass as a function of redhsift, and the right panel the distribution of errors on the source--frame chirp mass and symmetric mass ratio as a function of the total source--frame mass. The numbers reported on the dashed lines refer to the fraction of observed events lying inside the corresponding region.}
    \label{fig:NSBH_scatter}
\end{figure}

In \autoref{fig:NSBH_scatter} we show scatter plots for some combinations of parameters. In particular, in
the left and central panels we show
the scatter plots in the 
$(\Delta\Omega_{90\%},\Delta d_L/d_L)$ plane and in the $(\Delta{\cal M}_c/{\cal M}_c,\Delta d_L/d_L)$ plane, respectively, with a color code carrying information on the redshift. Also in the case of NSBHs, these quantities  are relevant when exploiting statistical techniques to extract cosmological and astrophysical information from GWs. NSBH systems could be particularly relevant for probing the so--called lower mass gap, whose identification would have significant implications \citep{Ezquiaga:2022zkx}. For this reason, in the right panels we show the relative errors in the reconstruction of the source--frame chirp mass and symmetric mass ratio as a function of the total source--frame mass. These parameters are in fact fundamental to identify the edges of the lower mass gap, as well as identifying the system.

The knowledge that we currently have on the distribution of NSBH systems is quite limited. The large  uncertainties characterising the astrophysical properties of the NSBH population are apparent looking, e.g., at Fig. 14 in \cite{10.1093/mnras/stab2716}, where the authors compare the results for 420 different models, and already the shapes of the resulting distributions can be significantly different. We thus  find useful to study how our results for this class of sources change with different assumptions. In  our fiducial NSBH model, discussed in \autoref{sect:populations}, the BH mass distribution  is relatively narrow, and we assumed  a time delay distribution $P(t_d)\propto1/t_d$, which favours small time delays. We compare it with two of the models presented in \cite{10.1093/mnras/stab2716}: their fiducial model, characterised by a broader distribution for the BH mass and a time delay distribution favouring higher values, and the model denoted by O\_111, which predicts smaller masses and low time delays. These can be seen, respectively, as an ‘‘optimistic'' and a ‘‘pessimistic'' scenario, since systems with higher masses and closer to the observer are louder and easier to detect. Observe that even more optimistic or pessimistic scenarios are possible, see e.g. \cite{10.1093/mnras/stab280}.

\begin{figure}[ht]
    \centering
    \includegraphics[width=1.02339\textwidth]{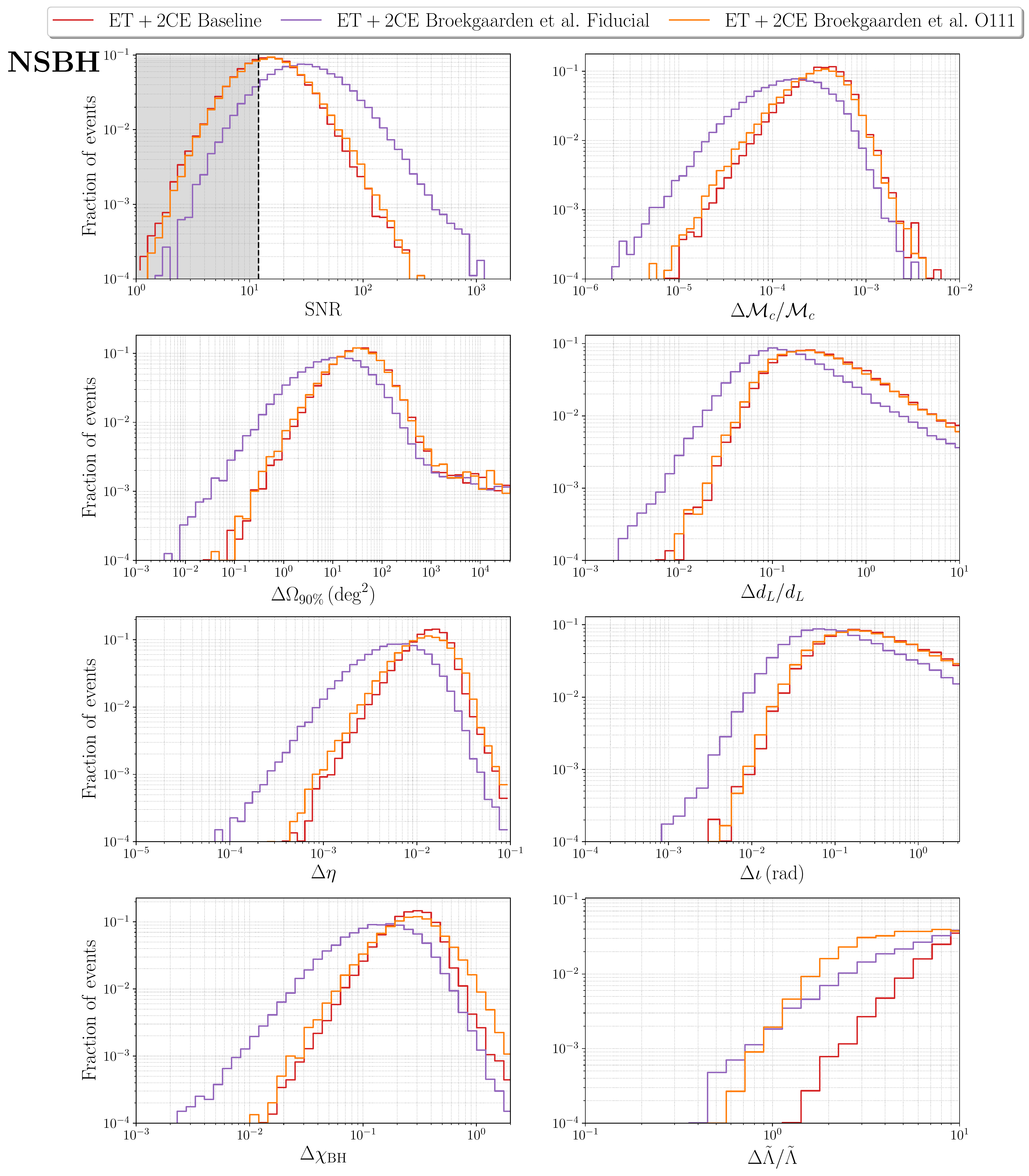}
    \caption{Histograms of the parameter errors for NSBHs with the ET+2CE network configuration, using the waveform model \texttt{IMRPhenomNSBH}, and sampling the BH mass and system redshift from our baseline distribution, and from two distributions taken from \citep{10.1093/mnras/stab2716}. The tails of the distributions featuring extremely large values of the relative errors, as in some of the panels, simply represents the contribution of events for which the Fisher matrix approximation cannot be trusted.
    }    
    \label{fig:HistBNScomp_BasevsBroek}
\end{figure}

To perform the comparison, we fitted the quantiles of the chosen distributions provided in \cite{10.1093/mnras/stab2716},\footnote{All data is publicly available at \url{https://zenodo.org/record/4574727\#.Ymv8jy8QN70}, in the \texttt{distributionQuantiles} folder.} 
with suitable metalog distributions \citep{doi:10.1287/deca.2016.0338}, capable of reproducing PDFs with irregular features. The comparison is only made using the ET + 2CE detector network, which has the highest number of detections, in order to avoid a proliferation of plots and for ease of readability. We show the results in \autoref{fig:HistBNScomp_BasevsBroek}, from which it is apparent that the uncertainties characterising the distribution of these events can strongly impact our results. In particular, as expected, the parameter errors for the events characterised by the narrower rate distribution can be much lower, even by more than one order of magnitude, as compared to the results obtained with our baseline choices, which, in contrast,  are in  good agreement with the broader distribution. Furthermore, also the SNRs obtained for the ‘‘optimistic'' population are on average much higher, resulting in an increased percentage of detections, from $66\%$ to $89\%$, while the percentage of detections in the ‘‘pessimistic'' case remains consistent with our fiducial choices. The results on NSBHs  shown in this work could thus change significantly in the future, when the population of NSBH systems will be better explored, and should only be considered as a first step toward  understanding the capabilities of GW detectors for NSBH systems, although our estimates appear to be more on the conservative side. Also in this case we summarise some of the main results presented in this section in \autoref{tab:NSBH_Summary}.

\begin{table}[t]
    \vspace{-.2cm}
    \centering\hspace{-2.5cm}
    \begin{tabular}{!{\vrule width .09em}c|c|c||c|c|c!{\vrule width .09em}}
    \toprule\midrule
    \multicolumn{6}{!{\vrule width .09em}c!{\vrule width .09em}}{\bf NSBH}\\
    \midrule\midrule
    Network & Detected & Analysed & $\rm SNR\geq 50$ & $\Delta d_L/d_L \leq 10\%$ & $\Delta\Omega_{90\%}\leq\SI{100}{\square\degfull}$\\
    \midrule\midrule
    \textbf{LVK--O4} & 3 & 3 & 0 & 0 & 1\\
    \midrule
    \textbf{ET} & 10792 & 9384 & 223 & 2 & 44\\
    \midrule
    \textbf{ET+2CE} & 29707 & 29462 & 1785 & 2687 & 22821 \\
    \midrule\bottomrule
    \end{tabular}
    \caption{A selection of  results from the analysis of the \num{4.5e4} NSBHs (corresponding to the full population in about \SI{1}{\year} with our choices for the parameters) at the considered networks.}
    \label{tab:NSBH_Summary}
    \vspace{-.2cm}
\end{table}

\newpage

\subsection{`Golden binaries' at 3G detectors}\label{sect:golden}

The detection rate of  compact binaries at 3G detectors is huge. As we have seen, for BBHs (keeping fixed the redshift dependence of the merger rate and varying only the local rate) the number of detections per year  should be in the range  $[\num{3.1e4}, \num{8.0e4}]$ at ET, raising to $[\num{4.2e4}, \num{1.1e5}]$  for ET+2CE.  Similarly, for BNSs,  again keeping fixed the redshift dependence of the merger rate and varying only the local rate, the  number of detections per year are expected to be in the range $[\num{9.1e2}, \num{1.6e5}]$ for ET, and $[\num{3.8e3}, \num{6.5e5}]$  for ET+2CE, and even larger numbers can be obtained with different models for the redshift dependence of the merger rate, as we discuss in \autoref{app:comparison}.
For NSBHs the detections should be of order $[\num{1.9e3}, \num{3.4e4}]$ for ET and $[\num{5.2e3}, \num{9.2e4}]$  for ET+2CE. 

These numbers are remarkable, and constitute one of the strength of 3G detectors. However, such large rates pose new challenges, compared to the situations at 2G detectors. For instance, a detection rate  of $\num{5e3}\, {\rm BNS/yr}$ corresponds to about a BNS every $\SI{2}{\hour}$, while a detection rate of $\num{8e4}$ BNS/yr means about one every \SI{7}{\minute}.  As we already mentioned, BNSs  can stay in the bandwidth of 3G detectors for hours, or up to order one day.  This means that we will have to deal with overlapping signals, as has been recognized already since several years~\citep{Regimbau:2009rk,2012PhRvD..86l2001R,2016PhRvD..93b4018M,Regimbau:2016ike}.  
In particular, in \cite{Samajdar:2021egv}, assuming values of the BBH local rate in the range $[15,37.9]~\si{\per\cubic\giga\parsec\per\year}$ and BNS local rates in the range $[80,810]~\si{\per\cubic\giga\parsec\per\year}$, consistent with the range of values that we have used, it is found that, at a network ET+2CE, a BNS signal will typically have tens of overlapping BBH and BNS signals, and it will happen up to $\order{10^4}$ times per year that two signals enters their final merger phase within seconds of each other (see also \cite{Himemoto:2021ukb} for similar estimates).\footnote{Overlapping signals also takes places for galactic white dwarf binaries in LISA. For 3G detectors, however, the BBH and BNS systems  are observed in their full inspiral--merger--ringdown phases, while white dwarf binaries in LISA are deep into their inspiral phase, so several aspects of the problem are different.} The bias that the overlap of signals can introduce in the parameter reconstruction of each separate event has been studied recently in
\cite{Samajdar:2021egv,Pizzati:2021apa,Himemoto:2021ukb,Antonelli:2021vwg}; these works show that the overlap of signals can indeed introduce a bias in parameter reconstruction, when two signals with comparable SNR have a difference in merger times which is no longer  large compared to the accuracy to which merger times themselves  can be measured. 

Beside  problems related to the overlap of signals, another practical problem related to the huge number of detections expected at 3G detectors  is that the computation of the full multi--dimensional posteriors of such a large number of events might be computationally too expensive (see \cite{Smith:2021bqc} for recent progress in this direction).

For some scientific questions, a possible solution to these problems, which might be sub--optimal but has the advantage of simplicity, is to restrict to events with a very large SNR, say ${\rm SNR}\geq100$;  we will refer to them as `golden events', or `golden binaries'. For some aspects of the science at 3G detectors such a strategy might have limited value. For instance,
for populations studies, the completeness of the sample of detections is a key element.
However, for precision studies, such as tests of General Relativity or cosmological observations, the result can be largely dominated by the very best events. For instance, in order to test gravity near the BH horizons and to discriminate the BH solution of General Relativity from other possible types of compact objects (see \cite{Cardoso:2016oxy} for general discussion)  one needs events with a sufficiently large SNR in the ringdown phase, which would allow the extraction of the frequency and damping time of at least two different quasi--normal modes, allowing us to perform `BH spectroscopy' \citep{Berti:2016lat}. For this kind of questions, only events with a sufficiently large SNR  are relevant. The same can happen in cosmological studies that make use of `dark sirens', i.e. coalescing binaries without an observed electromagnetic counterpart, by correlating them with galaxy catalogs. In this case, the results that can be obtained for $H_0$, or for the parameter $\Xi_0$ that characterizes modified GW propagation~\citep{Belgacem:2017ihm, Belgacem:2018lbp}, is largely dominated by the events with the smallest localization volume, that fall in a region where a galaxy catalog is sufficiently complete, see  \cite{Finke:2021aom}.
Another example is provided by multi--messenger observations;  given a BNS detection rate of order  one event per hour or more, only for a small fraction of detections it will be conceivable to perform a dedicated follow--up by electromagnetic observations. 

In all these cases, for each redshift range of interest, it will be natural to focus on the best characterized GW events. Generically, these are the ones with the highest SNR although, depending on the application that one has in mind, one might wish to perform the cut  by selecting directly the events with $\Delta d_L/d_L$ below a given value, or with an angular resolution below a given value (of course, these choices are correlated, as we also see from the scatter plots in \autoref{fig:BBH_scatter}).  It is therefore interesting to discuss in more detail the ensemble of detections with large SNR, or with small error on distance, or with good angular resolution,
that we generically call  `golden binaries'. In the left panel of \autoref{fig:ET_BBH_zhists} we showed the distribution in redshift of BBHs  at ET alone, for different cuts on the SNR, including very high thresholds (${\rm SNR}\geq 100$,  ${\rm SNR}\geq 150$ and ${\rm SNR}\geq 200$), while in the central panel we considered the events detected, according to the criterion ${\rm SNR}>12$, and we imposed further cuts based on  $\Delta d_L/d_L$, or (in the right panel), based on $\Delta\Omega_{90\%}$; \autoref{fig:ET2CE_BBH_zhists} showed the same result for BBH at ET+2CE. These plots show the number of events obtained in one year, for our reference value ${\cal R}_{0, {\rm BBH}}=\SI{17}{\per\cubic\giga\parsec\per\year}$ of the local BBH rate, taking into account the duty cycle, computed according to the assumptions in \autoref{sect:detnetworks}. As we already mentioned, it is important to keep in mind that, since we are looking for events in the tails of the distributions (whether with respect to SNR, $\Delta d_L/d_L$, or  $\Delta\Omega_{90\%}$), changes in the rate, or in the observation time, or in the duty cycle, can have significant effects. In particular, increasing one or the other, will in general lead to the appearance of rarer events, i.e. events with larger SNR, better distance measurements, and/or better angular localization.

From \autoref{fig:ET_BBH_zhists}, or from \autoref{tab:BBH_Summary}, we can appreciate that, already at ET alone (with our fiducial choices for the local rate, the redshift dependence of the merger rate, and the BBH mass function) there
will be $\order{800}$  BBH detections per year  with ${\rm SNR}\geq100$, of which a significant fraction is at  $z \gtrsim 1$, and  a few of them are even at redshifts as large as
$z\sim 3-4$;  in our sample realization, the farthest event with  ${\rm SNR}\geq100$ has $z\simeq 3.8$. In terms of accuracy on distance,  there are $\order{\num{e3}}$  events/yr with $\Delta d_L/d_L<0.1$, of which a large fraction is at $z>1$; furthermore, there are  $\order{200}$ events/yr with $d_L$ measured to better than $5\%$, again with a large fraction of them at $z>1$.

For ET+2CE, we see from \autoref{fig:ET2CE_BBH_zhists} or \autoref{tab:BBH_Summary} that these numbers raise by almost  one order of magnitude. There are now $\order{5000}$  events/yr with ${\rm SNR}\geq100$, of which $\order{150}$  events/yr have $z\gtrsim 3$. For the luminosity distance, we find $\order{1000}$  events/yr with $d_L$ measured better than $1\%$; again a significant fraction of them is at $z>1$, and, out of them,  $\order{40}$  events/yr are at redshifts as large as $z\sim 2-3$.
Thanks to triangulation, now also for the angular resolution we have remarkable results: there are $\order{1200}$   events/yr at $z\gtrsim 1$ localized to better than \SI{1}{\square\degfull}, and tens of events, still at cosmologically significant redshifts, localized to better than \SI{0.05}{\square\degfull}.  An especially remarkable result is that about 4  BBH systems per year at ET+2CE could be localised in a region as small as \SI{0.01}{\square\degfull} or lower, meaning that the true host galaxy could be identified even in absence of an EM counterpart. In fact, considering the COSMOS2020 survey \citep{Weaver:2021obz}, among the deepest and most complete galaxy surveys to date, we find that, on average, only a single galaxy is present in a region of \SI{0.01}{\square\degfull} and a redshift extension of 0.05 [in agreement with Eq. (7) of \cite{Singer:2016eax}, see also the discussion in \cite{Borhanian:2020vyr}].\footnote{We thank Nicola Borghi for providing this estimate.} Note that  the similarity of the shapes of  the distributions  for the SNR, for $\Delta d_L/d_L$ and for $\Delta\Omega_{90\%}$ confirms that the events with high SNR, good accuracy on $d_L$ and good angular localization are essentially the same, as was also visible, on the whole ensemble of detections, from the scatter plots in \autoref{fig:BBH_scatter}.

\begin{figure}[t]
    \centering
    \subfloat{
    \includegraphics[width=8cm]{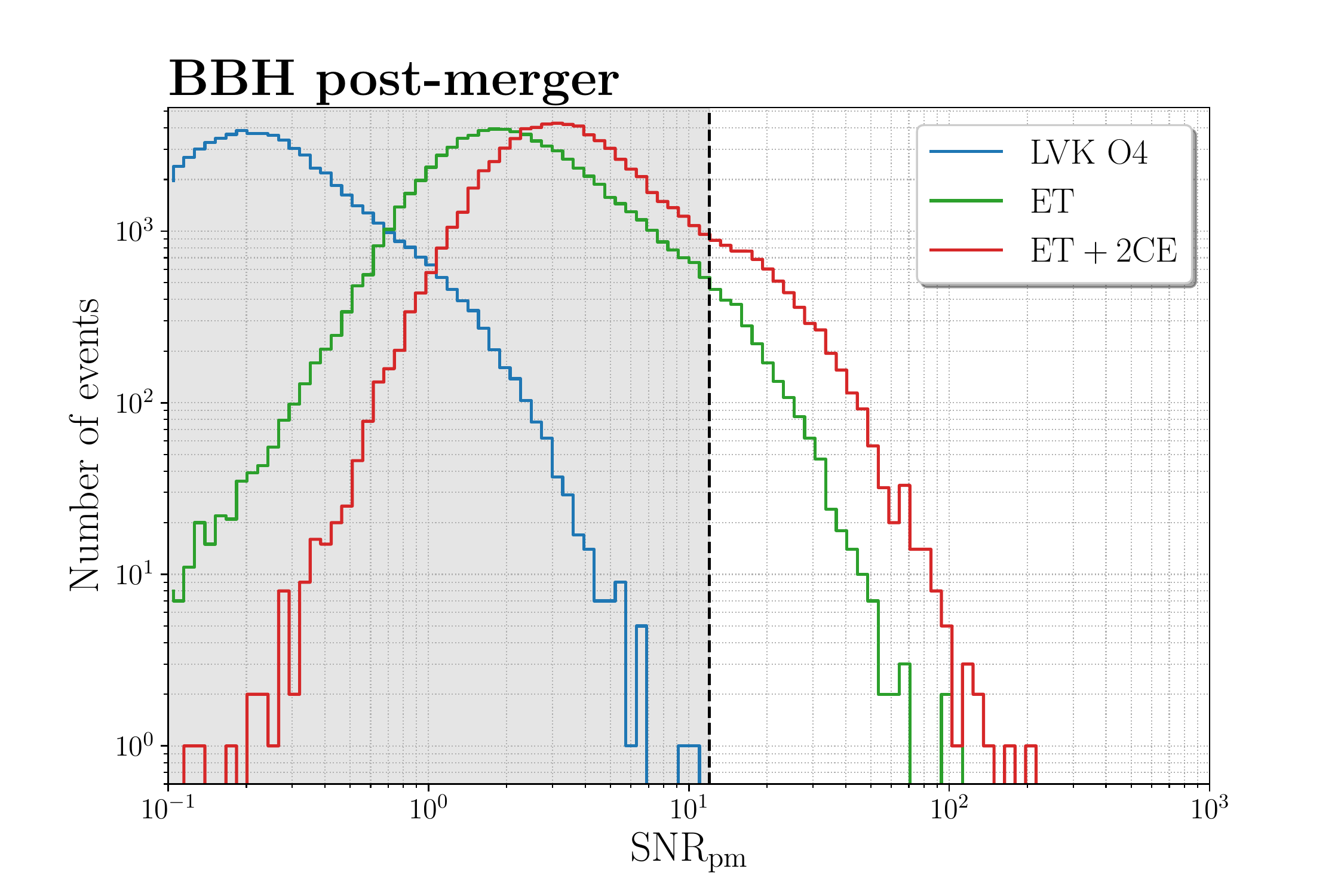}
    }
    \qquad
    \subfloat{
    \includegraphics[width=8cm]{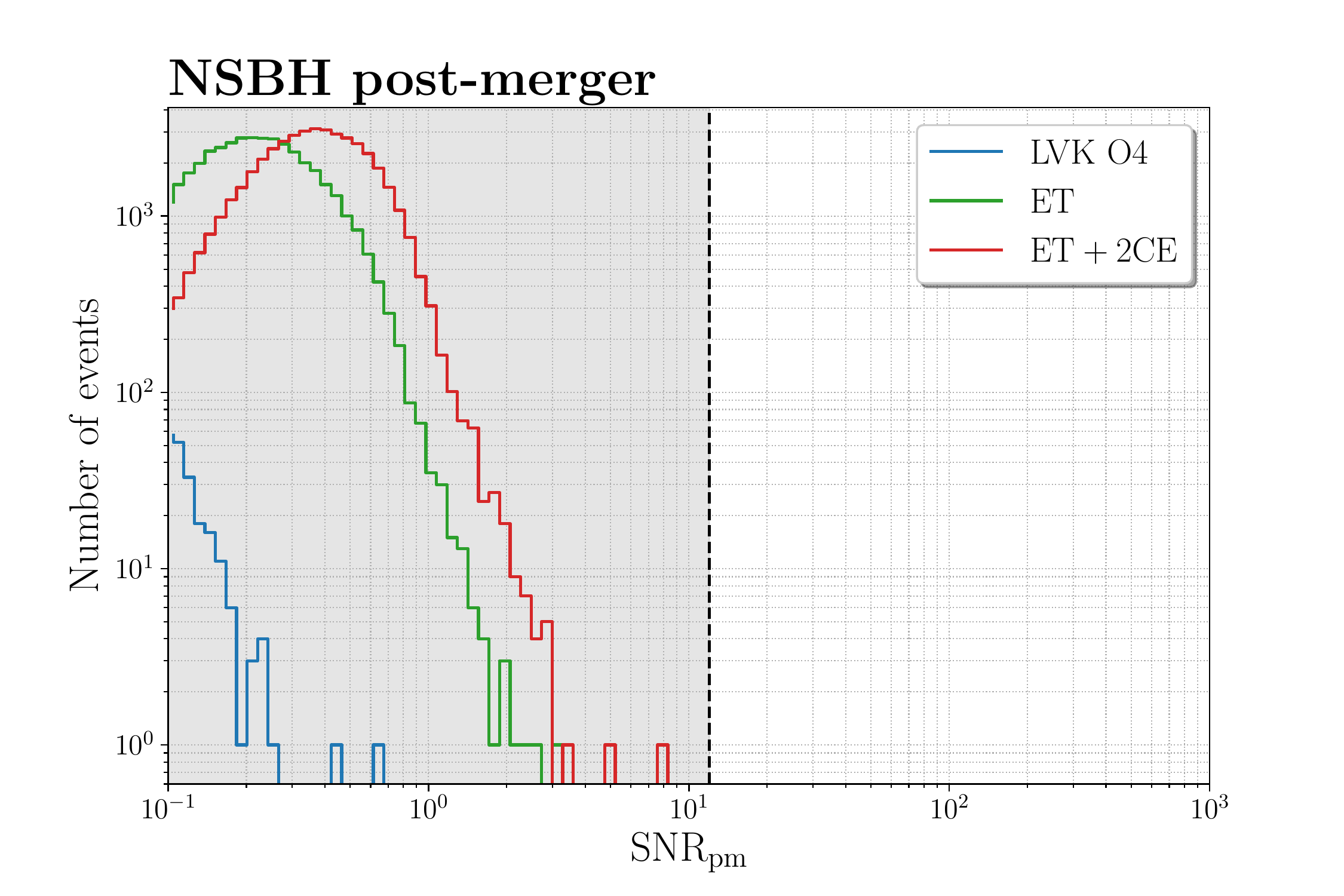}
    }
    \caption{Histograms of the SNR in the post--merger phase of BBH binaries (left panel) and NSBH binaries (right panel), for 1 year of data. For BBHs the computation is performed using the \texttt{IMRPhenomHM} waveform, while for NSBH systems we adopted \texttt{IMRPhenomNSBH}. 
    }    
    \label{fig:BBH_PostMerg_PhHM}
\end{figure}

For testing the near--horizon physics and the nature of BHs, as discussed above, it is crucial to have an accurate reconstruction of the ringdown phase. In the left panel of \autoref{fig:BBH_PostMerg_PhHM} we show the distribution of ${\rm SNR}_{\rm pm}$ for BBH systems,  where ${\rm SNR}_{\rm pm}$ is defined as the portion of the SNR obtained just from the post--merger phase, i.e. considering only the part of the  signal  above the peak frequency of the waveform.\footnote{This is the frequency at which the amplitude of the signal, normalised to the overall $f^{-\nicefrac{7}{6}}$ factor, reaches its maximum, determined internally by \texttt{IMRPhenomHM} according to Eq. (20) of \cite{Khan:2015jqa}.}
We see that, already at ET alone, there will be $\order{100}$  BBHs per year for which the SNR of the post--merger phase is larger than 30, with a few events per year reaching ${\rm SNR}_{\rm pm}\sim 80-100$. For ET+2CE these numbers further rise by a factor of a few and, for the best events, ${\rm SNR}_{\rm pm}$ can reach values of order $200$. In contrast, at LVK--O4, one can expect at most $\order{1}$ events with ${\rm SNR}_{\rm pm}\gtrsim 10$.

Some properties of `golden events' for BNS can be appreciated from
\autoref{fig:ET_BNS_zhists} for ET, and from \autoref{fig:ET2CE_BNS_zhists} for ET+2CE. Given that BNS are intrinsically less loud, now meaningful cuts  are at lower values of the SNR. With our fiducial choice for the population, at ET  there are $\order{800}$ events per year at  with ${\rm SNR}\geq30$, $\order{150}$ with  ${\rm SNR}\geq50$, and $\order{20}$ with
${\rm SNR}\geq100$, of which, in our sample, the farthest is at $z\simeq 0.19$. It is important to notice that these results have been obtained setting  ${\cal R}_{0, {\rm BNS}}=\SI{105.5}{\per\cubic\giga\parsec\per\year}$ and, as we repeatedly stressed, there is an uncertainty of one order of magnitude, in both directions, on this number. Similarly to the case of BBHs, for ET+2CE  there is a further  improvement by typically one order of magnitude, both for the number of sources with a given high value of the SNR and for those with a small value of $\Delta d_L/d_L$. For instance, the number of BNS/yr with ${\rm SNR}\geq50$ is $\order{1200}$. As usual, the improvement is even  more important on the localization, thanks to triangulation. While, for ET alone,  there are just about $\order{10}$ BNS/yr localized to better than \SI{50}{\square\degfull},  with ET+2CE there are $\order{2400}$  BNS/yr localized to better than \SI{10}{\square\degfull}, including events up to $z\sim 1.5$, and  $\order{100}$ events/yr localized to better than \SI{1}{\square\degfull}, including events  up to $z\sim 0.44$.
To put this result in the correct perspective, however, it is useful to recall, from the discussion in \autoref{sect:resBNS}, that, for multi--messenger observations, localization regions of the order of  hundreds of square degree, as can be obtained by ET alone, are already sufficient to trigger the search of a kilonova with large FOV telescopes, which are then needed to localize the source to the arcsec precision necessary to point instruments such as the ELT and perform spectroscopic studies. However, the remarkable improvement in localization between ET and ET+2CE will be very important in some cosmological applications, such as those making use of dark sirens by correlating them with galaxy catalogs.

Finally, \autoref{fig:ET_NSBH_zhists} and \ref{fig:ET2CE_NSBH_zhists} show the analogous results for NSBHs.
`Golden events' for NSBHs could be particularly interesting 
for cosmological applications, since these events, compared to BNSs, are louder because of the larger mass, while they could still have an electromagnetic counterpart, if the NS is tidally disrupted during the merger; furthermore, while the spin of NSs in binaries is expected to be negligible, the spin of the BH in a NSBH binary in general might not be small (see, however, \cite{Mandel:2020lhv,Broekgaarden:2021hlu,Mandel:2021ewy})
and, if it induces a precession of the orbital plane, it can lift the degeneracy between $d_L$ and $\cos\iota$, leading to an improved distance measurement~\citep{Vitale:2018wlg}.
The plots in \autoref{fig:ET_NSBH_zhists} and \ref{fig:ET2CE_NSBH_zhists} show a pattern intermediate between that for BBHs and that for BNSs. This is expected since,  as discussed in \autoref{sect:populations}, for the models that we have used, the typical BH mass in a NSBH binary is of order \SI{8}{\Msun}, so these systems are heavier than BNSs, but lighter than typical BBHs and, in the bandwidth of 3G detectors, they are louder than typical BNS, but  weaker than  typical BBHs. 
Finally, we observe that NSBH binaries look less promising than BBHs for testing the post--merger and ringdown phase of the final BH. 
In the right panel of \autoref{fig:BBH_PostMerg_PhHM} we show the distribution of ${\rm SNR}_{\rm pm}$ for these systems,
which shows that the signal--to--noise ratio in the post--merger phase never exceed 10, even for ET+2CE.\footnote{In this case the computation of the peak frequency is less straightforward, due to the various possible outcomes of the merger, depending on the possible disruption of the NS and the formation of an accretion torus \citep{1976ApJ...210..549L}. We thus proceeded according to the prescriptions in Sect. IV of \cite{PhysRevD.92.084050}, and checked with an actual numerical computation of the full waveform.} 

\section{Conclusions} \label{sec:conclusions}
 
In this paper we have performed  a comprehensive study of the capabilities of the Einstein Telescope, alone and in a network with two Cosmic Explorer detectors, and we have also provided  forecasts for the forthcoming LVK--O4 run. The results have been obtained using a 
new parameter estimation code, \codename{}, conceived especially for application to third--generation GW detectors. Our results broadly confirm the overall picture obtained in other recent studies of 3G detectors (see in particular
\cite{Maggiore:2019uih,Borhanian:2022czq,Harms:2022ymm,Ronchini:2022gwk}, and  references therein), on the extraordinary scientific potential of 3G detectors. We have applied our code to study populations of BBHs, BNSs and NSBH binaries. Compared to some of the other works devoted to forecast for 3G detectors, our results are more tuned toward ET, including detailed studies of what ET can do as a single detector; we used updated information on the astrophysical population from the LVK population studies based on the GWTC--3 catalog of detections; we used state--of--the--art waveforms for all these sources, and included the Earth's motion in the detectors response, which is especially relevant  for BNSs; we expanded the range of parameters on which inference is performed, including the spin of the BHs and the tidal deformability of NSs; we also presented the first systematic  study of parameter estimation for NSBH binaries at 3G detectors.

From the technical point of view our code, which is presented  in more detail in the companion paper \cite{Iacovelli:2022mbg}, has been designed so to optimally exploit the parallel nature of the problem of computing a large number of Fisher matrices for independent events, 
by vectorizing the evaluation 
of the  Fisher matrices even on a single CPU. This relies on the implementation of the waveforms in \texttt{Python}, 
resulting in a gain in computational speed, which motivates the name \codename{}. 
Furthermore, this allows the use of automatic differentiation with the library \texttt{JAX} for the accurate computation of derivatives.
Together with this paper, we publicly release the code \codename{} \raisebox{-1pt}{\href{https://github.com/CosmoStatGW/gwfast}{\includegraphics[width=10pt]{GitHub-Mark.pdf}}}, which is available at \url{https://github.com/CosmoStatGW/gwfast}. 
This paper is associated to version v1.0.1 which is archived on Zenodo \citep{gwfast_zenodo}.
We also release the library \wfname{} \raisebox{-1pt}{\href{https://github.com/CosmoStatGW/WF4Py}{\includegraphics[width=10pt]{GitHub-Mark.pdf}}}, available at \url{https://github.com/CosmoStatGW/WF4Py}, which contains state--of--the--art gravitational--wave waveforms in pure \texttt{Python}.
This paper is associated to version v1.0.0 which is archived on Zenodo \citep{WF4Py_zenodo}.

\begin{acknowledgments}
{\em Acknowledgments.} We are grateful to  Marica Branchesi, Jan Harms, Ulyana Dupletsa, Samuele Ronchini and the GSSI group for extremely useful discussions and comparisons of our respective codes and results,  all along this project. We also thank  Ssohrab Borhanian and B. S. Sathyaprakash for comparisons with their code and Andrea Maselli, Ik Siong Heng, Mauro Pieroni and the members of the ET OSB div.~9 for hosting discussions about comparisons of Fisher codes.
We thank Ed Porter for useful advice on the implementation of the code, Christopher Finlay for helpful discussion about \texttt{JAX} usage, and Floor Broekgaarden for a careful reading of the manuscript and extremely  useful comments.
The research leading to these results has been conceived and developed within the ET Observational Science Board (OSB).
Our research is supported by  the  Swiss National Science Foundation, grant 200020$\_$191957, and  by the SwissMap National Center for Competence in Research. Computations made use of the Yggdrasil cluster at the University of Geneva. 
\end{acknowledgments}

\software{\texttt{JAX} \citep{jax2018github}, \texttt{astropy} \citep{price2018astropy}, \texttt{LALSimulation} \citep{lalsuite}}, \texttt{HEALPix} \citep{Zonca2019}. 

\appendix
\section{Details of the adopted population distributions}\label{sec:appendix_adoptedDistr}
Here we report the shape and parameters of the population distributions mentioned in the main text. All these are available in the public code \texttt{MGCosmoPop}\footnote{\url{https://github.com/CosmoStatGW/MGCosmoPop}.} \citep{Mancarella:2021ecn}, together with sampling routines, which we used to obtain the catalogs.
\subsection{Redshift distribution}

For the BBH merger distribution with redshift we adopt a model that follows the star formation rate.  As a caveat, we mention that this is just a first approximation, and other effects could induce  departures from the star formation rate. In particular,  metallicity--dependent effects  might significantly boost the merger rate at low metallicities, see
\cite{Santoliquido:2020axb} (in particular their Fig.~2) and 
\cite{Chruslinska:2022ovf}.
Given the current uncertainties, a simple model of the merger rate following the star formation rate  can be considered a meaningful starting point that, however,  will have to be updated as our understanding of the merger rate of compact objects sharpens.

The Madau--Dickinson profile \citep{Madau:2014bja} for the star formation rate density has the functional form \citep{Madau:2016jbv}
\begin{equation}
    \psi(z| \alpha_z, \beta_z, z_p) \propto \frac{(1+z)^{\alpha_z}}{1 + \left(  \frac{1+z}{1+z_p}   \right)^{\alpha_z + \beta_z}} \, .
\end{equation}
Observationally, the merger rate evolution of BBHs is currently constrained only at small redshift; assuming a profile $\psi(z| \kappa) \propto (1+z)^{\kappa}$, 
the analysis of the first three LVK observing runs gives the median value $\kappa = 2.7$.\footnote{All the values of the population parameters given in this Appendix are taken from the results of the combined analysis of the events coming from the O1, O2 and O3 runs, available at \url{https://zenodo.org/record/5655785\#.YnUnPS8QN70}. All values are the medians of the estimations, contained in the file \texttt{PowerLawPeakObsOneTwoThree.json}\label{footnote:popGWTC3datarelease}.}
Here we use the Madau--Dickinson profile, with the low--redshift slope fixed to the value of $\alpha_z=2.7$. This is chosen to match the value of $\kappa$ since at low redshift the Madau--Dickinson profile reduces to $\sim(1+z)^{\alpha_z}$.
For the other parameters, we fix them to $\beta_z=3$ and $z_p=2$, which, within current uncertainties, are typical values used in the literature \citep{Madau:2014bja}.

For BNS systems, instead, we convolve the Madau--Dickinson profile with a time delay distribution $P(t_d)\propto1/t_d$, using a minimum time delay of \SI{20}{\mega\year}, as suggested in \cite{2012PhRvD..86l2001R}. To get a faster software implementation, we further re--fit  the curve obtained after the convolution to another Madau--Dickinson profile, with new  parameters $\alpha_z$, $\beta_z$ and $z_p$ now determined from the fit; we obtain the values  $\alpha_z = 1.42$, $\beta_z = 4.62$, $z_p = 1.84$, and this is what we use to generate the population (the overall rescaling to the local rate is discarded, since we always simulate a fixed amount of sources). A population with the same parameters is also adopted for NSBH systems. 

The distribution in redshift of the merger rate of compact binaries is finally 
\begin{equation}
    p_z(z) \propto \psi(z| \alpha_z, \beta_z, z_p) \frac{1}{1+z} \dfrac{{\rm d} V_c}{{\rm d}z}(z)\, ,
\end{equation}
where ${\rm d} V_c/{\rm d}z$ is the comoving volume element and the factor of $1+z$ relates the source times and the observer time. In a flat $\Lambda$CDM Universe, the redshift is related to the luminosity distance of the source by 
\begin{equation}
    d_L(z) = (1+z)\dfrac{c}{H_0} \int_0^z \dfrac{{\rm d}\tilde{z}}{\sqrt{\Omega_m (1+\tilde{z})^3+\Omega_{\Lambda}}}\, ,
\end{equation}
(the contribution of radiation is negligible even at the redshifts explored by 3G detectors),
and we use \textsc{Planck18} parameters \citep{Aghanim_Planck18}, i.e. $H_0 = \SI{67.66}{\kilo\meter\per\second\per\mega\parsec}$, $\Omega_m = 0.3097$ and $\Omega_{\Lambda} = 1 - \Omega_m = 0.6903$.\par\medskip
\subsection{Mass distribution}
The \textsc{Power Law + Peak} distribution \citep{LIGOScientific:2021psn}  used for the BBH mass function has the functional form, for the mass of the primary component (always taken as the heaviest),
\begin{fleqn}
\begin{equation}
     p(m_1|\lambda_{\rm peak}, \alpha, m_{\rm min},\delta_m,m_{\rm max},\mu_m,\sigma_m) = \left[(1 - \lambda_{\rm peak})\mathcal{P}(m_1|-\alpha,m_{\rm max}) + \lambda_{\rm peak}\mathcal{N}(m_1|\mu_m, \sigma_m)\right]S(m_1|m_{\rm min},\delta_m)
\end{equation}
\end{fleqn}
where $\mathcal{P}(m_1|-\alpha,m_{\rm max})$  is a normalized power--law distribution with spectral index $-\alpha$ and high--mass cut--off $m_{\rm max}$, $\mathcal{N}(m_1|\mu_m, \sigma_m)$ is a normalised Gaussian distribution with mean $\mu_m$ and standard deviation $\sigma_m$, while $S(m_1|m_{\rm min},\delta_m)$ is a smoothing function of the form
\begin{equation}
    S(m_1|m_{\rm min},\delta_m) = 
    \begin{dcases}
    0 & (m<m_{\rm min})\\
    [f(m-m_{\rm min},\delta_m)+1]^{-1} & (m_{\rm min}\leq m<m_{\rm min} + \delta_m) \\
    1 & (m\geq m_{\rm min} + \delta_m)
    \end{dcases}\,,
\end{equation}
with 
\begin{equation}
    f(m', \delta_m) = {\rm exp}\left\{\dfrac{\delta_m}{m'}+\dfrac{\delta_m}{m'-\delta_m}\right\}.
\end{equation}
The conditional mass ratio distribution of this model is 
\begin{equation}
    p(q|\beta_q,m_1,m_{\rm min},\delta_m)\propto q^{\beta_q} S(q m_1|m_{\rm min},\delta_m).
\end{equation}
The numerical values of the parameters we used for this distribution are $\lambda_{\rm peak} = 0.039,\ \alpha = 3.4,\ m_{\rm min} = 5.1,\ \delta_m = 4.8, \ m_{\rm max} = 87,\ \mu_m = 34,\ \sigma_m = 3.6,\ \beta_q = 1.1$.\textsuperscript{\ref{footnote:popGWTC3datarelease}}\par\medskip
\subsection{Spin distribution}
The \textsc{Default} distribution used to simulate the BBH spins [introduced in \cite{LIGOScientific:2018jsj}] assumes a Beta distribution for the spin magnitudes (taken to be independent) 
\begin{equation}
    p(\chi_{1,2}|\alpha_{\chi},\beta_{\chi}) = {\rm Beta}(\alpha_{\chi}, \beta_{\chi})\,,
\end{equation}
while the orientations are drawn from 
\begin{equation}
    p({\bf z}|\zeta,\sigma_t) = \zeta{\cal N}_t({\bf z}|0,\sigma_t) + (1-\zeta){\rm I}({\bf z})\,,
\end{equation}
where $z_i = {\rm cos}\theta_i$, $\theta_i$ is the tilt angle between the component spin and the binary's orbital angular momentum, ${\cal N}_t({\bf z}|0,\sigma_t)$ is a truncated Gaussian distribution centered at 0 and with standard deviation $\sigma_t$, while ${\rm I}({\bf z})$ denotes an isotropic distribution. Though this model could be used to sample all the 6 components of the spins of the two objects, we will always extract only the distribution of the spin component aligned with the orbital angular momentum, setting the others to zero. The numerical values of the parameters we used for this distribution are $\zeta=0.66,\ \sigma_t = 1.5,\ \mu_{\chi}= 0.28,\ \sigma_{\chi}= 0.03$,\textsuperscript{\ref{footnote:popGWTC3datarelease}} from which it follows $\alpha_{\chi} = 1.6,\ \beta_{\chi} = 4.12$.

\section{Comparison with other parameter estimation codes}
\label{app:comparison}
 
With the recent boost in activities for Einstein Telescope and Cosmic Explorer, parameter estimation for 3G detectors has become a particularly important topic, both as a tool to develop the science case for these detectors, and to inform the choices for  optimal detector's design and detector network configuration, so  several parameter estimation codes have been developed recently. In this appendix we compare our code and our results to those obtained with \texttt{GWBENCH}~\citep{Borhanian:2020ypi, Borhanian:2022czq}, developed at Penn  State, and with  \texttt{GWFISH}, developed by the GSSI group \citep{Harms:2022ymm} which, together with \codename{}, are among the most complete and advanced codes available, for parameter estimation at 3G detectors [see also  \cite{Chan:2018csa,Grimm:2020ivq,Nitz:2021pbr,Li:2021mbo,Pieroni:2022bbh}].

We begin by observing that, in the context of the activities of the ET Observational Science Board (OSB),
our group, the GSSI group and the Penn State group have analyzed
a small sample of common injections, using all the same waveform model [the \texttt{TaylorF2} waveform, developed in  \cite{Buonanno:2009zt}], finding excellent agreement, at the level of one part in $10^4-10^5$ for the SNR, and one part in $10^3$ on the parameter estimation of individual signals. This is an important consistency check for these rather complex codes, that had been developed totally independently. Further activity on this is in progress.

We next compare the results presented in our paper  with those presented in \cite{Borhanian:2022czq}, and  with those in \cite{Ronchini:2022gwk} (that  provides an extensive set of results for BNSs obtained with \texttt{GWFISH}).
The comparison  must take into account the fact that, currently, there are large uncertainties on several aspects, ranging from the detector configurations considered, the assumptions on the astrophysical models, as well as several differences in technical details, such as the waveform models used, the assumed threshold for the SNR, down to different implementations of the Fisher matrix inversion. To orient the reader, we therefore find useful first of all to give a summary of several different choices adopted in these works. We stress that all these different choices are  fully legitimate, and reflect current uncertainties. It is therefore instructive to see how the results change when using different assumptions. We also stress that, while the results, in our paper as well as in  \cite{Borhanian:2022czq} and in \cite{Harms:2022ymm,Ronchini:2022gwk},   are necessarily presented just for a  selection of few different choices, the codes themselves can be adapted to a larger variety of choices.
The main differences in the assumptions used in these works can be summarized as follows.

\vspace{2mm}\noindent
{\em Detector networks}. Our work is more focused on ET. To avoid a proliferation of plots, or of lines in each plot, we only presented the results for two 3G configurations: ET alone, and ET+2CE. Studying ET alone allows us to assess the strength of the ET Science Case, independently of decisions that will be taken by different funding agencies on CE. On the other hand, combining ET with two CE detectors allows us to examine the full strength of a 3G detector network.  The two CE detectors are assumed to be one with \SI{40}{\kilo\meter} arms, and one with \SI{20}{\kilo\meter} arms, which is the reference CE configuration~\citep{Evans:2021gyd}. We 
set both of them in the U.S. (the other options that is under consideration, with one of them in Australia, allows  better triangulation but, otherwise, is not very different). In contrast,
in \cite{Ronchini:2022gwk}  are studied ET alone, ET+CE (\SI{40}{\kilo\meter}) and ET+2CE, with both CE detectors with \SI{40}{\kilo\meter} arms, and  one of them located in Australia; compared to the ET+2CE network that we have studied, the latter network therefore has a greater horizon to coalescing binaries, because of the 40+40 km arms, and   better sky localization, because one of the CE detectors is set in Australia. 
The work in~\cite{Borhanian:2022czq} studies a larger set of combinations of 2G\textsuperscript{+} and 3G configurations, somewhat more tuned toward CE. ET alone is not considered, but is studied in a so--called ‘ECS' network, consisting of ET and 2CE, both having \SI{40}{\kilo\meter} arms, with a CE located in Australia.\footnote{For performing detailed comparison among our results, it should also be observed  that that the  PSDs of CE that we use, available at
\url{https://dcc.cosmicexplorer.org/CE-T2000017/public}
are an update of the sensitivity curves available at
\url{https://dcc.cosmicexplorer.org/public/0163/T2000007/005/}, that have been used in \cite{Borhanian:2022czq}.}

\vspace{2mm}\noindent
{\em Astrophysical populations}. In \cite{Borhanian:2022czq} are given predictions for BBHs and BNSs, and  \cite{Ronchini:2022gwk} focus on BNSs,  while we provide predictions for BBHs, BNSs and NSBHs.
As we repeatedly stressed in the text, there are currently large uncertainties on the BBH, BNS and NSBH populations, and the fiducial choices made in different works can be sensibly different, resulting potentially in differences as large as one order of magnitude. 

For BNSs, we assume  a redshift distribution of the merger rate given by a rather standard Madau--Dickinson distribution with a delay of 20~Myr between formation and merger, as
described in \autoref{sec:appendix_adoptedDistr};  the same choice, with the same time delay, is made in \cite{Borhanian:2022czq}, while \cite{Ronchini:2022gwk} use a different model, 
based on a catalog generated according to the recent study in \cite{Santoliquido:2020axb}. Furthermore, the overall normalizations are different. Our fiducial value for the local rate is
${\cal R}_{0, \rm BNS} = \SI{105.5}{\per\cubic\giga\parsec\per\year}$, which is the median value for 
${\cal R}_{0, \rm BNS}$  inferred  from the GWTC--3 catalog, using the same flat mass distribution that we are using~\citep{LIGOScientific:2021psn}. In contrast, \cite{Borhanian:2022czq} use 
${\cal R}_{0, \rm BNS} = \SI{320}{\per\cubic\giga\parsec\per\year}$, which is 
the median value  inferred  from the GWTC--2 catalog (actually, with a flat mass distribution, while \cite{Borhanian:2022czq} use a truncated Gaussian distribution), and \cite{Ronchini:2022gwk}
use ${\cal R}_{0, \rm BNS} = \SI{365}{\per\cubic\giga\parsec\per\year}$ (with the same  flat mass distribution between 1 and $\SI{2.5}{\Msun}$ that we also use). As a result of these differences, the rate of BNS mergers in our population model is \num{1e5} events/yr, in \cite{Borhanian:2022czq} is \num{4.7e5} events/yr, and in
\cite{Ronchini:2022gwk} is \num{9e5}. These numbers therefore span one full order of magnitude, and are the main reason for  the differences among our results for BNSs. Our estimates are the most conservative among these works, but we stress again that all these assumptions are consistent with current knowledge, and, in fact, none of them are on the extreme side. As we mentioned, values of  ${\cal R}_{0, \rm BNS}$ in the range $(10-1700)~\si{\per\cubic\giga\parsec\per\year}$ are still consistent with current observations, so one could consider even more conservative, or even more optimistic  models.\footnote{Earlier works, based on the GWTC--1 catalog \citep{LIGOScientific:2018mvr},   used even larger values of 
${\cal R}_{0, \rm BNS}$, close to  ${\cal R}_{0, \rm BNS} = \SI{1000}{\per\cubic\giga\parsec\per\year}$. For instance, in \cite{Belgacem:2019tbw} were used the two values  ${\cal R}_{0, \rm BNS} = \SI{662}{\per\cubic\giga\parsec\per\year}$ and
${\cal R}_{0, \rm BNS} = \SI{920}{\per\cubic\giga\parsec\per\year}$ given in \cite{LIGOScientific:2018mvr} 
for 
a flat or a Gaussian mass distribution,  respectively, while in  \cite{Sathyaprakash:2019rom} the value
${\cal R}_{0, \rm BNS} = \SI{1000}{\per\cubic\giga\parsec\per\year}$ was used.}

For BBHs, both us and  \cite{Borhanian:2022czq} use a  Madau--Dickinson distribution (with negligible delay  between formation and merger for us, and a delay of 10~Myr for \cite{Borhanian:2022czq}). For the local rate,
current constraints  are more stringent than for BNSs. Still, we use ${\cal R}_{0, {\rm BBH}}=\SI{17}{\per\cubic\giga\parsec\per\year}$, which is  the median value from GWTC--3, while  \cite{Borhanian:2022czq} use ${\cal R}_{0, {\rm BBH}}=\SI{24}{\per\cubic\giga\parsec\per\year}$ from GWTC--2. As a result, our BBH merger rate is 
\num{7.5e4} events/yr, while in  \cite{Borhanian:2022czq} is \num{1.2e5} events/yr.

\vspace{2mm}\noindent
{\em Waveforms and detection model}. A number of other differences are related to the waveforms and the detection model assumed.
For BNS, we use  the \texttt{IMRPhenomD\_NRTidalv2} waveform, as in  \cite{Borhanian:2022czq},
which includes inspiral, merger, ringdown, as well as tidal effects, while \cite{Ronchini:2022gwk}
uses \texttt{TaylorF2}, a simpler inspiral--only  waveform, and truncates it at a frequency of the order of 
$4f_{\rm ISCO}$, where  $f_{\rm ISCO}$ is the frequency at the innermost stable orbit [however,  \texttt{GWFISH} can use all waveforms models of the LALSimulation~\citep{lalsuite}]. We all include the effect of the Earth's rotation on the localization of the BNSs.
For BBHs both us and  \cite{Borhanian:2022czq} use \texttt{IMRPhenomHM}, which includes higher modes. There is, however, a difference on the number of parameters on which the inference is performed, both for BBHs and BNSs. In 
\cite{Borhanian:2022czq} the inference is performed on the standard set of 9 parameters
$\{{\cal M}_c,\ \eta,\ d_L,\ \theta,\ \phi,\ \iota,\ \psi,\ t_c,\ \Phi_c\}$, while we add to this list also the  two parallel spin components 
$\chi_{1,z}$ and $\chi_{2,z}$ and, for BNS and NSBH systems, we also perform the inference on  the tidal deformability parameters, see \autoref{sect:Results}.

Another difference concerns the threshold for the SNR. A lower threshold on the SNR increases the number of detections, but (beside raising the false--alarm rate in an actual run) at the level of our analysis it makes less reliable the parameter estimation made with the Fisher matrix. In this work, we present the bulk of our results with a detection threshold ${\rm SNR}\geq 12$, while
\cite{Borhanian:2022czq} use ${\rm SNR}\geq 10$; in 
\cite{Ronchini:2022gwk}, which is concerned with multi--messenger observation of BNS, ${\rm SNR}\geq 8$ is used for the detection in survey mode (which can be more appropriate to multi--messenger observations, given the temporal coincidence with the electromagnetic signal), but it has been checked that in pointing mode, where the sky localization is used, most of the events have ${\rm SNR}\geq 12$. 

Another significant point, especially for detector networks, is the assumption on the duty cycle. In this work, as in \cite{Ronchini:2022gwk}, for 3G detectors  we have assumed  an uncorrelated duty cycle of $85\%$ for each arm, while in \cite{Borhanian:2022czq} a duty cycle of $100\%$ is assumed. As we showed in 
 \autoref{fig:HistBBHcomp_PhenDvsHM}, including a realistic  duty cycle  results in long tails of  events with poor accuracy on  $\Delta d_L/d_L$ and, to lesser extent, of $\Delta\Omega_{90\%}$, which are missed assuming a $100\%$ duty cycle. The same remark applies to the results presented for BBHs in
\cite{Pieroni:2022bbh}, which agree with ours once we also assume a $100\%$  duty cycle.

As an example of the role of these different assumptions, we perform a more detailed comparison with the results reported in 
\cite{Ronchini:2022gwk} for the angular resolution of BNS at ET alone,  shown in their  Tab.~6, and in the cumulative distribution functions of their  Figs.~A1 and A2. In particular, they find that, at ET alone,  one would detect $370\, {\rm events/yr}$ with $\Delta\Omega_{90\%}<\SI{100}{\square\degfull}$, and 2~events/yr with $\Delta\Omega_{90\%}<\SI{1}{\square\degfull}$, while, from the results in \autoref{sect:resBNS}, 
we find 
$51\, {\rm events/yr}$ with $\Delta\Omega_{90\%}<\SI{100}{\square\degfull}$, and $2\, {\rm events/yr}$ event below $\SI{10}{\square\degfull}$. 
As a test, to isolate the effect of the choice of  population, we have performed further runs using ${\cal R}_{0, \rm BNS} = \SI{365}{\per\cubic\giga\parsec\per\year}$, ${\rm SNR}\geq 8$ and
the \texttt{TaylorF2} waveform,\footnote{However keeping the cut frequency of the waveform to $2 f_{\rm ISCO}$, rather than using $4 f_{\rm ISCO}$ as in \cite{Ronchini:2022gwk}.} so that the only difference is now in the model for the redshift evolution of the population. Under these conditions, 
in our model the population consists of \num{3.5e5} BNS mergers per year,  to be compared with \num{9.0e5} in \cite{Ronchini:2022gwk}, which is higher by a factor $\simeq 2.6$. Out of these mergers, we find that ET alone would detect  \num{8.6e4} events/yr, to be compared with  \num{1.4e5} events/yr in \cite{Ronchini:2022gwk}, which is now larger only by a factor $\simeq 1.7$. This is due to the fact that the excess in the population used in \cite{Ronchini:2022gwk}, compared to ours, is especially significant at large $z$ where, however, BNSs are more difficult to detect.
Then, selecting the events with $\Delta\Omega_{90\%}<\SI{1}{\square\degfull}$, $\SI{10}{\square\degfull}$ and $\SI{100}{\square\degfull}$ we find, respectively,
$(1,\,16,\,298)$ events/yr, to be compared with $(2,\,10,\,370)$ in \cite{Ronchini:2022gwk}.  Given the remaining difference in the redshift distribution of the population, we can then conclude that the results are fully consistent, once we make the same assumptions on local rate,  and we use the SNR threshold and the same waveform. This underlines that, within current uncertainties of the rate and on the population model, and different choices for technical aspects such as waveform, SNR, etc,   the results for BNSs can still vary sensibly.
A comparison with the results reported in   \cite{Borhanian:2022czq} for  their ECS configuration with our results for ET+2CE also show that our results broadly consistent, once taken into account the different rates, SNR threshold and network configuration.\footnote{In the same sense, our results are also broadly consistent with previous results presented in  \cite{Chan:2018csa,Maggiore:2019uih,Grimm:2020ivq,Nitz:2021pbr}. They also agree with those presented in \cite{Pieroni:2022bbh} (which are limited to BBHs and set a threshold ${\rm SNR}\geq 20$); indeed a number of cross--checks between our codes were performed prior to publication of \cite{Pieroni:2022bbh}.}

\vspace{2mm}\noindent
{\em Technical differences}. Among the various technical aspects of the code, one worth mentioning here 
is the  difference in the treatment of ill--conditioned Fisher matrices. In the present work these are  discarded, on the basis of a criterion on the inversion error discussed in \autoref{sec:singularities} and \autoref{sec:appendix_comparisonInv}. A similar strategy is adopted in \texttt{GWBENCH}, where the criterion is rather based on the condition number, i.e. the ratio of the largest to the smallest eigenvalue. In contrast,  in 
\texttt{GWFISH}, ill--conditioned Fisher matrices are conditioned and inverted (i.e., after passing to a diagonal basis, the lines and columns corresponding to the eigenvector with almost vanishing eigenvalue are deleted, and the remaining reduced--order matrix is inverted) \citep{Harms:2022ymm}. This leads to treating a larger number of events with the Fisher matrix technique; however, the vanishing, or nearly vanishing, of an eigenvalue means that the linear approximation on which the Fisher matrix technique is based becomes invalid, and one must  worry about the effect of  non--linear contributions; the basis in parameter space that diagonalizes the likelihood at quadratic order, in general, will no longer diagonalize it when cubic and higher--order terms are included; so the parameter that, at quadratic order, has vanishing eigenvalue and is discarded, beyond quadratic order could have non--trivial correlations with all other parameters, possibly leading to larger marginalized errors on them, compared to the answer obtained from the quadratic approximation, i.e. from the Fisher matrix.
For this reason, we have chosen to discard events with ill--conditioned Fisher matrices. In \autoref{sec:appendix_comparisonInv} we show a comparison of different techniques.
Note that, however, \codename{} offers different choices, including the conditioning of the singular values,
and the final choice is left to the user (see \cite{Iacovelli:2022mbg} for a detailed description). 

As we see from \autoref{tab:BBH_Summary}, for BBHs the number of events that we have discarded because the Fisher matrix was ill--conditioned is about $4\%$ of the total number of detected events at ET, and just $0.3\%$ for ET+2CE. Therefore, at the population level, the effect is quite marginal. For BNSs we see from \autoref{tab:BNS_Summary} that the number of events discarded is slightly higher but still consistent,
$4.5\%$ at ET and $1.3\%$ at ET+2CE. The effect of different treatments of the problem could then in principle be relevant. 
On the other hand,  discarding ill--conditioned matrices is not a perfect solution either, since it can potentially introduce some bias; for instance, events with $\cos\iota$ close to one are preferentially discarded, since in this limit the derivative with respect to $\iota$ vanishes. Basically, we are facing here  an intrinsic limitations of the Fisher matrix technique, and it is reassuring that codes employing different approaches reach consistent conclusions. At present, the uncertainty in the results is completely dominated by the astrophysical uncertainties.

\section{Comparison among different ways to quantify uncertainty from the Fisher matrix}\label{sec:appendix_comparisonInv}

\begin{table}[t]
    \centering
    \begin{tabular}{!{\vrule width .09em}c|c!{\vrule width .09em}}
    \toprule\midrule
        Parameter  & Prior range\\
        \midrule\midrule
         $\mathcal{M}_c,\ d_L$  & $(0,\, +\infty)$\\
         \midrule
         $\eta$ & $(0,\, 0.25]$\\
         \midrule
         $\phi,\ \Phi_c$ &  $[0,\, 2\pi]$ \\
         \midrule
         $\theta,\ \iota,\ \psi, $ & $[0,\, \pi]$\\
        \midrule
        $t_c$ & $[0,\, 1]$\\
        \midrule
        $\chi_{i,z}$ $(i=\{1,2\})$ &  $[-1,\,1]$ \\
        \midrule
        $\Lambda_{i}$ $(i=\{1,2\})$ & $[0,\, +\infty)$\\
  
    \midrule\bottomrule
    \end{tabular}
    \caption{Summary of the prior ranges used when sampling from the posterior probability.}
    \label{tab:priors}
\end{table}

In this appendix we provide a comparison among different methods of extracting the forecasted uncertainty on the waveform parameters from the Fisher matrix, which motivates our choice for this paper as discussed in \autoref{sec:singularities}.
As a test case we use the sample of BNSs, which is generally the most prone to showing large condition numbers.
%
In particular, discarding matrices with condition number higher than the inverse machine precision ($10^{-15}$ in our case) would lead to having only a few events left, which does not allow a statistical study of the population unless the even more stringent and less realistic assumption of discarding some parameters (e.g. neglecting the presence of spins and tidal deformabilities) is made.
The options that we consider are (see \cite{Iacovelli:2022mbg} for a detailed description of the methods and their implementation in \codename{}): 
\begin{enumerate}[noitemsep, label=\alph*)]
    
    \item we invert the Fisher matrix by a  Cholesky decomposition, compute the inversion error as defined in \eqref{eq:invError}, and exclude events with inversion error less than a chosen threshold $\epsilon_{\rm max}$. In particular, we compare the choices $\epsilon_{\rm max}=\num{e-3}$ and $\epsilon_{\rm max}=\num{5e-2}$;
    \item we invert all Fisher matrices by using the singular--value decomposition (SVD), and exclude from the inversion the singular values below a threshold of $10^{-10}$, to ensure that the numerical error does not propagate to well--measured parameters \citep{Harms:2022ymm}. Note that the inversion error is not defined in this case since the singular values of the matrix are regularized before inversion;
    \item We impose physically--motivated priors on top of the likelihood in \eqref{eq:LSAlikelihoodBayes}, draw samples from the corresponding posterior, and quantify the resulting forecasted uncertainty from the samples. This procedure is detailed below and does not require any inversion of the Fisher matrix, hence it is not sensitive to numerical instabilities and encodes the correct boundaries for the waveform parameters. On the other hand, this technique is computationally too expensive to be used for all catalogs; we only resort to it in this case for a validation of the forecasts.
\end{enumerate}

Let us describe the method used for the third option. The posterior probability for source parameters $\vb*{\theta}$ given data $s$ is defined as 
$p(\vb*{\theta} \,|\, s) \propto \pi({\vb*{\theta}}) \, \mathcal{L}(s \,|\, \vb*{\theta})$ where the likelihood $\mathcal{L}(s \,|\, \vb*{\theta})$ is defined in \eqref{eq:LSAlikelihoodBayes} and $\pi({\vb*{\theta}})$ is the chosen prior. We use flat priors with the boundaries summarized in \autoref{tab:priors} and also fix $\langle {\delta \theta}^i \rangle = 0$, i.e. we assume that the posterior mean coincides with the true value, that we denote by $\vb*{\mu}$. This is the same assumption made when quantifying the uncertainty directly from the Fisher matrix. 
Written in the form of \eqref{eq:LSAlikelihoodBayes}, the evaluation of the likelihood does not require the explicit inversion of the Fisher Matrix, and one can extract samples from this distribution.
For each event in the simulated catalogue of detections, we draw samples from the corresponding posterior by the following procedure: 
\begin{enumerate}[ label=\arabic*)]
    \item for variables whose $1\sigma$ uncertainty as forecasted by the Fisher analysis is larger than the prior range, we add a Gaussian prior with standard deviation equal to the prior range; this has only the purpose of improving the efficiency and numerical stability, and such prior is undone during the sampling procedure; 
    \item We sample from the likelihood \eqref{eq:LSAlikelihoodBayes} as follows. First, we generate a vector $\vb*{z} = (z_1,\ ...,\ z_{D})$ of independent, standard normal variables (with $D$ being the dimensionality of the Fisher matrix). Then, we find the Cholesky decomposition $\vb*{C}$ of the Fisher matrix $\vb*{\Gamma}$, such that $\vb*{C} \vb*{C}^{T} = \vb*{\Gamma}$. Note that this decomposition might not be found if the Fisher matrix is ill--conditioned, hence the need of the Gaussian prior discussed in the previous step, which ensures regularization.
    Then, the desired samples are given by $\vb*{\mu} + \vb*{X} $ where $\vb*{X}$ is the solution of the system $\vb*{C}^T \vb*{X} = \vb*{z}$, and $\vb*{\mu}$ the vector of the true values;
    \item We use rejection sampling to discard samples outside the prior range, as well as to undo the Gaussian prior previously imposed to regularize the Fisher matrix.
\end{enumerate}

%
%

The uncertainty is finally quantified by computing the standard deviation of the samples. The only caveat concerns the use of \eqref{eq:skyLoc} to determine the sky localization, since its derivation assumes a multivariate Gaussian distribution, while the use of priors can spoil this assumption. Hence we compute the $90\%$ sky localization by projecting the samples on the sphere with a \texttt{HEALPix} pixelization \citep{Zonca2019},\footnote{\url{https://healpix.sourceforge.io}.} locating the pixels within the $90\%$ localization region, and computing the corresponding sky area. This procedure is exact, but can require a very extensive and expensive sampling for events whose localization is not well determined. A second caveat is that the number of pixels used must be adapted on an event--by--event basis.
In order to reduce the computational cost, we only extract $2000$ samples/event, which can make this procedure underestimate the sky localization for poorly localized sources. However, this is enough for an accurate computation for well--localized events, in which case this provides an important numerical confirmation of the validity of \eqref{eq:skyLoc}.
As an alternative check, we compute the sky localization by binning the samples in $(\theta, \phi)$, finding the bins within the $90\%$ localization region, and computing the corresponding area on the sphere. This procedure does not require a large number of samples, but it is sensitive to the number of bins used, which is a quantity that has to be fixed \emph{a priori}, and whose mis--specification can lead to wrong estimates.
In summary, there is no fast and $100\%$ accurate way to compute the localization region, but comparing the techniques listed above we will show that the different techniques agree very well for events with sufficiently small ($\lesssim 10^{3}\  \rm deg^2$) localization region.

\begin{figure}
    \centering
    \includegraphics[width=.9\textwidth]{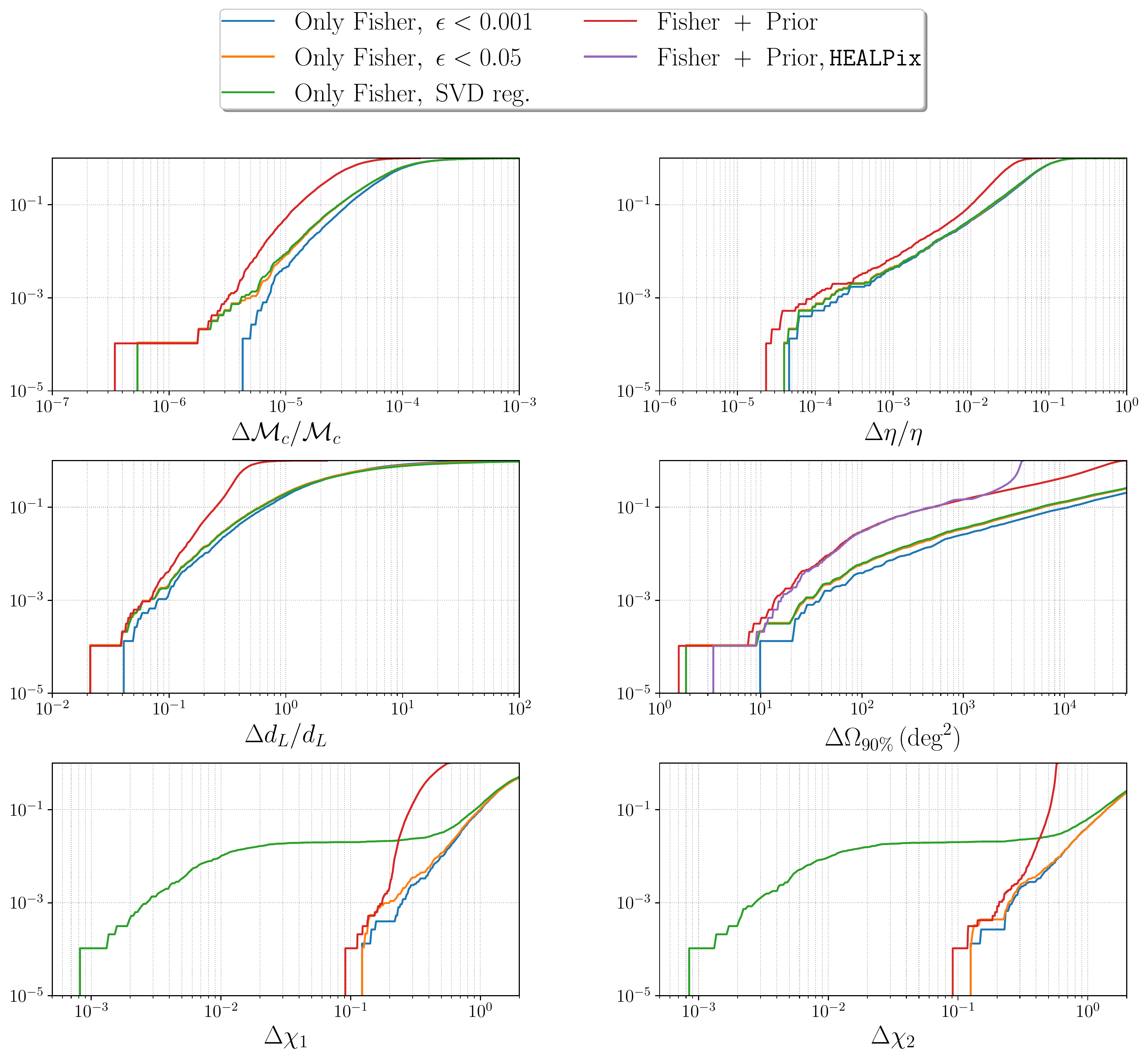}
    \caption{Comparison of cumulative distributions of the errors on the detector--frame chirp mass, symmetric mass ratio, luminosity distance, $90\%$ sky localization and spin components along the $z$ axis, for BNS observed at ET, with different choices to quantify the uncertainty. The default choice for this paper, $\epsilon<0.05$, corresponds to the orange curve. }
    \label{fig:comparison_inversion_ET}
\end{figure}

\begin{figure}
    \centering
    \includegraphics[width=.9\textwidth]{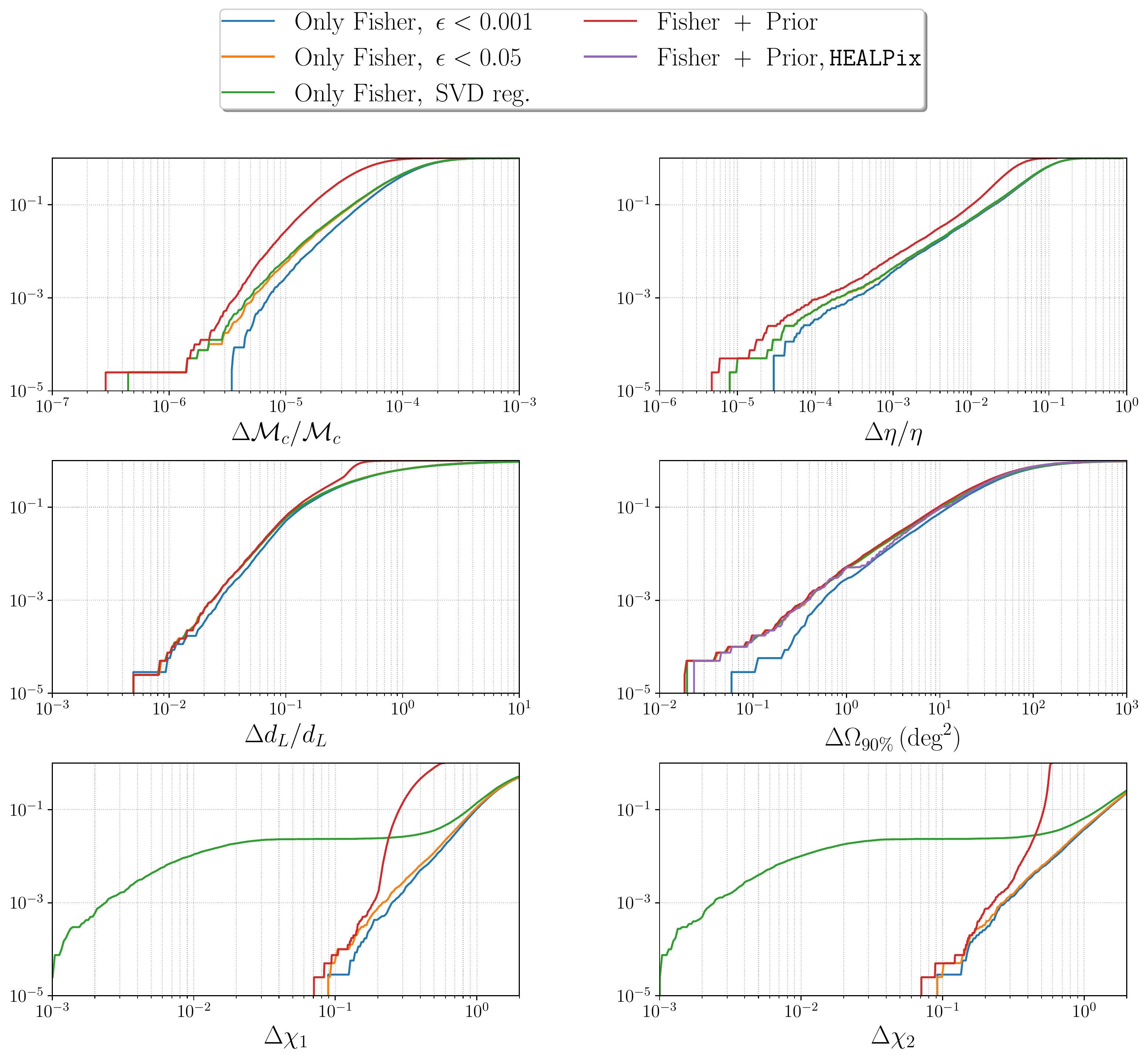}
    \caption{As in \autoref{fig:comparison_inversion_ET}, for a network of ET and two CE. }
    \label{fig:comparison_inversion_ET2CE}
\end{figure}

\autoref{fig:comparison_inversion_ET} and \autoref{fig:comparison_inversion_ET2CE} summarize our results. We show the cumulative distribution of the forecasted $1\sigma$ errors on the detector--frame chirp mass, symmetric mass ratio, luminosity distance, $90\%$ sky localization and spins along the $z$ axis, which are the most interesting for the discussion. \autoref{fig:comparison_inversion_ET} shows the results for ET alone, while \autoref{fig:comparison_inversion_ET2CE} for ET and two CE.
%
%
%
%
From these figures we can draw a number of interesting conclusions.

First, we see that for the left tails of the distributions, corresponding to well--resolved events, the result obtained with the SVD inversion and regularization (green curve) and the one obtained setting a threshold on the inversion error of $\epsilon_{\rm max}=\num{5e-2}$ (orange curve) agree almost perfectly for all variables except the spin, in which case the regularization of the singular values leads to severely underestimating the error. We conclude that degeneracies among the spin parameters are those that mostly affect the condition number and lead to the presence of very small singular values, which are consequently regularized. Hence, the regularization procedure should not be adopted if one is interested in forecasting errors on spin parameters.
Secondly we note that, adopting a threshold on the inversion error which is too stringent, such as $\epsilon_{\rm max}= \num{e-3}$ (blue curves), can lead to missing the tail of the best resolved events for some parameters, in particular the chirp mass. On the contrary, the choice $\epsilon_{\rm max}=\num{5e-2}$ ensures that all the lower tails of the distributions match perfectly both the distribution obtained with SVD and the one obtained by sampling (red curves). This motivates the choice of this threshold in this paper.

Finally, we comment on the errors obtained with the sampling procedure (red curves). As expected, the left tail of the distributions, corresponding to very--well measured parameters, match the results obtained by the inversion of the Fisher matrix. In general, the use of priors affects differently different parameters and different configurations.
For the mass parameters, we note a general improvement, which can be explained by the fact that most of the detected events lie very close to the threshold value $\eta=0.25$ for the symmetric mass ratio. This leads the finite prior range $\eta\leq0.25$ to play a significant role in reducing the forecasted uncertainty; in turn, it also affects the constraint on the chirp mass, due to the correlation of this parameter with $\eta$.
Similarly, the relative error on the luminosity distance is reduced at ET alone by the use of a prior because this reduces the allowed range for the inclination angle $\iota$, whose degeneracy with $d_L$ contributes significantly to the error on the latter. For ET in combination with two CE, this effects is mitigated because $\iota$ is generally better constrained without the need of a prior. 
As for the spins, these are generally poorly constrained and subject to strong degeneracies among the two components; this explains the improvement given by the prior.
We conclude with the sky localization. The red curve corresponds to the results obtained by the binning procedure described above, while we additionally show in violet the results obtained by the \texttt{HEALPix} projection. As anticipated, both methods match well the forecast obtained by the Fisher matrix alone for $\Delta \Omega_{90 \%} \lesssim 10^{3}\  \rm deg^2$, providing a confirmation of \eqref{eq:skyLoc}. 
The use of priors has a larger effect on the forecasts for ET alone, because for this configuration a precise localization is more difficult. In this sense, the Fisher forecast provides a conservative estimation of the localization capabilities.
The method based on \texttt{HEALPix} (violet lines) shows some disagreement for localization regions larger than $\sim 10^{3}\  \rm deg^2$, due to the limited number of samples available, which do not sufficiently cover all the pixels. Drawing more samples would lead to better convergence also of the upper tail.
This results confirm that the use of of \eqref{eq:skyLoc} is reliable for all events with sufficiently small localization regions.

\section{Comparison with realistic neutron star equations of state}\label{sec:appendix_comparisonNSEoS}
\begin{figure}
    \centering
    \includegraphics[width=0.91273\textwidth]{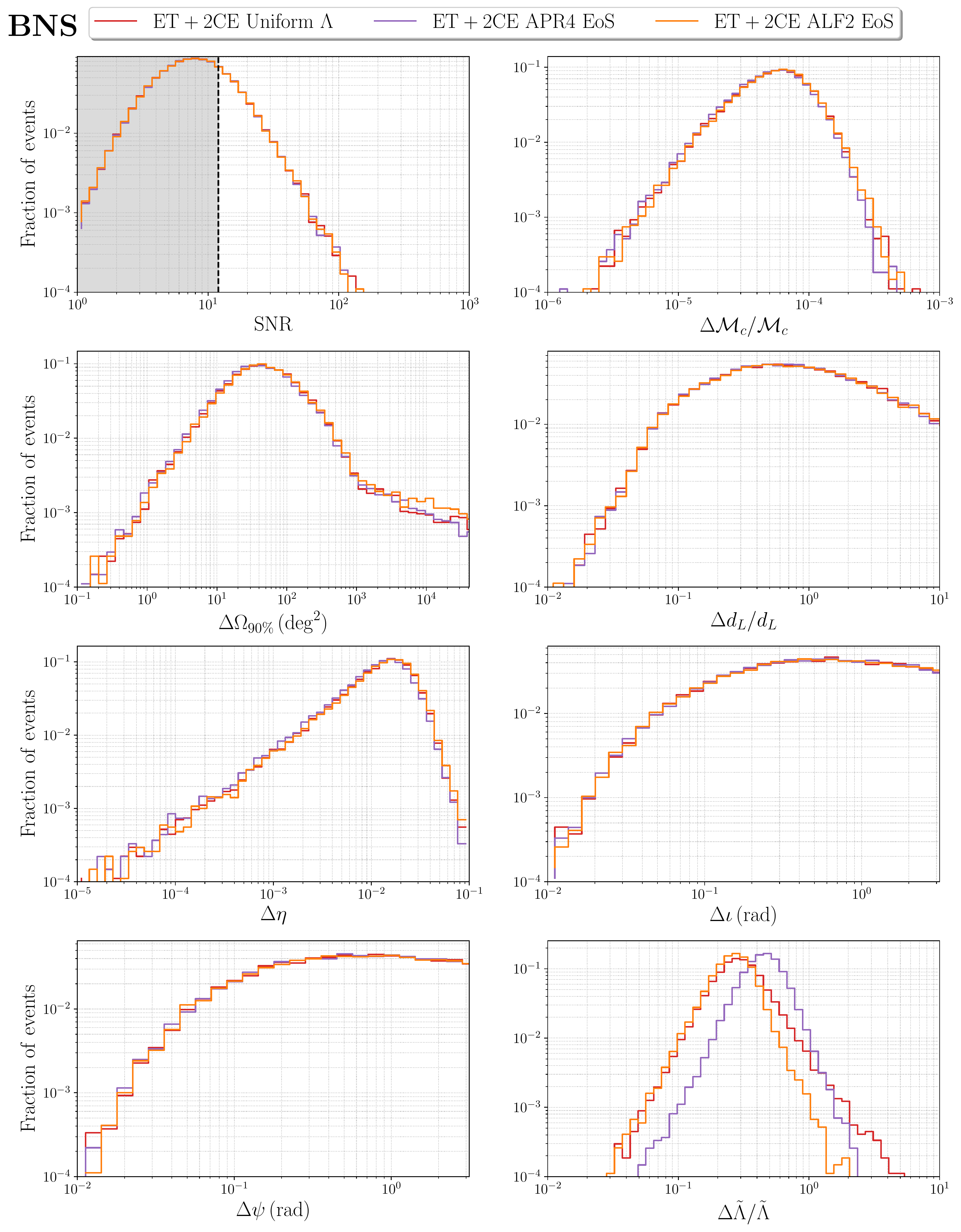}
    \caption{Histograms of the parameter errors for BNSs with the ET+2CE network configuration, using the waveform model \texttt{IMRPhenomD\_NRTidalv2}, and sampling the tidal deformability uniformly in the interval $[0,\,2000]$ (red line), or assuming for all BNSs a common equation of state, chosen to be  APR4 (violet) or ALF2 (orange).}
    \label{fig:HistBNScomp_randLamvsEoS}
\end{figure}

The equation of state (EoS) of dense matter above the nuclear density is still unknown, thus we cannot yet  reliably relate the mass of a neutron star to its adimensional tidal deformability. In the main text we made the agnostic choice of sampling the tidal deformabilities of the simulated BNS systems from a uniform distribution with broad limits, while  here we explore the impact of assuming a common equation of state for all BNS systems. In particular, we consider the so--called APR4 \citep{PhysRevC.58.1804} and ALF2 \citep{Alford_2005EoS} EoS, as representative of soft and hard EoS, respectively. The former thus predicts more compact stars with respect to the latter, leading to smaller values of the adimensional tidal deformability parameter for the same mass, which seems to be favoured by the observation of the BNS event GW170817 \citep{TheLIGOScientific:2017qsa}. The results, for the ET + 2CE network only, are shown in \autoref{fig:HistBNScomp_randLamvsEoS}, where we compare sources having the same parameters apart from the tidal deformability, which in the case of the green curve is sampled uniformly in the interval $[0,\, 2000]$, while for the others is computed from the mass of the two components assuming one of the chosen EoS.\footnote{The computation of $\Lambda(M)$ for the two EoS has been performed using the Tolman–Oppenheimer–Volkoff equations solver implemented in \texttt{LALSuite} \citep{lalsuite}.} Notice that the uniform distribution used in the main text for the BNS masses, ranging up to \SI{2.5}{\Msun}, is not suitable for performing this test, given that it exceeds the maximum masses of the chosen EoS.\footnote{For the APR4 EoS the maximum mass is \SI{2.21}{\Msun}, while for ALF2 it is \SI{2.09}{\Msun}.} To perform this analysis we thus adopted two independent Gaussian distribution for the masses of the components, ${\cal N}(1.33, 0.09)$ in units of \si{\Msun}. From the plot it is apparent that different choices for the tidal deformability distribution  have negligible impact on the distributions of the parameter errors apart from the tidal deformability itself. In particular, we find the stiff EoS to produce systems whose relative error on the adimensional tidal deformability parameter is smaller, as compared to the soft. This is in line with the fact that a stiffer EoS predicts higher values for the adimensional tidal deformability parameter, so the effect on  the waveform is larger, which makes the estimation easier. This  also provides a consistency check of our code.
We thus conclude that our results for BNS systems are robust with respect to different choices for the adimensional tidal deformability distribution.

\section{Comparison among different waveform models}\label{sec:appendix_comparisonWFmod}
We report here the comparison of the results for the SNR and parameter estimation errors for the three classes of sources, with waveform models different from the ones adopted in the main text. Again comparison is only made using the ET + 2CE detector network. \par\medskip

\begin{figure}
    \centering
    \includegraphics[width=.9\textwidth]{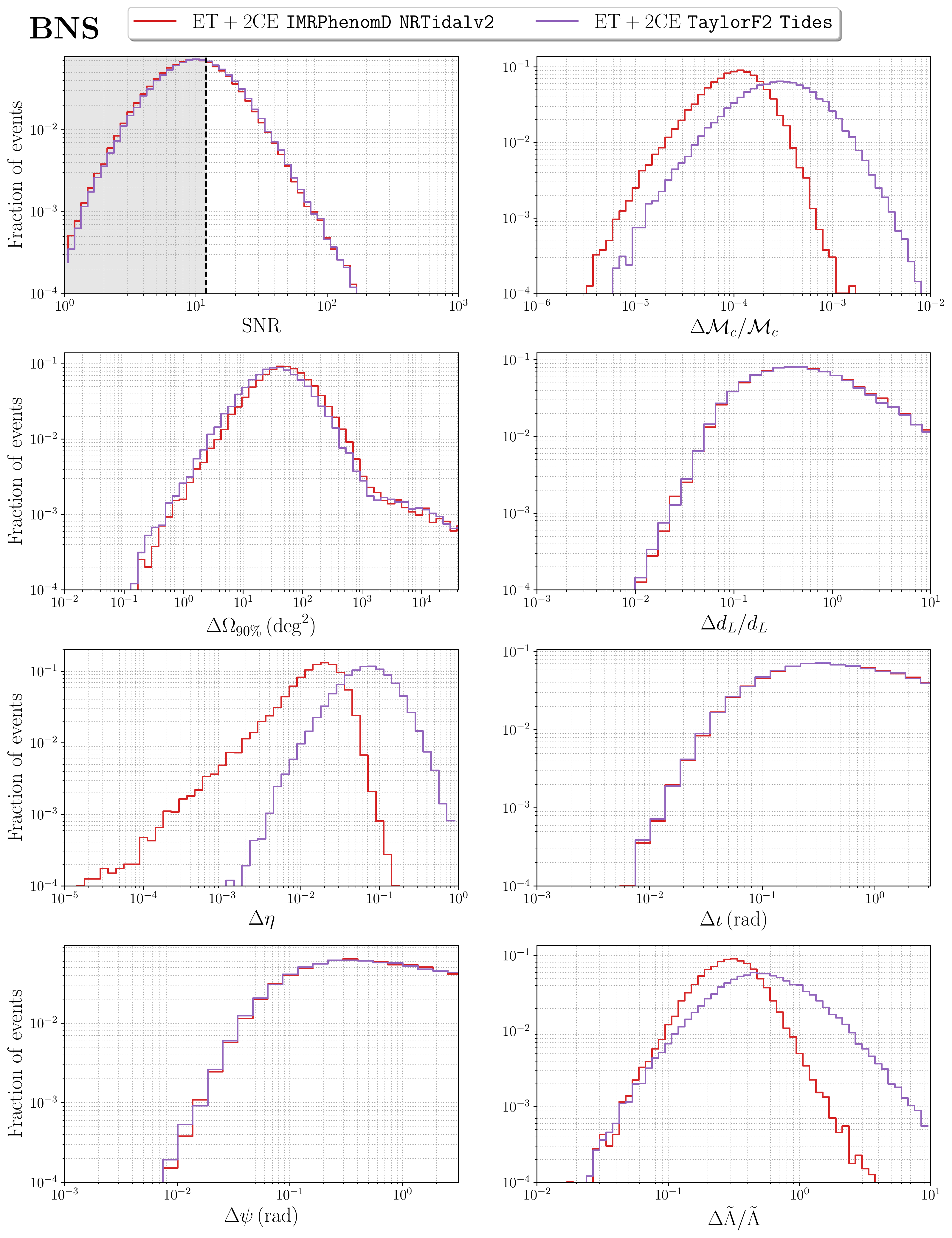}
    \caption{Histograms of the parameter errors for BNSs with the ET+2CE network configuration, using the full inspiral--merger--ringdown waveform model \texttt{IMRPhenomD\_NRTidalv2} and the 3.5 restricted PN model \texttt{TaylorF2} with tidal effects.}
    \label{fig:HistBNScomp_PhenDTidvsTF2Tid}
\end{figure}

\begin{figure}
    \centering
    \includegraphics[width=.9\textwidth]{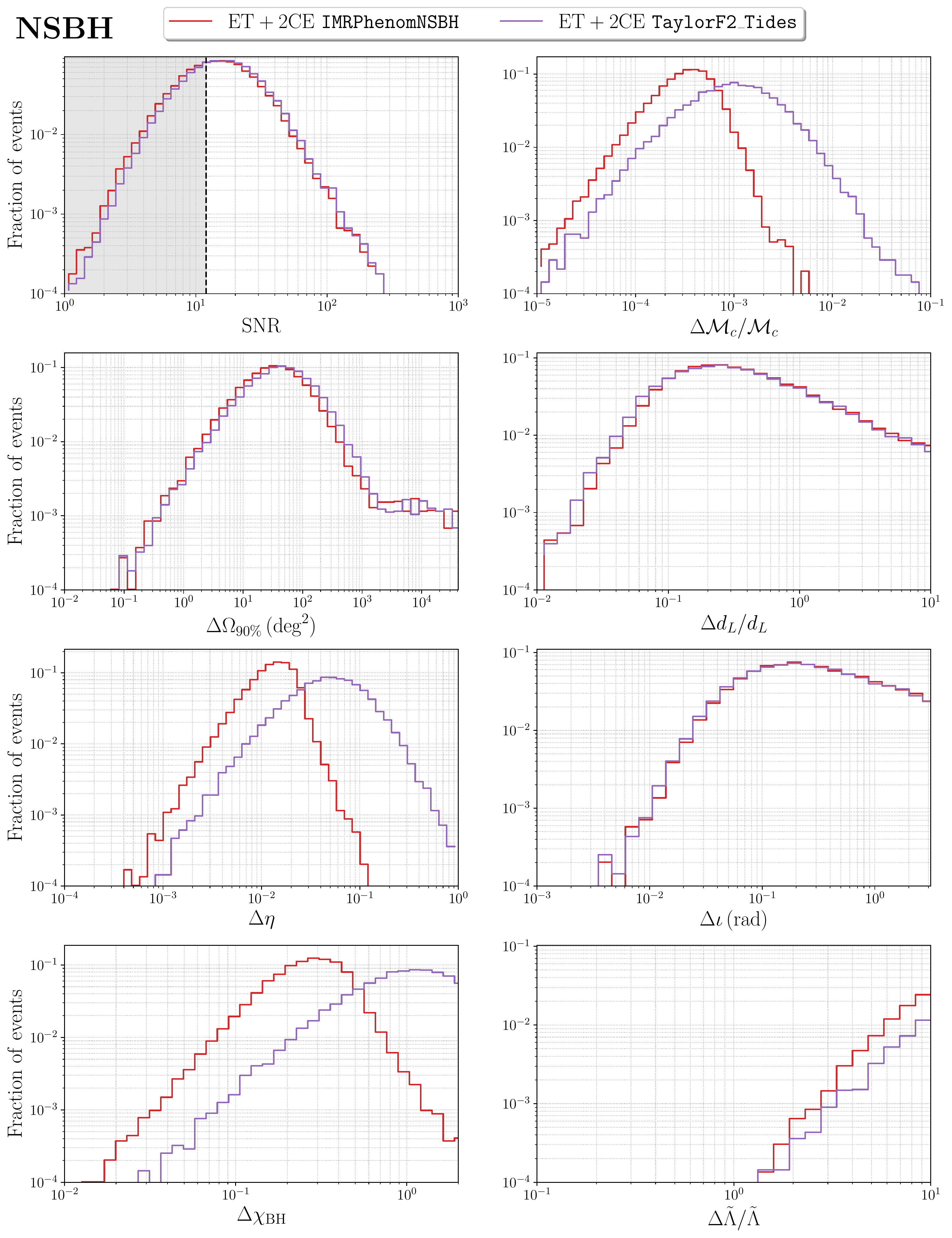}
    \caption{Histograms of the parameter errors for NSBH with the ET+2CE network configuration, using the full inspiral--merger--ringdown waveform model \texttt{IMRPhenomNSBH} and the 3.5 restricted PN model \texttt{TaylorF2} with tidal effects.}
    \label{fig:HistNSBHcomp_PhenNSBHvsTF2Tid}
\end{figure}

For BNS and NSBH systems, we compare the results obtained using the full inspiral--merger--ringdown models \texttt{IMRPhenomD\_NRTidalv2} and \texttt{IMRPhenomNSBH}, respectively, to the inspiral--only restricted PN model \texttt{TaylorF2} with terms up to 3.5 order \citep{PhysRevD.80.084043, PhysRevD.84.084037, PhysRevD.93.084054}, also including in it tidal effects (which enter at order 5 and 6 in the PN expansion) \citep{PhysRevD.89.103012}, and a cut at twice the Innermost Stable Circular Orbit frequency, $f_{\rm ISCO}$. As it is apparent from \autoref{fig:HistBNScomp_PhenDTidvsTF2Tid} and \ref{fig:HistNSBHcomp_PhenNSBHvsTF2Tid}, the change in the distribution of SNRs and extrinsic parameters errors, i.e. the sky location, luminosity distance, inclination angle and polarisation angle, is slight, as expected, since these parameters are not linked to the waveform model itself in absence of higher modes, which are not expected to give strong contributions for these classes of sources. The situation is different for intrinsic parameters, i.e. the chirp mass, symmetric mass ratio, spins and tidal deformability, whose impact on the GW signal is obviously better captured by the full inspiral--merger--ringdown models, resulting in significant improvements in the estimation.

\newpage
\bibliography{myrefs}{}
\bibliographystyle{aasjournal}

\end{document}